\documentclass[%
 prd,
 reprint,
superscriptaddress,
nofootinbib, 
 amsmath,amssymb,
 aps,
]{revtex4-2}

\usepackage{graphicx}
\usepackage{dcolumn}
\usepackage{bm}
\usepackage{hyperref}
\usepackage{url}
\usepackage{tabularx}

\usepackage{newtxtext,newtxmath}
\usepackage{bm}
\usepackage[T1]{fontenc}
\usepackage{xcolor}
\usepackage{tcolorbox}
\usepackage{orcidlink}
\usepackage{bigints}

\newcommand{\mnras}{Mon.~Not.~Roy.~Astron.~Soc.}
\newcommand{\physrep}{Phys.~Rep.}
\newcommand{\jcap}{JCAP}
\newcommand{\pasj}{Publ.~Astron.~Soc.~Japan}
\newcommand{\apjl}{\apj~Lett.}
\newcommand{\apjs}{\apj~Suppl.}
\newcommand{\aap}{Astronomy~\&~Astrophysics}
\newcommand{\aj}{The~Astronomical~J.}

\newcommand{\araa}{Ann.~Rev.~Astro.~Astrophys.}

\begin{document}

\title{Cosmology and Source Redshift Constraints from Galaxy Clustering and Tomographic Weak Lensing with HSC Y3 and SDSS using the Point-Mass Correction Model}

\author{Tianqing~Zhang\orcidlink{0000-0002-5596-198X}}
\thanks{tq.zhang@pitt.edu}
\affiliation{Department of Physics and Astronomy and PITT PACC, University of Pittsburgh, Pittsburgh, PA 15260, USA}
\affiliation{McWilliams Center for Cosmology and Astrophysics, Department of Physics, Carnegie
Mellon University, 5000 Forbes Ave, Pittsburgh, PA 15213, USA}

\author{Xiangchong~Li\orcidlink{0000-0003-2880-5102}}
\affiliation{McWilliams Center for Cosmology and Astrophysics, Department of Physics, Carnegie
Mellon University, 5000 Forbes Ave, Pittsburgh, PA 15213, USA}
\affiliation{Brookhaven National Laboratory, Bldg 510, Upton, New York 11973, USA}

\author{Sunao~Sugiyama\orcidlink{0000-0003-1153-6735}}
\affiliation{Department of Physics and Astronomy, University of Pennsylvania, Philadelphia, PA 19104, USA}

\author{Rachel~Mandelbaum\orcidlink{0000-0003-2271-1527}}
\affiliation{McWilliams Center for Cosmology and Astrophysics, Department of Physics, Carnegie
Mellon University, 5000 Forbes Ave, Pittsburgh, PA 15213, USA}

\author{Surhud~More\orcidlink{0000-0002-2986-2371}}
\affiliation{The Inter-University Centre for Astronomy and Astrophysics, Post
bag 4, Ganeshkhind, Pune 411007, India}
\affiliation{Kavli Institute for the Physics and Mathematics of the Universe (WPI), The University of Tokyo Institutes for Advanced Study (UTIAS), The University of Tokyo, Chiba 277-8583, Japan}

\author{Roohi~Dalal\orcidlink{0000-0002-7998-9899}}
\affiliation{American Astronomical Society, 1667 K St NW 800, Washington, D.C. 20006}

\author{Arun~Kannawadi\orcidlink{0000-0001-8783-6529}}
\affiliation{Department of Physics, Duke University, Durham, NC 27708, USA}

\author{Hironao~Miyatake\orcidlink{0000-0001-7964-9766}}
\affiliation{Kavli Institute for the Physics and Mathematics of the Universe (WPI), The University of Tokyo Institutes for Advanced Study (UTIAS), The University of Tokyo, Chiba 277-8583, Japan}
\affiliation{Kobayashi-Maskawa Institute for the Origin of Particles and the
Universe (KMI), Nagoya University, Nagoya, 464-8602, Japan}
\affiliation{Institute for Advanced Research, Nagoya University, Nagoya
464-8601, Japan}

\author{Atsushi~J.~Nishizawa\orcidlink{0000-0002-6109-2397}}
\affiliation{Kobayashi-Maskawa Institute for the Origin of Particles and the Universe (KMI), Nagoya University, Nagoya, 464-8602, Japan}
\affiliation{Gifu Shotoku Gakuen University, 1-1 Takakuwanishi, Yanaizu, Gifu, 501-6194, Japan}

 \author{Takahiro~Nishimichi\orcidlink{0000-0002-9664-0760}}
 \affiliation{Kavli Institute for the Physics and Mathematics of the Universe (WPI), The University of Tokyo Institutes for Advanced Study (UTIAS), The University of Tokyo, Chiba 277-8583, Japan}
 \affiliation{Center for Gravitational Physics, Yukawa Institute for Theoretical
Physics, Kyoto University, Kyoto 606-8502, Japan}
\affiliation{Department of Astrophysics and Atmospheric Sciences, Faculty of Science, Kyoto Sangyo University, Motoyama, Kamigamo, Kita-ku, Kyoto 603-8555, Japan}

\author{Masamune~Oguri\orcidlink{0000-0003-3484-399X}}
\affiliation{Kavli Institute for the Physics and Mathematics of the Universe (WPI), The University of Tokyo Institutes for Advanced Study (UTIAS), The University of Tokyo, Chiba 277-8583, Japan}
\affiliation{Center for Frontier Science, Chiba University, Chiba 263-8522, Japan}
\affiliation{Department of Physics, Graduate School of Science, Chiba University, Chiba 263-8522, Japan}

\author{Ken~Osato\orcidlink{0000-0002-7934-2569}}
\affiliation{Kavli Institute for the Physics and Mathematics of the Universe (WPI), The University of Tokyo Institutes for Advanced Study (UTIAS), The University of Tokyo, Chiba 277-8583, Japan}
\affiliation{Center for Frontier Science, Chiba University, Chiba 263-8522, Japan}
\affiliation{Department of Physics, Graduate School of Science, Chiba University, Chiba 263-8522, Japan}

\author{Markus~M.~Rau\orcidlink{0000-0003-3709-1324}}
\affiliation{School of Mathematics, Statistics and Physics,Newcastle University, Newcastle upon Tyne, NE17RU, United Kingdom}
\affiliation{High Energy Physics Division, Argonne National Laboratory, Lemont, IL 60439, USA}

\author{Masato~Shirasaki\orcidlink{0000-0002-1706-5797}}
\affiliation{National Astronomical Observatory of Japan, National Institutes of
Natural Sciences, Mitaka, Tokyo 181-8588, Japan}
\affiliation{The Institute of Statistical Mathematics, Tachikawa, Tokyo
190-8562, Japan}

\author{Tomomi~Sunayama\orcidlink{0009-0004-6387-5784}}
\affiliation{Kavli Institute for the Physics and Mathematics of the Universe (WPI), The University of Tokyo Institutes for Advanced Study (UTIAS), The University of Tokyo, Chiba 277-8583, Japan}
\affiliation{Kobayashi-Maskawa Institute for the Origin of Particles and the Universe (KMI), Nagoya University, Nagoya, 464-8602, Japan}
\affiliation{Academia Sinica Institute of Astronomy and Astrophysics (ASIAA), No.1, Sec. 4, Roosevelt Rd, Taipei 106319, Taiwan, R.O.C}

\author{Masahiro~Takada\orcidlink{0000-0002-5578-6472}}
\affiliation{Kavli Institute for the Physics and Mathematics of the Universe (WPI), The University of Tokyo Institutes for Advanced Study (UTIAS), The University of Tokyo, Chiba 277-8583, Japan}
\affiliation{Center for Data-Driven Discovery (CD3), Kavli IPMU (WPI), UTIAS, The University of Tokyo, Kashiwa, Chiba 277-8583, Japan}

\date{\today}

\begin{abstract}
The combination of galaxy clustering and weak lensing is a powerful probe of the cosmology model. We present a joint analysis of galaxy clustering and weak lensing cosmology using SDSS data as the tracer of dark matter (lens sample) and the HSC Y3 dataset as source galaxies. The analysis divides HSC Y3 galaxies into four tomographic bins for both galaxy-galaxy lensing and cosmic shear measurements, and employs a point-mass correction model to utilize galaxy-galaxy lensing signals down to 2$h^{-1}$Mpc, extending up to 70$h^{-1}$Mpc. These strategies enhance the signal-to-noise ratio of the galaxy-galaxy lensing data vector. Using a flat $\Lambda$CDM model, we find $S_8 = 0.780^{+0.029}_{-0.030}$, and using a $w$CDM model, we obtain $S_8 = 0.756^{+0.038}_{-0.036}$ with $w = -1.176^{+0.310}_{-0.346}$. We apply uninformative priors on the redshift mean-shift parameters for the third and fourth tomographic bins. Leveraging the self-calibration power of tomographic weak lensing, we measure $\Delta z_3 = -0.112^{+0.046}_{-0.049}$ and $\Delta z_4 = -0.185^{+0.071}_{-0.081}$, in agreement with previous HSC Y3 results. This demonstrates that weak lensing self-calibration can achieve redshift constraints comparable to other methods such as photometric and clustering redshift calibration.
\end{abstract}

\maketitle

\section{Introduction}

The standard cosmological model describes a universe composed of regular (baryonic) matter, dark matter, and dark energy. Together, regular and dark matter account for approximately 30\% of the total energy density, while dark energy makes up the remaining 70\% \cite{Planck2018Cosmology, DESY3_3x2_2022, DESI_BAO}. Dark matter and baryons drive the growth of large-scale structures, forming gravitationally bound dark matter halos in which galaxies form and evolve. In contrast, dark energy causes the accelerated expansion of the universe \cite{Riess2019, Scolnic2018}, which pushes galaxies and large-scale structures apart over cosmic time.

Such a cosmological model can be described using fewer than a dozen parameters, several of which are tightly constrained by measurements of galaxy clustering and weak gravitational lensing \cite{Hu1999, Dodelson2020}, due to their sensitivity to the matter power spectrum across different epochs. Galaxy clustering assumes that galaxies trace the underlying dark matter distribution with a bias, making the galaxy power spectrum a biased probe of the matter power spectrum. Weak gravitational lensing, on the other hand, leverages the deflection of light from background (source) galaxies to measure the projected matter distribution along the line of sight to foreground structures \cite{Kilbinger2015, Mandelbaum2018_review}.

By correlating the shapes of galaxy pairs at a given angular separation—referred to as cosmic shear -- one measures the integrated lensing effect caused by large-scale structure in the foreground \cite{Amon2022, Secco2022, Li2023, Dalal2023, wright2025}. Correlation of the shapes of source galaxies with the positions of foreground (lens) galaxies -- known as galaxy-galaxy lensing -- probes the average projected matter density around those lens galaxies \cite{Prat2018, Miyatake2022}. 
Galaxy clustering, galaxy-galaxy lensing, and cosmic shear each depend differently on galaxy bias. Therefore, a joint analysis of these three two-point correlation functions helps break the degeneracy between galaxy bias and the amplitude of the matter power spectrum. This combined approach is commonly referred to as the “three-by-two-point” (3×2pt) analysis \cite{DESY3_3x2_2022, Sugiyama2023, Miyatake2023, vanUitert2018}.

As previously mentioned, matter drives the formation of large-scale structure through gravitational attraction, while dark energy causes the accelerated expansion of space-time. As a result, the matter power spectrum evolves over cosmic time. To observe this evolution, both galaxy tracers and source galaxies must be divided into redshift ``slices'',  or tomographic bins \cite{Hu1999}. Specifically, galaxy clustering in a given tomographic bin probes the matter power spectrum within that redshift range; galaxy-galaxy lensing measures the projected matter profile around lens galaxies at those redshifts; and cosmic shear captures the cumulative lensing effect from all large-scale structures in front of the source galaxies \cite{Dodelson2020}.

Using tomographic source bins in galaxy-galaxy lensing provides additional self-calibration power on the mean redshift of the source galaxies. Since previous analysis based on HSC Y3 shear catalog found inconsistent redshift parameter from the cosmological analysis to the photo-$z$ redshift distribution, we utilize this self-calibration power to provide redshift parameter constraints for the third and fourth HSC bins. Furthermore, employing multiple source bins improves the signal-to-noise ratio of the galaxy-galaxy lensing measurement by increasing the number of tomographic lens–source bin pairs available.

In this work, a sample of SDSS DR11 galaxies is used as tracers of large-scale structure in the redshift range $0.15 < z_l < 0.7$. These galaxies also serve as the lens sample in the galaxy-galaxy lensing measurement and are divided into three tomographic bins. The HSC Y3 source galaxies, covering an area of 419 deg$^2$, are divided into four tomographic bins spanning $0.3 < z_s < 1.5$. The binning scheme of the source galaxies here is identical to that used in the HSC Y3 cosmic shear analysis \cite{Li2023, Dalal2023}.

The clustering measurements of the SDSS galaxies are taken from More et al. \cite{More2023}, and the tomographic galaxy-galaxy lensing measurements are adapted from \cite{2x2pt_paper}, which also established the $2\times2$pt analysis framework combining galaxy clustering and galaxy-galaxy lensing. The cosmic shear measurements are consistent with those in \cite{Li2023}, although a different multiplicative factor is applied for blinding purposes. Details of the $3\times2$pt data vector and its covariance matrix are provided in Section~\ref{sec:data:0}.

The theoretical forward model generates predicted $3\times2$pt data vectors to constrain cosmological and nuisance parameters via Markov Chain Monte-Carlo (MCMC) sampling. Most of the theoretical modeling used in this work follows \cite{Sugiyama2023} and \cite{2x2pt_paper}. We compute the nonlinear matter power spectrum with a single parameter controlling the strength of baryonic feedback. A single galaxy bias parameter is used to model the galaxy clustering signal $w_p$, and an additional point-mass correction parameter is introduced to account for small-scale modeling uncertainties in $\Delta \Sigma(R_p)$. For intrinsic alignment, we adopt the tidal alignment and tidal torquing model. Each tomographic bin of the source galaxies is assigned one nuisance parameter for residual multiplicative shear bias, redshift bias, and PSF systematics. We perform likelihood analyses on the full data vector as well as on its subsets to test internal consistency. Details of the theoretical modeling are provided in Section~\ref{sec:model:0}.

We present the cosmological constraints from our fiducial data vector, along with results obtained from various subsets of the data vector and alternative analysis choices. These tests help assess the robustness of the fiducial results. We also compare our fiducial cosmological constraints with results from other surveys. The main results are presented in Section~\ref{sec:results:0}, and the implications and conclusions are summarized in Section~\ref{sec:conclusion:0}. We summarize our findings in Section~\ref{sec:conclusion:0}.

\section{Datasets and Data Vectors}
\label{sec:data:0}

In this section, we describe the datasets and the 3×2pt data vector used in our analysis. The 3×2pt data vector is constructed by combining previously published measurements and our companion $2\times2$pt paper \cite{2x2pt_paper}. The galaxy clustering measurement is identical to that presented in \cite{More2023}, the galaxy-galaxy lensing measurement of $\Delta\Sigma(R_p)$ follows \cite{2x2pt_paper}, and the cosmic shear measurement is redone consistently with \cite{Li2023}, except for a global multiplicative factor applied for blinding purposes. We briefly summarize the measurement procedures and underlying datasets here and refer the reader to the respective works for further details.

\subsection{Datasets}

\subsubsection{SDSS DR11}
\label{sec:data:sdss}

We use the SDSS DR11 large-scale structure catalog to measure the galaxy clustering signal \cite{gunn2006, ABAZAJIAN2009}, and to define the lens galaxy sample for the galaxy-galaxy lensing measurements. The SDSS catalog covers an area of approximately 11,000 deg$^2$ and is divided into three redshift tomographic bins: $0.15 < z_l < 0.35$ for LOWZ, $0.43 < z_l < 0.55$ for CMASS1, and $0.55 < z_l < 0.7$ for CMASS2. To make the sample approximately volume-limited, we apply absolute magnitude cuts of $M_q - 5 \log h < -21.5$, $-21.9$, and $-22.2$ for the LOWZ, CMASS1, and CMASS2 bins, respectively. The comoving number densities of galaxies in these bins are $1.8$, $0.74$, and $0.45 \times 10^{-4}~(h^{-1} \mathrm{Mpc})^{-3}$, respectively.

In Fig.~\ref{fig:nz_lens_source}, we demonstrate the drastically different angular and redshift space coverage of the lens and source galaxies used in our $3\times2$pt analysis, plotted in (Dec, $z$) coordinates. The SDSS lens galaxies span a larger sky area and occupy a lower redshift range compared to the HSC source galaxies. The SDSS galaxy sample used in this work is identical to that used in the $2\times2$pt paper \cite{2x2pt_paper} and the previous HSC Y3 $3\times2$pt analyses \cite{Sugiyama2023, Miyatake2023, More2023}. For further details on the SDSS catalog used in this analysis, we refer readers to \cite{More2023} and \cite{2x2pt_paper}.

\subsubsection{HSC Y3}
\label{sec:data:hsc}

We use the HSC Year-3 shear catalog as the source sample for measuring the galaxy-galaxy lensing and cosmic shear signal. The catalog \cite{Li2022} \footnote{The shape catalog is publicly available as a catalog of the PDR3 \cite{Aihara2022} at \url{https://hsc-release.mtk.nao.ac.jp/doc/index.php/data-access\_\_pdr3/}.} is based on the S19A internal data release, which uses data between the second and third public releases of the HSC-SSP survey \citep{Aihara2019, Aihara2022}. It contains 35.7 million galaxies over a footprint of 433 deg$^2$, with an effective source galaxy number density of $19.9~\mathrm{arcmin}^{-2}$. The HSC Y3 footprint is divided into six fields: GAMA09H, GAMA15H, WIDE12H, XMM, VVDS, and HECTOMAP.

Galaxy shapes in the catalog are measured using the re-Gaussianization method \cite{Hirata2003, Mandelbaum2005}, and the shear response is calibrated using image simulations \cite{Mandelbaum2018_image, Li2022}, implemented with GalSim \cite{Rowe2014}. The multiplicative shear bias is calibrated to the 1\% level, and various systematics and null tests confirm that the catalog is free of significant additive shear bias \cite{Li2022, Zhang2023b}. Following \cite{Li2023}, we exclude a 20 deg$^2$ region in the GAMA09H field due to excessive B-mode power in the cosmic shear signal, yielding a total effective area of 416 deg$^2$ used in this work.
For further details on the shear catalog selection function and image simulation-based calibration, we refer readers to \cite{Li2022}. For PSF-related systematics and null tests, we refer to \cite{Zhang2023b}.

The redshift information in the HSC Y3 shape catalog is estimated using three photometric redshift (photo-$z$) algorithms \cite{Nishizawa2020}. \textsc{Mizuki} is a template-fitting based method, while \textsc{DNNz} and \textsc{DEmPz} are machine learning-based approaches. All three methods provide a redshift probability density function (PDF), $P(z_s)$, for each galaxy. These PDFs are used both to assign optimal weights in the galaxy-galaxy lensing measurement and to define priors on the redshift distributions of the tomographic source bins.

The optimal weights, which is expressed in Eq.~(14) in \cite{2x2pt_paper} are construct to (a) weight the lens-source pairs based on lensing effiency\cite{mandelbaum2013}; (b) weight lens sample based on observing condition\cite{sheldon2004,Mandelbaum2005}; (c) weight source sample based on shear SNR\cite{Hirata2003}. The optimal weights effectively suppress the IA contamination to GGL data vector, and increase the overall SNR for GGL. 

For the tomographic cosmological analysis, the HSC Y3 source galaxies are divided into four tomographic bins based on the \texttt{DNNz} photo-$z$ estimator's $z_{\rm best}$ values \cite{Rau2022}. The bin edges are defined as $[0.3, 0.6]$, $[0.6, 0.9]$, $[0.9, 1.2]$, and $[1.2, 1.5]$. To reduce contamination from galaxies with poorly constrained redshifts, we exclude galaxies with double-peaked PDFs in either \texttt{Mizuki} or \texttt{DNNz}, which removes 31\% of galaxies in bin 1 and 8\% in bin 2.

\begin{figure*}
    \includegraphics[width=2\columnwidth]{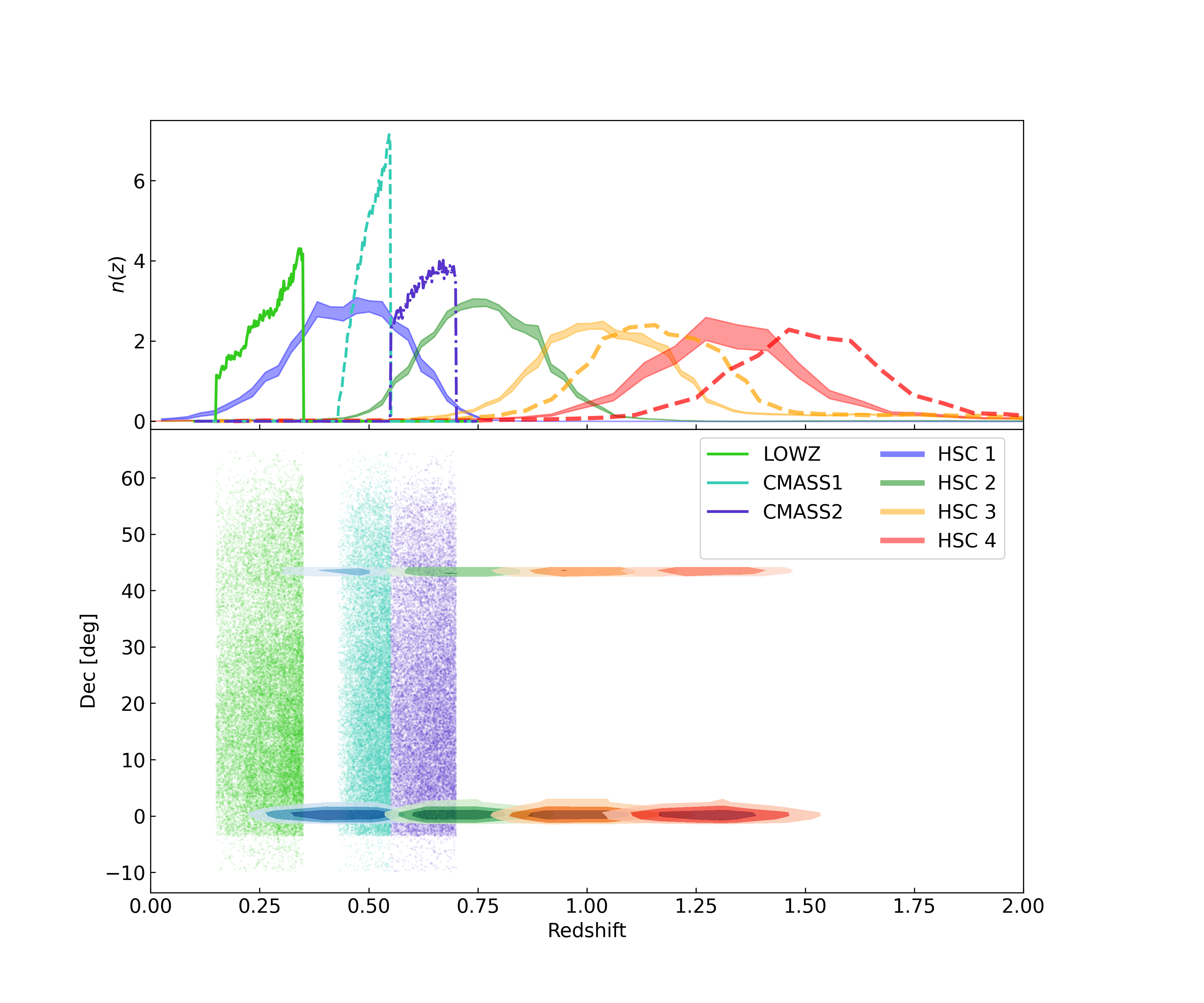}
    \caption{\label{fig:nz_lens_source} \textbf{Top}:Redshift distributions $n(z)$ for the three SDSS lens galaxy bins and the four HSC source bins used in this work. The dashed line for HSC bin 3 and 4 are the preferred $n(z)$ by the fiducial $3\times2$pt analysis. \textbf{bottom}: Declination-redshift distribution of galaxies. SDSS lens galaxies are shown as individual points, color-coded by lens bin, highlighting their broad angular coverage. HSC source galaxies are overlaid by the contour plots, which covers a larger redshift range. The HSC redshift of the galaxies are sampled by the corresponding $n(z)$. This figure demonstrates the complementary characteristic of the SDSS and HSC datasets. 
    }
\end{figure*}

The redshift distribution $n_i(z)$ for each tomographic bin is inferred by combining stacked photo-$z$ PDFs (used as priors) with clustering redshift measurements between the photometric galaxies and CAMIRA luminous red galaxies (CAMIRA-LRGs) \cite{CAMIRA_Oguri2014, CAMIRA3}, which serve as the likelihood term. The inference is performed using a logistic Gaussian process. It is important to note that the CAMIRA-LRG sample spans the redshift range $0.1 < z < 1.1$, so redshift information for $z > 1.1$ relies solely on photometric estimates. This limitation may explain the tension observed between $\Lambda$CDM-preferred and photometric-redshift-preferred redshift mean for the third and fourth redshift bins (see Section~\ref{sec:model:nz} for further discussion). In the upper panel of Fig.~\ref{fig:nz_lens_source}, we show the fiducial 67\% confidence interval of the $n_i(z)$ of the four HSC tomographic bins. 

The effective number densities in HSC bins 1 through 4 are 3.77, 5.07, 4.00, and 2.12 arcmin$^{-2}$, respectively, giving a total of 14.96 arcmin$^{-2}$. For more details on the redshift calibration of the source sample, we refer the reader to \cite{Rau2022}.

We note that the source catalog used in this work was jointly blinded with \cite{2x2pt_paper} to mitigate confirmation bias. The three blinded catalogs differ by $\sim$5\% offsets in their applied multiplicative shear bias. All internal consistency tests were performed independently on each of the three blinded catalogs, and we verified that the best-fitting models lie within the empirical $\chi^2$ distributions computed from mock realizations of the data vector. After all internal consistency and validation tests were passed, we unblinded and identified \texttt{blind\_id = 0} as the true catalog. For further details on the blinding procedure, we refer the reader to \cite{2x2pt_paper}.

\subsubsection{Mock Catalog}
\label{sec:data:mock}

Mock catalogs generated from high-fidelity simulations play a critical role in HSC cosmological analyses \cite{takahashi2017}. In this work, we utilize mock catalogs to compute the covariance matrix of the $3\times2$pt data vector.

We use the same suite of mock catalogs as those employed in the HSC Y3 analysis. For details of the underlying simulations, we refer the reader to \cite{takahashi2017}, and for their application in tomographic $2\times2$pt and $3\times2$pt analyses, to \cite{2x2pt_paper}. Briefly, the simulation suite consists of 108 full-sky gravitational lensing realizations, each divided into 13 non-overlapping HSC Y3-like fields, yielding a total of 1404 independent mock realizations.

SDSS-like lens galaxies are populated into the simulations using a halo occupation distribution (HOD) model. HSC-like source galaxies are randomly distributed in angular coordinates and assigned redshifts to match the observed redshift distribution. For each mock realization, we compute the full $3\times2$pt data vector using the same measurement pipeline applied to the real data. These mock data vectors are then used to estimate the covariance matrix (see Section~\ref{sec:data:cov}). 

\subsection{Data Vector}

\begin{figure*}
\includegraphics[width=1.8\columnwidth]{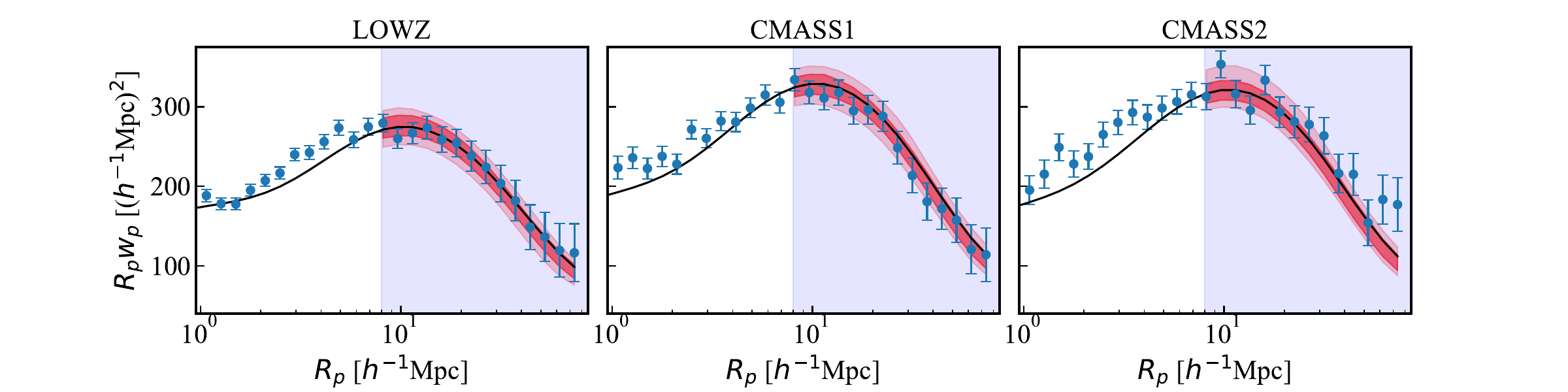}
\includegraphics[width=1.8\columnwidth]{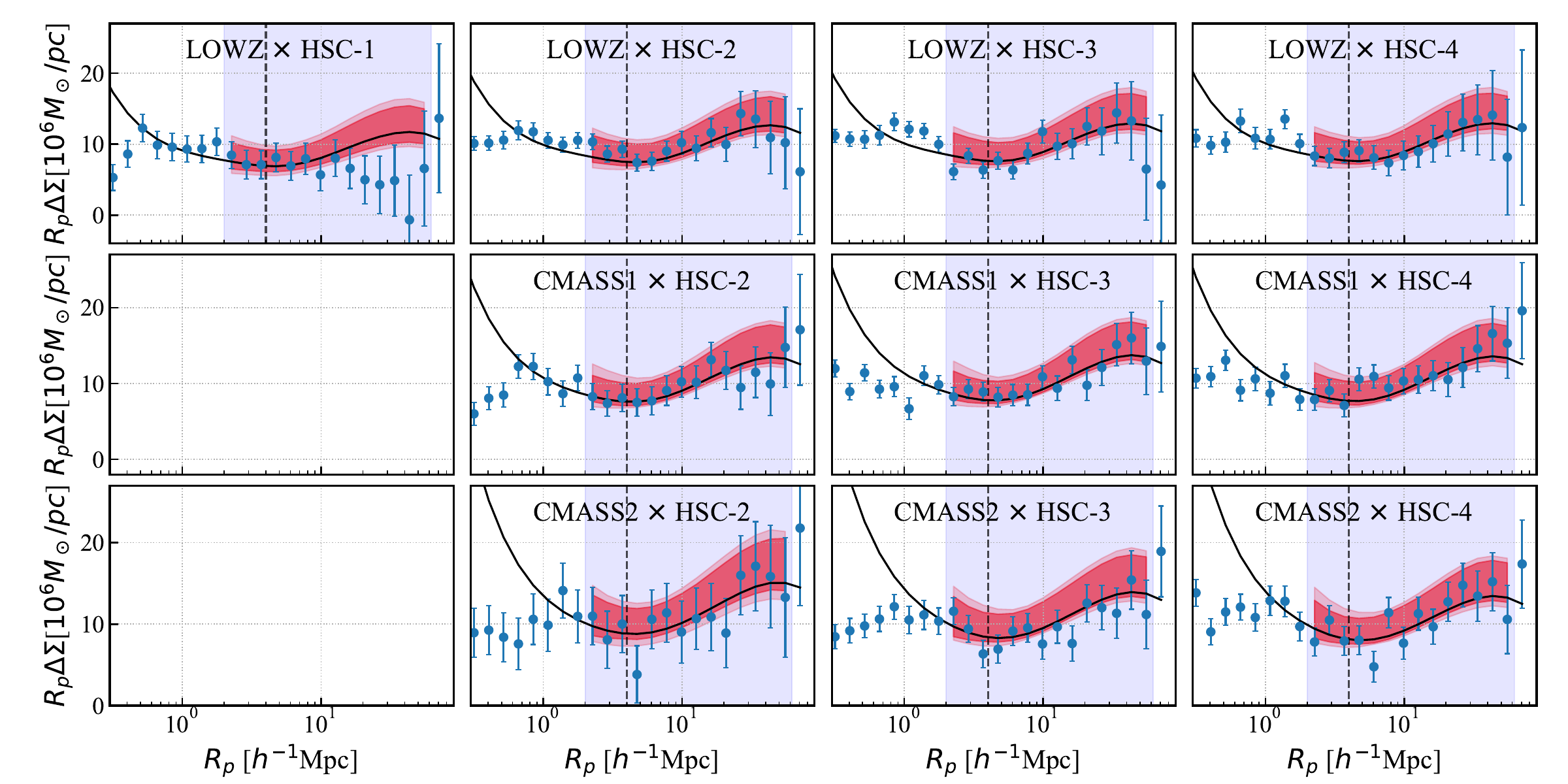}
\includegraphics[width=1.8\columnwidth]{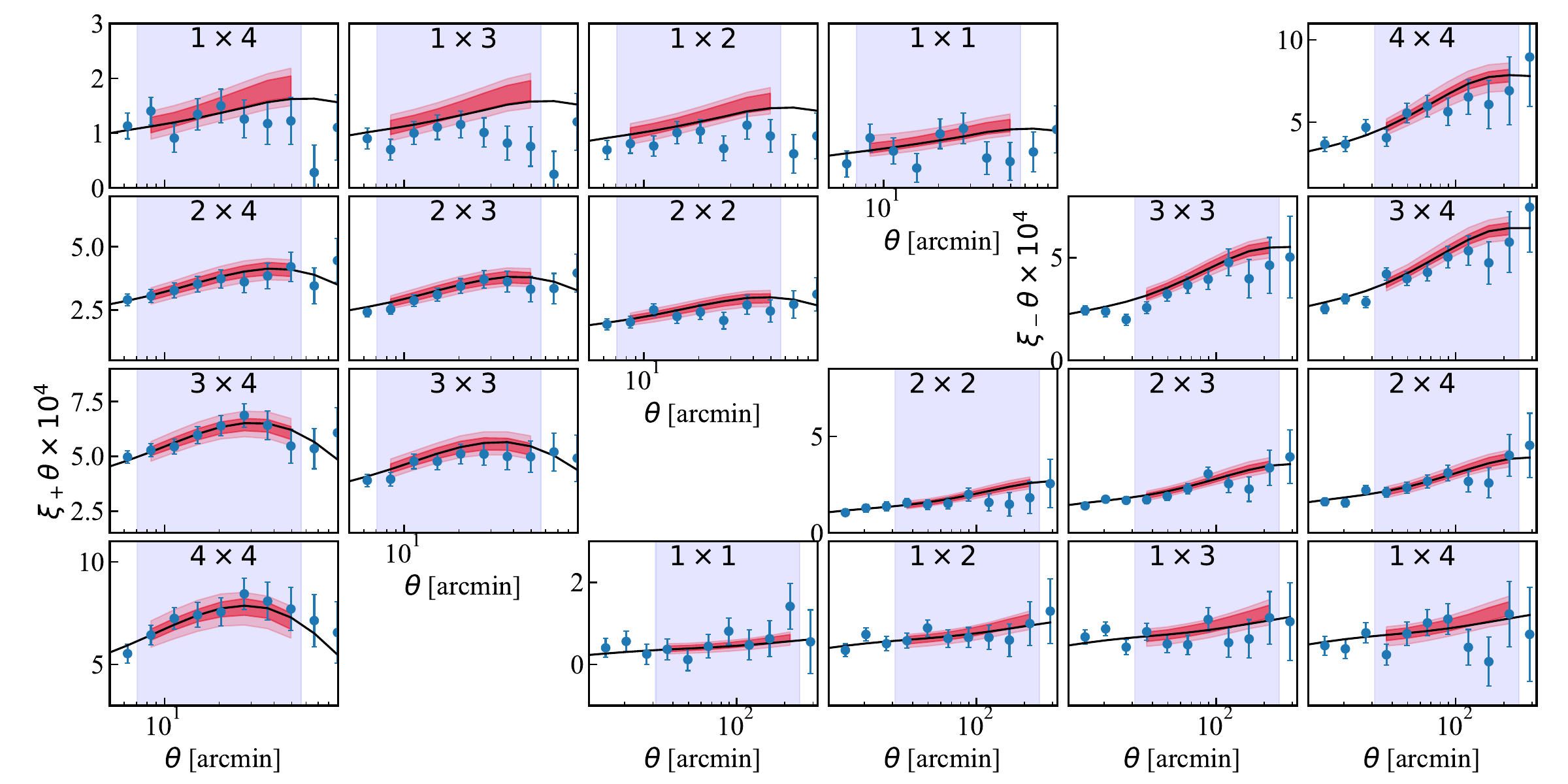}
\caption{\label{fig:measurement} Measurements (blue points), best-fit theoretical predictions (black lines), and the 68\% and 95\% confidence intervals of the model (red shaded regions) for the three components of the $3\times2$pt data vector: the projected galaxy clustering signal $w_p(R_p)$ (top), the galaxy-galaxy lensing signal $\Delta \Sigma(R_p)$ (middle), and the cosmic shear correlation functions $\xi_{+}$ and $\xi_{-}$ (bottom). The blue shaded regions indicate the radial scales used in the cosmological inference. In the $\Delta \Sigma$ panels, the dashed lines denote $R_0 = 4 h^{-1}{\rm Mpc }$, as discussed in Section~\ref{sec:model:ds}. } 
\end{figure*}

The full $3\times2$pt data vector is presented in Fig.~\ref{fig:measurement}, organized into three components: the galaxy clustering signal $w_p(R_p)$, the galaxy-galaxy lensing signal $\Delta \Sigma(R_p)$, and the cosmic shear correlation functions $\xi_{+/-}$. The measured two-point functions are shown as blue points with error bars. The solid black lines represent the theoretical predictions from the best-fit model, defined as the parameter set that minimizes the $\chi^2$. The red shaded bands denote the $68\%$ and $95\%$ confidence intervals of the model predictions, derived from the posterior distribution of the fiducial analysis in Section~\ref{sec:results:key}.

\subsubsection{Galaxy Clustering}
\label{sec:data:wp}


The galaxy clustering data vector is measured using the volume-limited sample of BOSS galaxies from SDSS DR11 \cite{sdss_dr11}. As the measurement procedure is identical to that described in \cite{More2023}, we only provide a brief summary here and refer readers to that work for full details.

The first step in measuring the clustering signal involves computing the three-dimensional correlation function using the Landy–Szalay estimator\cite{Landy1993}, evaluated as a function of projected separation $R_p$ and line-of-sight separation $\pi$:
\begin{equation}
\xi(R_p,\pi) = \frac{DD - 2DR + RR}{RR},
\end{equation}
where $DD$, $DR$, and $RR$ represent the number of data–data, data–random, and random–random galaxy pairs in a given $(R_p, \pi)$ bin, respectively. Here $R_p$ is the projected radius. To suppress shot noise in the estimator, the random catalog contains 50 times more points than the galaxy sample.

We further compress the measurement by integrating the three-dimensional correlation function along the line of sight to obtain the projected correlation function:
\begin{equation}
w_p(R_p) = 2 \int_0^{\pi_{\rm max}} \xi(R_p,\pi) d\pi,
\end{equation}
where $\pi_{\rm max} = 100h^{-1}$Mpc. The $w_p(R_p)$ measurement is binned into 30 logarithmically spaced bins from $0.5h^{-1}$Mpc to $80h^{-1}$Mpc, resulting in a total of 90 data points across the three SDSS tomographic bins. However, for cosmological inference, we restrict the analysis to large scales ($R_p > 8h^{-1}$Mpc) to avoid modeling uncertainties at small scales. This scale cut yields an effective clustering data vector length of 42. The measurements, along with the best-fit theoretical prediction and the $68\%$ and $95\%$ confidence intervals, are shown in the top three panels of Fig.~\ref{fig:measurement}.

\subsubsection{Galaxy-galaxy Lensing}
\label{sec:data:ggl}


Galaxy-galaxy lensing measures the gravitational lensing effect caused by the mass distribution around foreground (lens) galaxies, using the shapes of background (source) galaxies. In this work, we use the same galaxies as in the clustering measurement (Section~\ref{sec:data:wp}) as the lens sample, and the same galaxies as in the cosmic shear measurement (Section~\ref{sec:data:cs}) as the source sample. Using a consistent lens sample with the clustering analysis helps break the degeneracy between the matter power spectrum and galaxy bias. Similarly, using the same source sample as the cosmic shear measurement allows the galaxy-galaxy lensing to provide self-calibration power on the source redshift distribution.

The galaxy-galaxy lensing measurement used in this work is identical to that of ; therefore, we provide only a brief description here and refer the reader to \cite{2x2pt_paper} for further details.

The galaxy-galaxy lensing measurement in this work adopts the formalism of the excess surface mass density, $\Delta \Sigma(R_p)$ \cite{mandelbaum2013}, which is computed for a given lens-source pair at redshifts $z_l$ and $z_s$ as:
\begin{equation}
\label{eq:delta_sigma_obs}
\Delta \Sigma (R_p) = \gamma_t (R_p) \Sigma_{\mathrm{cr}}(z_l, z_s).
\end{equation}
Here, $R_p$ is the projected comoving separation between the lens and source, $\gamma_t$ \cite{sheldon2004,bartelmann2001} is the tangential shear, approximated by the tangential component of the galaxy shape $e_t$ with respect to the lens galaxies, defined as:
\begin{equation}
e_t = -e_1 \cos(2\phi) - e_2 \sin(2\phi),
\end{equation}
where $e_1 + i e_2$ is the complex ellipticity measured for the source galaxies, and $\phi$ is the position angle between the source and the lens. The critical surface density $\Sigma_{\mathrm{cr}}$ is given by:
\begin{equation}
\Sigma_{\mathrm{cr}}(z_l, z_s) = \frac{c^2}{4 \pi G} \frac{D_A(z_s)}{D_A(z_l) D_A(z_l, z_s) (1 + z_l)^2},
\end{equation}
where $D_A(z)$ denotes the angular diameter distance between redshifts.

Our full estimator computes the galaxy-galaxy lensing signal $\hat{\Delta \Sigma}(R_p)$ for the $q$-th lens bin and $i$-th source bin as:
\begin{equation}
\label{eq:dsigma_def}
\Delta \Sigma^{qi} (R_p) = \frac{1}{1+\hat{m}_i} \left( \frac{\sum_{l\in q, s\in i} w_{\text{ls}} e_{\text{t, ls}} \langle \Sigma_{\text{cr}}^{-1} \rangle_{\text{ls}}^{-1}}{ 2 \mathcal{R} \sum_{l\in q, s\in i} w_{\text{ls}}} \right),
\end{equation}
where $\langle \Sigma_{\text{cr}}^{-1} \rangle_{\text{ls}}$ is the expectation value of the inverse critical surface density. This is computed by integrating over the source redshift probability distribution to account for photometric redshift uncertainty. The lensing weight for each lens-source pair is defined as $w_{\text{ls}} = w_l w_s \langle \Sigma_{\text{cr}}^{-1} \rangle_{\text{ls}}$, which assigns zero weight to pairs with $z_s \leq z_l$, and higher weights to pairs with larger $z_s - z_l$.

The sums $\sum_{l\in q, s\in i}$ run over all lens-source pairs where the lens is in the $q$-th lens bin and the source is in the $i$-th source bin. We correct for shear calibration by applying the responsivity factor $\mathcal{R}$ and the tomographic-bin-dependent multiplicative bias $\hat{m}_i$. The multiplicative bias $m$ includes a blinding offset $dm_2$, which remains hidden from the analyst until the catalog is unblinded.

We additionally apply corrections for additive and multiplicative shear selection bias and for the boost factor, which accounts for contamination from physically associated sources due to photometric redshift uncertainty. To mitigate residual B-mode systematics in the lensing field, the same estimator is also computed using shapes measured around random points, and the resulting signal is subtracted from the lensing measurement.
We refer the reader to \cite{2x2pt_paper} for a detailed description of the galaxy-galaxy lensing measurement procedure used in this work.

The $\Delta \Sigma(R_p)$ signal is measured for all SDSS-HSC lens-source bin pairs, although we exclude the LOWZ-HSC1 and CMASS1-HSC1 pairs due to significant overlap in redshift space. To examine potential field dependence, we perform the $\Delta \Sigma(R_p)$ measurement separately in each of the six HSC fields. The measurements are binned into 30 logarithmically spaced bins between $0.05h^{-1}{\rm Mpc}$ and $80h^{-1}{\rm Mpc}$. The full data vector has a length of 360. For the cosmological analysis, we apply a scale cut of $2h^{-1}{\rm Mpc} < R < 70h^{-1}{\rm Mpc}$, resulting in an effective data vector length of 140. The lower limit of the scale cuts are motivated by the sensitivity to the offcentering of the lens samples (See Section~\textbf{VI C 2} of \cite{2x2pt_paper}). The measurements, best-fit model predictions, and the 68\% and 95\% confidence intervals for the $\Delta \Sigma(R_p)$ signal in the SDSS-HSC bin pairs included in the analysis are shown in the second to fourth row panels of Fig.~\ref{fig:measurement}.

\subsubsection{Cosmic Shear}
\label{sec:data:cs}


The cosmic shear measurement in this work follows the methodology of \cite{Li2023}. However, because the source catalog is blinded independently from other analyses, we remeasure the shear two-point functions using both the real and mock catalogs. The cosmic shear correlation functions $\xi_{\pm}^{ij}(\theta)$ \cite{Kilbinger2015} are computed for galaxy pairs in tomographic bins $i$ and $j$ as follows:
\begin{align}
\nonumber \hat{\xi}{\pm}^{ij}(\theta) &= \frac{\sum_x w(x)\gamma_+^i(x) w(x+\theta)\gamma_+^j(x+\theta)}{\sum_x w(x) w(x+\theta)} \\
&\pm \frac{\sum_x w(x)\gamma_\times^i(x) w(x+\theta)\gamma_\times^j(x+\theta)}{\sum_x w(x) w(x+\theta)}.
\end{align}
Here, the summation runs over all galaxy pairs with angular separation $\theta$ between tomographic bins $i$ and $j$. The quantities $\gamma_+$ and $\gamma_\times$ represent the tangential and cross shear components, respectively, defined with respect to the vector connecting each galaxy pair.

We use the open-source software \textsc{TreeCorr} to measure the shear-shear correlation functions. Following \cite{Li2023}, we apply angular scale cuts of $7.1 < \theta/\mathrm{arcmin} < 56.6$ for $\xi_+$ and $31.2 < \theta/\mathrm{arcmin} < 248$ for $\xi_-$. The small-scale cuts are chosen to minimize modeling uncertainties due to baryonic effects, while the large-scale cuts are determined by the presence of B-mode systematics. For all tomographic bin pairs, we adopt logarithmic angular binning with an interval of $\Delta \log(\theta) = 0.29$, resulting in 7 bins for both $\xi_+$ and $\xi_-$. In total, the cosmic shear data vector consists of 140 data points. The measurements, best-fit theoretical model, and the 68\% and 95\% confidence regions for the auto- and cross-correlations $\xi_{\pm}^{ij}(\theta)$ are shown in the bottom four rows of Fig.~\ref{fig:measurement}.

\begin{figure*}
\includegraphics[width=2\columnwidth]{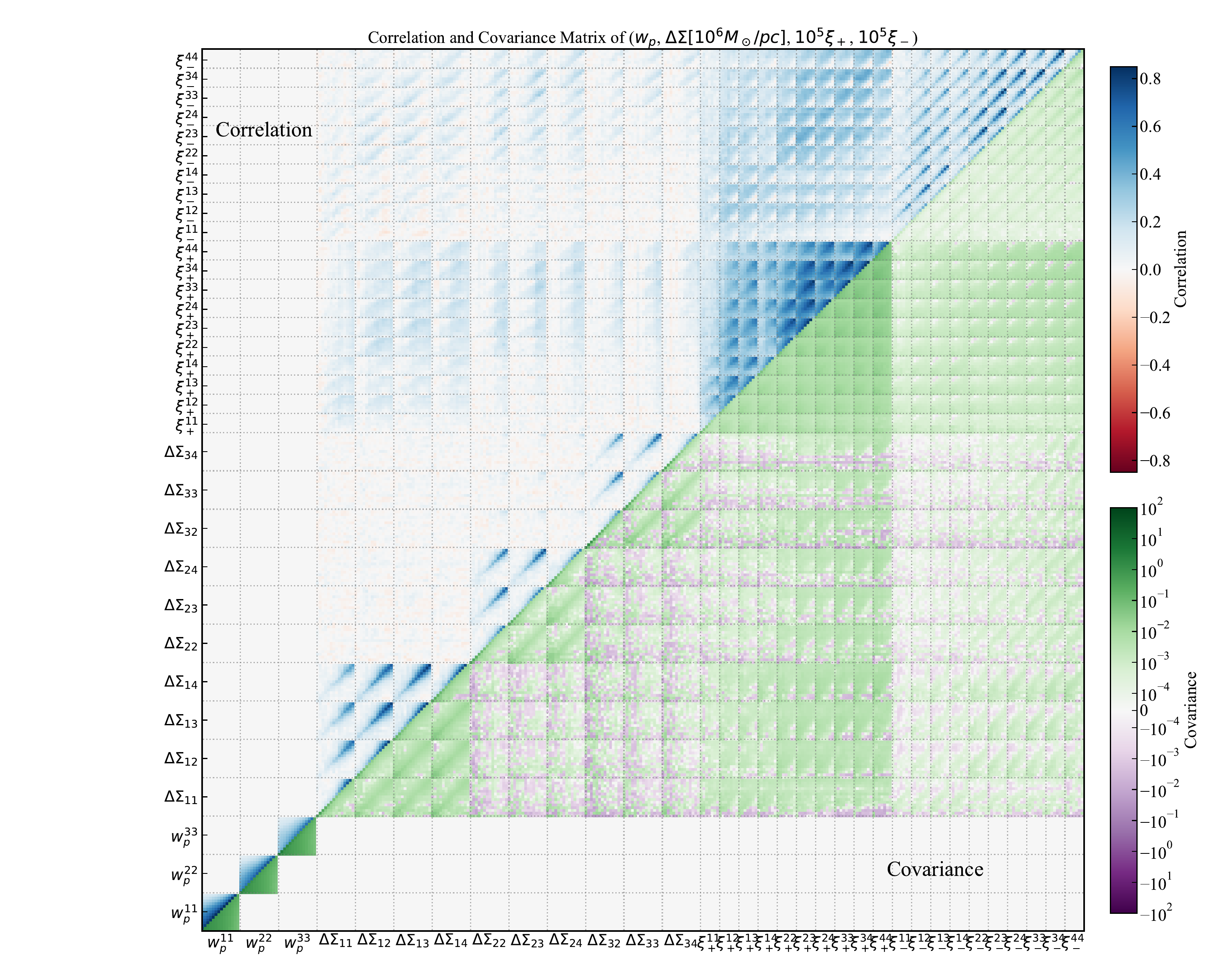}
\caption{\label{fig:correlation} The correlation and covariance matrix of the data vector used in the fiducial $3\times2$pt cosmological analysis, computed from measurements on 1404 mock SDSS and HSC realizations. $\xi_{\pm}^{ij}$ are multiplied by $10^5$ to match th The covariance matrix are shown in the lower triangular and the diagonal elements, while the correlation matrix are shown in as an upper triangular matrix. We assume no correlation between the galaxy clustering signal $w_p$ and the weak lensing observables. Significant correlations are observed among $\Delta \Sigma$ measurements that share the same lens tomographic bin, as well as among the shear-shear correlation functions $\xi_{\pm}^{ij}$. Additionally, we detect notable correlations between the galaxy-galaxy lensing and cosmic shear signals, which arise from their shared sensitivity to the same underlying large-scale structure.  
}
\end{figure*}

\subsection{Covariance Matrix}
\label{sec:data:cov}

The covariance matrix used for the $3\times2$pt analysis is estimated from 108 realizations of full-sky gravitational lensing simulations \cite{Shirasaki2017, Shirasaki2019, takahashi2017}. Each realization is populated with SDSS-like galaxies and subdivided into 13 non-overlapping regions to emulate the HSC-like footprint, yielding a total of 1404 mock realizations ($108 \times 13$). Given that the overlap between the SDSS DR11 and HSC-Y3 footprints is only about 5\% (i.e., 416 deg$^2$ / 8300 deg$^2$), we neglect the correlation between galaxy clustering ($w_p$) and weak lensing observables. However, since the HSC source catalog contributes to both galaxy-galaxy lensing ($\Delta \Sigma$) and cosmic shear ($\xi_{\pm}$), we expect a positive correlation between these two lensing signals. The correlation and covariance matrix of the full data vector is shown in Fig.~\ref{fig:correlation}. The mock measurements of $w_p$ and $\Delta \Sigma$ are described in \cite{2x2pt_paper}, and the mock cosmic shear measurements $\xi_{\pm}$ follow the methodology described in \cite{Li2023}.

From Fig.~\ref{fig:correlation}, we observe strong correlation between GGL data points that share the same lens bins. This suggest that using multiple source bin will not significantly increase the GGL signal-to-noise. We also observe strong correlation betwen the GGL and cosmic shear, highlighting the importance of measuring the mock data vector jointly. 

\section{Modeling and Inference}
\label{sec:model:0}

\begin{table}\centering
\begin{tabular}{ccc}
\hline
Parameter          &   & Prior          \\ \hline
\multicolumn{3}{l}{\hspace{-1em}\bf Cosmological parameters
(Section~\ref{sec:model:pk})}\\
$\log(A_s\times 10^{10})$   & \hspace{2cm}  & $U[1.0,5.0]$            \\
$\Omega_bh^2$          &    &  $U[0.1, 0.025]$       \\
$n_s$               &   & $U[0.94, 1.0]$    \\
$\Omega_ch^2$          &     & $U[0.0998, 0.1398]$     \\
$\Omega_m$          &  & $U[0.0906, 0.5406]$    \\
$\sum m_\nu [{\rm eV}]$       &   & $0.06$         \\
$w$                &   & $-1.0$/($U[-2,-0.33]$ wCDM)   \\
$w_a$              &   & $0.0$    \\
$\Omega_k$         &   & $0.0$     \\\hline
\multicolumn{3}{l}{\hspace{-1em}\bf Baryonic Feedback
(Section~\ref{sec:model:theory})}\\
$A_{b}$            &   & $U[1.5, 3.13]$ \\
\multicolumn{3}{l}{\hspace{-1em}\bf Intrinsic Alignment
(Section~\ref{sec:model:xipm})}\\
$A_{\rm IA, 1}$    &   & $U[-6,6]$\\
$A_{\rm IA, 2}$    &   & $U[-6,6]$\\
$\alpha_{\rm IA, 1}$    &   & $U[-6,6]$\\
$\alpha_{\rm IA, 2}$    &   & $U[-6,6]$\\
$b_{\rm TA}$    &   & $U[0,2]$\\\hline
\multicolumn{3}{l}{\hspace{-1em}\bf Linear galaxy bias
(Section~\ref{sec:model:wp} and Section~\ref{sec:model:ds})}\\
$b_1$              &   & $U[0.1,5.0]$ \\
$b_2$              &   & $U[0.1,5.0]$  \\
$b_3$              &   & $U[0.1,5.0]$  \\
\multicolumn{3}{l}{\hspace{-1em}\bf Point-mass correction
(Section~\ref{sec:model:ds})}\\
$\Delta \Sigma_{\rm PM,1} (4 {\rm Mpc/h})$ &   & $U[0.0,10.0]$            \\
$\Delta \Sigma_{\rm PM,2} (4 {\rm Mpc/h})$ &  & $U[0.0,10.0]$              \\
$\Delta \Sigma_{\rm PM,3} (4 {\rm Mpc/h})$ &   & $U[0.0,10.0]$              \\
\multicolumn{3}{l}{\hspace{-1em}\bf Magnification biases
(Section~\ref{sec:model:ds})}\\
$\alpha_{\rm mag, 1}$    &    & $\mathcal{N}(2.258,0.5)$          \\
$\alpha_{\rm mag, 2}$    &    & $\mathcal{N}(3.563,0.5)$         \\
$\alpha_{\rm mag, 3}$    &    & $\mathcal{N}(3.729,0.5)$         \\\hline
\multicolumn{3}{l}{\hspace{-1em}\bf Multiplicative shear biases
(Section~\ref{sec:model:m})}\\
$m_1$       &    & $\mathcal{N}(0.0,0.01)$  \\
$m_2$       &    & $\mathcal{N}(0.0,0.01)$ \\
$m_3$       &    & $\mathcal{N}(0.0,0.01)$ \\
$m_4$       &    & $\mathcal{N}(0.0,0.01)$  \\
\multicolumn{3}{l}{\hspace{-1em}\bf Photo-$z$ systematics
(Section~\ref{sec:model:nz})}\\
$\Delta z_{1}$    &    & $\mathcal{N}(0,0.024)$    \\
$\Delta z_{2}$     &      & $\mathcal{N}(0,0.022)$      \\
$\Delta z_{3}$     &     & $U[-1,1]$     \\
$\Delta z_{4}$     &    & $U[-1,1]$   \\
\multicolumn{3}{l}{\hspace{-1em}\bf PSF systematics
(Section~\ref{sec:model:psf})}\\
$\alpha^{\rm '(2)}$     &    &  $\mathcal{N}(0,1)$  \\
$\beta^{\rm '(2)}$     &    &  $\mathcal{N}(0,1)$  \\
$\alpha^{\rm '(2)}$     &    &  $\mathcal{N}(0,1)$  \\
$\beta^{\rm '(2)}$     &    &  $\mathcal{N}(0,1)$  \\

\hline
\end{tabular}
\caption{The prior distributions used in the fiducial Bayesian analysis. The first section lists the cosmological parameters: five parameters are varied in the fiducial $\Lambda$CDM analysis, and the dark energy equation-of-state parameter $w$ is additionally varied in the $w$CDM model; other cosmological parameters are fixed. The second section includes astrophysical parameters: one for baryonic feedback and five for the TATT intrinsic alignment model. The third section contains lens-related nuisance parameters, including linear galaxy biases, point-mass correction terms, and magnification bias parameters for each tomographic lens bin. The final section presents source-related nuisance parameters, which include one multiplicative shear biase and photometric redshift shift parameter per tomographic source bin, and four PSF systematics parameters.
}
\label{tab:prior}
\end{table}

In this section, we describe the forward model used to compute the $3\times2$pt data vector, which is parameterized by cosmological and nuisance parameters. This includes the theoretical modeling of galaxy clustering $w_p$, galaxy-galaxy lensing $\Delta \Sigma$ and cosmic shear $\xi_\pm$, treatments for systematic errors, the framework for Bayesian inference, and the validation and internal consistency tests performed to ensure the robustness of the results. The modeling of the cosmic shear correlation functions $\xi_\pm$ follows the methodology in \cite{Li2023}. Therefore, we only briefly here and refer the readers to \cite{Li2023}.

We use the open-source software package \textsc{CosmoSIS} \cite{Zuntz2015} for forward modeling and Bayesian inference. All modules employed in this work are publicly available through the \textsc{CosmoSIS} Standard Library\footnote{\url{https://github.com/joezuntz/cosmosis-standard-library}}.

\subsection{Cosmological and Astrophysical Modeling}
\label{sec:model:theory}

\subsubsection{Matter Power Spectrum}
\label{sec:model:pk}

We compute the matter power spectrum as the first step in the forward modeling process.
The linear matter power spectrum, $P_{\rm mm}(k; z_l)$, is calculated using the CAMB interface \cite{Lewis2000} within \textsc{CosmoSIS}. The shape and amplitude of the linear power spectrum are governed by the logarithmic amplitude of the primordial curvature perturbations, $\log(10^{10} A_s)$, and the spectral index $n_s$, as well as the physical densities of baryons and cold dark matter, $\Omega_b h^2$ and $\Omega_c h^2$, respectively, where $h$ is the dimensionless Hubble parameter. Throughout this work, we fix the total neutrino mass to $\sum m_\nu = 0.06$ eV. Throughout the survey, we assump a flat geometry for the universe, i.e., $\Omega_k = 0$.

At small scales, the power spectrum is not well described by linear perturbation theory. Therefore, we need to compute the non-linear power spectrum. We use \textsc{HMCode} \cite{Mead2016} 2016 to compute the nonlinear power spectrum through a customized interface in \textsc{CosmoSIS}. 

The clustering signal $w_p(R_p)$ and galaxy-galaxy lensing signal $\Delta \Sigma(R_p)$ are directly derived from the nonlinear matter power spectrum $P_{\rm nl}(k;z)$ (see Section~\ref{sec:model:wp} and Section~\ref{sec:model:ds}). In contrast, the cosmic shear two-point correlation functions $\xi_{\pm}^{ij}(\theta)$ are related to the angular power spectrum $C^{ij}_{\mathrm{GG}}(\ell)$\cite{Chisari2018}, which, under the assumption of a flat universe and the Limber approximation, can be written as:
\begin{equation}
\label{eq:cl_gg}
C^{ij}_{\mathrm{GG}}(\ell) = \int_0^{\chi_H} d\chi \, \frac{q_i(\chi) \, q_j(\chi)}{\chi^2} \, P_{\rm m}\left(k = \frac{\ell + 1/2}{\chi}, \chi\right),
\end{equation}
where $\chi$ is the comoving radial distance, $\chi_H$ is the comoving distance to the horizon, and $q_i(\chi)$ is the lensing efficiency kernel for the $i$-th tomographic bin, defined as:
\begin{equation}
q_i(\chi) = \frac{3}{2} \, \Omega_m \left( \frac{H_0}{c} \right)^2 \frac{\chi}{a(\chi)} \int_{\chi}^{\chi_H} d\chi' \, n_i(\chi') \, \frac{\chi' - \chi}{\chi'},
\end{equation}
where $n_i(\chi')$ is the comoving redshift distribution of the $i$-th tomographic bin, and $H_0 = 100\, h\, {\rm km\,s^{-1}\,Mpc^{-1}}$. For a single source galaxy at comoving distance $\chi$, the lensing efficiency peaks near $\chi/2$.

\subsubsection{Baryonic Feedback}
\label{sec:model:baryonic}

Baryonic effects such as feedback from supernovae and active galactic nuclei (AGN) are known to suppress the matter power spectrum at small scales \cite{Huang2021, Ryo2025}. To account for these effects, we follow \cite{Asgari2018} and adopt \textsc{HMCode} 2016 to model baryonic suppression. \textsc{HMCode} 2016 parameterizes baryonic feedback using the halo mass–concentration amplitude $A_b$ and a halo bloating parameter $\eta_b$, which depends on $A_b$ according to \cite{Joachimi2021}:
\begin{equation}
    \eta_b = 0.98 - 0.12 A_b.
\end{equation}
We place a uniform prior on $A_b$ over the range $U[1.5, 3.13]$, where $A_b = 3.13$ corresponds to the dark-matter-only scenario with no baryonic feedback. The parameter $A_b$ is marginalized over in our Bayesian inference. The lower bound $A_b = 1.5$ is chosen based on mock validation results, which showed that stronger feedback models below this threshold can bias the inferred $S_8$. We do not use the more recent \textsc{HMCode} 2020 model due to computational efficiency. The improvements in \textsc{HMCode} 2020 primarily affect models with large neutrino masses ($\sum m_\nu > 0.5\,\mathrm{eV}$), which are not relevant to this analysis.
In Appendix~\ref{ap:validation}, we validate that our fiducial model is robust against baryonic physics.

\subsubsection{Intrinsic Alignment}
\label{sec:model:ia}

Intrinsic alignment (IA) refers to the correlation between the shapes of galaxies due to physical effects unrelated to gravitational lensing. IA contaminates cosmic shear through two main mechanisms: (a) the intrinsic shapes of two galaxies can be correlated because they are physically associated and experience the same tidal field—this is known as the II (intrinsic-intrinsic) term; (b) the intrinsic shape of one galaxy is aligned with the local tidal field, while the other galaxy, located farther away, is gravitationally lensed by the same large-scale structure that generated the tidal field—this is known as the GI (gravitational-intrinsic) term \cite{Hirata2004}.

Under the Limber approximation\cite{Limber1953}, the angular power spectra for the II and GI terms between the $i$-th and $j$-th tomographic bins are given by:
\begin{align}
C_{II}^{ij} (\ell) &= \int_0^{\chi(z_{\rm max})} \!\! d\chi\, \frac{n_i(\chi)\, n_j(\chi)}{\chi^2}\, P_{\rm II}\!\left(k = \frac{\ell + 1/2}{\chi};\chi\right), \\
C_{GI}^{ij} (\ell) &= \int_0^{\chi(z_{\rm max})} \!\! d\chi\, \frac{q_i(\chi)\, n_j(\chi)}{\chi^2}\, P_{\rm GI}\!\left(k = \frac{\ell + 1/2}{\chi};\chi\right),
\end{align}
where $q_i(\chi)$ is the lensing efficiency of the $i$-th bin and $n_i(\chi)$ is the comoving number density. $P_{\rm II}$ and $P_{\rm GI}$ are the intrinsic alignment power spectra. $z_{\rm max}$ is set to $6$ across the analysis. 

In this work, we follow \cite{Li2023} and adopt the tidal alignment and tidal torque (TATT) model \cite{Blazek2019} to model IA contamination to the cosmic shear correlation functions $\xi_\pm$. The TATT model expresses the IA power spectra in terms of the following components:
\begin{align}
P^E_{\rm GI}(k)
    &= c_1 P_\delta(k) + b_{\rm TA} c_1 P_{0|0E}(k) + c_2 P_{0|E2}(k)\,, \\
P^E_{\rm II}(k)
    &= c_1^2 P_\delta(k) + 2 b_{\rm TA} c_1^2 P_{0|0E}(k) + b_{\rm TA}^2 c_1^2 P_{0E|0E}(k) \nonumber\\
    &+ c_2^2 P_{E2|E2}(k) + 2 c_1 c_2 P_{0|E2}(k) \nonumber\\ &+ 2 b_{\rm TA} c_1 c_2 P_{0E|E2}(k)\,, \\
P^B_{\rm II}(k)
    &= b_{\rm TA}^2 c_1^2 P_{0B|0B}(k) + c_2^2 P_{B2|B2}(k) \nonumber\\ &+2 b_{\rm TA} c_1 c_2 P_{0B|B2}(k)\,.
\end{align}
The terms on the right-hand side represent various one-loop contributions to the IA power spectrum. For details of the notation and derivation, we refer readers to \cite{Blazek2019}. These spectra are computed using \textsc{FAST-PT} v2.1 \cite{McEwen2016, Fang2017}, which we interface through a \textsc{CosmoSIS} module.

The coefficients $c_1$ and $c_2$ are redshift-dependent and parameterized as:
\begin{align}
c_1(z) &= - A_{\rm IA,1}  \frac{\bar{C} \rho_{\rm c}
    \Omega_\mathrm{m}}{D(z)} \left ( \frac{1+z}{1+z_0} \right )^{\eta_{\rm IA,1}},\\
c_2(z) &= 5 A_{\rm IA,2}  \frac{\bar{C} \rho_{\rm c}
    \Omega_\mathrm{m}}{D^2(z)} \left ( \frac{1+z}{1+z_0} \right )^{\eta_{\rm IA,2}}.
\end{align}
Here, $A_{\rm IA,1}$ and $A_{\rm IA,2}$ are the amplitudes of the tidal alignment and tidal torque components, $\eta_{\rm IA,1}$ and $\eta_{\rm IA,2}$ control their redshift evolution, $D(z)$ is the linear growth factor, $\rho_c$ is the critical density today, and $\bar{C} = 5 \times 10^{-14} h^{-2} M_\odot^{-1} \mathrm{Mpc}^3$ is a normalization constant.

In this work, we parameterize the TATT model using five free parameters: $A_{\rm IA,1}$, $A_{\rm IA,2}$, $\eta_{\rm IA,1}$, $\eta_{\rm IA,2}$, and $b_{\rm TA}$. We adopt wide, uninformative priors in the range $[-6, 6]$ for all five parameters due to the lack of reliable prior constraints. The pivot redshift is fixed at $z_0 = 0.62$ for all IA terms, and the normalization constant is set to $\bar{C} = 5 \times 10^{-14} h^{-2}\, M_\odot^{-1}\, \mathrm{Mpc}^3$. We note that the commonly used Nonlinear Alignment (NLA) model \cite{Bridle2007, Krause2010} is a special case of the TATT model obtained by setting $A_{\rm IA,2} = 0$ and $b_{\rm TA} = 0$.

Intrinsic alignment can also affect the galaxy-galaxy lensing (GGL) signal if the source galaxies are physically associated with the lens galaxies. However, our GGL measurements employ optimal weighting, which down-weights or eliminates lens-source pairs at the same redshift. As a result, the impact of IA on the GGL signal is expected to be negligible. We confirm this expectation through validation tests in which we artificially inject significant IA contamination into the mock GGL data vector. These tests show that the key cosmological constraints from the $3\times2$pt analysis remain unaffected. Details of these validation tests are provided in Appendix~\ref{ap:validation}.

\subsection{Correlation Functions}

\subsubsection{Projected Correlation Function $w_p (R_p)$}
\label{sec:model:wp}

The modeling of the projected correlation function $w_p(R_p)$ follows the formalism presented in Sugiyama et al. \cite{Sugiyama2023}, which adopts the ``minimal bias'' model. This model describes the galaxy number density field as a linear tracer of the underlying matter density field. Sugiyama et al. \cite{Sugiyama2020} demonstrated that the minimal bias model provides an accurate description of $w_p$ and allows unbiased inference of the cosmological parameters when applied to scales $R_p > 8\,h^{-1}\,\mathrm{Mpc}$. In this work, we use the implementation of the minimal bias model provided in the \texttt{CosmoSIS} framework to compute $w_p(R_p)$.

In the forward modeling, we first compute the 3D galaxy correlation function:
\begin{equation}
\xi_{gg}(r; z_l) = b_l(z_l)^2 \int_0^{\infty} \frac{k^2 \, \mathrm{d}k}{2\pi^2} P_{\rm mm}^{\rm NL}(k; z_l) j_0(kr),
\end{equation}
where $P_{\rm mm}^{\rm NL}(k; z_l)$ is the nonlinear matter power spectrum described in Sections~\ref{sec:model:pk} and \ref{sec:model:baryonic}, including baryonic feedback. The function $j_0(x)$ is the zeroth-order spherical Bessel function. The Hankel transform is performed using \texttt{FFTLog} \citep{Fang2020}. We introduce one linear galaxy bias parameter $b_l(z_l)$ for each lens redshift bin.

The projected correlation function $w_p$ is derived by integrating $\xi_{gg} (r;z_l)$ over the line-of-sight direction of $z_l$, 
\begin{equation}
w_p(R_p;z_l) = 2 f^{\rm RSD}_{\rm corr} (R_p; z_l) \int_0^{\Pi_{\rm max}} d\Pi \xi_{gg} (\sqrt{R_p^2 + \Pi^2};z_l).
\end{equation}
Here $\Pi$ is the line-of-sight distance. Following \cite{Zehavi2005, Sugiyama2023}, the maximum line-of-sight distance $\Pi_{\rm max}$ is set to $100h^{-1}$ Mpc, $f^{\rm RSD}$ is the correction factor for the Kaiser redshift space distortion effect (RSD) \cite{Kaiser1984}. 

As validated in More et al. \cite{More2023}, the redshift evolution of $w_p$ under the assumption of a redshift-independent galaxy bias $b_l(z_l)$ contributes at most 4\% of the statistical uncertainty. Therefore, this bias evolution can be safely neglected. In this work, we evaluate the clustering signal at the weighted average redshifts of each tomographic bin: $z_l = 0.26$, $0.51$, and $0.63$ for the LOWZ, CMASS1, and CMASS2 bins, respectively.

\subsubsection{Projected Surface Density $\Delta \Sigma(R_p)$}
\label{sec:model:ds}

The modeling of the projected surface mass density profile $\Delta \Sigma(R_p)$ from the nonlinear power spectrum follows the point-mass correction model described in the companion $2\times2$pt paper \cite{2x2pt_paper}. 

The galaxy-galaxy lensing signal is quantified by the excess surface density $\Delta \Sigma(R_p)$ at projected radius $R_p$. To model this signal across a wide range of scales ($2h^{-1}$Mpc to $80h^{-1}$Mpc), we combine the minimal bias model with a point-mass correction. The latter assumes that the additional mass within a small radius $R_0$ to what the minimal bias model predicts is concentrated at the center of the halo and can be approximated as a power-law contribution at larger radii. The model includes three components:
\begin{equation}
\label{eq:ggl_theory_all}
\Delta \Sigma (R_p) = \Delta \Sigma_{gG}(R_p) + \Delta \Sigma_{\rm PM}(R_p) + \Delta \Sigma_{\rm mag}(R_p),
\end{equation}
where $\Delta \Sigma_{gG}(R_p)$ is the predicted excess surface density from the minimal bias model assuming a single linear galaxy bias $b_l$ for all lens galaxies in a given tomographic bin:
\begin{equation}
\label{eq:ggl_first_term}
\Delta \Sigma_{gG}(R_p) = b_l(z_l) \bar{\rho}_{m_0} \int_0^\infty \frac{k\, {\rm d}k}{2 \pi} P_{\rm mm}^{\rm NL}(k;z_l) J_2(kR_p),
\end{equation}
where $\bar{\rho}_{m_0}$ is the present-day mean matter density, and $J_2(x)$ is the second-order Bessel function. Here, $b_l(z_l)$ denotes the galaxy bias in the $l$-th lens bin evaluated at redshift $z_l$. The Hankel transform is computed using \texttt{FFTLog} \citep{Fang2020}. This analysis uses the \texttt{CosmoSIS} modules for the minimal bias model, the point-mass (Upsilon) correction, and the magnification bias to compute the three terms in Eq.~\ref{eq:ggl_theory_all}.

The second term, $\Delta \Sigma_{\rm PM}(R_p)$, represents a point-mass correction to account for the small-scale behavior of the galaxy-matter correlation. It assumes that the mass profile within a radius $R_0$ can be approximated as a point mass located at the center of the halo, leading to a $R^{-2}$ dependence due to the inverse-square law. The reduced observable, originally introduced as the annular differential surface density (ADSD), is expressed using the $\Upsilon(R_p)$ statistic \cite{Baldauf2010, mandelbaum2013}:
\begin{equation}
\label{eq:upsilon}
\Upsilon(R_p) = \Delta \Sigma_{gG}(R_p) - \Delta \Sigma_{gG}(R_0) \left( \frac{R_p}{R_0} \right)^{-2}.
\end{equation}

In the point-mass correction model, we account for the missing 1-halo contribution, denoted $\Delta \Sigma_{\rm PM}(R_0)$, which is not captured by the \textsc{HaloFit} prediction. This leads to a modified model for $\Delta \Sigma(R_p)$:
\begin{align}
\Delta \Sigma'(R_p) &= \Upsilon(R_p) + \Delta \Sigma_{\rm PM}(R_0) \left( \frac{R_p}{R_0} \right)^{-2} \\
&= \Delta \Sigma_{gG}(R_p) + \left[ \Delta \Sigma_{\rm PM}(R_0) - \Delta \Sigma_{gG}(R_0) \right] \left( \frac{R_0}{R_p} \right)^2.
\end{align}
Here, $\Delta \Sigma_{gG}(R_p)$ is the prediction from the minimal bias model, and $\Delta \Sigma_{\rm PM}(R_0)$ is a free parameter in our analysis, defined per lens bin. We fix the inner scale radius to $R_0 = 4\,h^{-1}\mathrm{Mpc}$, and set the minimum scale used in the cosmological inference to $R_{p,\rm min} = 2\,h^{-1}\mathrm{Mpc}$. 

The third term, $\Delta \Sigma_{\rm mag} (R_p)$, accounts for the magnification bias in the lensing signal. This effect arises because gravitational lensing increases the apparent brightness of background (source) galaxies, making more of them detectable in flux-limited samples when being lensed. As a result, the observed galaxy-galaxy lensing signal is enhanced in regions of higher foreground mass density. The magnification bias contribution depends on the nonlinear matter power spectrum $P_{\rm mm}^{\rm NL}$, as well as the redshift distributions of the lens galaxies $n_l(z)$ and the source galaxies $n_s(z)$. 

We model the magnification bias using a single parameter $\alpha_{\rm mag, j}$ for each lens bin $j$, which characterizes the logarithmic slope of the number counts at the luminosity threshold used to define the lens sample. We adopt a conservative Gaussian prior on $\alpha_{\rm mag, j}$ with a standard deviation of $0.5$ to account for uncertainties in the slope estimation. We refer the reader to Sugiyama et al. \cite{Sugiyama2023} and \cite{2x2pt_paper} for a detailed description of the formalism used to calculate the contribution of the magnification bias.

\subsubsection{Shear-shear Correlation Function $\xi_{\pm} (\theta)$}
\label{sec:model:xipm}

The calculation of the shear-shear correlation function between the $i$-th and $j$-th tomographic bins, $\xi_{+/-}^{ij}(\theta)$, largely follows Li et al.\ \cite{Li2023}. Under the flat-sky approximation, the shear-shear two-point correlation function can be expressed as
\begin{equation}
    \xi_{+/-}^{ij} (\theta) = \frac{1}{2\pi} \int d\ell\, \ell\, J_{0/4}(\theta \ell) \left( C^{E;ij}(\ell) \pm C^{B;ij}(\ell)\right),
\end{equation}
where $J_{0/4}(\theta \ell)$ is the zeroth- or fourth-order Bessel function. The Hankel transform is performed using \texttt{FFTLog} \citep{Fang2020} as implemented in \textsc{CosmoSIS}.

$C^{E/B;ij}(\ell)$ denotes the total $E$- and $B$-mode angular power spectra of the shear-shear correlation, which are determined by both gravitational lensing of the large-scale structure (Section~\ref{sec:model:pk}) and intrinsic alignments (Section~\ref{sec:model:ia}). Specifically,
\begin{align}
C^{E;ij}(\ell) &= C^{E;ij}_{GG}(\ell) + C^{E;ij}_{II}(\ell) + C^{E;ij}_{GI}(\ell) + C^{E;ji}_{GI}(\ell), \\
C^{B;ij}(\ell) &= C^{B;ij}_{II}(\ell).
\end{align}
The $E$-mode spectrum includes contributions from lensing-lensing auto-spectra ($C^{E;ij}_{GG}$), intrinsic-intrinsic alignments ($C^{E;ij}_{II}$), and the galaxy-lensing cross terms ($C^{E;ij}_{GI}$ and $C^{E;ji}_{GI}$). The $B$-mode spectrum is assumed to arise solely from intrinsic-intrinsic correlations, as the $B$-mode contribution from lensing is neglected in this work \cite{Dodelson2010}. The $B$-mode signal in the HSC Y3 shear catalog is not statistically significant and is therefore set to zero throughout the analysis.

\subsection{Observational Systematics}

The systematics of the shear catalog need to be meticulously modeled to shield our results from systematic bias \cite{Mandelbaum2018_review, jefferson2025}.

In this section, we describe the parameters used to model observational systematics present in the HSC Y3 shear catalog, including uncertainties in the redshift distribution, multiplicative shear bias, and PSF-related systematics.

\subsubsection{Source Redshift Distribution}
\label{sec:model:nz}

The fiducial redshift distribution of the HSC Y3 shear catalog is shown as the shaded region in the upper panel of Fig.~\ref{fig:nz_lens_source}. For the fiducial analysis, the HSC Y3 shear catalog is divided into four tomographic bins using the \texttt{dNNz} $z_{\rm best}$ estimates. The redshift distribution of each tomographic bin is modeled as a joint probability distribution on a redshift grid spanning $z = 0$ to $4$, with a step size of $0.025$. Each distribution is inferred using a logistic Gaussian process \cite{Rau2022}, constrained by:
\begin{enumerate}
    \item The \texttt{dNNz} photo-$z$ estimates, with prior information that accounts for the cosmic variance of the photo-$z$ training set; and
    \item The clustering redshift between the photometric sample and the CAMIRA LRG reference sample \cite{CAMIRA_Oguri2014,Oguri2018_camira,CAMIRA3}, available in the range $z = 0.1$--$1.1$.
\end{enumerate}

As discovered in the HSC Y3 cosmic shear analysis \cite{Li2023, Dalal2023}, the mean redshifts of the third and fourth tomographic bins preferred by the $\Lambda$CDM model are inconsistent with those inferred from the fiducial redshift distributions based on photometric and clustering redshift calibrations. This discrepancy may arise because the clustering redshift calibration is limited to $z < 1.1$, which excludes part of bin~3 and all of bin~4. Independent $3\times2$pt analyses--though using only a single source bin--also suggest deviations from the fiducial mean redshifts for the source sample \cite{Sugiyama2023, Miyatake2023}. As highlighted in \cite{Li2023}, the inconsistency between the fiducial redshift distribution and the value preferred by cosmic shear is unlikely to be caused by statistical fluctuations, with statistical significance exceeding $95\%$.

We use a shift parameter $\Delta z_i$ for each source redshift distribution \cite{Zhang2023_nz}, which modifies the mean redshift of the $i$-th tomographic bin as:
\begin{equation}
\label{eq:model_nz_shift}
    n_i(z) \longrightarrow n_i(z + \Delta z_i).
\end{equation}
This shift affects both the cosmic shear and galaxy-galaxy lensing components of the data vector. In the case of cosmic shear, modifying $n_i(z)$ alters the lensing efficiency $q_i(\chi)$ defined in Eq.~\ref{eq:cl_gg}. A redshift distribution shifted to higher redshift corresponds to (a) larger volume of foreground matter field, (b) higher lensing efficiency and therefore tends to prefer a lower amplitude of the matter power spectrum, parameterized by $S_8 = \sigma_8 \sqrt{\Omega_m / 0.3}$.

For galaxy-galaxy lensing, the redshift distribution $n_i(z)$ of the source galaxies does not directly affect the theoretical value of $\Delta \Sigma(R_p)$ for a given lens bin $q$, since the model assumes fixed lens redshifts. However, the observed galaxy-galaxy lensing signal $\Delta \Sigma_{qi}$ between the $q$-th lens bin and the $i$-th source bin must be corrected for changes in the source redshift distribution $n_i(z)$. 

We define the correction factor based on the ratio of average inverse critical surface densities, weighted by the lens and source weights $w_l$ and $w_s$, over all galaxies in the $q$-th lens bin and $i$-th source bin:
\begin{equation}
\label{eq:redshift_correction}
f_{\Delta \Sigma}^{qi}(\Delta z_{i}) = \frac{\sum_{\rm l\in q, s\in i} w_{\rm l} w_{\rm s}  \langle \Sigma_{\rm crit}^{-1}\rangle^{\rm est}_{\rm ls} (\Delta z_i)}{\sum_{\rm l\in q, s\in i}  w_{\rm l} w_{\rm s}  \langle \Sigma_{\rm crit}^{-1}\rangle^{\rm fid}_{\rm ls}}.
\end{equation}
Here, $\langle \Sigma_{\rm crit}^{-1} \rangle^{\rm est}_{ls}(\Delta z_i)$ denotes the estimated average inverse critical surface density using the shifted source redshift distribution $n_i(z + \Delta z_i)$, while $\langle \Sigma_{\rm crit}^{-1} \rangle^{\rm fid}_{ls}$ uses the fiducial $n_i(z)$. The lens and source weights $w_l$ and $w_s$ follow the definitions in Section~III.B of \cite{2x2pt_paper}. Notice that $\langle \Sigma_{\rm crit}^{-1} \rangle^{\rm fid}_{ls}$ is computed by a cosmology with $\Omega_m = 0.279$ and $w_0 = -1$, while $\langle \Sigma_{\rm crit}^{-1} \rangle^{\rm est}_{ls}(\Delta z_i)$ is computed by the sampled $\Omega_m$ and $w_0$. Therefore, $f_{\Delta \Sigma, qi}(\Delta z_{i})$ also correct for the difference in comoving distance by the cosmology. It is shown that the impact of the assumed cosmology when measuring $\Delta \Sigma$ does not cause significant bias on the GGL inference if it is accounted for \cite{more2015}.

The corrected galaxy-galaxy lensing measurement is then:
\begin{equation} 
\label{eq:corrected_dsigma} 
\Delta \Sigma^{qi}_{\rm corr}(R_p, \Delta z_i) = f_{\Delta \Sigma}^{qi}(\Delta z_i) \Delta \Sigma^{qi}(R_p). 
\end{equation} 

Note that the mean lens redshift $z_q$ is assumed to be fixed, as the redshifts of the lens galaxies are spectroscopic and thus do not carry significant uncertainty. This correction factor applies to the entire theoretical prediction $\Delta \Sigma(R_p | z_q)$, including the main galaxy-galaxy lensing signal $\Delta \Sigma_{gG}(R_p)$, the magnification bias term $\Delta \Sigma_{\rm mag}(R_p)$, and the 1-halo correction term $\Delta \Sigma_{\rm PM}(R_p)$.

As a result of the redshift discrepancies identified in previous HSC Y3 cosmology analyses, we adopt informative priors for $\Delta z_1$ and $\Delta z_2$ based on the estimates from \cite{Rau2022}, and use flat priors in the range $[-1, 1]$ for $\Delta z_3$ and $\Delta z_4$. The full set of priors on $\Delta z_i$ is listed in Table~\ref{tab:prior}. We expect that the implicit lensing ratio encoded in the tomographic galaxy-galaxy lensing and cosmic shear data vectors can self-constrain the redshift parameters $\Delta z_3$ and $\Delta z_4$, and yield competitive cosmological constraints. We leave the calibration of the high-redshift source redshift distribution for future work, through methods such as clustering redshift inference \cite{Newman2008} or direct calibration \cite{Hildebrandt2017}, leveraging emerging high-redshift spectroscopic datasets overlapping the HSC footprint, including the DESI DR1 large-scale structure sample \cite{desi_dr1}.

\subsubsection{Multiplicative Calibration Bias}
\label{sec:model:m}

The shear multiplicative bias of the HSC Y3 shear catalog is calibrated using image simulations, as described in \cite{Li2022}. These simulations account for galaxy model bias, noise bias, selection bias, and detection bias, using realistic galaxy samples derived from HST F814W-band images.

In this work, we model the residual multiplicative shear bias with one nuisance parameter per source tomographic bin, denoted $m_i$. For the cosmic shear data vector, the correction is applied as
\begin{equation}
    \xi_{\pm}^{ij}(\theta) \longrightarrow (1 +  m_i)(1+ m_j) \xi_{\pm}^{ij}(\theta).
\end{equation}
For galaxy-galaxy lensing, as shown in Eq.~\ref{eq:delta_sigma_obs}, the $\Delta \Sigma$ signal is modified as
\begin{equation}
    \Delta \Sigma_{qi}(R_p) \longrightarrow (1 +  m_i) \Delta \Sigma_{qi}(R_p).
\end{equation}
The image simulations confirm that the multiplicative bias $m$ is controlled to within 1\% across the redshift range of the source sample. In our analysis, the priors on $m_i$ are Gaussian distributions centered at zero with a standard deviation of $0.01$.

\subsubsection{PSF Additive Bias}
\label{sec:model:psf}

The correlation between PSF modeling residuals and PSF shapes can contaminate the shear-shear correlation function\cite{jarvis2021,zhang2023a}. Following \cite{Zhang2023b}, the HSC Y3 shear catalog is impacted by PSF leakage and modeling residuals associated with the spin-2 components of the second moments $e_{\rm PSF}$ and the fourth moments $M^{(4)}_{\rm PSF}$. The observed shear can be modeled as
\begin{align}
\label{eq:PSF_additive}
\gamma_\mathrm{obs} &= \frac{1}{2\mathcal{R}} e_\mathrm{obs} \\\nonumber &= \frac{1}{2\mathcal{R}} \left( e_{\rm gal} + \alpha^{(2)} e^{(2)}_\text{psf} + \beta^{(2)} \Delta e^{(2)}_\text{psf}
    + \alpha^{(4)} M^{(4)}_\text{psf} + \beta^{(4)} \Delta M^{(4)}_\text{psf}\right),
\end{align}
where the second through fifth terms represent, respectively, second-moment leakage, second-moment modeling residuals, fourth-moment leakage, and fourth-moment modeling residuals. \cite{Zhang2023b} find significant additive bias in $\xi_+$, while the contamination to $\xi_-$ is negligible.

The additive PSF bias on $\xi_+$ is modeled as
\begin{equation}
\label{eq:expand-esys_esys}
\xi_+(\theta) \longrightarrow \xi_+(\theta) +
\sum_{k=1}^4 \sum_{q=1}^4 p_k p_q \langle S_k S_q \rangle,
\end{equation}
where $\vec{p} = [\alpha^{(2)}, \beta^{(2)}, \alpha^{(4)}, \beta^{(4)}]$ is the parameter vector, and $\vec{S} = [e^{(2)}_\text{psf}, \Delta e^{(2)}_\text{psf}, M^{(4)}_\text{psf}, \Delta M^{(4)}_\text{psf}]$ is the vector of PSF moments. The parameters $\vec{p}$ are constrained by the shear-PSF and PSF-PSF correlation functions; we refer the reader to \cite{Zhang2023b} for a detailed description.

To account for correlations between the components of $\vec{p}$, we re-parameterize the PSF additive bias using $\vec{p}' = [\alpha'^{(2)}, \beta'^{(2)}, \alpha'^{(4)}, \beta'^{(4)}]$. The parameters $\vec{p}$ are related to $\vec{p}'$ via an invertible linear transformation:
\begin{equation}
\vec{p} = \mathbf{T} \cdot \vec{p}' + \vec{\bar{p}},
\end{equation}
where $\mathbf{T}$ is constructed so that the components of $\vec{p}'$ are uncorrelated and follow standard normal distributions (mean 0 and standard deviation 1). The transformation matrix $\mathbf{T}$ is given by
\begin{equation}
    \mathbf{T} = \mathbf{V}^{1/2} \mathbf{U},
\end{equation}
where $\mathbf{V}$ is the diagonal matrix of eigenvalues of the covariance matrix of $\vec{p} - \vec{\bar{p}}$, and the columns of $\mathbf{U}$ are the corresponding eigenvectors. The Bayesian inference samples from the $\vec{p}'$ space with Gaussian priors centered at zero and standard deviation of one. 

We do not model the PSF systematics in the galaxy-galaxy lensing signal, since the PSF condition does not correlate with lens selection in this work, thus does not affect the two point statistics between lens and shape.

\subsection{Summary of the Theoretical Template}
\label{label:sec:inference}

\begin{figure*}
\includegraphics[width=1.5\columnwidth]{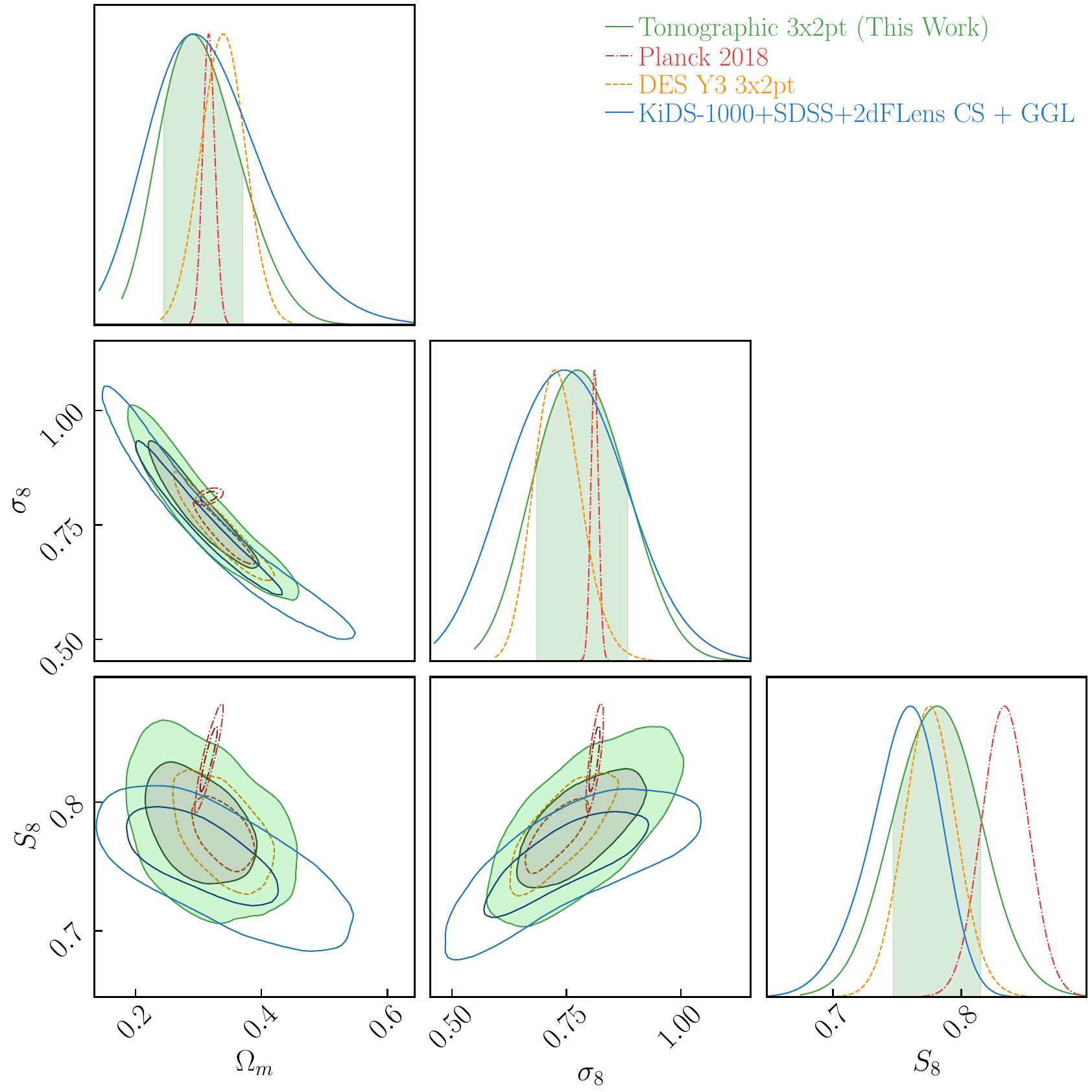}
\caption{\label{fig:external_datasets} Comparison of the fiducial cosmological constraints from this work (green contours) with external datasets. Shown for reference are the Planck 2018 TTTEEE+lowE CMB results (red contours; \cite{Planck2018Cosmology}), the Dark Energy Survey (DES) Year 3 3$\times$2pt analysis (orange contours; \cite{DESY3_3x2_2022}), and the Kilo-Degree Survey (KiDS)-1000 joint cosmic shear and galaxy-galaxy lensing analysis (blue contours; \cite{Heymans2021}). Our 3$\times$2pt constraints from HSC Y3 are consistent with those from other clustering and weak lensing surveys, while providing improved redshift calibration through self-consistent tomographic lensing. 
}
\end{figure*}

\begin{figure*}
\includegraphics[width=1.5\columnwidth]{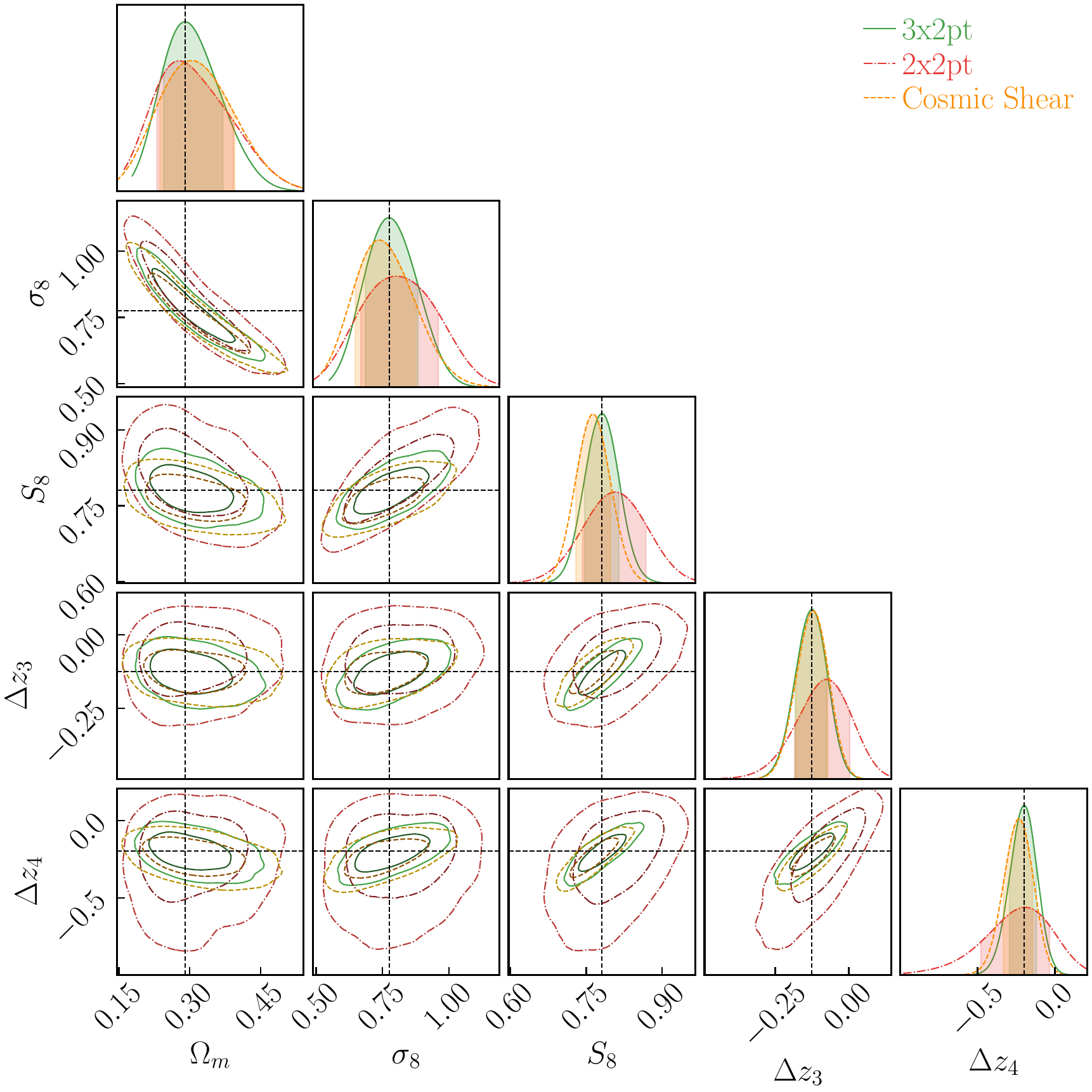}
\caption{\label{fig:fiducial_probes} Marginalized constraints on $\Omega_m$, $\sigma_8$, $S_8$, $\Delta z_3$, and $\Delta z_4$ from the 3$\times$2pt (clustering + galaxy-galaxy lensing + cosmic shear), 2$\times$2pt (clustering + galaxy-galaxy lensing), and cosmic shear analyses presented in this work. The constraints on $\Omega_m$ and $\sigma_8$ are jointly informed by the 2$\times$2pt and cosmic shear measurements, while the constraints on $S_8$, $\Delta z_3$, and $\Delta z_4$ are primarily driven by the cosmic shear component. The 2$\times$2pt analysis yields results consistent with the 3$\times$2pt constraints across all parameters.}
\end{figure*}

We use both public and custom modules within \texttt{CosmoSIS} to construct the theoretical model for the $3\times2$pt data vector. The linear matter power spectrum is computed using \texttt{CAMB}, with the parameters $\log(10^{10} A_s)$, $n_s$, $\Omega_b h^2$, $\Omega_c h^2$, and $\Omega_c$ as free parameters for the $\Lambda$CDM model. For the $w$CDM model, we additionally free the dark energy equation-of-state parameter $w_0$. We assign flat prior on $\log(10^{10} A_s)$ following \cite{Planck2018Cosmology}.

Nonlinear evolution and baryonic effects are modeled using \texttt{HMCode} 2016, parameterized by a single baryonic feedback parameter $A_b$. Intrinsic alignments are modeled using the tidal alignment and tidal torquing (TATT) model, with five free parameters: $A_{\rm IA,1}$, $A_{\rm IA,2}$, $\eta_{\rm IA,1}$, $\eta_{\rm IA,2}$, and $b_{\rm TA}$.

The clustering signal $w_p$ and galaxy-galaxy lensing signal $\Delta \Sigma$ are computed using the minimal bias model, and are modulated by the linear galaxy bias parameters $b_1$, $b_2$, and $b_3$ for each lens bin. Magnification bias is modeled with one parameter $\alpha_{\rm mag, q}$ per lens bin, with informative priors derived from the slope of the SDSS galaxy number counts near the luminosity cut.

We implement a customized module in \texttt{CosmoSIS} for the point-mass correction term, with one free parameter $\Delta \Sigma_{\rm PM, q}(4\,h^{-1}\mathrm{Mpc})$ per lens bin and uninformative priors. For the shear-shear correlation function $\xi_\pm$, we compute the angular power spectra including both lensing and intrinsic alignment terms and transform them into configuration space via Hankel transforms.

The shear observables are modified with one multiplicative shear bias parameter $m_i$ per source bin, calibrated using image simulations. We adopt Gaussian priors centered at zero with standard deviation $\sigma = 0.01$. For the source redshift distribution uncertainties, we use informative priors on $\Delta z_1$ and $\Delta z_2$ from photometric and clustering redshift calibration, and adopt flat uninformative priors $\Delta z_{3,4} \sim \mathcal{U}[-1, 1]$ due to empirical evidence for a non-negligible bias in $\Delta z_{3,4}$.

We model the PSF additive bias with four parameters $\alpha'^{(2)}$, $\beta'^{(2)}$, $\alpha'^{(4)}$, and $\beta'^{(4)}$, using informative priors constrained by PSF-shape cross-correlations.

In total, the fiducial Bayesian inference involves 32 (or 33) free parameters depending on whether the $w$CDM model is used. We collectively denote the full parameter vector as $\mathbf{p}$.

For the $3\times2$pt data vector $\mathbf{d}$, we arrange the probes in the following order: 
\begin{equation}
\mathbf{d} = \{ \mathbf{w}_{p}, \mathbf{\Delta \Sigma},  \mathbf{\xi_+}, \mathbf{\xi_-} \}.
\end{equation}
Throughout this work, the elements of the data vector $d^{ij}(R_k)$ are ordered lexicographically in $ijk$.

For the galaxy clustering, we define $\mathbf{w}_{p} = w_{qq}(R_k)$, where $q = 0, 1, 2$ corresponds to the LOWZ, CMASS1, and CMASS2 tomographic bins, respectively. We use the projected clustering signal in the range $R_p = [8, 80]\,h^{-1}\mathrm{Mpc}$, sampled at 14 logarithmically spaced points per tomographic bin. The lower limit of the scale cut is set by the validity range of the minimal bias model \cite{Sugiyama2022, Sugiyama2023}, while the upper limit is chosen to avoid the baryon acoustic oscillation (BAO) feature. In total, the clustering data vector $\mathbf{w}_{p}$ consists of 42 data points.

For galaxy-galaxy lensing, we define $\mathbf{\Delta \Sigma} = \Delta \Sigma_{qi}(R_k)$, where $q$ indexes the lens bins and $i$ indexes the source bins. We include all four HSC source bins for the LOWZ lens bin, and use the second through fourth HSC bins for CMASS1 and CMASS2, in order to avoid significant redshift overlap between the lens and source populations. For each lens-source bin pair, we use the $\Delta \Sigma(R_p)$ signal in the range $R_p = [2, 70]\,h^{-1}\mathrm{Mpc}$ for Bayesian inference. The lower scale cut is chosen based on the off-centering effect in the lens sample, as characterized in mock validation tests \cite{2x2pt_paper}, while the upper scale cut is determined from the null test of the cross-component $\Delta \Sigma_\times$. Each lens-source pair contributes 14 logarithmically spaced data points, resulting in a total of 140 data points for the galaxy-galaxy lensing data vector.

For cosmic shear, we define $\mathbf{\xi_\pm} = \xi^{ij}_\pm(\theta_k)$, where $i$ and $j$ label the tomographic bins. We use the angular range $\theta \in [7.1, 56.6]$ arcmin for all $\xi_+$ bin pairs, and $\theta \in [31.2, 248]$ arcmin for all $\xi_-$ bin pairs. The small-scale cuts are imposed to avoid the modeling uncertainties due to baryonic feedback, while the large-scale cuts are set based on the B-mode null test of the observed data.

\subsection{Bayesian Inference}
\label{label:sec:inference}

We conduct Bayesian inference to estimate cosmological parameter constraints using the data vector measured from the HSC and SDSS datasets. We define the terminology relevant to the inference, as well as the parameter space and data vector used in the analysis.

We denote the parameter vector as $\mathbf{p}$, the observed data vector as $\mathbf{d}^{\rm obs}$, the theoretical data vector as $\mathbf{d}^{\rm theory}(\mathbf{p})$, and the covariance matrix of $\mathbf{d}^{\rm obs}$ as $\mathbf{C}$. For each analysis, we sample $\mathbf{p}$ and compute the log-likelihood function as:
\begin{equation}
\label{eq:log_likelihood}
\log \mathcal{L}(\mathbf{d}|\mathbf{p}) = -\frac{1}{2} \left[\mathbf{d}^{\rm obs} - \mathbf{d}^{\rm theory}(\mathbf{p})\right]^{T} \mathbf{C}^{-1} \left[\mathbf{d}^{\rm obs} - \mathbf{d}^{\rm theory}(\mathbf{p})\right].
\end{equation}
Since our covariance matrix is estimated by finite realizations of the mock catalogs, we multiply the $\log$ likelihood by the Hartlap factor \cite{Hartlap2007} $\mathcal{P}$, defined by,
\begin{equation}
\label{eq:hartlap_factor}
\mathcal{P} = \frac{N_{\rm mock} - n_{\rm DV} - 2}{N_{\rm mock} - 1}.
\end{equation}
For our fiducial analysis, $N_{\rm mock} = 1404$, and $n_{\rm DV} = 322$, so $\mathcal{P}_{\rm fid} = 0.77$. The Hartlap factor inflates our constraining power by $\sim 0.1\sigma$.


The priors for all parameters are summarized in Table~\ref{tab:prior}. We sample the posterior using the nested sampling algorithm \texttt{MultiNest}, implemented within the \texttt{CosmoSIS} framework~\cite{Zuntz2015}. For the fiducial analysis, we set the number of live points to 500, the efficiency parameter to 0.3, and the tolerance parameter to 1.0. We additionally run a \texttt{MultiNest} analysis with the tolerance parameter reduced to 0.05 and find no significant differences in the results (see Section~\ref{ap:bayesian}).

We note that physically $\Delta \Sigma_{\rm PM,q}(4 {\rm Mpc}/h)$ and $b_q$ should be correlated due to their dependence on halo mass. Due to software constraints, we model the priors as independent parameters and leave the study with correlated priors to future studies.

For the fiducial analysis, we report the results of the Bayesian inference using the $1\sigma$ and $2\sigma$ confidence contours, along with summary statistics in the format:
\begin{equation}
\mathrm{mode}^{+34\%\ \mathrm{upper}}_{-34\%\ \mathrm{lower}} \ (\mathrm{MAP},\ \mathrm{mean}),
\end{equation}
where the mode refers to the peak of the one-dimensional marginalized posterior distribution of the parameter. We also compute the Maximum A Posteriori (MAP) estimate for the fiducial analysis by performing optimization from 100 random starting points in the parameter space. However, we do not compute the MAP for the validation tests or internal consistency analyses.

\section{Results and Comparison}
\label{sec:results:0}

\begin{figure*}
\includegraphics[width=1.5\columnwidth]{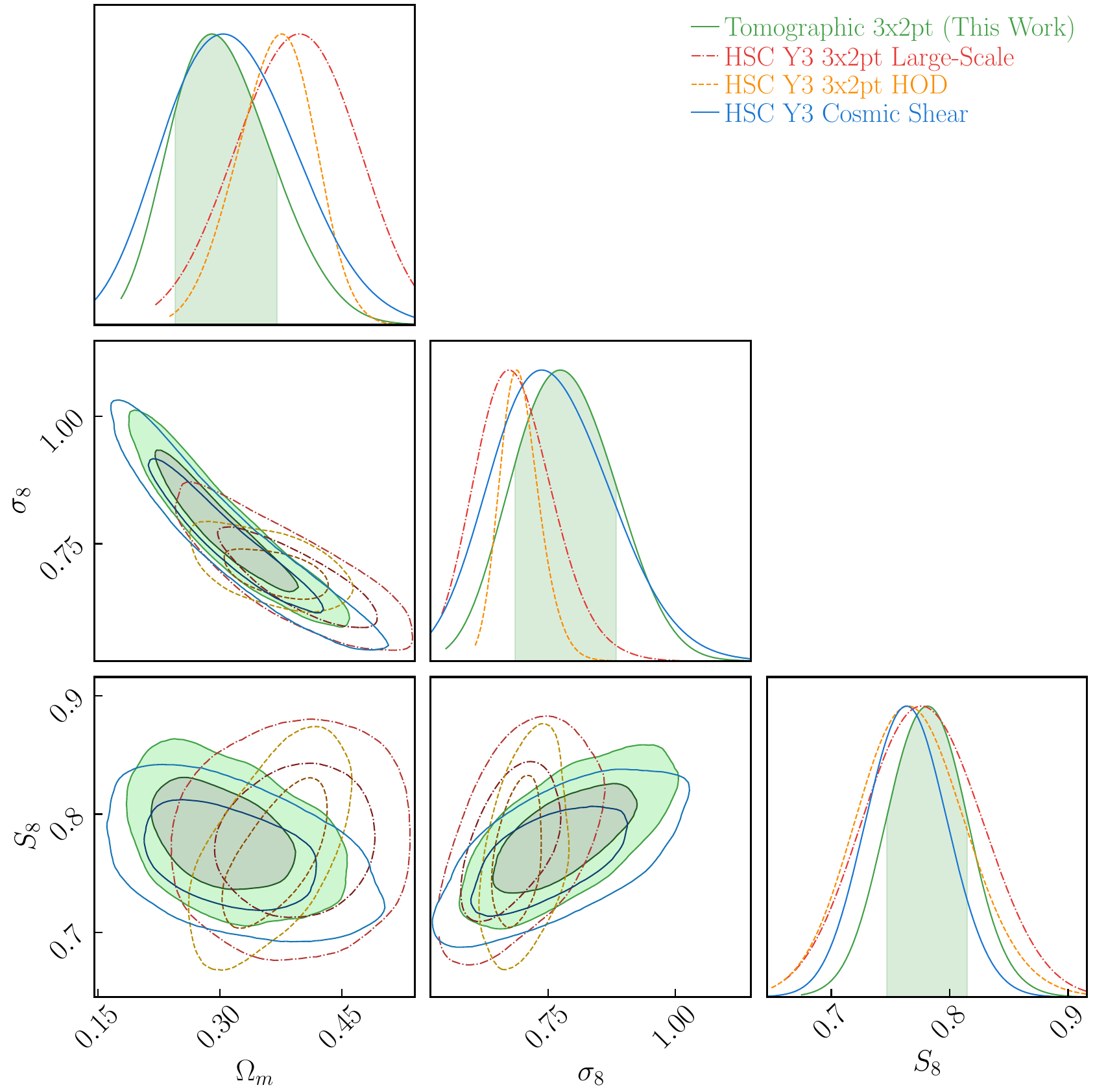}
\caption{\label{fig:compare_hsc_analyses} Comparison of the 3$\times$2pt analysis results from this work (black contours) to previous HSC Y3 results: the 3$\times$2pt analysis using a single source bin at large scales \citep[red contours;][]{Sugiyama2023}, the 3$\times$2pt analysis using a single source bin including small scales \citep[orange contours;][]{Miyatake2023}, and the cosmic shear analysis \citep[blue contours;][]{Li2023}. Relative to previous 3$\times$2pt results, our analysis yields slightly lower $\Omega_m$ and higher $\sigma_8$ values, while the $S_8$ constraints remain statistically consistent across all analyses.}
\end{figure*}

\begin{figure}
\includegraphics[width=1\columnwidth]{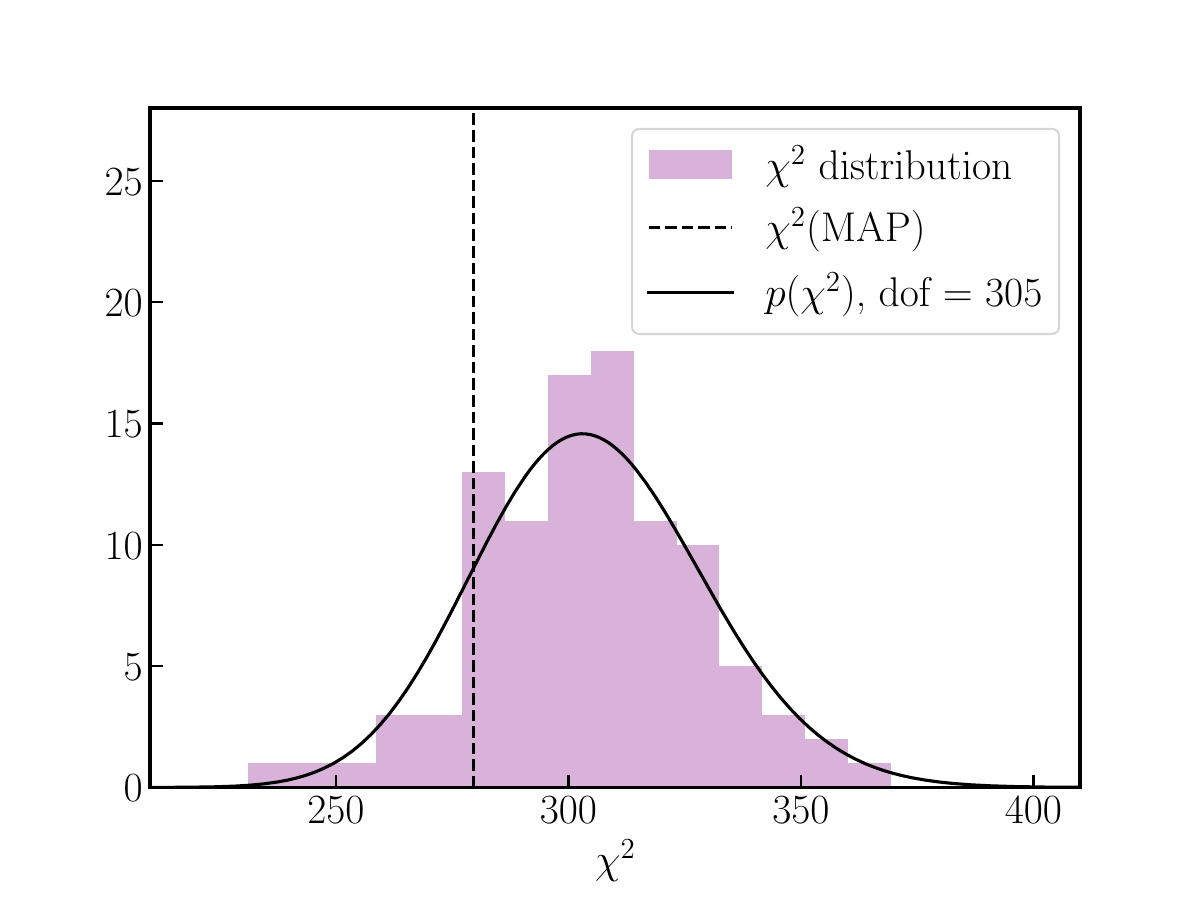}
\caption{\label{fig:chi2_dist} The empirical $\chi^2$ distribution (purple) of the 3$\times$2pt data vector, estimated from the minimum $\chi^2$ values of the MAP parameters fitted to 100 noisy mock data vectors. The $\chi^2$ value corresponding to the MAP fit of the real data vector is shown as the black dotted line. The resulting $p$-value for the goodness-of-fit of our fiducial analysis is 0.91, indicating good agreement between the model and the data.
}
\end{figure}

\subsection{Key Results}
\label{sec:results:key}

We present the key findings of this work. The fiducial analysis, defined in Section~\ref{sec:model:0}, is a Bayesian inference of 32 free parameters using a $3\times2$pt data vector that includes galaxy clustering, galaxy-galaxy lensing, and cosmic shear measurements. After successfully passing all null tests, internal consistency checks, and validation tests across the three blinded catalogs, we unblinded the results and identified the \texttt{blind\_id = 0} catalog as the true dataset. Importantly, no modifications were made to the fiducial analysis after unblinding.

The fiducial analysis under the $\Lambda$CDM model yields the following constraints on key cosmological and redshift parameters:
\begin{align}
   \mathbf{S_8}   = \sigma_8 \left(\frac{\Omega_m}{0.3}\right)^{0.5} &= 0.780 ^{+0.029}_{-0.030}~(0.782,~0.781),\\
    \mathbf{\Delta z_3} &= -0.112^{+0.046}_{-0.049}~(-0.128,~{-0.129}),\\
    \mathbf{\Delta z_4} &= -0.185^{+0.071}_{-0.081}~(-0.192,~{-0.206}),
\end{align}
where the values are presented in the format ${\rm mode} ^{+34\%}_{-34\%}~({\rm MAP,~mean})$. When optimizing along the principal degeneracy between $\Omega_m$ and $\sigma_8$, the fiducial analysis yields:
\begin{equation}
    \tilde{S}_8 = \sigma_8 \left(\frac{\Omega_m}{0.3}\right)^{0.6} = 0.781 ^{+0.027}_{-0.027}~(0.783,~0.781).
\end{equation}
We also obtain constraints on parameters that derive $S_8$
\begin{align}
    \Omega_m  &= 0.284^{+0.061}_{-0.050}~(0.300,~0.306),\\
    \sigma_8  &= 0.774^{+0.093}_{-0.086}~(0.782,~0.783).\\ 
\end{align}

The $\tilde{S}_8$ constraint achieves a precision of $3.5\%$, representing the most precise $S_8$ measurement obtained to date from HSC cosmological analyses. In Fig.~\ref{fig:external_datasets}, we compare the fiducial results on $\Omega_m$, $\sigma_8$, and $S_8$ with external datasets, including the Planck TTTEEE+lowE CMB results~\cite{Planck2018Cosmology}, the DES Y3 $3\times2$pt analysis\cite{DESY3_3x2_2022}, and the KiDS-1000 + SDSS + 2dFLenS cosmic shear and galaxy-galaxy lensing analysis\cite{Heymans2021}. The HSC Y3 tomographic $3\times2$pt analysis shows consistency with other clustering and weak lensing analyses.

We do not find significant tension between our 3$\times$2pt results and the Planck 2018 constraints. To quantify this comparison, we compute the eigen-tension metric introduced by \cite{Park2020}. Specifically, we diagonalize the posterior covariance matrix of our fiducial analysis and identify the two principal eigenvectors:
\begin{align}
\mathbf{e}_1 &= \sigma_8 \Omega_m^{0.6},\\
\mathbf{e}_2 &= \Omega_m \sigma_8^{-0.6}.
\end{align}
The first eigenmode $\mathbf{e}_1$ corresponds to the most constrained direction in the 3$\times$2pt data vector—closely aligned with $S_8$—while $\mathbf{e}_2$ is approximately orthogonal and reflects the degeneracy direction. We find that the probability that our fiducial 3$\times$2pt posterior and the Planck 2018 posterior are statistically inconsistent in the $(\mathbf{e}_1, \mathbf{e}_2)$ space is 80.1\%, corresponding to a $1.31\sigma$ tension.

Fig.~\ref{fig:fiducial_probes} shows the 1D marginalized constraints on $\Omega_m$, $\sigma_8$, $S_8$, $\Delta z_3$, and $\Delta z_4$ from our 3$\times$2pt, 2$\times$2pt, and cosmic shear analyses. We observe that the constraints on $S_8$, $\Delta z_3$, and $\Delta z_4$ are primarily driven by cosmic shear, while the constraints on $\Omega_m$ and $\sigma_8$ are jointly informed by both 2$\times$2pt and cosmic shear probes.

The $S_8$ result from our analysis is statistically consistent with previous HSC Y3 3$\times$2pt analyses presented in \cite{Sugiyama2023, Miyatake2023}, as well as with cosmic shear results in \cite{Li2023, Dalal2023}. In Fig.~\ref{fig:compare_hsc_analyses}, we compare our constraints with those from the HSC Y3 3$\times$2pt large-scale analysis \cite{Sugiyama2023}, the 3$\times$2pt small-scale halo-based analysis \cite{Miyatake2023}, and the HSC Y3 cosmic shear analysis \cite{Li2023}. Our analysis provides the most precise constraint on $S_8$ among the HSC Y3 cosmology results to date.

We also find that the redshift distribution parameters $\Delta z_3$ and $\Delta z_4$ are constrained by the 3$\times$2pt analysis and are statistically consistent with the results from the cosmic shear analysis \cite{Li2023, Dalal2023}. By computing the covariance matrix of the joint posterior of $(\Delta z_3, \Delta z_4)$, we determine that the probability of inconsistency with the fiducial value $(0, 0)$ is 98.3\%, corresponding to a $2.4\sigma$ tension between the $\Lambda$CDM-preferred redshift distributions and those inferred from photometric redshifts. This emphasizes the importance of accurately calibrating the redshift distributions of high-redshift source bins for HSC, and, in the future, for Rubin Observatory and Roman, using external spectroscopic datasets such as DESI DR1 \cite{desi_dr1}.

In Fig.~\ref{fig:chi2_dist}, we show the empirical $\chi^2$ distribution of the 3$\times$2pt data vector (purple histogram), compared to the $\chi^2$ value of the best-fit parameters from the real data (black dotted line). The empirical distribution is estimated from 100 synthetic data vectors generated by sampling from the fiducial covariance matrix, which has been corrected by the Hartlap factor \cite{Hartlap2007} to account for the noise due to the limited number of simulations used in its estimation. For each synthetic data vector, we compute the $\chi^2$ of the best-fit parameter set, defined as the maximum a posteriori (MAP) solution obtained by running the optimizer from 10 different starting points to mitigate the risk of converging to local maxima in the likelihood surface. The resulting $\chi^2$ distribution corresponds to a model with 305 effective degrees of freedom. Given the minimum $\chi^2$ value of $279.6$ from the real data, we find a $p$-value of 0.91, indicating that the fiducial theoretical model provides an excellent fit to the data vector.

\subsection{$w$CDM Analysis} 
\label{sec:res:wcdm}

\begin{figure}
\includegraphics[width=1\columnwidth]{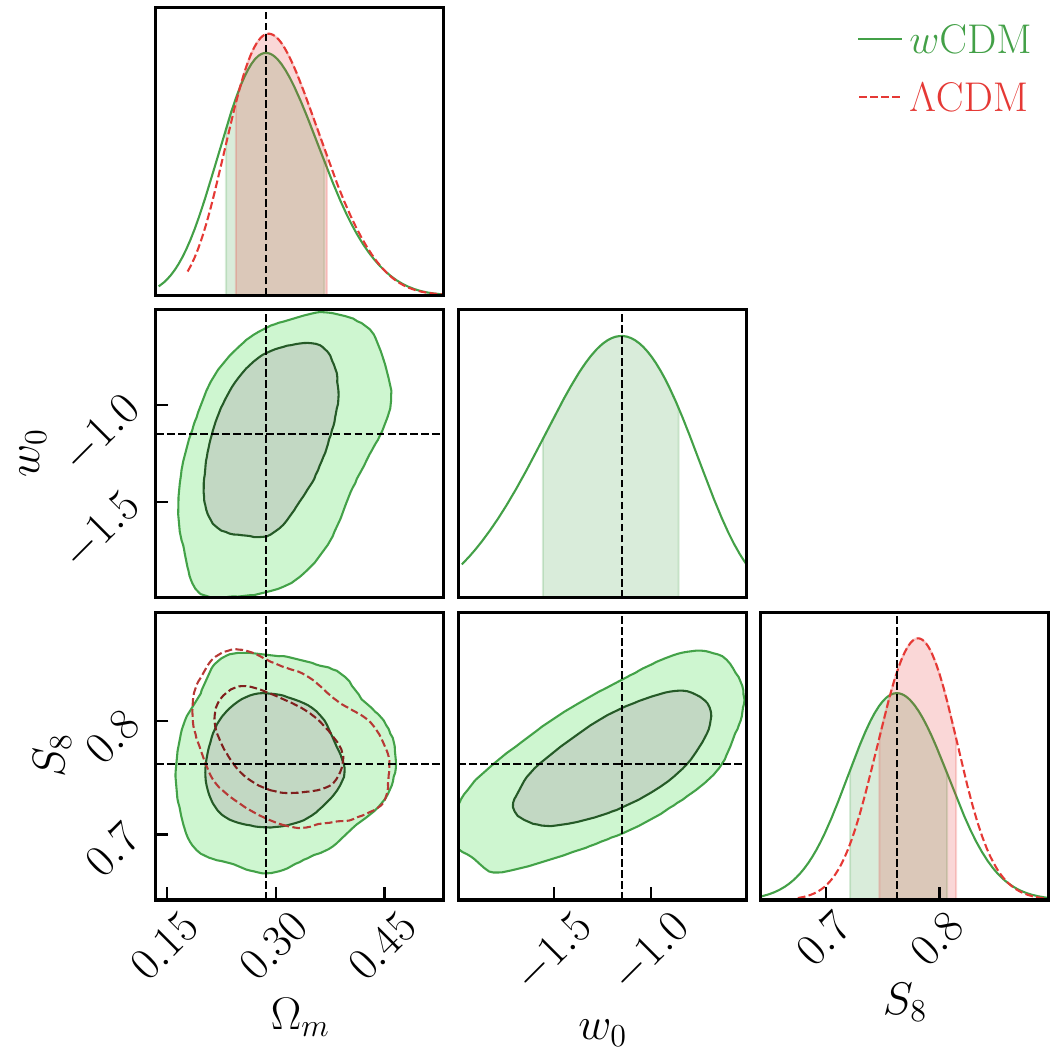}
\caption{\label{fig:wcdm} Comparison of the 68\% and 95\% confidence contours between the $\Lambda$CDM (red) and $w$CDM (green) analyses in the $(w_0, \Omega_m, S_8)$ parameter space. The $w$CDM model allows for a free dark energy equation of state $w_0$, while $w_a$ is fixed to zero. The two models yield consistent results, with $w$CDM exhibiting slightly broader contours and a marginally lower mode value of $S_8$. The dotted lines indicate the mode of the marginalized 1D posterior distributions in the $w$CDM model.
 }
\end{figure}

In addition to the $\Lambda$CDM results, we perform a $w$CDM analysis in which the dark energy equation of state parameter $w_0$ is treated as a free parameter with a uniform prior $U[-2, -0.33]$, while the time variation parameter $w_a$ is fixed to zero. In Fig.~\ref{fig:wcdm}, we show the 68\% and 95\% confidence contours for the $w$CDM and $\Lambda$CDM analyses in the $\Omega_m$–$w_0$–$S_8$ parameter space. Under the $w$CDM model, we recover the following parameter constraints:
\begin{align}
    S_8  &= \sigma_8 \left(\frac{\Omega_m}{0.3}\right)^{0.5} = 0.756^{+0.038}_{-0.036} \quad (0.773, 0.764),\\
    w_0  &= -1.17^{+0.31}_{-0.35} \quad (-1.15, -1.21),\\ 
    \Delta z_3 &= -0.120^{+0.048}_{-0.051} \quad (-0.121, -0.124),\\
    \Delta z_4 &= -0.199^{+0.077}_{-0.079} \quad (-0.164, -0.201).
\end{align}
Compared to the $\Lambda$CDM results, we observe a slightly lower $S_8$ value under the $w$CDM model. The recovered $w_0$ value is consistent with the cosmological constant ($w_0 = -1$) within $1\sigma$. Importantly, the constraining power on the source redshift shift parameters $\Delta z_3$ and $\Delta z_4$ remains robust, indicating that the tomographic lensing ratio information is largely insensitive to the expansion history parameterized by $w_0$.

\subsection{Internal Consistency Tests}
\label{sec:results:internal}

\begin{figure*}
\includegraphics[width=2\columnwidth]{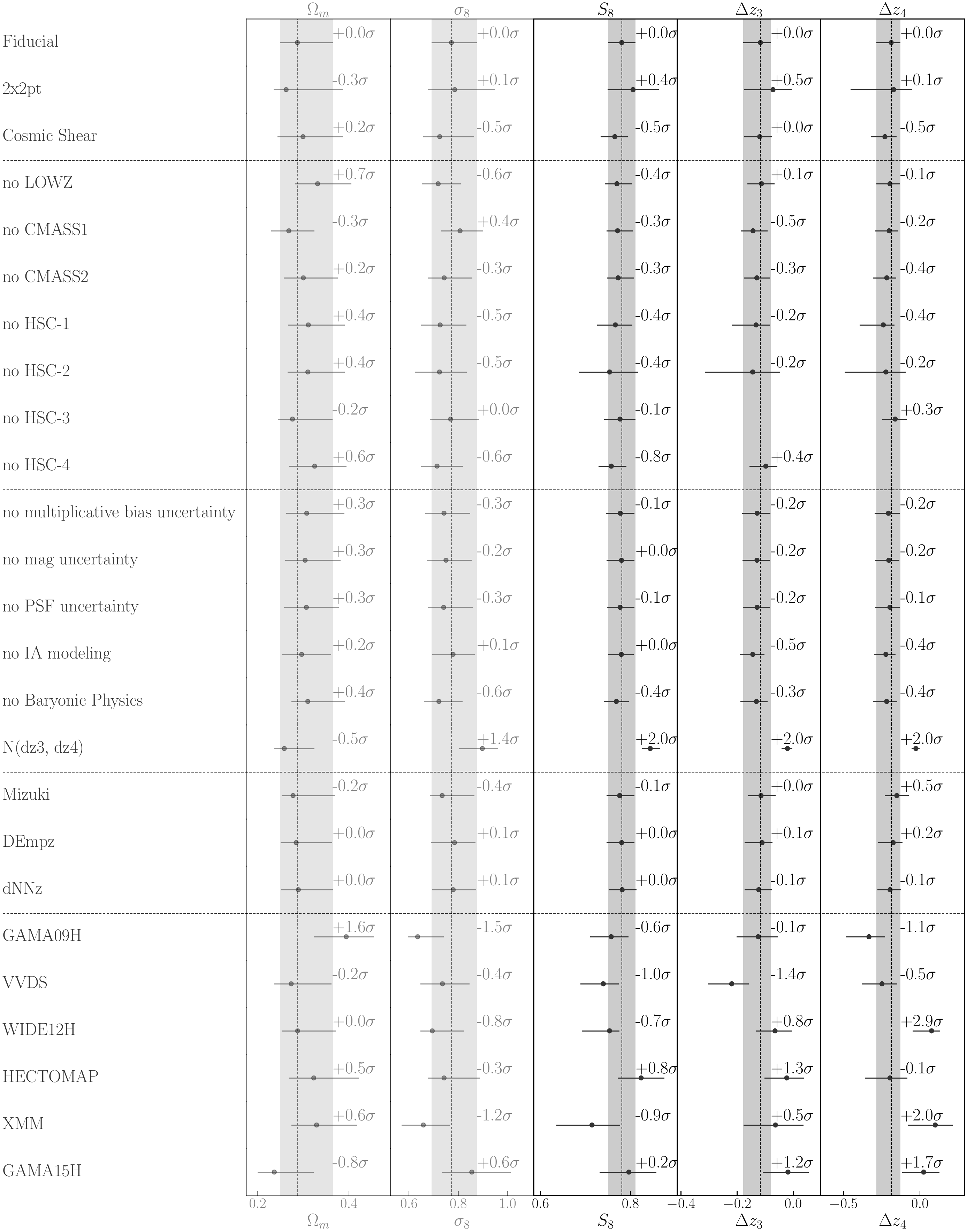}
\caption{\label{fig:internal_consistency} Summary of internal consistency results for five key parameters. Each point represents the mode of the marginalized 1D posterior distribution from a different test, with error bars denoting the 34\% upper and lower confidence intervals. The verticle dashed lines and shaded bands mark the mode and $1\sigma$ region of the fiducial analysis for comparison. We find that the inferred parameters are consistent across all subsets of the data vector and insensitive to variations in systematic modeling, except for the case where we apply Gaussian prior to $\Delta z_3$ and $\Delta z_4$. For systematics where results show sensitivity (e.g., baryonic feedback and redshift distribution), we adopt conservative modeling choices in the fiducial analysis. The bias of the mode compared to the fiducial analysis are printed onto the figure. }
\end{figure*}

\begin{figure}
    \centering
    \includegraphics[width=1\linewidth]{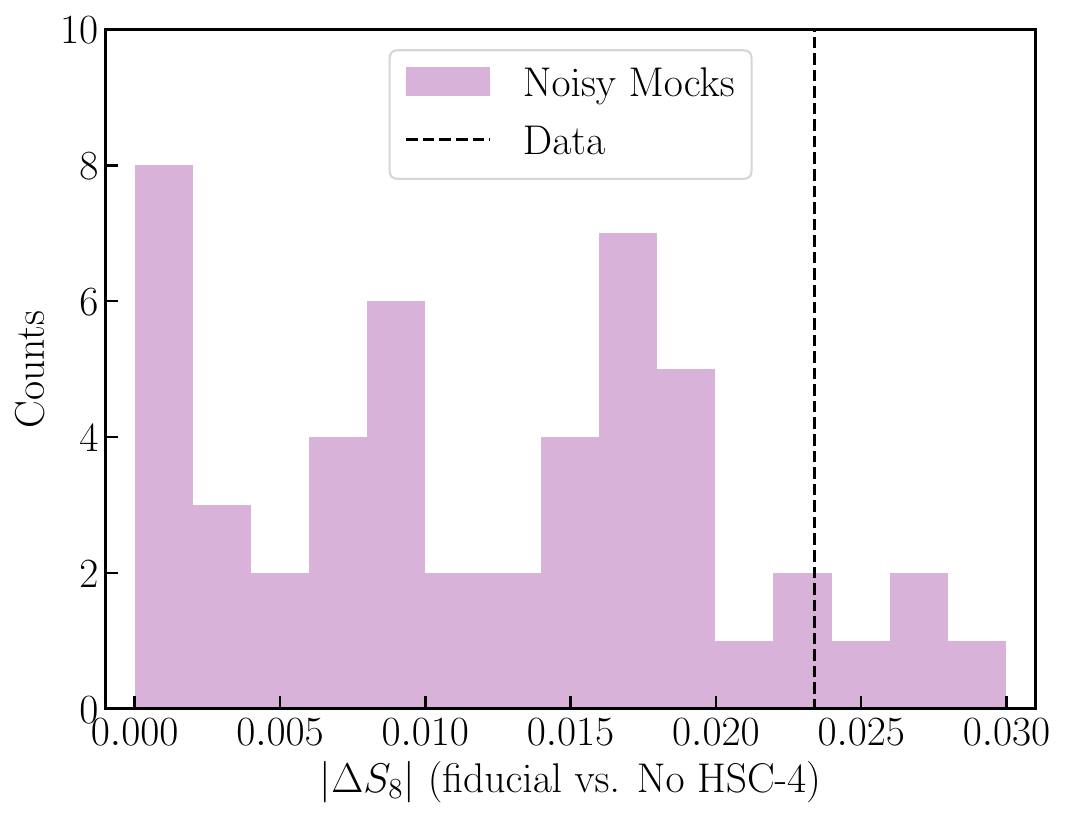}
    \caption{The distribution of $\Delta S_8$ between the fiducial and ``no HSC-4'' analyses of 50 noisy data vectors. The dashed line show the level of bias in Fig.~\ref{fig:internal_consistency}, i.e., $0.8\sigma$. The $p$-value that this bias is drawn from the distribution is $0.08$, which is not a conclusive evidence that the bias is significant. }
    \label{fig:noz4_validation}
\end{figure}

We conduct extended internal consistency tests to ensure that our cosmological results are unbiased. These tests include analyses performed on subsets of the data vector or with slightly different analysis choices. In Fig.~\ref{fig:internal_consistency}, we show the 1D constraints on $\Omega_m$, $\sigma_8$, $S_8$, $\Delta z_3$, and $\Delta z_4$ from all internal consistency tests and the bias compared to the fiducial analysis.

The first set of tests (``$2\times2$pt'' and ``cosmic shear'') compares results from individual cosmological probes within the $3\times2$pt data vector. As discussed in Section~\ref{sec:results:key}, both the $2\times2$pt and cosmic shear results are consistent with the fiducial analysis, although the $2\times2$pt analysis yields a slightly higher $S_8$ value than cosmic shear. This trend is also reported in \cite{Sugiyama2023} and \cite{Miyatake2023}.

The second set of tests (``no LOWZ'', ``no CMASS1'', ``no CMASS2'', ``no HSC-1'', ``no HSC-2'', ``no HSC-3'', ``no HSC-4'') assesses the impact of removing individual lens or source tomographic bins. We observe no significant bias in $S_8$ from the removal of a single tomographic bin, except for the ``no HSC-4'' case. To test whether the $-0.8\sigma$ bias is statistically significant, we run fiducial and ``no HSC-4'' analyses on 50 noisy data vectors generated for the $\chi^2$ calculation, and computed the distribution of $|\Delta S_8|$ with/without the HSC-4 bin, shown in Fig.~\ref{fig:noz4_validation}. We find that the $p$-value that the $-0.8\sigma$ is drawn from the uncertainty to be $0.08$, which is not a strong enough evidence to conclude that the bias is significant. However, this is worth further notice in future analyses. 
We also note the constraints on $S_8$ and $(\Delta z_3, \Delta z_4)$ rely strongly on the second source bin, which serves as an anchor point for redshift self-calibration. We find that removing the 3rd and 4th HSC bins merely increase the $S_8$ uncertainties, suggesting that the cosmological parameters, the point-mass correction term, and the galaxy bias are mainly constrained by the 1st and 2nd HSC bins, due to their degeneracy with the redshift parameters.

The third set of internal consistency tests applies alternative analysis choices, including removing the uncertainty associated with nuisance parameters (``no magnification uncertainty'', ``no multiplicative bias uncertainty'', ``no PSF uncertainty''), removing systematic modeling altogether (``no IA modeling'', ``no Baryonic Physics''), and using Gaussian informative priors on the redshift shift parameters (``$\mathcal{N}(\Delta z_3), \mathcal{N}(\Delta z_4)$''). We find that removing the uncertainty of nuisance parameters does not significantly alter the results. Similarly, omitting the intrinsic alignment modeling does not induce a significant shift in the constraints. However, turning off baryonic feedback modeling (by setting $A_{\rm bary} = 3.13$) leads to a lower inferred value of $S_8$ by $0.4\sigma$.

When adopting the Gaussian informative priors on $\Delta z_3$ and $\Delta z_4$ inferred from photometric redshift and clustering redshift cross-correlation calibrations \cite{Rau2022}, we observe significant shifts in the posterior distributions of $S_8$, $\Delta z_3$, and $\Delta z_4$. This is consistent with findings from \cite{Li2023, Dalal2023}, which motivated the use of uninformative priors in our fiducial analysis. Notably, this test also demonstrates the potential constraining power of the HSC $3\times2$pt analysis if the redshift distribution were accurately calibrated—yielding a precision of approximately $2\%$ on $S_8$.

We also test the robustness of our results against the choice of photometric redshift (photo-$z$) estimation method (``\texttt{DNNz}'', ``\texttt{DEmPz}'', ``\texttt{Mizuki}''). In these tests, we retain the same tomographic bin assignments as in the fiducial analysis but use each specific photo-$z$ method to compute the critical surface density $\Sigma_{\rm cr}^{-1}(z_l, z_s)$ in the galaxy-galaxy lensing measurement, as well as the redshift distribution $n_i(z)$ used in the theoretical modeling. We find that our cosmological results remain consistent across all photo-$z$ methods tested.

In the final set of internal consistency tests, we measure the $3\times2$pt data vector and its covariance matrix independently on each HSC field, while continue using the entire SDSS catalog. We then perform the fiducial analysis on these individual field datasets. The resulting parameter constraints show scatter consistent with their expected uncertainties. Given that the fields are largely independent—except for their shared clustering signal—this level of variation is expected.

We note that the parameter constraints in the internal consistency tests are likely to be correlated to the fiducial analysis. Therefore, the significance of the parameter difference $\mathbf{\theta_1} - \mathbf{\theta_2}$ relates to their correlation matrix, and should not be judged only based on their own uncertainty. Without the correlation of $\mathbf{\theta_1}$ and $ \mathbf{\theta_2}$, we would not draw conclusion of whether the parameter difference is statistically significant. 

Overall, we do not find significant bias in our cosmological results when using subsets of the data. The only cases where notable shifts occur are when making more aggressive assumptions about baryonic feedback or applying informative redshift priors. As a result, we adopt conservative modeling choices for these components in the fiducial analysis.

\section{Discussion and Conclusion}
\label{sec:conclusion:0}

This work presents a cosmological analysis based on the measurement of clustering and weak lensing two-point correlation functions using the HSC Y3 shear catalog and the SDSS DR11 large-scale structure catalog. The analysis combines three types of two-point functions (commonly referred to as 3$\times$2pt): the projected correlation function $w_p$ for galaxy clustering; the excess surface mass density $\Delta \Sigma$ for galaxy-galaxy lensing; and the angular shear-shear correlation functions $\xi_\pm$ for cosmic shear. This work introduces two major improvements over previous cosmological analyses with HSC:

\begin{enumerate}
    \item We employ tomographic binning for the source galaxies in both galaxy-galaxy lensing and cosmic shear measurements. This approach increases the overall signal-to-noise ratio and enables joint constraints on the redshift distribution parameters $\Delta z_3$ and $\Delta z_4$, which are critical due to the limited redshift calibration in the third and fourth tomographic bins.
    \item We extend the modeling of the galaxy-galaxy lensing signal down to $2\,h^{-1}\,\mathrm{Mpc}$ by adopting a point-mass correction term in additional to the linear galaxy bias model in the $\Delta \Sigma$ formalism. The point-mass correction accounts for additional lensing from a hypothetical mass concentration at the halo center and is parameterized by one free parameter per lens bin. This allows us to include small-scale information without explicitly modeling higher-order galaxy bias terms.
\end{enumerate}

We measure the galaxy clustering signal using the auto-correlation function in three SDSS tomographic bins: LOWZ, CMASS1, and CMASS2. Each bin contains 14 data points, logarithmically spaced in projected radius over the range $8$--$80\,h^{-1}\,\mathrm{Mpc}$, yielding a total of 42 clustering data points. The galaxy-galaxy lensing signal is measured for all lens-source bin pairs, except for CMASS1-HSC1 and CMASS2-HSC1, which are excluded due to significant redshift overlap. Each valid lens-source pair contains 14 measurements in projected radius bins, logarithmically spaced from $2$ to $70\,h^{-1}\,\mathrm{Mpc}$, resulting in a total of 140 galaxy-galaxy lensing data points.
For cosmic shear, we measure the angular shear-shear correlation functions $\xi_+$ and $\xi_-$ in 7 logarithmically spaced angular bins. The angular scale ranges from $7.1$ to $56.6$ arcmin for $\xi_+$, and from $31.2$ to $248$ arcmin for $\xi_-$. This yields a total of 140 data points for the cosmic shear measurement.
Altogether, the 3$\times$2pt data vector contains 322 data points. To estimate its covariance matrix, the full 3$\times$2pt data vector is measured on 1,404 mock realizations generated from simulations.

The theoretical templates are constructed using the open-source software \texttt{CosmoSIS} \cite{Zuntz2015}. We use five cosmological parameters to compute the $\Lambda$CDM model, assuming a flat universe, constant dark energy equation of state, and cold (non-relativistic) dark matter. Baryonic physics is modeled using \texttt{HMCode} 2016 \cite{Mead2016}, with a single free parameter. Intrinsic alignment for the cosmic shear signal is modeled using the TATT framework, introducing five additional free parameters. In addition, we run a $w$CDM model analysis by setting a free prior on the dark energy equation of state parameter $w_0$.
For each lens tomographic bin, we include one parameter each for the magnification bias, galaxy bias, and point-mass correction. For each source tomographic bin, we introduce one parameter for the redshift distribution bias and one for the multiplicative shear bias. Additionally, we include four parameters to account for PSF additive bias in the cosmic shear measurements. In total, the model includes 32(33) free parameters for the $\Lambda(w)$CDM  Bayesian inference.
To sample the posterior distribution of these parameters, we use \texttt{MultiNest}, a nested sampling algorithm implemented within \texttt{CosmoSIS}.

Our fiducial analysis yields a large-scale structure clustering amplitude of $S_8 = \sigma_8 (\Omega_m/0.3)^{0.5} = 0.780 ^{+0.029}_{-0.030}$, consistent with previous HSC cosmology results \cite{Sugiyama2023,Miyatake2023,Li2023,Dalal2023} as well as the DES Y3 cosmic shear and $3\times2$pt analyses \cite{Amon2022,DESY3_3x2_2022}. The optimized parameter yield $\tilde{S}_8 = \sigma_8 (\Omega_m/0.3)^{0.6} = 0.781 ^{+0.027}_{-0.027}$.  Our $S_8$ result shows a mild $1.3\sigma$ tension with the Planck CMB constraint \cite{Planck2018Cosmology}, although the discrepancy is not statistically significant. Additionally, we report constraints on $\Omega_m = 0.284^{+0.061}_{-0.050}$ and $\sigma_8 = 0.774^{+0.093}_{-0.086}$. Our measurement of $\Omega_m$ is in good agreement with independent cosmological probes, including BAO measurements and the Planck CMB results. In the $w$CDM analysis, we obtain $S_8 = 0.756^{+0.038}_{-0.036}$ and $w_0 = -1.17^{+0.31}_{-0.35}$, consistent with a cosmological constant and slightly lower $S_8$ than in $\Lambda$CDM.

One important feature of our analysis is that we do not fully rely on photometric redshift estimates or external redshift calibration for the third and fourth source bins. Instead, we adopt flat (uninformative) priors on the redshift shift parameters $\Delta z_3$ and $\Delta z_4$, allowing the $3\times2$pt data vector to self-calibrate these parameters. Our fiducial analysis yields $\Delta z_3 = -0.112^{+0.046}_{-0.049}$ and $\Delta z_4 = -0.185^{+0.071}_{-0.081}$. The joint posterior on $(\Delta z_3, \Delta z_4)$ is in $2.3\sigma$ tension with $(0,0)$, highlighting the critical importance of accurate redshift calibration in weak lensing cosmology.

We conduct extensive internal consistency tests by analyzing partial datasets and exploring different modeling choices for systematic errors. We find no significant evidence that our modeling assumptions bias the results. In addition, we carry out comprehensive validation tests to assess the robustness of our analysis framework against a variety of systematic effects and find no significant impact on the $S_8$ constraints.

This work motivates future improvements in several directions. As tomographic galaxy-galaxy lensing and cosmic shear measurements extract higher signal-to-noise ratios from the catalog, the corresponding increase in the size of the data vector brings it closer to the number of simulations used to estimate the covariance matrix. As a result, the Hartlap correction becomes significant in the likelihood evaluation, reducing the constraining power of the analysis. To mitigate this issue, one can generate more simulations or consider alternative methods for estimating the covariance matrix, such as analytic covariance modeling. Additionally, data compression techniques may be explored [e.g., \cite{Ferreira2021}] to reduce the dimensionality of the data vector without losing cosmological information.

Looking ahead, this work lays the groundwork for future HSC $3\times2$pt analyses that aim to maximize signal-to-noise across a broad range of redshifts and angular scales, particularly in galaxy-galaxy lensing. It also provides a methodological foundation for joint DESI–LSST analyses. Importantly, we demonstrate that $3\times2$pt analyses can still deliver competitive cosmological constraints even when redshift calibration is uncertain, provided that the source catalog is defined consistently for both cosmic shear and galaxy-galaxy lensing. Nonetheless, this work further underscores the critical importance of precise redshift calibration in the current era of weak lensing cosmology.

\begin{acknowledgments}

TZ and RM are supported by Schmidt Sciences. TZ thanks SLAC National Accelerator Laboratory for providing hospitality and an excellent research environment during the course of this study. RM is supported in part by a grant from the Simons Foundation (Simons Investigator in Astrophysics, Award ID 620789). AJN s supported by JSPS Kakenhi Grant Numbers: JP22K21349, JP23H00108 and JP25H0155. HM is supported by JSPS Kakenhi Grant Numbers: JP22K21349, JP23H00108, and JP24KK0065. MS is supported by JSPS Kakenhi Grant Numbers: JP24H00215 and JP24H00221. TS is supported by JSPS Kakenhi Grant Number: 24K17067. KO is supported by JSPS KAKENHI Grant Number JP24H00215, JP25K17380, JP25H01513, and JP25H00662. TN is supported by JSPS KAKENHI Grant Numbers: JP20H05861, JP23K20844, JP22K03634, JP24H00215, and JP24H00221. 

Work at Argonne National Laboratory was supported by the U.S. Department of Energy, Office of High Energy Physics. Argonne, a U.S. Department of Energy Office of Science Laboratory, is operated by UChicago Argonne LLC under contract no. DE-AC02-06CH11357. MMR acknowledges the Laboratory Directed Research and Development (LDRD) funding from Argonne National Laboratory, provided by the Director, Office of Science, of the U.S. Department of Energy under Contract No. DE-AC02-06CH11357. Work at Argonne National Laboratory was also supported
under the U.S. Department of Energy contract DE-AC02-06CH11357.

This work was supported in part by World Premier International Research Center Initiative (WPI Initiative), MEXT, Japan, and JSPS KAKENHI Grant Number 24H00215.

The Hyper Suprime-Cam (HSC) collaboration includes the astronomical communities of Japan and Taiwan, and Princeton University. The HSC instrumentation and software were developed by the National Astronomical Observatory of Japan (NAOJ), the Kavli Institute for the Physics and Mathematics of the Universe (Kavli IPMU), the University of Tokyo, the High Energy Accelerator Research Organization (KEK), the Academia Sinica Institute for Astronomy and Astrophysics in Taiwan (ASIAA), and Princeton University. Funding was contributed by the FIRST program from the Japanese Cabinet Office, the Ministry of Education, Culture, Sports, Science and Technology (MEXT), the Japan Society for the Promotion of Science (JSPS), Japan Science and Technology Agency (JST), the Toray Science Foundation, NAOJ, Kavli IPMU, KEK, ASIAA, and Princeton University.

This paper makes use of software developed for Vera C. Rubin Observatory. We thank the Rubin Observatory for making their code available as free software at http://pipelines.lsst.io/.

This paper is based on data collected at the Subaru Telescope and retrieved from the HSC data archive system, which is operated by the Subaru Telescope and Astronomy Data Center (ADC) at NAOJ. Data analysis was in part carried out with the cooperation of Center for Computational Astrophysics (CfCA), NAOJ. We are honored and grateful for the opportunity of observing the Universe from Maunakea, which has the cultural, historical and natural significance in Hawaii.

The Pan-STARRS1 Surveys (PS1) and the PS1 public science archive have been made possible through contributions by the Institute for Astronomy, the University of Hawaii, the Pan-STARRS Project Office, the Max Planck Society and its participating institutes, the Max Planck Institute for Astronomy, Heidelberg, and the Max Planck Institute for Extraterrestrial Physics, Garching, The Johns Hopkins University, Durham University, the University of Edinburgh, the Queen’s University Belfast, the Harvard-Smithsonian Center for Astrophysics, the Las Cumbres Observatory Global Telescope Network Incorporated, the National Central University of Taiwan, the Space Telescope Science Institute, the National Aeronautics and Space Administration under grant No. NNX08AR22G issued through the Planetary Science Division of the NASA Science Mission Directorate, the National Science Foundation grant No. AST-1238877, the University of Maryland, Eotvos Lorand University (ELTE), the Los Alamos National Laboratory, and the Gordon and Betty Moore Foundation.
\end{acknowledgments}

\appendix

\section{Validation Analysis}
\label{ap:validation}

We perform 3×2pt analyses on a suite of mock catalogs incorporating various features to validate that our analysis pipeline and methodological choices are robust against systematic errors. 
These mock data vectors are computed by halo-based mock catalogs generated in \cite{Miyatake2022}.
In this section, we describe the validation tests and present their results.

\begin{figure*}
\includegraphics[width=2\columnwidth]{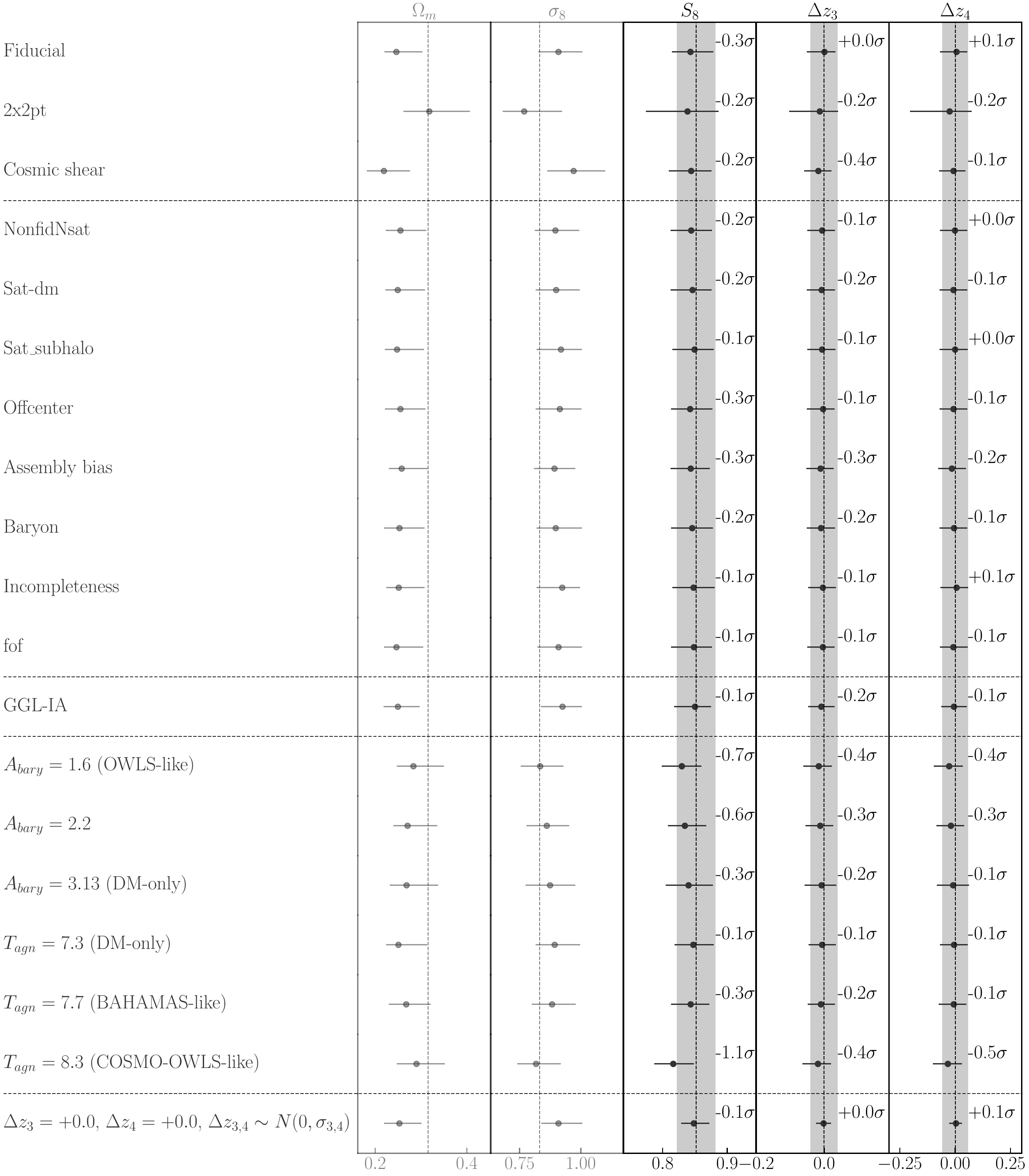}
\includegraphics[width=2\columnwidth]{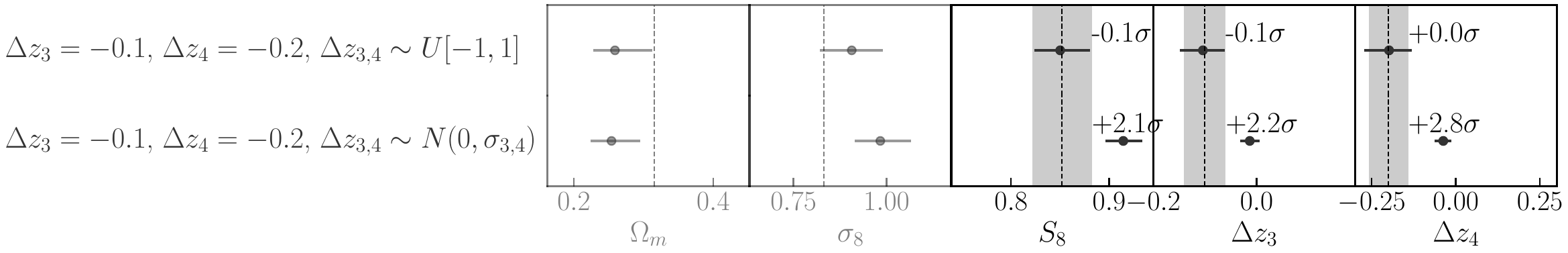}
\caption{\label{fig:validation_test} The validation results are summarized for five key parameters. The points indicate the mode of the marginalized one-dimensional posterior distributions, while the error bars represent the central $68\%$ credible intervals. The dashed lines and shaded gray regions denote the true parameter values and the $68\%$ confidence interval from the fiducial analysis, respectively, for comparison. Overall, we find that our pipeline is robust against a wide range of astrophysical systematics and successfully recovers the true values of $S_8$, $\Delta z_3$, and $\Delta z_4$. In particular, we demonstrate that the analysis framework can accurately recover redshift parameter biases when they are present in the data.
}
\end{figure*}

Fig.~\ref{fig:validation_test} displays the mode and 1$\sigma$ uncertainty of the one-dimensional posterior distributions for $\Omega_m$, $\sigma_8$, $S_8$, $\Delta z_3$, and $\Delta z_4$ across all validation tests. The true input values of the data vector are indicated by dashed lines. We emphasize that our validation focuses on $S_8$, $\Delta z_3$, and $\Delta z_4$, which are the primary parameters constrained in this work. The shaded regions represent the 1$\sigma$ uncertainty of the fiducial analysis, centered on the true parameter values. While deviations from the fiducial result quantify the relative impact of a specific systematic effect, the offset from the true values reflects a combination of model imperfections and projection effects \cite{Krause2010}.

The first three tests—labeled "fiducial," "2×2pt," and "cosmic shear"—analyze the 3×2pt, 2×2pt, and cosmic shear probes, respectively. We observe no significant biases in $S_8$, $\Delta z_3$, or $\Delta z_4$ for any of these configurations. However, the 3×2pt and cosmic shear analyses exhibit notable projection effects in the inferred values of $\Omega_m$ and $\sigma_8$, potentially driven by the imposed priors on the parameter space.

In the second set of validation tests, we assess the robustness of the analysis against various astrophysical effects. 
\texttt{nonfidNsat}, \texttt{sat-DM}, and \texttt{sat-subhalo} represent different methods for populating satellite galaxies. 
\texttt{offcenter} refers to simulations where central galaxies are displaced from the halo centers. 
\texttt{assembly\_bias} modifies the assembly history relative to the fiducial model. 
\texttt{baryon} includes the effects of baryonic feedback in the simulations, while \texttt{incompleteness} introduces incompleteness in the galaxy population to emulate observational limitations in SDSS. 
\texttt{fof} assigns galaxies using halos identified via the friends-of-friends algorithm. 
We find that none of these astrophysical effects introduces a significant bias in the key parameters.

We introduce a significant intrinsic alignment (IA) signal into the galaxy-galaxy lensing measurement based on the overlapping redshift distributions and prediction of  nonlinear alignment (NLA) model, and perform an analysis without modeling this effect (\texttt{GGL-IA}). We find no significant bias in the inferred parameters, as galaxy-galaxy lensing is not the dominant probe for the constraints on $S_8$, $\Delta z_3$, or $\Delta z_4$. Furthermore, we do not expect the real IA contamination to our GGL data vector to be severe, because our lens-source pair weights $w_{\rm ls}$ effectively eliminate lens-source pair that are physically associated.

The third set of validation tests focuses on the impact of baryonic effects on the cosmic shear data vector. We apply baryonic feedback consistent with the OWLS \cite{Schaye2010}, BAHAMAS \cite{mccarthy2017}, and COSMO-OWLS \cite{LeBrun2014} simulations by setting the baryonic parameters to $A_{\rm bary} = 1.6$, $\log(T_{\rm agn}) = 7.7$, and $\log(T_{\rm agn}) = 8.3$, respectively. We also test a case with $A_{\rm bary} = 2.2$, which approximately corresponds to the mode value inferred from our fiducial analysis.

We find no significant bias in $S_8$ for the OWLS-like, BAHAMAS-like, or $A_{\rm bary} = 2.2$ mock data vectors. However, the COSMO-OWLS-like data vector exhibits a $\sim1\sigma$ downward bias in $S_8$. While this bias is marginal, we note that $\log(T_{\rm agn}) = 8.3$ corresponds to a relatively extreme AGN feedback scenario and may not reflect realistic astrophysical conditions.

\begin{figure}
\includegraphics[width=1\columnwidth]{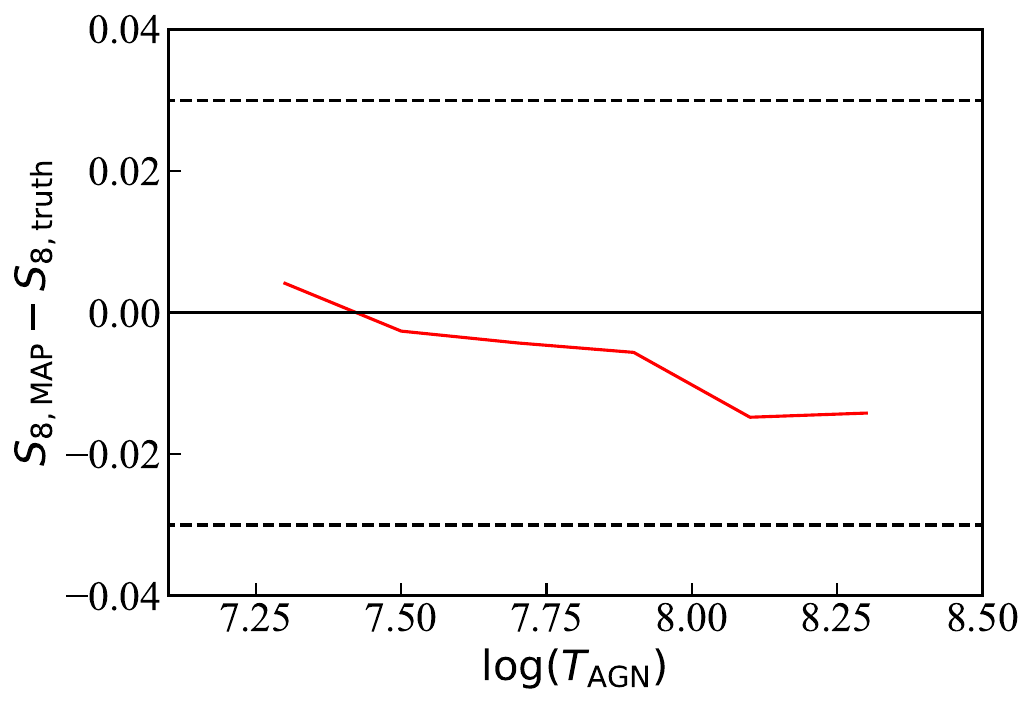}
\caption{\label{fig:map_s8} 
Bias in the $S_8$ parameter at the MAP estimate as a function of $\log(T_{\rm agn})$, used to model varying levels of baryonic feedback. The dashed horizontal line indicates the $1\sigma$ uncertainty from the fiducial analysis. Across the full range of $\log(T_{\rm agn}) = 7.3$–$8.3$, the $S_8$ bias remains well within $1\sigma$. This demonstrates that baryonic effects do not induce significant bias in cosmological parameter inference.}
\end{figure}

To further explore this trend, we perform MAP estimations on a suite of data vectors constructed with baryonic feedback models spanning $\log(T_{\rm agn}) = 7.3$ to $8.3$. Fig.~\ref{fig:map_s8} shows the $S_8$ bias at the MAP relative to the true input value, illustrating that the bias induced by baryonic feedback remains below $1\sigma$ across this range. The mild bias in the posterior mode observed in the full Bayesian inference is likely attributable to projection effects in the high-dimensional parameter space.

The final set of validation tests ensures that our pipeline is robust against potential systematics in the redshift distribution. In the test labeled ``$\Delta z_3 = -0.1$, $\Delta z_4 = -0.2$, $\Delta z_{3,4} \sim U[-1,1]$'', we use the Gaussian prior centered around zero for $\Delta z_{3,4}$. In the test labeled ``$\Delta z_3 = -0.1$, $\Delta z_4 = -0.2$, $\Delta z_{3,4} \sim U[-1,1]$'', we introduce redshift biases to the HSC-3 and HSC-4 bins and demonstrate that our inference correctly recovers these input shifts. In the test ``$\Delta z_3 = -0.1$, $\Delta z_4 = -0.2$, $\Delta z_{3,4} \sim \mathcal{N}(0, \sigma_{3,4})$'', we apply the same redshift bias but assume a Gaussian prior on $\Delta z_3$ and $\Delta z_4$ centered at zero during the analysis. In this case, we observe a significant bias in the inferred value of $S_8$, which highlights the importance of accurate prior modeling for redshift uncertainties.

\section{Bayesian inference parameters}
\label{ap:bayesian}

\begin{figure}
\includegraphics[width=1\columnwidth]{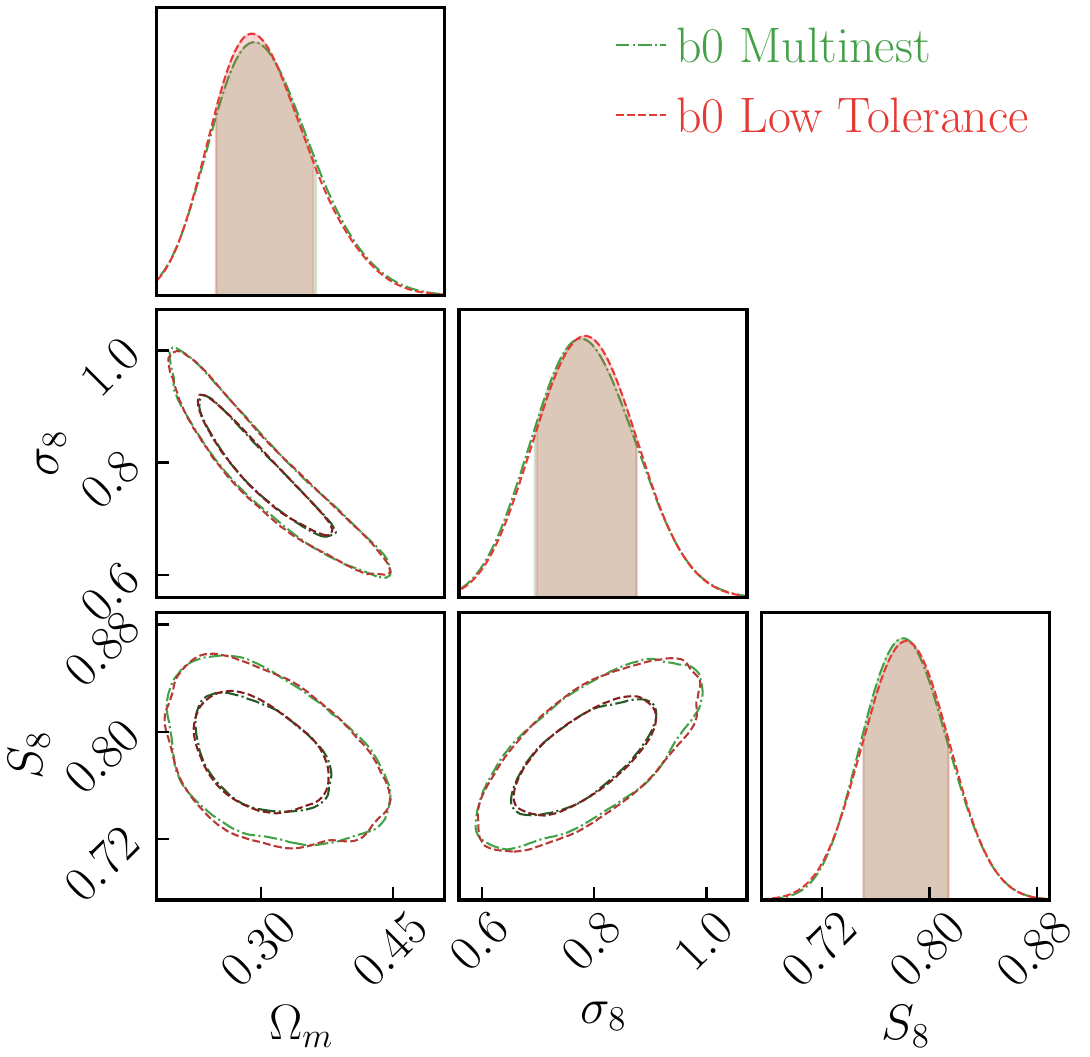}
\caption{\label{fig:low_tolorance_check} The 1 and $2-\sigma$ contour $\Omega_m, \sigma_8$, and $S_8$ constriants of the fiducial analysis (green contours), and of the analysis with low tolerance (red contours). There is no visible difference in the results. }
\end{figure}

The default tolerance parameter in \textit{MultiNest} is set to $1.0$ in this work. To validate this choice, we perform an additional \textit{MultiNest} run with the tolerance set to $0.05$, which corresponds to the default setting used in previous works such as \cite{Li2023,Dalal2023}. In Fig.~\ref{fig:low_tolorance_check}, we compare the posterior contours from the low-tolerance run to our fiducial results. We find no visible differences between the two, indicating that our parameter constraints are not sensitive to this setting. Given the significantly longer runtime required for convergence at lower tolerance, we adopt $1.0$ as the default tolerance value for our analysis.

\section{Full corner plot}
\label{ap:full_corner_plot}

\begin{figure*}
\includegraphics[width=2.0\columnwidth]{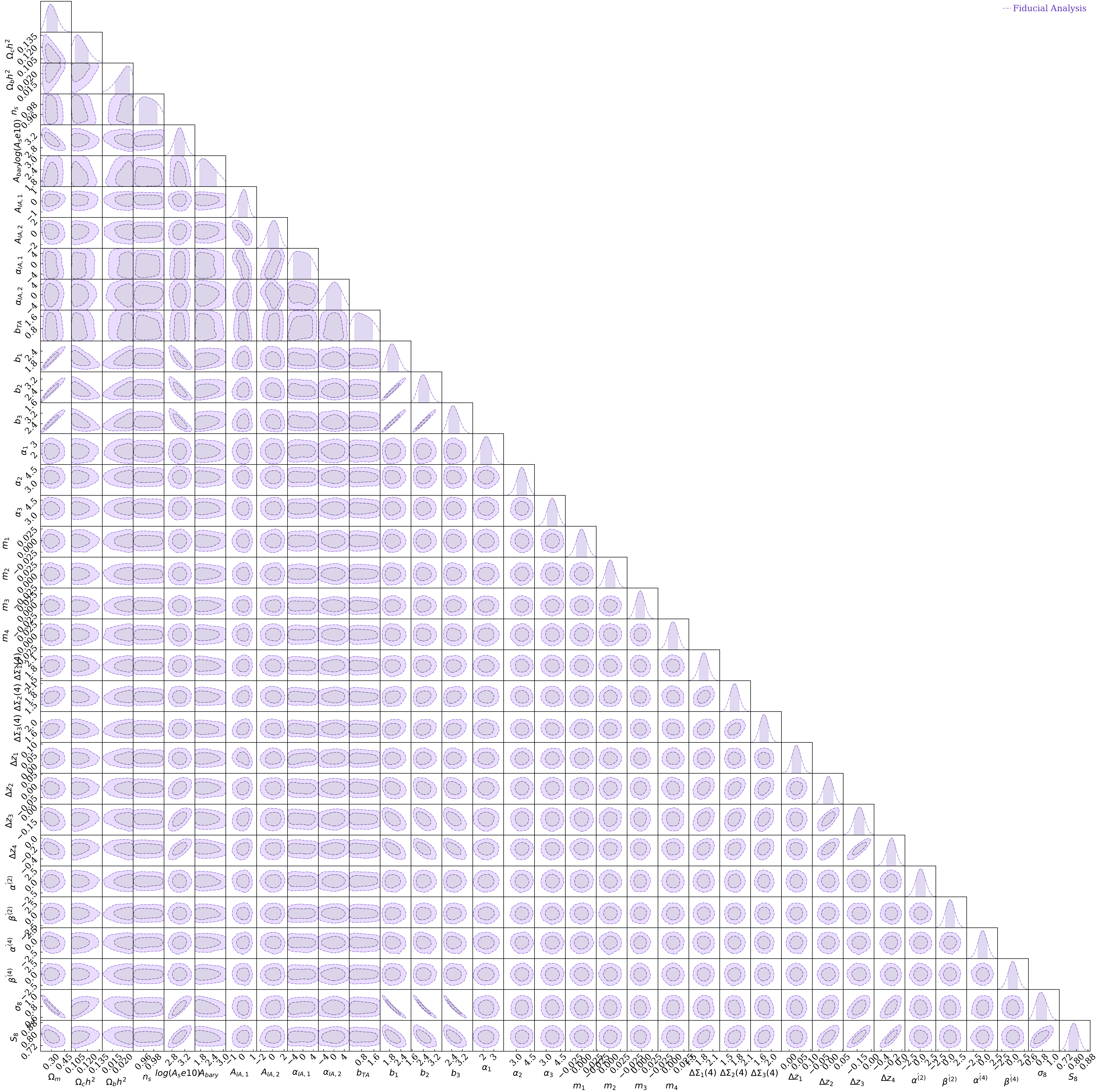}
\caption{\label{fig:full_corner_plot} The posterior distribution of the fiducial $3\times2$pt analysis on 32 free parameters and two derived parameters $\sigma_8$, and $S_8$.}
\end{figure*}

Fig~\ref{fig:full_corner_plot} shows the full corner plot of the 32 sampled parameters and two derived parameters $\sigma_8$, and $S_8$. The parameter constraints are obtained by the fiducial $3\times2$pt analysis.

\bibliography{apssamp}

\begin{thebibliography}{81}%
\makeatletter
\providecommand \@ifxundefined [1]{%
 \@ifx{#1\undefined}
}%
\providecommand \@ifnum [1]{%
 \ifnum #1\expandafter \@firstoftwo
 \else \expandafter \@secondoftwo
 \fi
}%
\providecommand \@ifx [1]{%
 \ifx #1\expandafter \@firstoftwo
 \else \expandafter \@secondoftwo
 \fi
}%
\providecommand \natexlab [1]{#1}%
\providecommand \enquote  [1]{``#1''}%
\providecommand \bibnamefont  [1]{#1}%
\providecommand \bibfnamefont [1]{#1}%
\providecommand \citenamefont [1]{#1}%
\providecommand \href@noop [0]{\@secondoftwo}%
\providecommand \href [0]{\begingroup \@sanitize@url \@href}%
\providecommand \@href[1]{\@@startlink{#1}\@@href}%
\providecommand \@@href[1]{\endgroup#1\@@endlink}%
\providecommand \@sanitize@url [0]{\catcode `\\12\catcode `\$12\catcode `\&12\catcode `\#12\catcode `\^12\catcode `\_12\catcode `\%12\relax}%
\providecommand \@@startlink[1]{}%
\providecommand \@@endlink[0]{}%
\providecommand \url  [0]{\begingroup\@sanitize@url \@url }%
\providecommand \@url [1]{\endgroup\@href {#1}{\urlprefix }}%
\providecommand \urlprefix  [0]{URL }%
\providecommand \Eprint [0]{\href }%
\providecommand \doibase [0]{https://doi.org/}%
\providecommand \selectlanguage [0]{\@gobble}%
\providecommand \bibinfo  [0]{\@secondoftwo}%
\providecommand \bibfield  [0]{\@secondoftwo}%
\providecommand \translation [1]{[#1]}%
\providecommand \BibitemOpen [0]{}%
\providecommand \bibitemStop [0]{}%
\providecommand \bibitemNoStop [0]{.\EOS\space}%
\providecommand \EOS [0]{\spacefactor3000\relax}%
\providecommand \BibitemShut  [1]{\csname bibitem#1\endcsname}%
\let\auto@bib@innerbib\@empty
\bibitem [{\citenamefont {{Planck Collaboration}}\ \emph {et~al.}(2020)\citenamefont {{Planck Collaboration}}, \citenamefont {{Aghanim}}, \citenamefont {{Akrami}}, \citenamefont {{Ashdown}}, \citenamefont {{Aumont}}, \citenamefont {{Baccigalupi}}, \citenamefont {{Ballardini}}, \citenamefont {{Banday}}, \citenamefont {{Barreiro}}, \citenamefont {{Bartolo}} \emph {et~al.}}]{Planck2018Cosmology}%
  \BibitemOpen
  \bibfield  {author} {\bibinfo {author} {\bibnamefont {{Planck Collaboration}}}, \bibinfo {author} {\bibfnamefont {N.}~\bibnamefont {{Aghanim}}}, \bibinfo {author} {\bibfnamefont {Y.}~\bibnamefont {{Akrami}}}, \bibinfo {author} {\bibfnamefont {M.}~\bibnamefont {{Ashdown}}}, \bibinfo {author} {\bibfnamefont {J.}~\bibnamefont {{Aumont}}}, \bibinfo {author} {\bibfnamefont {C.}~\bibnamefont {{Baccigalupi}}}, \bibinfo {author} {\bibfnamefont {M.}~\bibnamefont {{Ballardini}}}, \bibinfo {author} {\bibfnamefont {A.~J.}\ \bibnamefont {{Banday}}}, \bibinfo {author} {\bibfnamefont {R.~B.}\ \bibnamefont {{Barreiro}}}, \bibinfo {author} {\bibfnamefont {N.}~\bibnamefont {{Bartolo}}}, \emph {et~al.},\ }\href {https://doi.org/10.1051/0004-6361/201833910} {\bibfield  {journal} {\bibinfo  {journal} {\aap}\ }\textbf {\bibinfo {volume} {641}},\ \bibinfo {eid} {A6} (\bibinfo {year} {2020})},\ \Eprint {https://arxiv.org/abs/1807.06209} {arXiv:1807.06209 [astro-ph.CO]} \BibitemShut {NoStop}%
\bibitem [{\citenamefont {{Abbott}}\ \emph {et~al.}(2022)\citenamefont {{Abbott}}, \citenamefont {{Aguena}}, \citenamefont {{Alarcon}}, \citenamefont {{Allam}}, \citenamefont {{Alves}}, \citenamefont {{Amon}}, \citenamefont {{Andrade-Oliveira}}, \citenamefont {{Annis}}, \citenamefont {{Avila}}, \citenamefont {{Bacon}} \emph {et~al.}}]{DESY3_3x2_2022}%
  \BibitemOpen
  \bibfield  {author} {\bibinfo {author} {\bibfnamefont {T.~M.~C.}\ \bibnamefont {{Abbott}}}, \bibinfo {author} {\bibfnamefont {M.}~\bibnamefont {{Aguena}}}, \bibinfo {author} {\bibfnamefont {A.}~\bibnamefont {{Alarcon}}}, \bibinfo {author} {\bibfnamefont {S.}~\bibnamefont {{Allam}}}, \bibinfo {author} {\bibfnamefont {O.}~\bibnamefont {{Alves}}}, \bibinfo {author} {\bibfnamefont {A.}~\bibnamefont {{Amon}}}, \bibinfo {author} {\bibfnamefont {F.}~\bibnamefont {{Andrade-Oliveira}}}, \bibinfo {author} {\bibfnamefont {J.}~\bibnamefont {{Annis}}}, \bibinfo {author} {\bibfnamefont {S.}~\bibnamefont {{Avila}}}, \bibinfo {author} {\bibfnamefont {D.}~\bibnamefont {{Bacon}}}, \emph {et~al.},\ }\href {https://doi.org/10.1103/PhysRevD.105.023520} {\bibfield  {journal} {\bibinfo  {journal} {\prd}\ }\textbf {\bibinfo {volume} {105}},\ \bibinfo {eid} {023520} (\bibinfo {year} {2022})},\ \Eprint {https://arxiv.org/abs/2105.13549} {arXiv:2105.13549 [astro-ph.CO]} \BibitemShut {NoStop}%
\bibitem [{\citenamefont {{Adame}}\ \emph {et~al.}(2025)\citenamefont {{Adame}}, \citenamefont {{Aguilar}}, \citenamefont {{Ahlen}}, \citenamefont {{Alam}}, \citenamefont {{Alexander}}, \citenamefont {{Alvarez}}, \citenamefont {{Alves}}, \citenamefont {{Anand}}, \citenamefont {{Andrade}}, \citenamefont {{Armengaud}}, \citenamefont {{Avila}}, \citenamefont {{Aviles}}, \citenamefont {{Awan}}, \citenamefont {{Bahr-Kalus}}, \citenamefont {{Bailey}}, \citenamefont {{Baltay}}, \citenamefont {{Bault}}, \citenamefont {{Behera}}, \citenamefont {{BenZvi}}, \citenamefont {{Bera}}, \citenamefont {{Beutler}}, \citenamefont {{Bianchi}}, \citenamefont {{Blake}}, \citenamefont {{Blum}}, \citenamefont {{Brieden}}, \citenamefont {{Brodzeller}}, \citenamefont {{Brooks}}, \citenamefont {{Buckley-Geer}}, \citenamefont {{Burtin}}, \citenamefont {{Calderon}}, \citenamefont {{Canning}}, \citenamefont {{Carnero Rosell}}, \citenamefont {{Cereskaite}}, \citenamefont {{Cervantes-Cota}}, \citenamefont {{Chabanier}}, \citenamefont
  {{Chaussidon}}, \citenamefont {{Chaves-Montero}}, \citenamefont {{Chen}}, \citenamefont {{Chen}}, \citenamefont {{Claybaugh}}, \citenamefont {{Cole}}, \citenamefont {{Cuceu}}, \citenamefont {{Davis}}, \citenamefont {{Dawson}}, \citenamefont {{de la Macorra}}, \citenamefont {{de Mattia}}, \citenamefont {{Deiosso}}, \citenamefont {{Dey}}, \citenamefont {{Dey}}, \citenamefont {{Ding}}, \citenamefont {{Doel}}, \citenamefont {{Edelstein}}, \citenamefont {{Eftekharzadeh}}, \citenamefont {{Eisenstein}}, \citenamefont {{Elliott}}, \citenamefont {{Fagrelius}}, \citenamefont {{Fanning}}, \citenamefont {{Ferraro}}, \citenamefont {{Ereza}}, \citenamefont {{Findlay}}, \citenamefont {{Flaugher}}, \citenamefont {{Font-Ribera}}, \citenamefont {{Forero-S{\'a}nchez}}, \citenamefont {{Forero-Romero}}, \citenamefont {{Frenk}}, \citenamefont {{Garcia-Quintero}}, \citenamefont {{Gazta{\~n}aga}}, \citenamefont {{Gil-Mar{\'\i}n}}, \citenamefont {{Gontcho a Gontcho}}, \citenamefont {{Gonzalez-Morales}}, \citenamefont
  {{Gonzalez-Perez}}, \citenamefont {{Gordon}}, \citenamefont {{Green}}, \citenamefont {{Gruen}}, \citenamefont {{Gsponer}}, \citenamefont {{Gutierrez}}, \citenamefont {{Guy}}, \citenamefont {{Hadzhiyska}}, \citenamefont {{Hahn}}, \citenamefont {{Hanif}}, \citenamefont {{Herrera-Alcantar}}, \citenamefont {{Honscheid}}, \citenamefont {{Howlett}}, \citenamefont {{Huterer}}, \citenamefont {{Ir{\v{s}}i{\v{c}}}}, \citenamefont {{Ishak}}, \citenamefont {{Juneau}}, \citenamefont {{Kara{\c{c}}ayl{\i}}}, \citenamefont {{Kehoe}}, \citenamefont {{Kent}}, \citenamefont {{Kirkby}}, \citenamefont {{Kremin}}, \citenamefont {{Krolewski}}, \citenamefont {{Lai}}, \citenamefont {{Lan}}, \citenamefont {{Landriau}}, \citenamefont {{Lang}}, \citenamefont {{Lasker}}, \citenamefont {{Le Goff}}, \citenamefont {{Le Guillou}}, \citenamefont {{Leauthaud}}, \citenamefont {{Levi}}, \citenamefont {{Li}}, \citenamefont {{Linder}}, \citenamefont {{Lodha}}, \citenamefont {{Magneville}}, \citenamefont {{Manera}}, \citenamefont {{Margala}},
  \citenamefont {{Martini}}, \citenamefont {{Maus}}, \citenamefont {{McDonald}}, \citenamefont {{Medina-Varela}}, \citenamefont {{Meisner}}, \citenamefont {{Mena-Fern{\'a}ndez}}, \citenamefont {{Miquel}}, \citenamefont {{Moon}}, \citenamefont {{Moore}}, \citenamefont {{Moustakas}}, \citenamefont {{Mueller}}, \citenamefont {{Mu{\~n}oz-Guti{\'e}rrez}}, \citenamefont {{Myers}}, \citenamefont {{Nadathur}}, \citenamefont {{Napolitano}}, \citenamefont {{Neveux}}, \citenamefont {{Newman}}, \citenamefont {{Nguyen}}, \citenamefont {{Nie}}, \citenamefont {{Niz}}, \citenamefont {{Noriega}}, \citenamefont {{Padmanabhan}}, \citenamefont {{Paillas}}, \citenamefont {{Palanque-Delabrouille}}, \citenamefont {{Pan}}, \citenamefont {{Penmetsa}}, \citenamefont {{Percival}}, \citenamefont {{Pieri}}, \citenamefont {{Pinon}}, \citenamefont {{Poppett}}, \citenamefont {{Porredon}}, \citenamefont {{Prada}}, \citenamefont {{P{\'e}rez-Fern{\'a}ndez}}, \citenamefont {{P{\'e}rez-R{\`a}fols}}, \citenamefont {{Rabinowitz}}, \citenamefont
  {{Raichoor}}, \citenamefont {{Ram{\'\i}rez-P{\'e}rez}}, \citenamefont {{Ramirez-Solano}}, \citenamefont {{Rashkovetskyi}}, \citenamefont {{Ravoux}}, \citenamefont {{Rezaie}}, \citenamefont {{Rich}}, \citenamefont {{Rocher}}, \citenamefont {{Rockosi}}, \citenamefont {{Roe}}, \citenamefont {{Rosado-Marin}}, \citenamefont {{Ross}}, \citenamefont {{Rossi}}, \citenamefont {{Ruggeri}}, \citenamefont {{Ruhlmann-Kleider}}, \citenamefont {{Samushia}}, \citenamefont {{Sanchez}}, \citenamefont {{Saulder}}, \citenamefont {{Schlafly}}, \citenamefont {{Schlegel}}, \citenamefont {{Schubnell}}, \citenamefont {{Seo}}, \citenamefont {{Shafieloo}}, \citenamefont {{Sharples}}, \citenamefont {{Silber}}, \citenamefont {{Slosar}}, \citenamefont {{Smith}}, \citenamefont {{Sprayberry}}, \citenamefont {{Tan}}, \citenamefont {{Tarl{\'e}}}, \citenamefont {{Taylor}}, \citenamefont {{Trusov}}, \citenamefont {{Ure{\~n}a-L{\'o}pez}}, \citenamefont {{Vaisakh}}, \citenamefont {{Valcin}}, \citenamefont {{Valdes}}, \citenamefont
  {{Vargas-Maga{\~n}a}}, \citenamefont {{Verde}}, \citenamefont {{Walther}}, \citenamefont {{Wang}}, \citenamefont {{Wang}}, \citenamefont {{Weaver}}, \citenamefont {{Weaverdyck}}, \citenamefont {{Wechsler}}, \citenamefont {{Weinberg}}, \citenamefont {{White}}, \citenamefont {{Yu}}, \citenamefont {{Yu}}, \citenamefont {{Yuan}}, \citenamefont {{Y{\`e}che}}, \citenamefont {{Zaborowski}}, \citenamefont {{Zarrouk}}, \citenamefont {{Zhang}}, \citenamefont {{Zhao}}, \citenamefont {{Zhao}}, \citenamefont {{Zhou}},\ and\ \citenamefont {{Zhuang}}}]{DESI_BAO}%
  \BibitemOpen
  \bibfield  {author} {\bibinfo {author} {\bibfnamefont {A.~G.}\ \bibnamefont {{Adame}}}, \bibinfo {author} {\bibfnamefont {J.}~\bibnamefont {{Aguilar}}}, \bibinfo {author} {\bibfnamefont {S.}~\bibnamefont {{Ahlen}}}, \bibinfo {author} {\bibfnamefont {S.}~\bibnamefont {{Alam}}}, \bibinfo {author} {\bibfnamefont {D.~M.}\ \bibnamefont {{Alexander}}}, \bibinfo {author} {\bibfnamefont {M.}~\bibnamefont {{Alvarez}}}, \bibinfo {author} {\bibfnamefont {O.}~\bibnamefont {{Alves}}}, \bibinfo {author} {\bibfnamefont {A.}~\bibnamefont {{Anand}}}, \bibinfo {author} {\bibfnamefont {U.}~\bibnamefont {{Andrade}}}, \bibinfo {author} {\bibfnamefont {E.}~\bibnamefont {{Armengaud}}}, \bibinfo {author} {\bibfnamefont {S.}~\bibnamefont {{Avila}}}, \bibinfo {author} {\bibfnamefont {A.}~\bibnamefont {{Aviles}}}, \bibinfo {author} {\bibfnamefont {H.}~\bibnamefont {{Awan}}}, \bibinfo {author} {\bibfnamefont {B.}~\bibnamefont {{Bahr-Kalus}}}, \bibinfo {author} {\bibfnamefont {S.}~\bibnamefont {{Bailey}}}, \bibinfo {author}
  {\bibfnamefont {C.}~\bibnamefont {{Baltay}}}, \bibinfo {author} {\bibfnamefont {A.}~\bibnamefont {{Bault}}}, \bibinfo {author} {\bibfnamefont {J.}~\bibnamefont {{Behera}}}, \bibinfo {author} {\bibfnamefont {S.}~\bibnamefont {{BenZvi}}}, \bibinfo {author} {\bibfnamefont {A.}~\bibnamefont {{Bera}}}, \bibinfo {author} {\bibfnamefont {F.}~\bibnamefont {{Beutler}}}, \bibinfo {author} {\bibfnamefont {D.}~\bibnamefont {{Bianchi}}}, \bibinfo {author} {\bibfnamefont {C.}~\bibnamefont {{Blake}}}, \bibinfo {author} {\bibfnamefont {R.}~\bibnamefont {{Blum}}}, \bibinfo {author} {\bibfnamefont {S.}~\bibnamefont {{Brieden}}}, \bibinfo {author} {\bibfnamefont {A.}~\bibnamefont {{Brodzeller}}}, \bibinfo {author} {\bibfnamefont {D.}~\bibnamefont {{Brooks}}}, \bibinfo {author} {\bibfnamefont {E.}~\bibnamefont {{Buckley-Geer}}}, \bibinfo {author} {\bibfnamefont {E.}~\bibnamefont {{Burtin}}}, \bibinfo {author} {\bibfnamefont {R.}~\bibnamefont {{Calderon}}}, \bibinfo {author} {\bibfnamefont {R.}~\bibnamefont {{Canning}}},
  \bibinfo {author} {\bibfnamefont {A.}~\bibnamefont {{Carnero Rosell}}}, \bibinfo {author} {\bibfnamefont {R.}~\bibnamefont {{Cereskaite}}}, \bibinfo {author} {\bibfnamefont {J.~L.}\ \bibnamefont {{Cervantes-Cota}}}, \bibinfo {author} {\bibfnamefont {S.}~\bibnamefont {{Chabanier}}}, \bibinfo {author} {\bibfnamefont {E.}~\bibnamefont {{Chaussidon}}}, \bibinfo {author} {\bibfnamefont {J.}~\bibnamefont {{Chaves-Montero}}}, \bibinfo {author} {\bibfnamefont {S.}~\bibnamefont {{Chen}}}, \bibinfo {author} {\bibfnamefont {X.}~\bibnamefont {{Chen}}}, \bibinfo {author} {\bibfnamefont {T.}~\bibnamefont {{Claybaugh}}}, \bibinfo {author} {\bibfnamefont {S.}~\bibnamefont {{Cole}}}, \bibinfo {author} {\bibfnamefont {A.}~\bibnamefont {{Cuceu}}}, \bibinfo {author} {\bibfnamefont {T.~M.}\ \bibnamefont {{Davis}}}, \bibinfo {author} {\bibfnamefont {K.}~\bibnamefont {{Dawson}}}, \bibinfo {author} {\bibfnamefont {A.}~\bibnamefont {{de la Macorra}}}, \bibinfo {author} {\bibfnamefont {A.}~\bibnamefont {{de Mattia}}}, \bibinfo
  {author} {\bibfnamefont {N.}~\bibnamefont {{Deiosso}}}, \bibinfo {author} {\bibfnamefont {A.}~\bibnamefont {{Dey}}}, \bibinfo {author} {\bibfnamefont {B.}~\bibnamefont {{Dey}}}, \bibinfo {author} {\bibfnamefont {Z.}~\bibnamefont {{Ding}}}, \bibinfo {author} {\bibfnamefont {P.}~\bibnamefont {{Doel}}}, \bibinfo {author} {\bibfnamefont {J.}~\bibnamefont {{Edelstein}}}, \bibinfo {author} {\bibfnamefont {S.}~\bibnamefont {{Eftekharzadeh}}}, \bibinfo {author} {\bibfnamefont {D.~J.}\ \bibnamefont {{Eisenstein}}}, \bibinfo {author} {\bibfnamefont {A.}~\bibnamefont {{Elliott}}}, \bibinfo {author} {\bibfnamefont {P.}~\bibnamefont {{Fagrelius}}}, \bibinfo {author} {\bibfnamefont {K.}~\bibnamefont {{Fanning}}}, \bibinfo {author} {\bibfnamefont {S.}~\bibnamefont {{Ferraro}}}, \bibinfo {author} {\bibfnamefont {J.}~\bibnamefont {{Ereza}}}, \bibinfo {author} {\bibfnamefont {N.}~\bibnamefont {{Findlay}}}, \bibinfo {author} {\bibfnamefont {B.}~\bibnamefont {{Flaugher}}}, \bibinfo {author} {\bibfnamefont {A.}~\bibnamefont
  {{Font-Ribera}}}, \bibinfo {author} {\bibfnamefont {D.}~\bibnamefont {{Forero-S{\'a}nchez}}}, \bibinfo {author} {\bibfnamefont {J.~E.}\ \bibnamefont {{Forero-Romero}}}, \bibinfo {author} {\bibfnamefont {C.~S.}\ \bibnamefont {{Frenk}}}, \bibinfo {author} {\bibfnamefont {C.}~\bibnamefont {{Garcia-Quintero}}}, \bibinfo {author} {\bibfnamefont {E.}~\bibnamefont {{Gazta{\~n}aga}}}, \bibinfo {author} {\bibfnamefont {H.}~\bibnamefont {{Gil-Mar{\'\i}n}}}, \bibinfo {author} {\bibfnamefont {S.}~\bibnamefont {{Gontcho a Gontcho}}}, \bibinfo {author} {\bibfnamefont {A.~X.}\ \bibnamefont {{Gonzalez-Morales}}}, \bibinfo {author} {\bibfnamefont {V.}~\bibnamefont {{Gonzalez-Perez}}}, \bibinfo {author} {\bibfnamefont {C.}~\bibnamefont {{Gordon}}}, \bibinfo {author} {\bibfnamefont {D.}~\bibnamefont {{Green}}}, \bibinfo {author} {\bibfnamefont {D.}~\bibnamefont {{Gruen}}}, \bibinfo {author} {\bibfnamefont {R.}~\bibnamefont {{Gsponer}}}, \bibinfo {author} {\bibfnamefont {G.}~\bibnamefont {{Gutierrez}}}, \bibinfo {author}
  {\bibfnamefont {J.}~\bibnamefont {{Guy}}}, \bibinfo {author} {\bibfnamefont {B.}~\bibnamefont {{Hadzhiyska}}}, \bibinfo {author} {\bibfnamefont {C.}~\bibnamefont {{Hahn}}}, \bibinfo {author} {\bibfnamefont {M.~M.~S.}\ \bibnamefont {{Hanif}}}, \bibinfo {author} {\bibfnamefont {H.~K.}\ \bibnamefont {{Herrera-Alcantar}}}, \bibinfo {author} {\bibfnamefont {K.}~\bibnamefont {{Honscheid}}}, \bibinfo {author} {\bibfnamefont {C.}~\bibnamefont {{Howlett}}}, \bibinfo {author} {\bibfnamefont {D.}~\bibnamefont {{Huterer}}}, \bibinfo {author} {\bibfnamefont {V.}~\bibnamefont {{Ir{\v{s}}i{\v{c}}}}}, \bibinfo {author} {\bibfnamefont {M.}~\bibnamefont {{Ishak}}}, \bibinfo {author} {\bibfnamefont {S.}~\bibnamefont {{Juneau}}}, \bibinfo {author} {\bibfnamefont {N.~G.}\ \bibnamefont {{Kara{\c{c}}ayl{\i}}}}, \bibinfo {author} {\bibfnamefont {R.}~\bibnamefont {{Kehoe}}}, \bibinfo {author} {\bibfnamefont {S.}~\bibnamefont {{Kent}}}, \bibinfo {author} {\bibfnamefont {D.}~\bibnamefont {{Kirkby}}}, \bibinfo {author} {\bibfnamefont
  {A.}~\bibnamefont {{Kremin}}}, \bibinfo {author} {\bibfnamefont {A.}~\bibnamefont {{Krolewski}}}, \bibinfo {author} {\bibfnamefont {Y.}~\bibnamefont {{Lai}}}, \bibinfo {author} {\bibfnamefont {T.~W.}\ \bibnamefont {{Lan}}}, \bibinfo {author} {\bibfnamefont {M.}~\bibnamefont {{Landriau}}}, \bibinfo {author} {\bibfnamefont {D.}~\bibnamefont {{Lang}}}, \bibinfo {author} {\bibfnamefont {J.}~\bibnamefont {{Lasker}}}, \bibinfo {author} {\bibfnamefont {J.~M.}\ \bibnamefont {{Le Goff}}}, \bibinfo {author} {\bibfnamefont {L.}~\bibnamefont {{Le Guillou}}}, \bibinfo {author} {\bibfnamefont {A.}~\bibnamefont {{Leauthaud}}}, \bibinfo {author} {\bibfnamefont {M.~E.}\ \bibnamefont {{Levi}}}, \bibinfo {author} {\bibfnamefont {T.~S.}\ \bibnamefont {{Li}}}, \bibinfo {author} {\bibfnamefont {E.}~\bibnamefont {{Linder}}}, \bibinfo {author} {\bibfnamefont {K.}~\bibnamefont {{Lodha}}}, \bibinfo {author} {\bibfnamefont {C.}~\bibnamefont {{Magneville}}}, \bibinfo {author} {\bibfnamefont {M.}~\bibnamefont {{Manera}}}, \bibinfo
  {author} {\bibfnamefont {D.}~\bibnamefont {{Margala}}}, \bibinfo {author} {\bibfnamefont {P.}~\bibnamefont {{Martini}}}, \bibinfo {author} {\bibfnamefont {M.}~\bibnamefont {{Maus}}}, \bibinfo {author} {\bibfnamefont {P.}~\bibnamefont {{McDonald}}}, \bibinfo {author} {\bibfnamefont {L.}~\bibnamefont {{Medina-Varela}}}, \bibinfo {author} {\bibfnamefont {A.}~\bibnamefont {{Meisner}}}, \bibinfo {author} {\bibfnamefont {J.}~\bibnamefont {{Mena-Fern{\'a}ndez}}}, \bibinfo {author} {\bibfnamefont {R.}~\bibnamefont {{Miquel}}}, \bibinfo {author} {\bibfnamefont {J.}~\bibnamefont {{Moon}}}, \bibinfo {author} {\bibfnamefont {S.}~\bibnamefont {{Moore}}}, \bibinfo {author} {\bibfnamefont {J.}~\bibnamefont {{Moustakas}}}, \bibinfo {author} {\bibfnamefont {E.}~\bibnamefont {{Mueller}}}, \bibinfo {author} {\bibfnamefont {A.}~\bibnamefont {{Mu{\~n}oz-Guti{\'e}rrez}}}, \bibinfo {author} {\bibfnamefont {A.~D.}\ \bibnamefont {{Myers}}}, \bibinfo {author} {\bibfnamefont {S.}~\bibnamefont {{Nadathur}}}, \bibinfo {author}
  {\bibfnamefont {L.}~\bibnamefont {{Napolitano}}}, \bibinfo {author} {\bibfnamefont {R.}~\bibnamefont {{Neveux}}}, \bibinfo {author} {\bibfnamefont {J.~A.}\ \bibnamefont {{Newman}}}, \bibinfo {author} {\bibfnamefont {N.~M.}\ \bibnamefont {{Nguyen}}}, \bibinfo {author} {\bibfnamefont {J.}~\bibnamefont {{Nie}}}, \bibinfo {author} {\bibfnamefont {G.}~\bibnamefont {{Niz}}}, \bibinfo {author} {\bibfnamefont {H.~E.}\ \bibnamefont {{Noriega}}}, \bibinfo {author} {\bibfnamefont {N.}~\bibnamefont {{Padmanabhan}}}, \bibinfo {author} {\bibfnamefont {E.}~\bibnamefont {{Paillas}}}, \bibinfo {author} {\bibfnamefont {N.}~\bibnamefont {{Palanque-Delabrouille}}}, \bibinfo {author} {\bibfnamefont {J.}~\bibnamefont {{Pan}}}, \bibinfo {author} {\bibfnamefont {S.}~\bibnamefont {{Penmetsa}}}, \bibinfo {author} {\bibfnamefont {W.~J.}\ \bibnamefont {{Percival}}}, \bibinfo {author} {\bibfnamefont {M.~M.}\ \bibnamefont {{Pieri}}}, \bibinfo {author} {\bibfnamefont {M.}~\bibnamefont {{Pinon}}}, \bibinfo {author} {\bibfnamefont
  {C.}~\bibnamefont {{Poppett}}}, \bibinfo {author} {\bibfnamefont {A.}~\bibnamefont {{Porredon}}}, \bibinfo {author} {\bibfnamefont {F.}~\bibnamefont {{Prada}}}, \bibinfo {author} {\bibfnamefont {A.}~\bibnamefont {{P{\'e}rez-Fern{\'a}ndez}}}, \bibinfo {author} {\bibfnamefont {I.}~\bibnamefont {{P{\'e}rez-R{\`a}fols}}}, \bibinfo {author} {\bibfnamefont {D.}~\bibnamefont {{Rabinowitz}}}, \bibinfo {author} {\bibfnamefont {A.}~\bibnamefont {{Raichoor}}}, \bibinfo {author} {\bibfnamefont {C.}~\bibnamefont {{Ram{\'\i}rez-P{\'e}rez}}}, \bibinfo {author} {\bibfnamefont {S.}~\bibnamefont {{Ramirez-Solano}}}, \bibinfo {author} {\bibfnamefont {M.}~\bibnamefont {{Rashkovetskyi}}}, \bibinfo {author} {\bibfnamefont {C.}~\bibnamefont {{Ravoux}}}, \bibinfo {author} {\bibfnamefont {M.}~\bibnamefont {{Rezaie}}}, \bibinfo {author} {\bibfnamefont {J.}~\bibnamefont {{Rich}}}, \bibinfo {author} {\bibfnamefont {A.}~\bibnamefont {{Rocher}}}, \bibinfo {author} {\bibfnamefont {C.}~\bibnamefont {{Rockosi}}}, \bibinfo {author}
  {\bibfnamefont {N.~A.}\ \bibnamefont {{Roe}}}, \bibinfo {author} {\bibfnamefont {A.}~\bibnamefont {{Rosado-Marin}}}, \bibinfo {author} {\bibfnamefont {A.~J.}\ \bibnamefont {{Ross}}}, \bibinfo {author} {\bibfnamefont {G.}~\bibnamefont {{Rossi}}}, \bibinfo {author} {\bibfnamefont {R.}~\bibnamefont {{Ruggeri}}}, \bibinfo {author} {\bibfnamefont {V.}~\bibnamefont {{Ruhlmann-Kleider}}}, \bibinfo {author} {\bibfnamefont {L.}~\bibnamefont {{Samushia}}}, \bibinfo {author} {\bibfnamefont {E.}~\bibnamefont {{Sanchez}}}, \bibinfo {author} {\bibfnamefont {C.}~\bibnamefont {{Saulder}}}, \bibinfo {author} {\bibfnamefont {E.~F.}\ \bibnamefont {{Schlafly}}}, \bibinfo {author} {\bibfnamefont {D.}~\bibnamefont {{Schlegel}}}, \bibinfo {author} {\bibfnamefont {M.}~\bibnamefont {{Schubnell}}}, \bibinfo {author} {\bibfnamefont {H.}~\bibnamefont {{Seo}}}, \bibinfo {author} {\bibfnamefont {A.}~\bibnamefont {{Shafieloo}}}, \bibinfo {author} {\bibfnamefont {R.}~\bibnamefont {{Sharples}}}, \bibinfo {author} {\bibfnamefont
  {J.}~\bibnamefont {{Silber}}}, \bibinfo {author} {\bibfnamefont {A.}~\bibnamefont {{Slosar}}}, \bibinfo {author} {\bibfnamefont {A.}~\bibnamefont {{Smith}}}, \bibinfo {author} {\bibfnamefont {D.}~\bibnamefont {{Sprayberry}}}, \bibinfo {author} {\bibfnamefont {T.}~\bibnamefont {{Tan}}}, \bibinfo {author} {\bibfnamefont {G.}~\bibnamefont {{Tarl{\'e}}}}, \bibinfo {author} {\bibfnamefont {P.}~\bibnamefont {{Taylor}}}, \bibinfo {author} {\bibfnamefont {S.}~\bibnamefont {{Trusov}}}, \bibinfo {author} {\bibfnamefont {L.~A.}\ \bibnamefont {{Ure{\~n}a-L{\'o}pez}}}, \bibinfo {author} {\bibfnamefont {R.}~\bibnamefont {{Vaisakh}}}, \bibinfo {author} {\bibfnamefont {D.}~\bibnamefont {{Valcin}}}, \bibinfo {author} {\bibfnamefont {F.}~\bibnamefont {{Valdes}}}, \bibinfo {author} {\bibfnamefont {M.}~\bibnamefont {{Vargas-Maga{\~n}a}}}, \bibinfo {author} {\bibfnamefont {L.}~\bibnamefont {{Verde}}}, \bibinfo {author} {\bibfnamefont {M.}~\bibnamefont {{Walther}}}, \bibinfo {author} {\bibfnamefont {B.}~\bibnamefont {{Wang}}},
  \bibinfo {author} {\bibfnamefont {M.~S.}\ \bibnamefont {{Wang}}}, \bibinfo {author} {\bibfnamefont {B.~A.}\ \bibnamefont {{Weaver}}}, \bibinfo {author} {\bibfnamefont {N.}~\bibnamefont {{Weaverdyck}}}, \bibinfo {author} {\bibfnamefont {R.~H.}\ \bibnamefont {{Wechsler}}}, \bibinfo {author} {\bibfnamefont {D.~H.}\ \bibnamefont {{Weinberg}}}, \bibinfo {author} {\bibfnamefont {M.}~\bibnamefont {{White}}}, \bibinfo {author} {\bibfnamefont {J.}~\bibnamefont {{Yu}}}, \bibinfo {author} {\bibfnamefont {Y.}~\bibnamefont {{Yu}}}, \bibinfo {author} {\bibfnamefont {S.}~\bibnamefont {{Yuan}}}, \bibinfo {author} {\bibfnamefont {C.}~\bibnamefont {{Y{\`e}che}}}, \bibinfo {author} {\bibfnamefont {E.~A.}\ \bibnamefont {{Zaborowski}}}, \bibinfo {author} {\bibfnamefont {P.}~\bibnamefont {{Zarrouk}}}, \bibinfo {author} {\bibfnamefont {H.}~\bibnamefont {{Zhang}}}, \bibinfo {author} {\bibfnamefont {C.}~\bibnamefont {{Zhao}}}, \bibinfo {author} {\bibfnamefont {R.}~\bibnamefont {{Zhao}}}, \bibinfo {author} {\bibfnamefont
  {R.}~\bibnamefont {{Zhou}}},\ and\ \bibinfo {author} {\bibfnamefont {T.}~\bibnamefont {{Zhuang}}},\ }\href {https://doi.org/10.1088/1475-7516/2025/02/021} {\bibfield  {journal} {\bibinfo  {journal} {\jcap}\ }\textbf {\bibinfo {volume} {2025}},\ \bibinfo {eid} {021} (\bibinfo {year} {2025})},\ \Eprint {https://arxiv.org/abs/2404.03002} {arXiv:2404.03002 [astro-ph.CO]} \BibitemShut {NoStop}%
\bibitem [{\citenamefont {{Riess}}\ \emph {et~al.}(2019)\citenamefont {{Riess}}, \citenamefont {{Casertano}}, \citenamefont {{Yuan}}, \citenamefont {{Macri}},\ and\ \citenamefont {{Scolnic}}}]{Riess2019}%
  \BibitemOpen
  \bibfield  {author} {\bibinfo {author} {\bibfnamefont {A.~G.}\ \bibnamefont {{Riess}}}, \bibinfo {author} {\bibfnamefont {S.}~\bibnamefont {{Casertano}}}, \bibinfo {author} {\bibfnamefont {W.}~\bibnamefont {{Yuan}}}, \bibinfo {author} {\bibfnamefont {L.~M.}\ \bibnamefont {{Macri}}},\ and\ \bibinfo {author} {\bibfnamefont {D.}~\bibnamefont {{Scolnic}}},\ }\href {https://doi.org/10.3847/1538-4357/ab1422} {\bibfield  {journal} {\bibinfo  {journal} {\apj}\ }\textbf {\bibinfo {volume} {876}},\ \bibinfo {eid} {85} (\bibinfo {year} {2019})},\ \Eprint {https://arxiv.org/abs/1903.07603} {arXiv:1903.07603 [astro-ph.CO]} \BibitemShut {NoStop}%
\bibitem [{\citenamefont {{Scolnic}}\ \emph {et~al.}(2018)\citenamefont {{Scolnic}}, \citenamefont {{Jones}}, \citenamefont {{Rest}}, \citenamefont {{Pan}}, \citenamefont {{Chornock}}, \citenamefont {{Foley}}, \citenamefont {{Huber}}, \citenamefont {{Kessler}}, \citenamefont {{Narayan}}, \citenamefont {{Riess}} \emph {et~al.}}]{Scolnic2018}%
  \BibitemOpen
  \bibfield  {author} {\bibinfo {author} {\bibfnamefont {D.~M.}\ \bibnamefont {{Scolnic}}}, \bibinfo {author} {\bibfnamefont {D.~O.}\ \bibnamefont {{Jones}}}, \bibinfo {author} {\bibfnamefont {A.}~\bibnamefont {{Rest}}}, \bibinfo {author} {\bibfnamefont {Y.~C.}\ \bibnamefont {{Pan}}}, \bibinfo {author} {\bibfnamefont {R.}~\bibnamefont {{Chornock}}}, \bibinfo {author} {\bibfnamefont {R.~J.}\ \bibnamefont {{Foley}}}, \bibinfo {author} {\bibfnamefont {M.~E.}\ \bibnamefont {{Huber}}}, \bibinfo {author} {\bibfnamefont {R.}~\bibnamefont {{Kessler}}}, \bibinfo {author} {\bibfnamefont {G.}~\bibnamefont {{Narayan}}}, \bibinfo {author} {\bibfnamefont {A.~G.}\ \bibnamefont {{Riess}}}, \emph {et~al.},\ }\href {https://doi.org/10.3847/1538-4357/aab9bb} {\bibfield  {journal} {\bibinfo  {journal} {\apj}\ }\textbf {\bibinfo {volume} {859}},\ \bibinfo {eid} {101} (\bibinfo {year} {2018})},\ \Eprint {https://arxiv.org/abs/1710.00845} {arXiv:1710.00845 [astro-ph.CO]} \BibitemShut {NoStop}%
\bibitem [{\citenamefont {{Hu}}(1999)}]{Hu1999}%
  \BibitemOpen
  \bibfield  {author} {\bibinfo {author} {\bibfnamefont {W.}~\bibnamefont {{Hu}}},\ }\href {https://doi.org/10.1086/312210} {\bibfield  {journal} {\bibinfo  {journal} {\apjl}\ }\textbf {\bibinfo {volume} {522}},\ \bibinfo {pages} {L21} (\bibinfo {year} {1999})},\ \Eprint {https://arxiv.org/abs/astro-ph/9904153} {arXiv:astro-ph/9904153 [astro-ph]} \BibitemShut {NoStop}%
\bibitem [{\citenamefont {{Dodelson}}\ and\ \citenamefont {{Schmidt}}(2020)}]{Dodelson2020}%
  \BibitemOpen
  \bibfield  {author} {\bibinfo {author} {\bibfnamefont {S.}~\bibnamefont {{Dodelson}}}\ and\ \bibinfo {author} {\bibfnamefont {F.}~\bibnamefont {{Schmidt}}},\ }\href {https://doi.org/10.1016/C2017-0-01943-2} {\emph {\bibinfo {title} {{Modern Cosmology}}}}\ (\bibinfo {year} {2020})\BibitemShut {NoStop}%
\bibitem [{\citenamefont {{Kilbinger}}(2015)}]{Kilbinger2015}%
  \BibitemOpen
  \bibfield  {author} {\bibinfo {author} {\bibfnamefont {M.}~\bibnamefont {{Kilbinger}}},\ }\href {https://doi.org/10.1088/0034-4885/78/8/086901} {\bibfield  {journal} {\bibinfo  {journal} {Reports on Progress in Physics}\ }\textbf {\bibinfo {volume} {78}},\ \bibinfo {eid} {086901} (\bibinfo {year} {2015})},\ \Eprint {https://arxiv.org/abs/1411.0115} {arXiv:1411.0115 [astro-ph.CO]} \BibitemShut {NoStop}%
\bibitem [{\citenamefont {{Mandelbaum}}(2018)}]{Mandelbaum2018_review}%
  \BibitemOpen
  \bibfield  {author} {\bibinfo {author} {\bibfnamefont {R.}~\bibnamefont {{Mandelbaum}}},\ }\href {https://doi.org/10.1146/annurev-astro-081817-051928} {\bibfield  {journal} {\bibinfo  {journal} {\araa}\ }\textbf {\bibinfo {volume} {56}},\ \bibinfo {pages} {393} (\bibinfo {year} {2018})},\ \Eprint {https://arxiv.org/abs/1710.03235} {arXiv:1710.03235 [astro-ph.CO]} \BibitemShut {NoStop}%
\bibitem [{\citenamefont {{Amon}}\ \emph {et~al.}(2022)\citenamefont {{Amon}}, \citenamefont {{Gruen}}, \citenamefont {{Troxel}}, \citenamefont {{MacCrann}}, \citenamefont {{Dodelson}}, \citenamefont {{Choi}}, \citenamefont {{Doux}}, \citenamefont {{Secco}}, \citenamefont {{Samuroff}}, \citenamefont {{Krause}} \emph {et~al.}}]{Amon2022}%
  \BibitemOpen
  \bibfield  {author} {\bibinfo {author} {\bibfnamefont {A.}~\bibnamefont {{Amon}}}, \bibinfo {author} {\bibfnamefont {D.}~\bibnamefont {{Gruen}}}, \bibinfo {author} {\bibfnamefont {M.~A.}\ \bibnamefont {{Troxel}}}, \bibinfo {author} {\bibfnamefont {N.}~\bibnamefont {{MacCrann}}}, \bibinfo {author} {\bibfnamefont {S.}~\bibnamefont {{Dodelson}}}, \bibinfo {author} {\bibfnamefont {A.}~\bibnamefont {{Choi}}}, \bibinfo {author} {\bibfnamefont {C.}~\bibnamefont {{Doux}}}, \bibinfo {author} {\bibfnamefont {L.~F.}\ \bibnamefont {{Secco}}}, \bibinfo {author} {\bibfnamefont {S.}~\bibnamefont {{Samuroff}}}, \bibinfo {author} {\bibfnamefont {E.}~\bibnamefont {{Krause}}}, \emph {et~al.},\ }\href {https://doi.org/10.1103/PhysRevD.105.023514} {\bibfield  {journal} {\bibinfo  {journal} {\prd}\ }\textbf {\bibinfo {volume} {105}},\ \bibinfo {eid} {023514} (\bibinfo {year} {2022})},\ \Eprint {https://arxiv.org/abs/2105.13543} {arXiv:2105.13543 [astro-ph.CO]} \BibitemShut {NoStop}%
\bibitem [{\citenamefont {{Secco}}\ \emph {et~al.}(2022)\citenamefont {{Secco}}, \citenamefont {{Samuroff}}, \citenamefont {{Krause}}, \citenamefont {{Jain}}, \citenamefont {{Blazek}}, \citenamefont {{Raveri}}, \citenamefont {{Campos}}, \citenamefont {{Amon}}, \citenamefont {{Chen}}, \citenamefont {{Doux}} \emph {et~al.}}]{Secco2022}%
  \BibitemOpen
  \bibfield  {author} {\bibinfo {author} {\bibfnamefont {L.~F.}\ \bibnamefont {{Secco}}}, \bibinfo {author} {\bibfnamefont {S.}~\bibnamefont {{Samuroff}}}, \bibinfo {author} {\bibfnamefont {E.}~\bibnamefont {{Krause}}}, \bibinfo {author} {\bibfnamefont {B.}~\bibnamefont {{Jain}}}, \bibinfo {author} {\bibfnamefont {J.}~\bibnamefont {{Blazek}}}, \bibinfo {author} {\bibfnamefont {M.}~\bibnamefont {{Raveri}}}, \bibinfo {author} {\bibfnamefont {A.}~\bibnamefont {{Campos}}}, \bibinfo {author} {\bibfnamefont {A.}~\bibnamefont {{Amon}}}, \bibinfo {author} {\bibfnamefont {A.}~\bibnamefont {{Chen}}}, \bibinfo {author} {\bibfnamefont {C.}~\bibnamefont {{Doux}}}, \emph {et~al.},\ }\href {https://doi.org/10.1103/PhysRevD.105.023515} {\bibfield  {journal} {\bibinfo  {journal} {\prd}\ }\textbf {\bibinfo {volume} {105}},\ \bibinfo {eid} {023515} (\bibinfo {year} {2022})},\ \Eprint {https://arxiv.org/abs/2105.13544} {arXiv:2105.13544 [astro-ph.CO]} \BibitemShut {NoStop}%
\bibitem [{\citenamefont {{Li}}\ \emph {et~al.}(2023)\citenamefont {{Li}}, \citenamefont {{Zhang}}, \citenamefont {{Sugiyama}}, \citenamefont {{Dalal}}, \citenamefont {{Rau}}, \citenamefont {{Mandelbaum}}, \citenamefont {{Takada}}, \citenamefont {{More}}, \citenamefont {{Strauss}}, \citenamefont {{Miyatake}} \emph {et~al.}}]{Li2023}%
  \BibitemOpen
  \bibfield  {author} {\bibinfo {author} {\bibfnamefont {X.}~\bibnamefont {{Li}}}, \bibinfo {author} {\bibfnamefont {T.}~\bibnamefont {{Zhang}}}, \bibinfo {author} {\bibfnamefont {S.}~\bibnamefont {{Sugiyama}}}, \bibinfo {author} {\bibfnamefont {R.}~\bibnamefont {{Dalal}}}, \bibinfo {author} {\bibfnamefont {M.~M.}\ \bibnamefont {{Rau}}}, \bibinfo {author} {\bibfnamefont {R.}~\bibnamefont {{Mandelbaum}}}, \bibinfo {author} {\bibfnamefont {M.}~\bibnamefont {{Takada}}}, \bibinfo {author} {\bibfnamefont {S.}~\bibnamefont {{More}}}, \bibinfo {author} {\bibfnamefont {M.~A.}\ \bibnamefont {{Strauss}}}, \bibinfo {author} {\bibfnamefont {H.}~\bibnamefont {{Miyatake}}}, \emph {et~al.},\ }\href@noop {} {\bibfield  {journal} {\bibinfo  {journal} {arXiv e-prints}\ ,\ \bibinfo {eid} {arXiv:2304.00702}} (\bibinfo {year} {2023})},\ \Eprint {https://arxiv.org/abs/2304.00702} {arXiv:2304.00702 [astro-ph.CO]} \BibitemShut {NoStop}%
\bibitem [{\citenamefont {{Dalal}}\ \emph {et~al.}(2023)\citenamefont {{Dalal}}, \citenamefont {{Li}}, \citenamefont {{Nicola}}, \citenamefont {{Zuntz}}, \citenamefont {{Strauss}}, \citenamefont {{Sugiyama}}, \citenamefont {{Zhang}}, \citenamefont {{Rau}}, \citenamefont {{Mandelbaum}}, \citenamefont {{Takada}}, \citenamefont {{More}}, \citenamefont {{Miyatake}}, \citenamefont {{Kannawadi}}, \citenamefont {{Shirasaki}}, \citenamefont {{Taniguchi}}, \citenamefont {{Takahashi}}, \citenamefont {{Osato}}, \citenamefont {{Hamana}}, \citenamefont {{Oguri}}, \citenamefont {{Nishizawa}}, \citenamefont {{Malag{\'o}n}}, \citenamefont {{Sunayama}}, \citenamefont {{Alonso}}, \citenamefont {{Slosar}}, \citenamefont {{Luo}}, \citenamefont {{Armstrong}}, \citenamefont {{Bosch}}, \citenamefont {{Hsieh}}, \citenamefont {{Komiyama}}, \citenamefont {{Lupton}}, \citenamefont {{Lust}}, \citenamefont {{MacArthur}}, \citenamefont {{Miyazaki}}, \citenamefont {{Murayama}}, \citenamefont {{Nishimichi}}, \citenamefont {{Okura}},
  \citenamefont {{Price}}, \citenamefont {{Tait}}, \citenamefont {{Tanaka}},\ and\ \citenamefont {{Wang}}}]{Dalal2023}%
  \BibitemOpen
  \bibfield  {author} {\bibinfo {author} {\bibfnamefont {R.}~\bibnamefont {{Dalal}}}, \bibinfo {author} {\bibfnamefont {X.}~\bibnamefont {{Li}}}, \bibinfo {author} {\bibfnamefont {A.}~\bibnamefont {{Nicola}}}, \bibinfo {author} {\bibfnamefont {J.}~\bibnamefont {{Zuntz}}}, \bibinfo {author} {\bibfnamefont {M.~A.}\ \bibnamefont {{Strauss}}}, \bibinfo {author} {\bibfnamefont {S.}~\bibnamefont {{Sugiyama}}}, \bibinfo {author} {\bibfnamefont {T.}~\bibnamefont {{Zhang}}}, \bibinfo {author} {\bibfnamefont {M.~M.}\ \bibnamefont {{Rau}}}, \bibinfo {author} {\bibfnamefont {R.}~\bibnamefont {{Mandelbaum}}}, \bibinfo {author} {\bibfnamefont {M.}~\bibnamefont {{Takada}}}, \bibinfo {author} {\bibfnamefont {S.}~\bibnamefont {{More}}}, \bibinfo {author} {\bibfnamefont {H.}~\bibnamefont {{Miyatake}}}, \bibinfo {author} {\bibfnamefont {A.}~\bibnamefont {{Kannawadi}}}, \bibinfo {author} {\bibfnamefont {M.}~\bibnamefont {{Shirasaki}}}, \bibinfo {author} {\bibfnamefont {T.}~\bibnamefont {{Taniguchi}}}, \bibinfo {author}
  {\bibfnamefont {R.}~\bibnamefont {{Takahashi}}}, \bibinfo {author} {\bibfnamefont {K.}~\bibnamefont {{Osato}}}, \bibinfo {author} {\bibfnamefont {T.}~\bibnamefont {{Hamana}}}, \bibinfo {author} {\bibfnamefont {M.}~\bibnamefont {{Oguri}}}, \bibinfo {author} {\bibfnamefont {A.~J.}\ \bibnamefont {{Nishizawa}}}, \bibinfo {author} {\bibfnamefont {A.~A.~P.}\ \bibnamefont {{Malag{\'o}n}}}, \bibinfo {author} {\bibfnamefont {T.}~\bibnamefont {{Sunayama}}}, \bibinfo {author} {\bibfnamefont {D.}~\bibnamefont {{Alonso}}}, \bibinfo {author} {\bibfnamefont {A.}~\bibnamefont {{Slosar}}}, \bibinfo {author} {\bibfnamefont {W.}~\bibnamefont {{Luo}}}, \bibinfo {author} {\bibfnamefont {R.}~\bibnamefont {{Armstrong}}}, \bibinfo {author} {\bibfnamefont {J.}~\bibnamefont {{Bosch}}}, \bibinfo {author} {\bibfnamefont {B.-C.}\ \bibnamefont {{Hsieh}}}, \bibinfo {author} {\bibfnamefont {Y.}~\bibnamefont {{Komiyama}}}, \bibinfo {author} {\bibfnamefont {R.~H.}\ \bibnamefont {{Lupton}}}, \bibinfo {author} {\bibfnamefont {N.~B.}\
  \bibnamefont {{Lust}}}, \bibinfo {author} {\bibfnamefont {L.~A.}\ \bibnamefont {{MacArthur}}}, \bibinfo {author} {\bibfnamefont {S.}~\bibnamefont {{Miyazaki}}}, \bibinfo {author} {\bibfnamefont {H.}~\bibnamefont {{Murayama}}}, \bibinfo {author} {\bibfnamefont {T.}~\bibnamefont {{Nishimichi}}}, \bibinfo {author} {\bibfnamefont {Y.}~\bibnamefont {{Okura}}}, \bibinfo {author} {\bibfnamefont {P.~A.}\ \bibnamefont {{Price}}}, \bibinfo {author} {\bibfnamefont {P.~J.}\ \bibnamefont {{Tait}}}, \bibinfo {author} {\bibfnamefont {M.}~\bibnamefont {{Tanaka}}},\ and\ \bibinfo {author} {\bibfnamefont {S.-Y.}\ \bibnamefont {{Wang}}},\ }\href {https://doi.org/10.1103/PhysRevD.108.123519} {\bibfield  {journal} {\bibinfo  {journal} {\prd}\ }\textbf {\bibinfo {volume} {108}},\ \bibinfo {eid} {123519} (\bibinfo {year} {2023})},\ \Eprint {https://arxiv.org/abs/2304.00701} {arXiv:2304.00701 [astro-ph.CO]} \BibitemShut {NoStop}%
\bibitem [{\citenamefont {{Wright}}\ \emph {et~al.}(2025)\citenamefont {{Wright}}, \citenamefont {{St{\"o}lzner}}, \citenamefont {{Asgari}}, \citenamefont {{Bilicki}}, \citenamefont {{Giblin}}, \citenamefont {{Heymans}}, \citenamefont {{Hildebrandt}}, \citenamefont {{Hoekstra}}, \citenamefont {{Joachimi}}, \citenamefont {{Kuijken}}, \citenamefont {{Li}}, \citenamefont {{Reischke}}, \citenamefont {{von Wietersheim-Kramsta}}, \citenamefont {{Yoon}}, \citenamefont {{Burger}}, \citenamefont {{Chisari}}, \citenamefont {{de Jong}}, \citenamefont {{Dvornik}}, \citenamefont {{Georgiou}}, \citenamefont {{Harnois-D{\'e}raps}}, \citenamefont {{Jalan}}, \citenamefont {{William}}, \citenamefont {{Joudaki}}, \citenamefont {{Lesci}}, \citenamefont {{Linke}}, \citenamefont {{Loureiro}}, \citenamefont {{Mahony}}, \citenamefont {{Maturi}}, \citenamefont {{Miller}}, \citenamefont {{Moscardini}}, \citenamefont {{Napolitano}}, \citenamefont {{Porth}}, \citenamefont {{Radovich}}, \citenamefont {{Schneider}}, \citenamefont
  {{Tr{\"o}ster}}, \citenamefont {{Wittje}}, \citenamefont {{Yan}},\ and\ \citenamefont {{Zhang}}}]{wright2025}%
  \BibitemOpen
  \bibfield  {author} {\bibinfo {author} {\bibfnamefont {A.~H.}\ \bibnamefont {{Wright}}}, \bibinfo {author} {\bibfnamefont {B.}~\bibnamefont {{St{\"o}lzner}}}, \bibinfo {author} {\bibfnamefont {M.}~\bibnamefont {{Asgari}}}, \bibinfo {author} {\bibfnamefont {M.}~\bibnamefont {{Bilicki}}}, \bibinfo {author} {\bibfnamefont {B.}~\bibnamefont {{Giblin}}}, \bibinfo {author} {\bibfnamefont {C.}~\bibnamefont {{Heymans}}}, \bibinfo {author} {\bibfnamefont {H.}~\bibnamefont {{Hildebrandt}}}, \bibinfo {author} {\bibfnamefont {H.}~\bibnamefont {{Hoekstra}}}, \bibinfo {author} {\bibfnamefont {B.}~\bibnamefont {{Joachimi}}}, \bibinfo {author} {\bibfnamefont {K.}~\bibnamefont {{Kuijken}}}, \bibinfo {author} {\bibfnamefont {S.-S.}\ \bibnamefont {{Li}}}, \bibinfo {author} {\bibfnamefont {R.}~\bibnamefont {{Reischke}}}, \bibinfo {author} {\bibfnamefont {M.}~\bibnamefont {{von Wietersheim-Kramsta}}}, \bibinfo {author} {\bibfnamefont {M.}~\bibnamefont {{Yoon}}}, \bibinfo {author} {\bibfnamefont {P.}~\bibnamefont {{Burger}}},
  \bibinfo {author} {\bibfnamefont {N.~E.}\ \bibnamefont {{Chisari}}}, \bibinfo {author} {\bibfnamefont {J.}~\bibnamefont {{de Jong}}}, \bibinfo {author} {\bibfnamefont {A.}~\bibnamefont {{Dvornik}}}, \bibinfo {author} {\bibfnamefont {C.}~\bibnamefont {{Georgiou}}}, \bibinfo {author} {\bibfnamefont {J.}~\bibnamefont {{Harnois-D{\'e}raps}}}, \bibinfo {author} {\bibfnamefont {P.}~\bibnamefont {{Jalan}}}, \bibinfo {author} {\bibfnamefont {A.~J.}\ \bibnamefont {{William}}}, \bibinfo {author} {\bibfnamefont {S.}~\bibnamefont {{Joudaki}}}, \bibinfo {author} {\bibfnamefont {G.~F.}\ \bibnamefont {{Lesci}}}, \bibinfo {author} {\bibfnamefont {L.}~\bibnamefont {{Linke}}}, \bibinfo {author} {\bibfnamefont {A.}~\bibnamefont {{Loureiro}}}, \bibinfo {author} {\bibfnamefont {C.}~\bibnamefont {{Mahony}}}, \bibinfo {author} {\bibfnamefont {M.}~\bibnamefont {{Maturi}}}, \bibinfo {author} {\bibfnamefont {L.}~\bibnamefont {{Miller}}}, \bibinfo {author} {\bibfnamefont {L.}~\bibnamefont {{Moscardini}}}, \bibinfo {author}
  {\bibfnamefont {N.~R.}\ \bibnamefont {{Napolitano}}}, \bibinfo {author} {\bibfnamefont {L.}~\bibnamefont {{Porth}}}, \bibinfo {author} {\bibfnamefont {M.}~\bibnamefont {{Radovich}}}, \bibinfo {author} {\bibfnamefont {P.}~\bibnamefont {{Schneider}}}, \bibinfo {author} {\bibfnamefont {T.}~\bibnamefont {{Tr{\"o}ster}}}, \bibinfo {author} {\bibfnamefont {A.}~\bibnamefont {{Wittje}}}, \bibinfo {author} {\bibfnamefont {Z.}~\bibnamefont {{Yan}}},\ and\ \bibinfo {author} {\bibfnamefont {Y.-H.}\ \bibnamefont {{Zhang}}},\ }\href {https://doi.org/10.48550/arXiv.2503.19441} {\bibfield  {journal} {\bibinfo  {journal} {arXiv e-prints}\ ,\ \bibinfo {eid} {arXiv:2503.19441}} (\bibinfo {year} {2025})},\ \Eprint {https://arxiv.org/abs/2503.19441} {arXiv:2503.19441 [astro-ph.CO]} \BibitemShut {NoStop}%
\bibitem [{\citenamefont {{Prat}}\ \emph {et~al.}(2018)\citenamefont {{Prat}}, \citenamefont {{S{\'a}nchez}}, \citenamefont {{Fang}}, \citenamefont {{Gruen}}, \citenamefont {{Elvin-Poole}}, \citenamefont {{Kokron}}, \citenamefont {{Secco}}, \citenamefont {{Jain}}, \citenamefont {{Miquel}}, \citenamefont {{MacCrann}}, \citenamefont {{Troxel}}, \citenamefont {{Alarcon}}, \citenamefont {{Bacon}}, \citenamefont {{Bernstein}}, \citenamefont {{Blazek}}, \citenamefont {{Cawthon}}, \citenamefont {{Chang}}, \citenamefont {{Crocce}}, \citenamefont {{Davis}}, \citenamefont {{De Vicente}}, \citenamefont {{Dietrich}}, \citenamefont {{Drlica-Wagner}}, \citenamefont {{Friedrich}}, \citenamefont {{Gatti}}, \citenamefont {{Hartley}}, \citenamefont {{Hoyle}}, \citenamefont {{Huff}}, \citenamefont {{Jarvis}}, \citenamefont {{Rau}}, \citenamefont {{Rollins}}, \citenamefont {{Ross}}, \citenamefont {{Rozo}}, \citenamefont {{Rykoff}}, \citenamefont {{Samuroff}}, \citenamefont {{Sheldon}}, \citenamefont {{Varga}}, \citenamefont
  {{Vielzeuf}}, \citenamefont {{Zuntz}}, \citenamefont {{Abbott}}, \citenamefont {{Abdalla}}, \citenamefont {{Allam}}, \citenamefont {{Annis}}, \citenamefont {{Bechtol}}, \citenamefont {{Benoit-L{\'e}vy}}, \citenamefont {{Bertin}}, \citenamefont {{Brooks}}, \citenamefont {{Buckley-Geer}}, \citenamefont {{Burke}}, \citenamefont {{Carnero Rosell}}, \citenamefont {{Carrasco Kind}}, \citenamefont {{Carretero}}, \citenamefont {{Castander}}, \citenamefont {{Cunha}}, \citenamefont {{D'Andrea}}, \citenamefont {{da Costa}}, \citenamefont {{Desai}}, \citenamefont {{Diehl}}, \citenamefont {{Dodelson}}, \citenamefont {{Eifler}}, \citenamefont {{Fernandez}}, \citenamefont {{Flaugher}}, \citenamefont {{Fosalba}}, \citenamefont {{Frieman}}, \citenamefont {{Garc{\'\i}a-Bellido}}, \citenamefont {{Gaztanaga}}, \citenamefont {{Gerdes}}, \citenamefont {{Giannantonio}}, \citenamefont {{Goldstein}}, \citenamefont {{Gruendl}}, \citenamefont {{Gschwend}}, \citenamefont {{Gutierrez}}, \citenamefont {{Honscheid}}, \citenamefont
  {{James}}, \citenamefont {{Jeltema}}, \citenamefont {{Johnson}}, \citenamefont {{Johnson}}, \citenamefont {{Kirk}}, \citenamefont {{Krause}}, \citenamefont {{Kuehn}}, \citenamefont {{Kuhlmann}}, \citenamefont {{Lahav}}, \citenamefont {{Li}}, \citenamefont {{Lima}}, \citenamefont {{Maia}}, \citenamefont {{March}}, \citenamefont {{Marshall}}, \citenamefont {{Martini}}, \citenamefont {{Melchior}}, \citenamefont {{Menanteau}}, \citenamefont {{Mohr}}, \citenamefont {{Nichol}}, \citenamefont {{Nord}}, \citenamefont {{Plazas}}, \citenamefont {{Romer}}, \citenamefont {{Roodman}}, \citenamefont {{Sako}}, \citenamefont {{Sanchez}}, \citenamefont {{Scarpine}}, \citenamefont {{Schindler}}, \citenamefont {{Schubnell}}, \citenamefont {{Sevilla-Noarbe}}, \citenamefont {{Smith}}, \citenamefont {{Smith}}, \citenamefont {{Soares-Santos}}, \citenamefont {{Sobreira}}, \citenamefont {{Suchyta}}, \citenamefont {{Swanson}}, \citenamefont {{Tarle}}, \citenamefont {{Thomas}}, \citenamefont {{Tucker}}, \citenamefont {{Vikram}},
  \citenamefont {{Walker}}, \citenamefont {{Wechsler}}, \citenamefont {{Yanny}}, \citenamefont {{Zhang}},\ and\ \citenamefont {{DES Collaboration}}}]{Prat2018}%
  \BibitemOpen
  \bibfield  {author} {\bibinfo {author} {\bibfnamefont {J.}~\bibnamefont {{Prat}}}, \bibinfo {author} {\bibfnamefont {C.}~\bibnamefont {{S{\'a}nchez}}}, \bibinfo {author} {\bibfnamefont {Y.}~\bibnamefont {{Fang}}}, \bibinfo {author} {\bibfnamefont {D.}~\bibnamefont {{Gruen}}}, \bibinfo {author} {\bibfnamefont {J.}~\bibnamefont {{Elvin-Poole}}}, \bibinfo {author} {\bibfnamefont {N.}~\bibnamefont {{Kokron}}}, \bibinfo {author} {\bibfnamefont {L.~F.}\ \bibnamefont {{Secco}}}, \bibinfo {author} {\bibfnamefont {B.}~\bibnamefont {{Jain}}}, \bibinfo {author} {\bibfnamefont {R.}~\bibnamefont {{Miquel}}}, \bibinfo {author} {\bibfnamefont {N.}~\bibnamefont {{MacCrann}}}, \bibinfo {author} {\bibfnamefont {M.~A.}\ \bibnamefont {{Troxel}}}, \bibinfo {author} {\bibfnamefont {A.}~\bibnamefont {{Alarcon}}}, \bibinfo {author} {\bibfnamefont {D.}~\bibnamefont {{Bacon}}}, \bibinfo {author} {\bibfnamefont {G.~M.}\ \bibnamefont {{Bernstein}}}, \bibinfo {author} {\bibfnamefont {J.}~\bibnamefont {{Blazek}}}, \bibinfo {author}
  {\bibfnamefont {R.}~\bibnamefont {{Cawthon}}}, \bibinfo {author} {\bibfnamefont {C.}~\bibnamefont {{Chang}}}, \bibinfo {author} {\bibfnamefont {M.}~\bibnamefont {{Crocce}}}, \bibinfo {author} {\bibfnamefont {C.}~\bibnamefont {{Davis}}}, \bibinfo {author} {\bibfnamefont {J.}~\bibnamefont {{De Vicente}}}, \bibinfo {author} {\bibfnamefont {J.~P.}\ \bibnamefont {{Dietrich}}}, \bibinfo {author} {\bibfnamefont {A.}~\bibnamefont {{Drlica-Wagner}}}, \bibinfo {author} {\bibfnamefont {O.}~\bibnamefont {{Friedrich}}}, \bibinfo {author} {\bibfnamefont {M.}~\bibnamefont {{Gatti}}}, \bibinfo {author} {\bibfnamefont {W.~G.}\ \bibnamefont {{Hartley}}}, \bibinfo {author} {\bibfnamefont {B.}~\bibnamefont {{Hoyle}}}, \bibinfo {author} {\bibfnamefont {E.~M.}\ \bibnamefont {{Huff}}}, \bibinfo {author} {\bibfnamefont {M.}~\bibnamefont {{Jarvis}}}, \bibinfo {author} {\bibfnamefont {M.~M.}\ \bibnamefont {{Rau}}}, \bibinfo {author} {\bibfnamefont {R.~P.}\ \bibnamefont {{Rollins}}}, \bibinfo {author} {\bibfnamefont {A.~J.}\
  \bibnamefont {{Ross}}}, \bibinfo {author} {\bibfnamefont {E.}~\bibnamefont {{Rozo}}}, \bibinfo {author} {\bibfnamefont {E.~S.}\ \bibnamefont {{Rykoff}}}, \bibinfo {author} {\bibfnamefont {S.}~\bibnamefont {{Samuroff}}}, \bibinfo {author} {\bibfnamefont {E.}~\bibnamefont {{Sheldon}}}, \bibinfo {author} {\bibfnamefont {T.~N.}\ \bibnamefont {{Varga}}}, \bibinfo {author} {\bibfnamefont {P.}~\bibnamefont {{Vielzeuf}}}, \bibinfo {author} {\bibfnamefont {J.}~\bibnamefont {{Zuntz}}}, \bibinfo {author} {\bibfnamefont {T.~M.~C.}\ \bibnamefont {{Abbott}}}, \bibinfo {author} {\bibfnamefont {F.~B.}\ \bibnamefont {{Abdalla}}}, \bibinfo {author} {\bibfnamefont {S.}~\bibnamefont {{Allam}}}, \bibinfo {author} {\bibfnamefont {J.}~\bibnamefont {{Annis}}}, \bibinfo {author} {\bibfnamefont {K.}~\bibnamefont {{Bechtol}}}, \bibinfo {author} {\bibfnamefont {A.}~\bibnamefont {{Benoit-L{\'e}vy}}}, \bibinfo {author} {\bibfnamefont {E.}~\bibnamefont {{Bertin}}}, \bibinfo {author} {\bibfnamefont {D.}~\bibnamefont {{Brooks}}}, \bibinfo
  {author} {\bibfnamefont {E.}~\bibnamefont {{Buckley-Geer}}}, \bibinfo {author} {\bibfnamefont {D.~L.}\ \bibnamefont {{Burke}}}, \bibinfo {author} {\bibfnamefont {A.}~\bibnamefont {{Carnero Rosell}}}, \bibinfo {author} {\bibfnamefont {M.}~\bibnamefont {{Carrasco Kind}}}, \bibinfo {author} {\bibfnamefont {J.}~\bibnamefont {{Carretero}}}, \bibinfo {author} {\bibfnamefont {F.~J.}\ \bibnamefont {{Castander}}}, \bibinfo {author} {\bibfnamefont {C.~E.}\ \bibnamefont {{Cunha}}}, \bibinfo {author} {\bibfnamefont {C.~B.}\ \bibnamefont {{D'Andrea}}}, \bibinfo {author} {\bibfnamefont {L.~N.}\ \bibnamefont {{da Costa}}}, \bibinfo {author} {\bibfnamefont {S.}~\bibnamefont {{Desai}}}, \bibinfo {author} {\bibfnamefont {H.~T.}\ \bibnamefont {{Diehl}}}, \bibinfo {author} {\bibfnamefont {S.}~\bibnamefont {{Dodelson}}}, \bibinfo {author} {\bibfnamefont {T.~F.}\ \bibnamefont {{Eifler}}}, \bibinfo {author} {\bibfnamefont {E.}~\bibnamefont {{Fernandez}}}, \bibinfo {author} {\bibfnamefont {B.}~\bibnamefont {{Flaugher}}}, \bibinfo
  {author} {\bibfnamefont {P.}~\bibnamefont {{Fosalba}}}, \bibinfo {author} {\bibfnamefont {J.}~\bibnamefont {{Frieman}}}, \bibinfo {author} {\bibfnamefont {J.}~\bibnamefont {{Garc{\'\i}a-Bellido}}}, \bibinfo {author} {\bibfnamefont {E.}~\bibnamefont {{Gaztanaga}}}, \bibinfo {author} {\bibfnamefont {D.~W.}\ \bibnamefont {{Gerdes}}}, \bibinfo {author} {\bibfnamefont {T.}~\bibnamefont {{Giannantonio}}}, \bibinfo {author} {\bibfnamefont {D.~A.}\ \bibnamefont {{Goldstein}}}, \bibinfo {author} {\bibfnamefont {R.~A.}\ \bibnamefont {{Gruendl}}}, \bibinfo {author} {\bibfnamefont {J.}~\bibnamefont {{Gschwend}}}, \bibinfo {author} {\bibfnamefont {G.}~\bibnamefont {{Gutierrez}}}, \bibinfo {author} {\bibfnamefont {K.}~\bibnamefont {{Honscheid}}}, \bibinfo {author} {\bibfnamefont {D.~J.}\ \bibnamefont {{James}}}, \bibinfo {author} {\bibfnamefont {T.}~\bibnamefont {{Jeltema}}}, \bibinfo {author} {\bibfnamefont {M.~W.~G.}\ \bibnamefont {{Johnson}}}, \bibinfo {author} {\bibfnamefont {M.~D.}\ \bibnamefont {{Johnson}}},
  \bibinfo {author} {\bibfnamefont {D.}~\bibnamefont {{Kirk}}}, \bibinfo {author} {\bibfnamefont {E.}~\bibnamefont {{Krause}}}, \bibinfo {author} {\bibfnamefont {K.}~\bibnamefont {{Kuehn}}}, \bibinfo {author} {\bibfnamefont {S.}~\bibnamefont {{Kuhlmann}}}, \bibinfo {author} {\bibfnamefont {O.}~\bibnamefont {{Lahav}}}, \bibinfo {author} {\bibfnamefont {T.~S.}\ \bibnamefont {{Li}}}, \bibinfo {author} {\bibfnamefont {M.}~\bibnamefont {{Lima}}}, \bibinfo {author} {\bibfnamefont {M.~A.~G.}\ \bibnamefont {{Maia}}}, \bibinfo {author} {\bibfnamefont {M.}~\bibnamefont {{March}}}, \bibinfo {author} {\bibfnamefont {J.~L.}\ \bibnamefont {{Marshall}}}, \bibinfo {author} {\bibfnamefont {P.}~\bibnamefont {{Martini}}}, \bibinfo {author} {\bibfnamefont {P.}~\bibnamefont {{Melchior}}}, \bibinfo {author} {\bibfnamefont {F.}~\bibnamefont {{Menanteau}}}, \bibinfo {author} {\bibfnamefont {J.~J.}\ \bibnamefont {{Mohr}}}, \bibinfo {author} {\bibfnamefont {R.~C.}\ \bibnamefont {{Nichol}}}, \bibinfo {author} {\bibfnamefont
  {B.}~\bibnamefont {{Nord}}}, \bibinfo {author} {\bibfnamefont {A.~A.}\ \bibnamefont {{Plazas}}}, \bibinfo {author} {\bibfnamefont {A.~K.}\ \bibnamefont {{Romer}}}, \bibinfo {author} {\bibfnamefont {A.}~\bibnamefont {{Roodman}}}, \bibinfo {author} {\bibfnamefont {M.}~\bibnamefont {{Sako}}}, \bibinfo {author} {\bibfnamefont {E.}~\bibnamefont {{Sanchez}}}, \bibinfo {author} {\bibfnamefont {V.}~\bibnamefont {{Scarpine}}}, \bibinfo {author} {\bibfnamefont {R.}~\bibnamefont {{Schindler}}}, \bibinfo {author} {\bibfnamefont {M.}~\bibnamefont {{Schubnell}}}, \bibinfo {author} {\bibfnamefont {I.}~\bibnamefont {{Sevilla-Noarbe}}}, \bibinfo {author} {\bibfnamefont {M.}~\bibnamefont {{Smith}}}, \bibinfo {author} {\bibfnamefont {R.~C.}\ \bibnamefont {{Smith}}}, \bibinfo {author} {\bibfnamefont {M.}~\bibnamefont {{Soares-Santos}}}, \bibinfo {author} {\bibfnamefont {F.}~\bibnamefont {{Sobreira}}}, \bibinfo {author} {\bibfnamefont {E.}~\bibnamefont {{Suchyta}}}, \bibinfo {author} {\bibfnamefont {M.~E.~C.}\ \bibnamefont
  {{Swanson}}}, \bibinfo {author} {\bibfnamefont {G.}~\bibnamefont {{Tarle}}}, \bibinfo {author} {\bibfnamefont {D.}~\bibnamefont {{Thomas}}}, \bibinfo {author} {\bibfnamefont {D.~L.}\ \bibnamefont {{Tucker}}}, \bibinfo {author} {\bibfnamefont {V.}~\bibnamefont {{Vikram}}}, \bibinfo {author} {\bibfnamefont {A.~R.}\ \bibnamefont {{Walker}}}, \bibinfo {author} {\bibfnamefont {R.~H.}\ \bibnamefont {{Wechsler}}}, \bibinfo {author} {\bibfnamefont {B.}~\bibnamefont {{Yanny}}}, \bibinfo {author} {\bibfnamefont {Y.}~\bibnamefont {{Zhang}}},\ and\ \bibinfo {author} {\bibnamefont {{DES Collaboration}}},\ }\href {https://doi.org/10.1103/PhysRevD.98.042005} {\bibfield  {journal} {\bibinfo  {journal} {\prd}\ }\textbf {\bibinfo {volume} {98}},\ \bibinfo {eid} {042005} (\bibinfo {year} {2018})},\ \Eprint {https://arxiv.org/abs/1708.01537} {arXiv:1708.01537 [astro-ph.CO]} \BibitemShut {NoStop}%
\bibitem [{\citenamefont {{Miyatake}}\ \emph {et~al.}(2022)\citenamefont {{Miyatake}}, \citenamefont {{Kobayashi}}, \citenamefont {{Takada}}, \citenamefont {{Nishimichi}}, \citenamefont {{Shirasaki}}, \citenamefont {{Sugiyama}}, \citenamefont {{Takahashi}}, \citenamefont {{Osato}}, \citenamefont {{More}},\ and\ \citenamefont {{Park}}}]{Miyatake2022}%
  \BibitemOpen
  \bibfield  {author} {\bibinfo {author} {\bibfnamefont {H.}~\bibnamefont {{Miyatake}}}, \bibinfo {author} {\bibfnamefont {Y.}~\bibnamefont {{Kobayashi}}}, \bibinfo {author} {\bibfnamefont {M.}~\bibnamefont {{Takada}}}, \bibinfo {author} {\bibfnamefont {T.}~\bibnamefont {{Nishimichi}}}, \bibinfo {author} {\bibfnamefont {M.}~\bibnamefont {{Shirasaki}}}, \bibinfo {author} {\bibfnamefont {S.}~\bibnamefont {{Sugiyama}}}, \bibinfo {author} {\bibfnamefont {R.}~\bibnamefont {{Takahashi}}}, \bibinfo {author} {\bibfnamefont {K.}~\bibnamefont {{Osato}}}, \bibinfo {author} {\bibfnamefont {S.}~\bibnamefont {{More}}},\ and\ \bibinfo {author} {\bibfnamefont {Y.}~\bibnamefont {{Park}}},\ }\href {https://doi.org/10.1103/PhysRevD.106.083519} {\bibfield  {journal} {\bibinfo  {journal} {\prd}\ }\textbf {\bibinfo {volume} {106}},\ \bibinfo {eid} {083519} (\bibinfo {year} {2022})},\ \Eprint {https://arxiv.org/abs/2101.00113} {arXiv:2101.00113 [astro-ph.CO]} \BibitemShut {NoStop}%
\bibitem [{\citenamefont {{Sugiyama}}\ \emph {et~al.}(2023)\citenamefont {{Sugiyama}}, \citenamefont {{Miyatake}}, \citenamefont {{More}}, \citenamefont {{Li}}, \citenamefont {{Shirasaki}}, \citenamefont {{Takada}}, \citenamefont {{Kobayashi}}, \citenamefont {{Takahashi}}, \citenamefont {{Nishimichi}}, \citenamefont {{Nishizawa}}, \citenamefont {{Rau}}, \citenamefont {{Zhang}}, \citenamefont {{Dalal}}, \citenamefont {{Mandelbaum}}, \citenamefont {{Strauss}}, \citenamefont {{Hamana}}, \citenamefont {{Oguri}}, \citenamefont {{Osato}}, \citenamefont {{Kannawadi}}, \citenamefont {{Hsieh}}, \citenamefont {{Luo}}, \citenamefont {{Armstrong}}, \citenamefont {{Bosch}}, \citenamefont {{Komiyama}}, \citenamefont {{Lupton}}, \citenamefont {{Lust}}, \citenamefont {{Miyazaki}}, \citenamefont {{Murayama}}, \citenamefont {{Okura}}, \citenamefont {{Price}}, \citenamefont {{Tait}}, \citenamefont {{Tanaka}},\ and\ \citenamefont {{Wang}}}]{Sugiyama2023}%
  \BibitemOpen
  \bibfield  {author} {\bibinfo {author} {\bibfnamefont {S.}~\bibnamefont {{Sugiyama}}}, \bibinfo {author} {\bibfnamefont {H.}~\bibnamefont {{Miyatake}}}, \bibinfo {author} {\bibfnamefont {S.}~\bibnamefont {{More}}}, \bibinfo {author} {\bibfnamefont {X.}~\bibnamefont {{Li}}}, \bibinfo {author} {\bibfnamefont {M.}~\bibnamefont {{Shirasaki}}}, \bibinfo {author} {\bibfnamefont {M.}~\bibnamefont {{Takada}}}, \bibinfo {author} {\bibfnamefont {Y.}~\bibnamefont {{Kobayashi}}}, \bibinfo {author} {\bibfnamefont {R.}~\bibnamefont {{Takahashi}}}, \bibinfo {author} {\bibfnamefont {T.}~\bibnamefont {{Nishimichi}}}, \bibinfo {author} {\bibfnamefont {A.~J.}\ \bibnamefont {{Nishizawa}}}, \bibinfo {author} {\bibfnamefont {M.~M.}\ \bibnamefont {{Rau}}}, \bibinfo {author} {\bibfnamefont {T.}~\bibnamefont {{Zhang}}}, \bibinfo {author} {\bibfnamefont {R.}~\bibnamefont {{Dalal}}}, \bibinfo {author} {\bibfnamefont {R.}~\bibnamefont {{Mandelbaum}}}, \bibinfo {author} {\bibfnamefont {M.~A.}\ \bibnamefont {{Strauss}}}, \bibinfo
  {author} {\bibfnamefont {T.}~\bibnamefont {{Hamana}}}, \bibinfo {author} {\bibfnamefont {M.}~\bibnamefont {{Oguri}}}, \bibinfo {author} {\bibfnamefont {K.}~\bibnamefont {{Osato}}}, \bibinfo {author} {\bibfnamefont {A.}~\bibnamefont {{Kannawadi}}}, \bibinfo {author} {\bibfnamefont {B.-C.}\ \bibnamefont {{Hsieh}}}, \bibinfo {author} {\bibfnamefont {W.}~\bibnamefont {{Luo}}}, \bibinfo {author} {\bibfnamefont {R.}~\bibnamefont {{Armstrong}}}, \bibinfo {author} {\bibfnamefont {J.}~\bibnamefont {{Bosch}}}, \bibinfo {author} {\bibfnamefont {Y.}~\bibnamefont {{Komiyama}}}, \bibinfo {author} {\bibfnamefont {R.~H.}\ \bibnamefont {{Lupton}}}, \bibinfo {author} {\bibfnamefont {N.~B.}\ \bibnamefont {{Lust}}}, \bibinfo {author} {\bibfnamefont {S.}~\bibnamefont {{Miyazaki}}}, \bibinfo {author} {\bibfnamefont {H.}~\bibnamefont {{Murayama}}}, \bibinfo {author} {\bibfnamefont {Y.}~\bibnamefont {{Okura}}}, \bibinfo {author} {\bibfnamefont {P.~A.}\ \bibnamefont {{Price}}}, \bibinfo {author} {\bibfnamefont {P.~J.}\ \bibnamefont
  {{Tait}}}, \bibinfo {author} {\bibfnamefont {M.}~\bibnamefont {{Tanaka}}},\ and\ \bibinfo {author} {\bibfnamefont {S.-Y.}\ \bibnamefont {{Wang}}},\ }\href {https://doi.org/10.1103/PhysRevD.108.123521} {\bibfield  {journal} {\bibinfo  {journal} {\prd}\ }\textbf {\bibinfo {volume} {108}},\ \bibinfo {eid} {123521} (\bibinfo {year} {2023})},\ \Eprint {https://arxiv.org/abs/2304.00705} {arXiv:2304.00705 [astro-ph.CO]} \BibitemShut {NoStop}%
\bibitem [{\citenamefont {{Miyatake}}\ \emph {et~al.}(2023)\citenamefont {{Miyatake}}, \citenamefont {{Sugiyama}}, \citenamefont {{Takada}}, \citenamefont {{Nishimichi}}, \citenamefont {{Li}}, \citenamefont {{Shirasaki}}, \citenamefont {{More}}, \citenamefont {{Kobayashi}}, \citenamefont {{Nishizawa}}, \citenamefont {{Rau}} \emph {et~al.}}]{Miyatake2023}%
  \BibitemOpen
  \bibfield  {author} {\bibinfo {author} {\bibfnamefont {H.}~\bibnamefont {{Miyatake}}}, \bibinfo {author} {\bibfnamefont {S.}~\bibnamefont {{Sugiyama}}}, \bibinfo {author} {\bibfnamefont {M.}~\bibnamefont {{Takada}}}, \bibinfo {author} {\bibfnamefont {T.}~\bibnamefont {{Nishimichi}}}, \bibinfo {author} {\bibfnamefont {X.}~\bibnamefont {{Li}}}, \bibinfo {author} {\bibfnamefont {M.}~\bibnamefont {{Shirasaki}}}, \bibinfo {author} {\bibfnamefont {S.}~\bibnamefont {{More}}}, \bibinfo {author} {\bibfnamefont {Y.}~\bibnamefont {{Kobayashi}}}, \bibinfo {author} {\bibfnamefont {A.~J.}\ \bibnamefont {{Nishizawa}}}, \bibinfo {author} {\bibfnamefont {M.~M.}\ \bibnamefont {{Rau}}}, \emph {et~al.},\ }\href@noop {} {\bibfield  {journal} {\bibinfo  {journal} {arXiv e-prints}\ ,\ \bibinfo {eid} {arXiv:2304.00704}} (\bibinfo {year} {2023})},\ \Eprint {https://arxiv.org/abs/2304.00704} {arXiv:2304.00704 [astro-ph.CO]} \BibitemShut {NoStop}%
\bibitem [{\citenamefont {{van Uitert}}\ \emph {et~al.}(2018)\citenamefont {{van Uitert}}, \citenamefont {{Joachimi}}, \citenamefont {{Joudaki}}, \citenamefont {{Amon}}, \citenamefont {{Heymans}}, \citenamefont {{K{\"o}hlinger}}, \citenamefont {{Asgari}}, \citenamefont {{Blake}}, \citenamefont {{Choi}}, \citenamefont {{Erben}} \emph {et~al.}}]{vanUitert2018}%
  \BibitemOpen
  \bibfield  {author} {\bibinfo {author} {\bibfnamefont {E.}~\bibnamefont {{van Uitert}}}, \bibinfo {author} {\bibfnamefont {B.}~\bibnamefont {{Joachimi}}}, \bibinfo {author} {\bibfnamefont {S.}~\bibnamefont {{Joudaki}}}, \bibinfo {author} {\bibfnamefont {A.}~\bibnamefont {{Amon}}}, \bibinfo {author} {\bibfnamefont {C.}~\bibnamefont {{Heymans}}}, \bibinfo {author} {\bibfnamefont {F.}~\bibnamefont {{K{\"o}hlinger}}}, \bibinfo {author} {\bibfnamefont {M.}~\bibnamefont {{Asgari}}}, \bibinfo {author} {\bibfnamefont {C.}~\bibnamefont {{Blake}}}, \bibinfo {author} {\bibfnamefont {A.}~\bibnamefont {{Choi}}}, \bibinfo {author} {\bibfnamefont {T.}~\bibnamefont {{Erben}}}, \emph {et~al.},\ }\href {https://doi.org/10.1093/mnras/sty551} {\bibfield  {journal} {\bibinfo  {journal} {\mnras}\ }\textbf {\bibinfo {volume} {476}},\ \bibinfo {pages} {4662} (\bibinfo {year} {2018})},\ \Eprint {https://arxiv.org/abs/1706.05004} {arXiv:1706.05004 [astro-ph.CO]} \BibitemShut {NoStop}%
\bibitem [{\citenamefont {{More}}\ \emph {et~al.}(2023)\citenamefont {{More}}, \citenamefont {{Sugiyama}}, \citenamefont {{Miyatake}}, \citenamefont {{Rau}}, \citenamefont {{Shirasaki}}, \citenamefont {{Li}}, \citenamefont {{Nishizawa}}, \citenamefont {{Osato}}, \citenamefont {{Zhang}}, \citenamefont {{Takada}} \emph {et~al.}}]{More2023}%
  \BibitemOpen
  \bibfield  {author} {\bibinfo {author} {\bibfnamefont {S.}~\bibnamefont {{More}}}, \bibinfo {author} {\bibfnamefont {S.}~\bibnamefont {{Sugiyama}}}, \bibinfo {author} {\bibfnamefont {H.}~\bibnamefont {{Miyatake}}}, \bibinfo {author} {\bibfnamefont {M.~M.}\ \bibnamefont {{Rau}}}, \bibinfo {author} {\bibfnamefont {M.}~\bibnamefont {{Shirasaki}}}, \bibinfo {author} {\bibfnamefont {X.}~\bibnamefont {{Li}}}, \bibinfo {author} {\bibfnamefont {A.~J.}\ \bibnamefont {{Nishizawa}}}, \bibinfo {author} {\bibfnamefont {K.}~\bibnamefont {{Osato}}}, \bibinfo {author} {\bibfnamefont {T.}~\bibnamefont {{Zhang}}}, \bibinfo {author} {\bibfnamefont {M.}~\bibnamefont {{Takada}}}, \emph {et~al.},\ }\href@noop {} {\bibfield  {journal} {\bibinfo  {journal} {arXiv e-prints}\ ,\ \bibinfo {eid} {arXiv:2304.00703}} (\bibinfo {year} {2023})},\ \Eprint {https://arxiv.org/abs/2304.00703} {arXiv:2304.00703 [astro-ph.CO]} \BibitemShut {NoStop}%
\bibitem [{\citenamefont {{Zhang}}\ and\ \citenamefont {{Sugiyama}}(2025)}]{2x2pt_paper}%
  \BibitemOpen
  \bibfield  {author} {\bibinfo {author} {\bibfnamefont {T.}~\bibnamefont {{Zhang}}}\ and\ \bibinfo {author} {\bibfnamefont {S.}~\bibnamefont {{Sugiyama}}},\ }\href@noop {} {\bibfield  {journal} {\bibinfo  {journal} {arXiv e-prints}\ } (\bibinfo {year} {2025})}\BibitemShut {NoStop}%
\bibitem [{\citenamefont {{Gunn}}\ \emph {et~al.}(2006)\citenamefont {{Gunn}}, \citenamefont {{Siegmund}}, \citenamefont {{Mannery}}, \citenamefont {{Owen}}, \citenamefont {{Hull}}, \citenamefont {{Leger}}, \citenamefont {{Carey}}, \citenamefont {{Knapp}}, \citenamefont {{York}}, \citenamefont {{Boroski}}, \citenamefont {{Kent}}, \citenamefont {{Lupton}}, \citenamefont {{Rockosi}}, \citenamefont {{Evans}}, \citenamefont {{Waddell}}, \citenamefont {{Anderson}}, \citenamefont {{Annis}}, \citenamefont {{Barentine}}, \citenamefont {{Bartoszek}}, \citenamefont {{Bastian}}, \citenamefont {{Bracker}}, \citenamefont {{Brewington}}, \citenamefont {{Briegel}}, \citenamefont {{Brinkmann}}, \citenamefont {{Brown}}, \citenamefont {{Carr}}, \citenamefont {{Czarapata}}, \citenamefont {{Drennan}}, \citenamefont {{Dombeck}}, \citenamefont {{Federwitz}}, \citenamefont {{Gillespie}}, \citenamefont {{Gonzales}}, \citenamefont {{Hansen}}, \citenamefont {{Harvanek}}, \citenamefont {{Hayes}}, \citenamefont {{Jordan}}, \citenamefont
  {{Kinney}}, \citenamefont {{Klaene}}, \citenamefont {{Kleinman}}, \citenamefont {{Kron}}, \citenamefont {{Kresinski}}, \citenamefont {{Lee}}, \citenamefont {{Limmongkol}}, \citenamefont {{Lindenmeyer}}, \citenamefont {{Long}}, \citenamefont {{Loomis}}, \citenamefont {{McGehee}}, \citenamefont {{Mantsch}}, \citenamefont {{Neilsen}}, \citenamefont {{Neswold}}, \citenamefont {{Newman}}, \citenamefont {{Nitta}}, \citenamefont {{Peoples}}, \citenamefont {{Pier}}, \citenamefont {{Prieto}}, \citenamefont {{Prosapio}}, \citenamefont {{Rivetta}}, \citenamefont {{Schneider}}, \citenamefont {{Snedden}},\ and\ \citenamefont {{Wang}}}]{gunn2006}%
  \BibitemOpen
  \bibfield  {author} {\bibinfo {author} {\bibfnamefont {J.~E.}\ \bibnamefont {{Gunn}}}, \bibinfo {author} {\bibfnamefont {W.~A.}\ \bibnamefont {{Siegmund}}}, \bibinfo {author} {\bibfnamefont {E.~J.}\ \bibnamefont {{Mannery}}}, \bibinfo {author} {\bibfnamefont {R.~E.}\ \bibnamefont {{Owen}}}, \bibinfo {author} {\bibfnamefont {C.~L.}\ \bibnamefont {{Hull}}}, \bibinfo {author} {\bibfnamefont {R.~F.}\ \bibnamefont {{Leger}}}, \bibinfo {author} {\bibfnamefont {L.~N.}\ \bibnamefont {{Carey}}}, \bibinfo {author} {\bibfnamefont {G.~R.}\ \bibnamefont {{Knapp}}}, \bibinfo {author} {\bibfnamefont {D.~G.}\ \bibnamefont {{York}}}, \bibinfo {author} {\bibfnamefont {W.~N.}\ \bibnamefont {{Boroski}}}, \bibinfo {author} {\bibfnamefont {S.~M.}\ \bibnamefont {{Kent}}}, \bibinfo {author} {\bibfnamefont {R.~H.}\ \bibnamefont {{Lupton}}}, \bibinfo {author} {\bibfnamefont {C.~M.}\ \bibnamefont {{Rockosi}}}, \bibinfo {author} {\bibfnamefont {M.~L.}\ \bibnamefont {{Evans}}}, \bibinfo {author} {\bibfnamefont {P.}~\bibnamefont
  {{Waddell}}}, \bibinfo {author} {\bibfnamefont {J.~E.}\ \bibnamefont {{Anderson}}}, \bibinfo {author} {\bibfnamefont {J.}~\bibnamefont {{Annis}}}, \bibinfo {author} {\bibfnamefont {J.~C.}\ \bibnamefont {{Barentine}}}, \bibinfo {author} {\bibfnamefont {L.~M.}\ \bibnamefont {{Bartoszek}}}, \bibinfo {author} {\bibfnamefont {S.}~\bibnamefont {{Bastian}}}, \bibinfo {author} {\bibfnamefont {S.~B.}\ \bibnamefont {{Bracker}}}, \bibinfo {author} {\bibfnamefont {H.~J.}\ \bibnamefont {{Brewington}}}, \bibinfo {author} {\bibfnamefont {C.~I.}\ \bibnamefont {{Briegel}}}, \bibinfo {author} {\bibfnamefont {J.}~\bibnamefont {{Brinkmann}}}, \bibinfo {author} {\bibfnamefont {Y.~J.}\ \bibnamefont {{Brown}}}, \bibinfo {author} {\bibfnamefont {M.~A.}\ \bibnamefont {{Carr}}}, \bibinfo {author} {\bibfnamefont {P.~C.}\ \bibnamefont {{Czarapata}}}, \bibinfo {author} {\bibfnamefont {C.~C.}\ \bibnamefont {{Drennan}}}, \bibinfo {author} {\bibfnamefont {T.}~\bibnamefont {{Dombeck}}}, \bibinfo {author} {\bibfnamefont {G.~R.}\
  \bibnamefont {{Federwitz}}}, \bibinfo {author} {\bibfnamefont {B.~A.}\ \bibnamefont {{Gillespie}}}, \bibinfo {author} {\bibfnamefont {C.}~\bibnamefont {{Gonzales}}}, \bibinfo {author} {\bibfnamefont {S.~U.}\ \bibnamefont {{Hansen}}}, \bibinfo {author} {\bibfnamefont {M.}~\bibnamefont {{Harvanek}}}, \bibinfo {author} {\bibfnamefont {J.}~\bibnamefont {{Hayes}}}, \bibinfo {author} {\bibfnamefont {W.}~\bibnamefont {{Jordan}}}, \bibinfo {author} {\bibfnamefont {E.}~\bibnamefont {{Kinney}}}, \bibinfo {author} {\bibfnamefont {M.}~\bibnamefont {{Klaene}}}, \bibinfo {author} {\bibfnamefont {S.~J.}\ \bibnamefont {{Kleinman}}}, \bibinfo {author} {\bibfnamefont {R.~G.}\ \bibnamefont {{Kron}}}, \bibinfo {author} {\bibfnamefont {J.}~\bibnamefont {{Kresinski}}}, \bibinfo {author} {\bibfnamefont {G.}~\bibnamefont {{Lee}}}, \bibinfo {author} {\bibfnamefont {S.}~\bibnamefont {{Limmongkol}}}, \bibinfo {author} {\bibfnamefont {C.~W.}\ \bibnamefont {{Lindenmeyer}}}, \bibinfo {author} {\bibfnamefont {D.~C.}\ \bibnamefont
  {{Long}}}, \bibinfo {author} {\bibfnamefont {C.~L.}\ \bibnamefont {{Loomis}}}, \bibinfo {author} {\bibfnamefont {P.~M.}\ \bibnamefont {{McGehee}}}, \bibinfo {author} {\bibfnamefont {P.~M.}\ \bibnamefont {{Mantsch}}}, \bibinfo {author} {\bibfnamefont {J.}~\bibnamefont {{Neilsen}}, \bibfnamefont {Eric~H.}}, \bibinfo {author} {\bibfnamefont {R.~M.}\ \bibnamefont {{Neswold}}}, \bibinfo {author} {\bibfnamefont {P.~R.}\ \bibnamefont {{Newman}}}, \bibinfo {author} {\bibfnamefont {A.}~\bibnamefont {{Nitta}}}, \bibinfo {author} {\bibfnamefont {J.}~\bibnamefont {{Peoples}}, \bibfnamefont {John}}, \bibinfo {author} {\bibfnamefont {J.~R.}\ \bibnamefont {{Pier}}}, \bibinfo {author} {\bibfnamefont {P.~S.}\ \bibnamefont {{Prieto}}}, \bibinfo {author} {\bibfnamefont {A.}~\bibnamefont {{Prosapio}}}, \bibinfo {author} {\bibfnamefont {C.}~\bibnamefont {{Rivetta}}}, \bibinfo {author} {\bibfnamefont {D.~P.}\ \bibnamefont {{Schneider}}}, \bibinfo {author} {\bibfnamefont {S.}~\bibnamefont {{Snedden}}},\ and\ \bibinfo {author}
  {\bibfnamefont {S.-i.}\ \bibnamefont {{Wang}}},\ }\href {https://doi.org/10.1086/500975} {\bibfield  {journal} {\bibinfo  {journal} {\aj}\ }\textbf {\bibinfo {volume} {131}},\ \bibinfo {pages} {2332} (\bibinfo {year} {2006})},\ \Eprint {https://arxiv.org/abs/astro-ph/0602326} {arXiv:astro-ph/0602326 [astro-ph]} \BibitemShut {NoStop}%
\bibitem [{\citenamefont {{Abazajian}}\ \emph {et~al.}(2009)\citenamefont {{Abazajian}}, \citenamefont {{Adelman-McCarthy}}, \citenamefont {{Ag{\"u}eros}}, \citenamefont {{Allam}}, \citenamefont {{Allende Prieto}}, \citenamefont {{An}}, \citenamefont {{Anderson}}, \citenamefont {{Anderson}}, \citenamefont {{Annis}}, \citenamefont {{Bahcall}}, \citenamefont {{Bailer-Jones}}, \citenamefont {{Barentine}}, \citenamefont {{Bassett}}, \citenamefont {{Becker}}, \citenamefont {{Beers}}, \citenamefont {{Bell}}, \citenamefont {{Belokurov}}, \citenamefont {{Berlind}}, \citenamefont {{Berman}}, \citenamefont {{Bernardi}}, \citenamefont {{Bickerton}}, \citenamefont {{Bizyaev}}, \citenamefont {{Blakeslee}}, \citenamefont {{Blanton}}, \citenamefont {{Bochanski}}, \citenamefont {{Boroski}}, \citenamefont {{Brewington}}, \citenamefont {{Brinchmann}}, \citenamefont {{Brinkmann}}, \citenamefont {{Brunner}}, \citenamefont {{Budav{\'a}ri}}, \citenamefont {{Carey}}, \citenamefont {{Carliles}}, \citenamefont {{Carr}}, \citenamefont
  {{Castander}}, \citenamefont {{Cinabro}}, \citenamefont {{Connolly}}, \citenamefont {{Csabai}}, \citenamefont {{Cunha}}, \citenamefont {{Czarapata}}, \citenamefont {{Davenport}}, \citenamefont {{de Haas}}, \citenamefont {{Dilday}}, \citenamefont {{Doi}}, \citenamefont {{Eisenstein}}, \citenamefont {{Evans}}, \citenamefont {{Evans}}, \citenamefont {{Fan}}, \citenamefont {{Friedman}}, \citenamefont {{Frieman}}, \citenamefont {{Fukugita}}, \citenamefont {{G{\"a}nsicke}}, \citenamefont {{Gates}}, \citenamefont {{Gillespie}}, \citenamefont {{Gilmore}}, \citenamefont {{Gonzalez}}, \citenamefont {{Gonzalez}}, \citenamefont {{Grebel}}, \citenamefont {{Gunn}}, \citenamefont {{Gy{\"o}ry}}, \citenamefont {{Hall}}, \citenamefont {{Harding}}, \citenamefont {{Harris}}, \citenamefont {{Harvanek}}, \citenamefont {{Hawley}}, \citenamefont {{Hayes}}, \citenamefont {{Heckman}}, \citenamefont {{Hendry}}, \citenamefont {{Hennessy}}, \citenamefont {{Hindsley}}, \citenamefont {{Hoblitt}}, \citenamefont {{Hogan}}, \citenamefont
  {{Hogg}}, \citenamefont {{Holtzman}}, \citenamefont {{Hyde}}, \citenamefont {{Ichikawa}}, \citenamefont {{Ichikawa}}, \citenamefont {{Im}}, \citenamefont {{Ivezi{\'c}}}, \citenamefont {{Jester}}, \citenamefont {{Jiang}}, \citenamefont {{Johnson}}, \citenamefont {{Jorgensen}}, \citenamefont {{Juri{\'c}}}, \citenamefont {{Kent}}, \citenamefont {{Kessler}}, \citenamefont {{Kleinman}}, \citenamefont {{Knapp}}, \citenamefont {{Konishi}}, \citenamefont {{Kron}}, \citenamefont {{Krzesinski}}, \citenamefont {{Kuropatkin}}, \citenamefont {{Lampeitl}}, \citenamefont {{Lebedeva}}, \citenamefont {{Lee}}, \citenamefont {{Lee}}, \citenamefont {{French Leger}}, \citenamefont {{L{\'e}pine}}, \citenamefont {{Li}}, \citenamefont {{Lima}}, \citenamefont {{Lin}}, \citenamefont {{Long}}, \citenamefont {{Loomis}}, \citenamefont {{Loveday}}, \citenamefont {{Lupton}}, \citenamefont {{Magnier}}, \citenamefont {{Malanushenko}}, \citenamefont {{Malanushenko}}, \citenamefont {{Mandelbaum}}, \citenamefont {{Margon}}, \citenamefont
  {{Marriner}}, \citenamefont {{Mart{\'\i}nez-Delgado}}, \citenamefont {{Matsubara}}, \citenamefont {{McGehee}}, \citenamefont {{McKay}}, \citenamefont {{Meiksin}}, \citenamefont {{Morrison}}, \citenamefont {{Mullally}}, \citenamefont {{Munn}}, \citenamefont {{Murphy}}, \citenamefont {{Nash}}, \citenamefont {{Nebot}}, \citenamefont {{Neilsen}}, \citenamefont {{Newberg}}, \citenamefont {{Newman}}, \citenamefont {{Nichol}}, \citenamefont {{Nicinski}}, \citenamefont {{Nieto-Santisteban}}, \citenamefont {{Nitta}}, \citenamefont {{Okamura}}, \citenamefont {{Oravetz}}, \citenamefont {{Ostriker}}, \citenamefont {{Owen}}, \citenamefont {{Padmanabhan}}, \citenamefont {{Pan}}, \citenamefont {{Park}}, \citenamefont {{Pauls}}, \citenamefont {{Peoples}}, \citenamefont {{Percival}}, \citenamefont {{Pier}}, \citenamefont {{Pope}}, \citenamefont {{Pourbaix}}, \citenamefont {{Price}}, \citenamefont {{Purger}}, \citenamefont {{Quinn}}, \citenamefont {{Raddick}}, \citenamefont {{Re Fiorentin}}, \citenamefont {{Richards}},
  \citenamefont {{Richmond}}, \citenamefont {{Riess}}, \citenamefont {{Rix}}, \citenamefont {{Rockosi}}, \citenamefont {{Sako}}, \citenamefont {{Schlegel}}, \citenamefont {{Schneider}}, \citenamefont {{Scholz}}, \citenamefont {{Schreiber}}, \citenamefont {{Schwope}}, \citenamefont {{Seljak}}, \citenamefont {{Sesar}}, \citenamefont {{Sheldon}}, \citenamefont {{Shimasaku}}, \citenamefont {{Sibley}}, \citenamefont {{Simmons}}, \citenamefont {{Sivarani}}, \citenamefont {{Allyn Smith}}, \citenamefont {{Smith}}, \citenamefont {{Smol{\v{c}}i{\'c}}}, \citenamefont {{Snedden}}, \citenamefont {{Stebbins}}, \citenamefont {{Steinmetz}}, \citenamefont {{Stoughton}}, \citenamefont {{Strauss}}, \citenamefont {{SubbaRao}}, \citenamefont {{Suto}}, \citenamefont {{Szalay}}, \citenamefont {{Szapudi}}, \citenamefont {{Szkody}}, \citenamefont {{Tanaka}}, \citenamefont {{Tegmark}}, \citenamefont {{Teodoro}}, \citenamefont {{Thakar}}, \citenamefont {{Tremonti}}, \citenamefont {{Tucker}}, \citenamefont {{Uomoto}}, \citenamefont
  {{Vanden Berk}}, \citenamefont {{Vandenberg}}, \citenamefont {{Vidrih}}, \citenamefont {{Vogeley}}, \citenamefont {{Voges}}, \citenamefont {{Vogt}}, \citenamefont {{Wadadekar}}, \citenamefont {{Watters}}, \citenamefont {{Weinberg}}, \citenamefont {{West}}, \citenamefont {{White}}, \citenamefont {{Wilhite}}, \citenamefont {{Wonders}}, \citenamefont {{Yanny}}, \citenamefont {{Yocum}}, \citenamefont {{York}}, \citenamefont {{Zehavi}}, \citenamefont {{Zibetti}},\ and\ \citenamefont {{Zucker}}}]{ABAZAJIAN2009}%
  \BibitemOpen
  \bibfield  {author} {\bibinfo {author} {\bibfnamefont {K.~N.}\ \bibnamefont {{Abazajian}}}, \bibinfo {author} {\bibfnamefont {J.~K.}\ \bibnamefont {{Adelman-McCarthy}}}, \bibinfo {author} {\bibfnamefont {M.~A.}\ \bibnamefont {{Ag{\"u}eros}}}, \bibinfo {author} {\bibfnamefont {S.~S.}\ \bibnamefont {{Allam}}}, \bibinfo {author} {\bibfnamefont {C.}~\bibnamefont {{Allende Prieto}}}, \bibinfo {author} {\bibfnamefont {D.}~\bibnamefont {{An}}}, \bibinfo {author} {\bibfnamefont {K.~S.~J.}\ \bibnamefont {{Anderson}}}, \bibinfo {author} {\bibfnamefont {S.~F.}\ \bibnamefont {{Anderson}}}, \bibinfo {author} {\bibfnamefont {J.}~\bibnamefont {{Annis}}}, \bibinfo {author} {\bibfnamefont {N.~A.}\ \bibnamefont {{Bahcall}}}, \bibinfo {author} {\bibfnamefont {C.~A.~L.}\ \bibnamefont {{Bailer-Jones}}}, \bibinfo {author} {\bibfnamefont {J.~C.}\ \bibnamefont {{Barentine}}}, \bibinfo {author} {\bibfnamefont {B.~A.}\ \bibnamefont {{Bassett}}}, \bibinfo {author} {\bibfnamefont {A.~C.}\ \bibnamefont {{Becker}}}, \bibinfo {author}
  {\bibfnamefont {T.~C.}\ \bibnamefont {{Beers}}}, \bibinfo {author} {\bibfnamefont {E.~F.}\ \bibnamefont {{Bell}}}, \bibinfo {author} {\bibfnamefont {V.}~\bibnamefont {{Belokurov}}}, \bibinfo {author} {\bibfnamefont {A.~A.}\ \bibnamefont {{Berlind}}}, \bibinfo {author} {\bibfnamefont {E.~F.}\ \bibnamefont {{Berman}}}, \bibinfo {author} {\bibfnamefont {M.}~\bibnamefont {{Bernardi}}}, \bibinfo {author} {\bibfnamefont {S.~J.}\ \bibnamefont {{Bickerton}}}, \bibinfo {author} {\bibfnamefont {D.}~\bibnamefont {{Bizyaev}}}, \bibinfo {author} {\bibfnamefont {J.~P.}\ \bibnamefont {{Blakeslee}}}, \bibinfo {author} {\bibfnamefont {M.~R.}\ \bibnamefont {{Blanton}}}, \bibinfo {author} {\bibfnamefont {J.~J.}\ \bibnamefont {{Bochanski}}}, \bibinfo {author} {\bibfnamefont {W.~N.}\ \bibnamefont {{Boroski}}}, \bibinfo {author} {\bibfnamefont {H.~J.}\ \bibnamefont {{Brewington}}}, \bibinfo {author} {\bibfnamefont {J.}~\bibnamefont {{Brinchmann}}}, \bibinfo {author} {\bibfnamefont {J.}~\bibnamefont {{Brinkmann}}}, \bibinfo
  {author} {\bibfnamefont {R.~J.}\ \bibnamefont {{Brunner}}}, \bibinfo {author} {\bibfnamefont {T.}~\bibnamefont {{Budav{\'a}ri}}}, \bibinfo {author} {\bibfnamefont {L.~N.}\ \bibnamefont {{Carey}}}, \bibinfo {author} {\bibfnamefont {S.}~\bibnamefont {{Carliles}}}, \bibinfo {author} {\bibfnamefont {M.~A.}\ \bibnamefont {{Carr}}}, \bibinfo {author} {\bibfnamefont {F.~J.}\ \bibnamefont {{Castander}}}, \bibinfo {author} {\bibfnamefont {D.}~\bibnamefont {{Cinabro}}}, \bibinfo {author} {\bibfnamefont {A.~J.}\ \bibnamefont {{Connolly}}}, \bibinfo {author} {\bibfnamefont {I.}~\bibnamefont {{Csabai}}}, \bibinfo {author} {\bibfnamefont {C.~E.}\ \bibnamefont {{Cunha}}}, \bibinfo {author} {\bibfnamefont {P.~C.}\ \bibnamefont {{Czarapata}}}, \bibinfo {author} {\bibfnamefont {J.~R.~A.}\ \bibnamefont {{Davenport}}}, \bibinfo {author} {\bibfnamefont {E.}~\bibnamefont {{de Haas}}}, \bibinfo {author} {\bibfnamefont {B.}~\bibnamefont {{Dilday}}}, \bibinfo {author} {\bibfnamefont {M.}~\bibnamefont {{Doi}}}, \bibinfo {author}
  {\bibfnamefont {D.~J.}\ \bibnamefont {{Eisenstein}}}, \bibinfo {author} {\bibfnamefont {M.~L.}\ \bibnamefont {{Evans}}}, \bibinfo {author} {\bibfnamefont {N.~W.}\ \bibnamefont {{Evans}}}, \bibinfo {author} {\bibfnamefont {X.}~\bibnamefont {{Fan}}}, \bibinfo {author} {\bibfnamefont {S.~D.}\ \bibnamefont {{Friedman}}}, \bibinfo {author} {\bibfnamefont {J.~A.}\ \bibnamefont {{Frieman}}}, \bibinfo {author} {\bibfnamefont {M.}~\bibnamefont {{Fukugita}}}, \bibinfo {author} {\bibfnamefont {B.~T.}\ \bibnamefont {{G{\"a}nsicke}}}, \bibinfo {author} {\bibfnamefont {E.}~\bibnamefont {{Gates}}}, \bibinfo {author} {\bibfnamefont {B.}~\bibnamefont {{Gillespie}}}, \bibinfo {author} {\bibfnamefont {G.}~\bibnamefont {{Gilmore}}}, \bibinfo {author} {\bibfnamefont {B.}~\bibnamefont {{Gonzalez}}}, \bibinfo {author} {\bibfnamefont {C.~F.}\ \bibnamefont {{Gonzalez}}}, \bibinfo {author} {\bibfnamefont {E.~K.}\ \bibnamefont {{Grebel}}}, \bibinfo {author} {\bibfnamefont {J.~E.}\ \bibnamefont {{Gunn}}}, \bibinfo {author}
  {\bibfnamefont {Z.}~\bibnamefont {{Gy{\"o}ry}}}, \bibinfo {author} {\bibfnamefont {P.~B.}\ \bibnamefont {{Hall}}}, \bibinfo {author} {\bibfnamefont {P.}~\bibnamefont {{Harding}}}, \bibinfo {author} {\bibfnamefont {F.~H.}\ \bibnamefont {{Harris}}}, \bibinfo {author} {\bibfnamefont {M.}~\bibnamefont {{Harvanek}}}, \bibinfo {author} {\bibfnamefont {S.~L.}\ \bibnamefont {{Hawley}}}, \bibinfo {author} {\bibfnamefont {J.~J.~E.}\ \bibnamefont {{Hayes}}}, \bibinfo {author} {\bibfnamefont {T.~M.}\ \bibnamefont {{Heckman}}}, \bibinfo {author} {\bibfnamefont {J.~S.}\ \bibnamefont {{Hendry}}}, \bibinfo {author} {\bibfnamefont {G.~S.}\ \bibnamefont {{Hennessy}}}, \bibinfo {author} {\bibfnamefont {R.~B.}\ \bibnamefont {{Hindsley}}}, \bibinfo {author} {\bibfnamefont {J.}~\bibnamefont {{Hoblitt}}}, \bibinfo {author} {\bibfnamefont {C.~J.}\ \bibnamefont {{Hogan}}}, \bibinfo {author} {\bibfnamefont {D.~W.}\ \bibnamefont {{Hogg}}}, \bibinfo {author} {\bibfnamefont {J.~A.}\ \bibnamefont {{Holtzman}}}, \bibinfo {author}
  {\bibfnamefont {J.~B.}\ \bibnamefont {{Hyde}}}, \bibinfo {author} {\bibfnamefont {S.-i.}\ \bibnamefont {{Ichikawa}}}, \bibinfo {author} {\bibfnamefont {T.}~\bibnamefont {{Ichikawa}}}, \bibinfo {author} {\bibfnamefont {M.}~\bibnamefont {{Im}}}, \bibinfo {author} {\bibfnamefont {{\v{Z}}.}~\bibnamefont {{Ivezi{\'c}}}}, \bibinfo {author} {\bibfnamefont {S.}~\bibnamefont {{Jester}}}, \bibinfo {author} {\bibfnamefont {L.}~\bibnamefont {{Jiang}}}, \bibinfo {author} {\bibfnamefont {J.~A.}\ \bibnamefont {{Johnson}}}, \bibinfo {author} {\bibfnamefont {A.~M.}\ \bibnamefont {{Jorgensen}}}, \bibinfo {author} {\bibfnamefont {M.}~\bibnamefont {{Juri{\'c}}}}, \bibinfo {author} {\bibfnamefont {S.~M.}\ \bibnamefont {{Kent}}}, \bibinfo {author} {\bibfnamefont {R.}~\bibnamefont {{Kessler}}}, \bibinfo {author} {\bibfnamefont {S.~J.}\ \bibnamefont {{Kleinman}}}, \bibinfo {author} {\bibfnamefont {G.~R.}\ \bibnamefont {{Knapp}}}, \bibinfo {author} {\bibfnamefont {K.}~\bibnamefont {{Konishi}}}, \bibinfo {author} {\bibfnamefont
  {R.~G.}\ \bibnamefont {{Kron}}}, \bibinfo {author} {\bibfnamefont {J.}~\bibnamefont {{Krzesinski}}}, \bibinfo {author} {\bibfnamefont {N.}~\bibnamefont {{Kuropatkin}}}, \bibinfo {author} {\bibfnamefont {H.}~\bibnamefont {{Lampeitl}}}, \bibinfo {author} {\bibfnamefont {S.}~\bibnamefont {{Lebedeva}}}, \bibinfo {author} {\bibfnamefont {M.~G.}\ \bibnamefont {{Lee}}}, \bibinfo {author} {\bibfnamefont {Y.~S.}\ \bibnamefont {{Lee}}}, \bibinfo {author} {\bibfnamefont {R.}~\bibnamefont {{French Leger}}}, \bibinfo {author} {\bibfnamefont {S.}~\bibnamefont {{L{\'e}pine}}}, \bibinfo {author} {\bibfnamefont {N.}~\bibnamefont {{Li}}}, \bibinfo {author} {\bibfnamefont {M.}~\bibnamefont {{Lima}}}, \bibinfo {author} {\bibfnamefont {H.}~\bibnamefont {{Lin}}}, \bibinfo {author} {\bibfnamefont {D.~C.}\ \bibnamefont {{Long}}}, \bibinfo {author} {\bibfnamefont {C.~P.}\ \bibnamefont {{Loomis}}}, \bibinfo {author} {\bibfnamefont {J.}~\bibnamefont {{Loveday}}}, \bibinfo {author} {\bibfnamefont {R.~H.}\ \bibnamefont {{Lupton}}},
  \bibinfo {author} {\bibfnamefont {E.}~\bibnamefont {{Magnier}}}, \bibinfo {author} {\bibfnamefont {O.}~\bibnamefont {{Malanushenko}}}, \bibinfo {author} {\bibfnamefont {V.}~\bibnamefont {{Malanushenko}}}, \bibinfo {author} {\bibfnamefont {R.}~\bibnamefont {{Mandelbaum}}}, \bibinfo {author} {\bibfnamefont {B.}~\bibnamefont {{Margon}}}, \bibinfo {author} {\bibfnamefont {J.~P.}\ \bibnamefont {{Marriner}}}, \bibinfo {author} {\bibfnamefont {D.}~\bibnamefont {{Mart{\'\i}nez-Delgado}}}, \bibinfo {author} {\bibfnamefont {T.}~\bibnamefont {{Matsubara}}}, \bibinfo {author} {\bibfnamefont {P.~M.}\ \bibnamefont {{McGehee}}}, \bibinfo {author} {\bibfnamefont {T.~A.}\ \bibnamefont {{McKay}}}, \bibinfo {author} {\bibfnamefont {A.}~\bibnamefont {{Meiksin}}}, \bibinfo {author} {\bibfnamefont {H.~L.}\ \bibnamefont {{Morrison}}}, \bibinfo {author} {\bibfnamefont {F.}~\bibnamefont {{Mullally}}}, \bibinfo {author} {\bibfnamefont {J.~A.}\ \bibnamefont {{Munn}}}, \bibinfo {author} {\bibfnamefont {T.}~\bibnamefont {{Murphy}}},
  \bibinfo {author} {\bibfnamefont {T.}~\bibnamefont {{Nash}}}, \bibinfo {author} {\bibfnamefont {A.}~\bibnamefont {{Nebot}}}, \bibinfo {author} {\bibfnamefont {J.}~\bibnamefont {{Neilsen}}, \bibfnamefont {Eric~H.}}, \bibinfo {author} {\bibfnamefont {H.~J.}\ \bibnamefont {{Newberg}}}, \bibinfo {author} {\bibfnamefont {P.~R.}\ \bibnamefont {{Newman}}}, \bibinfo {author} {\bibfnamefont {R.~C.}\ \bibnamefont {{Nichol}}}, \bibinfo {author} {\bibfnamefont {T.}~\bibnamefont {{Nicinski}}}, \bibinfo {author} {\bibfnamefont {M.}~\bibnamefont {{Nieto-Santisteban}}}, \bibinfo {author} {\bibfnamefont {A.}~\bibnamefont {{Nitta}}}, \bibinfo {author} {\bibfnamefont {S.}~\bibnamefont {{Okamura}}}, \bibinfo {author} {\bibfnamefont {D.~J.}\ \bibnamefont {{Oravetz}}}, \bibinfo {author} {\bibfnamefont {J.~P.}\ \bibnamefont {{Ostriker}}}, \bibinfo {author} {\bibfnamefont {R.}~\bibnamefont {{Owen}}}, \bibinfo {author} {\bibfnamefont {N.}~\bibnamefont {{Padmanabhan}}}, \bibinfo {author} {\bibfnamefont {K.}~\bibnamefont {{Pan}}},
  \bibinfo {author} {\bibfnamefont {C.}~\bibnamefont {{Park}}}, \bibinfo {author} {\bibfnamefont {G.}~\bibnamefont {{Pauls}}}, \bibinfo {author} {\bibfnamefont {J.}~\bibnamefont {{Peoples}}, \bibfnamefont {John}}, \bibinfo {author} {\bibfnamefont {W.~J.}\ \bibnamefont {{Percival}}}, \bibinfo {author} {\bibfnamefont {J.~R.}\ \bibnamefont {{Pier}}}, \bibinfo {author} {\bibfnamefont {A.~C.}\ \bibnamefont {{Pope}}}, \bibinfo {author} {\bibfnamefont {D.}~\bibnamefont {{Pourbaix}}}, \bibinfo {author} {\bibfnamefont {P.~A.}\ \bibnamefont {{Price}}}, \bibinfo {author} {\bibfnamefont {N.}~\bibnamefont {{Purger}}}, \bibinfo {author} {\bibfnamefont {T.}~\bibnamefont {{Quinn}}}, \bibinfo {author} {\bibfnamefont {M.~J.}\ \bibnamefont {{Raddick}}}, \bibinfo {author} {\bibfnamefont {P.}~\bibnamefont {{Re Fiorentin}}}, \bibinfo {author} {\bibfnamefont {G.~T.}\ \bibnamefont {{Richards}}}, \bibinfo {author} {\bibfnamefont {M.~W.}\ \bibnamefont {{Richmond}}}, \bibinfo {author} {\bibfnamefont {A.~G.}\ \bibnamefont {{Riess}}},
  \bibinfo {author} {\bibfnamefont {H.-W.}\ \bibnamefont {{Rix}}}, \bibinfo {author} {\bibfnamefont {C.~M.}\ \bibnamefont {{Rockosi}}}, \bibinfo {author} {\bibfnamefont {M.}~\bibnamefont {{Sako}}}, \bibinfo {author} {\bibfnamefont {D.~J.}\ \bibnamefont {{Schlegel}}}, \bibinfo {author} {\bibfnamefont {D.~P.}\ \bibnamefont {{Schneider}}}, \bibinfo {author} {\bibfnamefont {R.-D.}\ \bibnamefont {{Scholz}}}, \bibinfo {author} {\bibfnamefont {M.~R.}\ \bibnamefont {{Schreiber}}}, \bibinfo {author} {\bibfnamefont {A.~D.}\ \bibnamefont {{Schwope}}}, \bibinfo {author} {\bibfnamefont {U.}~\bibnamefont {{Seljak}}}, \bibinfo {author} {\bibfnamefont {B.}~\bibnamefont {{Sesar}}}, \bibinfo {author} {\bibfnamefont {E.}~\bibnamefont {{Sheldon}}}, \bibinfo {author} {\bibfnamefont {K.}~\bibnamefont {{Shimasaku}}}, \bibinfo {author} {\bibfnamefont {V.~C.}\ \bibnamefont {{Sibley}}}, \bibinfo {author} {\bibfnamefont {A.~E.}\ \bibnamefont {{Simmons}}}, \bibinfo {author} {\bibfnamefont {T.}~\bibnamefont {{Sivarani}}}, \bibinfo
  {author} {\bibfnamefont {J.}~\bibnamefont {{Allyn Smith}}}, \bibinfo {author} {\bibfnamefont {M.~C.}\ \bibnamefont {{Smith}}}, \bibinfo {author} {\bibfnamefont {V.}~\bibnamefont {{Smol{\v{c}}i{\'c}}}}, \bibinfo {author} {\bibfnamefont {S.~A.}\ \bibnamefont {{Snedden}}}, \bibinfo {author} {\bibfnamefont {A.}~\bibnamefont {{Stebbins}}}, \bibinfo {author} {\bibfnamefont {M.}~\bibnamefont {{Steinmetz}}}, \bibinfo {author} {\bibfnamefont {C.}~\bibnamefont {{Stoughton}}}, \bibinfo {author} {\bibfnamefont {M.~A.}\ \bibnamefont {{Strauss}}}, \bibinfo {author} {\bibfnamefont {M.}~\bibnamefont {{SubbaRao}}}, \bibinfo {author} {\bibfnamefont {Y.}~\bibnamefont {{Suto}}}, \bibinfo {author} {\bibfnamefont {A.~S.}\ \bibnamefont {{Szalay}}}, \bibinfo {author} {\bibfnamefont {I.}~\bibnamefont {{Szapudi}}}, \bibinfo {author} {\bibfnamefont {P.}~\bibnamefont {{Szkody}}}, \bibinfo {author} {\bibfnamefont {M.}~\bibnamefont {{Tanaka}}}, \bibinfo {author} {\bibfnamefont {M.}~\bibnamefont {{Tegmark}}}, \bibinfo {author}
  {\bibfnamefont {L.~F.~A.}\ \bibnamefont {{Teodoro}}}, \bibinfo {author} {\bibfnamefont {A.~R.}\ \bibnamefont {{Thakar}}}, \bibinfo {author} {\bibfnamefont {C.~A.}\ \bibnamefont {{Tremonti}}}, \bibinfo {author} {\bibfnamefont {D.~L.}\ \bibnamefont {{Tucker}}}, \bibinfo {author} {\bibfnamefont {A.}~\bibnamefont {{Uomoto}}}, \bibinfo {author} {\bibfnamefont {D.~E.}\ \bibnamefont {{Vanden Berk}}}, \bibinfo {author} {\bibfnamefont {J.}~\bibnamefont {{Vandenberg}}}, \bibinfo {author} {\bibfnamefont {S.}~\bibnamefont {{Vidrih}}}, \bibinfo {author} {\bibfnamefont {M.~S.}\ \bibnamefont {{Vogeley}}}, \bibinfo {author} {\bibfnamefont {W.}~\bibnamefont {{Voges}}}, \bibinfo {author} {\bibfnamefont {N.~P.}\ \bibnamefont {{Vogt}}}, \bibinfo {author} {\bibfnamefont {Y.}~\bibnamefont {{Wadadekar}}}, \bibinfo {author} {\bibfnamefont {S.}~\bibnamefont {{Watters}}}, \bibinfo {author} {\bibfnamefont {D.~H.}\ \bibnamefont {{Weinberg}}}, \bibinfo {author} {\bibfnamefont {A.~A.}\ \bibnamefont {{West}}}, \bibinfo {author}
  {\bibfnamefont {S.~D.~M.}\ \bibnamefont {{White}}}, \bibinfo {author} {\bibfnamefont {B.~C.}\ \bibnamefont {{Wilhite}}}, \bibinfo {author} {\bibfnamefont {A.~C.}\ \bibnamefont {{Wonders}}}, \bibinfo {author} {\bibfnamefont {B.}~\bibnamefont {{Yanny}}}, \bibinfo {author} {\bibfnamefont {D.~R.}\ \bibnamefont {{Yocum}}}, \bibinfo {author} {\bibfnamefont {D.~G.}\ \bibnamefont {{York}}}, \bibinfo {author} {\bibfnamefont {I.}~\bibnamefont {{Zehavi}}}, \bibinfo {author} {\bibfnamefont {S.}~\bibnamefont {{Zibetti}}},\ and\ \bibinfo {author} {\bibfnamefont {D.~B.}\ \bibnamefont {{Zucker}}},\ }\href {https://doi.org/10.1088/0067-0049/182/2/543} {\bibfield  {journal} {\bibinfo  {journal} {\apjs}\ }\textbf {\bibinfo {volume} {182}},\ \bibinfo {pages} {543} (\bibinfo {year} {2009})},\ \Eprint {https://arxiv.org/abs/0812.0649} {arXiv:0812.0649 [astro-ph]} \BibitemShut {NoStop}%
\bibitem [{\citenamefont {{Li}}\ \emph {et~al.}(2022)\citenamefont {{Li}}, \citenamefont {{Miyatake}}, \citenamefont {{Luo}}, \citenamefont {{More}}, \citenamefont {{Oguri}}, \citenamefont {{Hamana}}, \citenamefont {{Mandelbaum}}, \citenamefont {{Shirasaki}}, \citenamefont {{Takada}}, \citenamefont {{Armstrong}} \emph {et~al.}}]{Li2022}%
  \BibitemOpen
  \bibfield  {author} {\bibinfo {author} {\bibfnamefont {X.}~\bibnamefont {{Li}}}, \bibinfo {author} {\bibfnamefont {H.}~\bibnamefont {{Miyatake}}}, \bibinfo {author} {\bibfnamefont {W.}~\bibnamefont {{Luo}}}, \bibinfo {author} {\bibfnamefont {S.}~\bibnamefont {{More}}}, \bibinfo {author} {\bibfnamefont {M.}~\bibnamefont {{Oguri}}}, \bibinfo {author} {\bibfnamefont {T.}~\bibnamefont {{Hamana}}}, \bibinfo {author} {\bibfnamefont {R.}~\bibnamefont {{Mandelbaum}}}, \bibinfo {author} {\bibfnamefont {M.}~\bibnamefont {{Shirasaki}}}, \bibinfo {author} {\bibfnamefont {M.}~\bibnamefont {{Takada}}}, \bibinfo {author} {\bibfnamefont {R.}~\bibnamefont {{Armstrong}}}, \emph {et~al.},\ }\href {https://doi.org/10.1093/pasj/psac006} {\bibfield  {journal} {\bibinfo  {journal} {\pasj}\ }\textbf {\bibinfo {volume} {74}},\ \bibinfo {pages} {421} (\bibinfo {year} {2022})},\ \Eprint {https://arxiv.org/abs/2107.00136} {arXiv:2107.00136 [astro-ph.CO]} \BibitemShut {NoStop}%
\bibitem [{\citenamefont {{Aihara}}\ \emph {et~al.}(2022)\citenamefont {{Aihara}}, \citenamefont {{AlSayyad}}, \citenamefont {{Ando}}, \citenamefont {{Armstrong}}, \citenamefont {{Bosch}}, \citenamefont {{Egami}}, \citenamefont {{Furusawa}}, \citenamefont {{Furusawa}}, \citenamefont {{Harasawa}}, \citenamefont {{Harikane}} \emph {et~al.}}]{Aihara2022}%
  \BibitemOpen
  \bibfield  {author} {\bibinfo {author} {\bibfnamefont {H.}~\bibnamefont {{Aihara}}}, \bibinfo {author} {\bibfnamefont {Y.}~\bibnamefont {{AlSayyad}}}, \bibinfo {author} {\bibfnamefont {M.}~\bibnamefont {{Ando}}}, \bibinfo {author} {\bibfnamefont {R.}~\bibnamefont {{Armstrong}}}, \bibinfo {author} {\bibfnamefont {J.}~\bibnamefont {{Bosch}}}, \bibinfo {author} {\bibfnamefont {E.}~\bibnamefont {{Egami}}}, \bibinfo {author} {\bibfnamefont {H.}~\bibnamefont {{Furusawa}}}, \bibinfo {author} {\bibfnamefont {J.}~\bibnamefont {{Furusawa}}}, \bibinfo {author} {\bibfnamefont {S.}~\bibnamefont {{Harasawa}}}, \bibinfo {author} {\bibfnamefont {Y.}~\bibnamefont {{Harikane}}}, \emph {et~al.},\ }\href {https://doi.org/10.1093/pasj/psab122} {\bibfield  {journal} {\bibinfo  {journal} {\pasj}\ }\textbf {\bibinfo {volume} {74}},\ \bibinfo {pages} {247} (\bibinfo {year} {2022})},\ \Eprint {https://arxiv.org/abs/2108.13045} {arXiv:2108.13045 [astro-ph.IM]} \BibitemShut {NoStop}%
\bibitem [{\citenamefont {{Aihara}}\ \emph {et~al.}(2019)\citenamefont {{Aihara}}, \citenamefont {{AlSayyad}}, \citenamefont {{Ando}}, \citenamefont {{Armstrong}}, \citenamefont {{Bosch}}, \citenamefont {{Egami}}, \citenamefont {{Furusawa}}, \citenamefont {{Furusawa}}, \citenamefont {{Goulding}}, \citenamefont {{Harikane}} \emph {et~al.}}]{Aihara2019}%
  \BibitemOpen
  \bibfield  {author} {\bibinfo {author} {\bibfnamefont {H.}~\bibnamefont {{Aihara}}}, \bibinfo {author} {\bibfnamefont {Y.}~\bibnamefont {{AlSayyad}}}, \bibinfo {author} {\bibfnamefont {M.}~\bibnamefont {{Ando}}}, \bibinfo {author} {\bibfnamefont {R.}~\bibnamefont {{Armstrong}}}, \bibinfo {author} {\bibfnamefont {J.}~\bibnamefont {{Bosch}}}, \bibinfo {author} {\bibfnamefont {E.}~\bibnamefont {{Egami}}}, \bibinfo {author} {\bibfnamefont {H.}~\bibnamefont {{Furusawa}}}, \bibinfo {author} {\bibfnamefont {J.}~\bibnamefont {{Furusawa}}}, \bibinfo {author} {\bibfnamefont {A.}~\bibnamefont {{Goulding}}}, \bibinfo {author} {\bibfnamefont {Y.}~\bibnamefont {{Harikane}}}, \emph {et~al.},\ }\href {https://doi.org/10.1093/pasj/psz103} {\bibfield  {journal} {\bibinfo  {journal} {\pasj}\ }\textbf {\bibinfo {volume} {71}},\ \bibinfo {eid} {114} (\bibinfo {year} {2019})},\ \Eprint {https://arxiv.org/abs/1905.12221} {arXiv:1905.12221 [astro-ph.IM]} \BibitemShut {NoStop}%
\bibitem [{\citenamefont {{Hirata}}\ and\ \citenamefont {{Seljak}}(2003)}]{Hirata2003}%
  \BibitemOpen
  \bibfield  {author} {\bibinfo {author} {\bibfnamefont {C.}~\bibnamefont {{Hirata}}}\ and\ \bibinfo {author} {\bibfnamefont {U.}~\bibnamefont {{Seljak}}},\ }\href {https://doi.org/10.1046/j.1365-8711.2003.06683.x} {\bibfield  {journal} {\bibinfo  {journal} {\mnras}\ }\textbf {\bibinfo {volume} {343}},\ \bibinfo {pages} {459} (\bibinfo {year} {2003})},\ \Eprint {https://arxiv.org/abs/astro-ph/0301054} {arXiv:astro-ph/0301054 [astro-ph]} \BibitemShut {NoStop}%
\bibitem [{\citenamefont {Mandelbaum}\ \emph {et~al.}(2005)\citenamefont {Mandelbaum}, \citenamefont {Hirata}, \citenamefont {Seljak}, \citenamefont {Guzik}, \citenamefont {Padmanabhan}, \citenamefont {Blake}, \citenamefont {Blanton}, \citenamefont {Lupton},\ and\ \citenamefont {Brinkmann}}]{Mandelbaum2005}%
  \BibitemOpen
  \bibfield  {author} {\bibinfo {author} {\bibfnamefont {R.}~\bibnamefont {Mandelbaum}}, \bibinfo {author} {\bibfnamefont {C.~M.}\ \bibnamefont {Hirata}}, \bibinfo {author} {\bibfnamefont {U.}~\bibnamefont {Seljak}}, \bibinfo {author} {\bibfnamefont {J.}~\bibnamefont {Guzik}}, \bibinfo {author} {\bibfnamefont {N.}~\bibnamefont {Padmanabhan}}, \bibinfo {author} {\bibfnamefont {C.}~\bibnamefont {Blake}}, \bibinfo {author} {\bibfnamefont {M.~R.}\ \bibnamefont {Blanton}}, \bibinfo {author} {\bibfnamefont {R.}~\bibnamefont {Lupton}},\ and\ \bibinfo {author} {\bibfnamefont {J.}~\bibnamefont {Brinkmann}},\ }\href {https://doi.org/10.1111/j.1365-2966.2005.09282.x} {\bibfield  {journal} {\bibinfo  {journal} {Mon. Not. Roy. Astron. Soc.}\ }\textbf {\bibinfo {volume} {361}},\ \bibinfo {pages} {1287} (\bibinfo {year} {2005})},\ \Eprint {https://arxiv.org/abs/astro-ph/0501201} {arXiv:astro-ph/0501201} \BibitemShut {NoStop}%
\bibitem [{\citenamefont {{Mandelbaum}}\ \emph {et~al.}(2018)\citenamefont {{Mandelbaum}}, \citenamefont {{Lanusse}}, \citenamefont {{Leauthaud}}, \citenamefont {{Armstrong}}, \citenamefont {{Simet}}, \citenamefont {{Miyatake}}, \citenamefont {{Meyers}}, \citenamefont {{Bosch}}, \citenamefont {{Murata}}, \citenamefont {{Miyazaki}} \emph {et~al.}}]{Mandelbaum2018_image}%
  \BibitemOpen
  \bibfield  {author} {\bibinfo {author} {\bibfnamefont {R.}~\bibnamefont {{Mandelbaum}}}, \bibinfo {author} {\bibfnamefont {F.}~\bibnamefont {{Lanusse}}}, \bibinfo {author} {\bibfnamefont {A.}~\bibnamefont {{Leauthaud}}}, \bibinfo {author} {\bibfnamefont {R.}~\bibnamefont {{Armstrong}}}, \bibinfo {author} {\bibfnamefont {M.}~\bibnamefont {{Simet}}}, \bibinfo {author} {\bibfnamefont {H.}~\bibnamefont {{Miyatake}}}, \bibinfo {author} {\bibfnamefont {J.~E.}\ \bibnamefont {{Meyers}}}, \bibinfo {author} {\bibfnamefont {J.}~\bibnamefont {{Bosch}}}, \bibinfo {author} {\bibfnamefont {R.}~\bibnamefont {{Murata}}}, \bibinfo {author} {\bibfnamefont {S.}~\bibnamefont {{Miyazaki}}}, \emph {et~al.},\ }\href {https://doi.org/10.1093/mnras/sty2420} {\bibfield  {journal} {\bibinfo  {journal} {\mnras}\ }\textbf {\bibinfo {volume} {481}},\ \bibinfo {pages} {3170} (\bibinfo {year} {2018})},\ \Eprint {https://arxiv.org/abs/1710.00885} {arXiv:1710.00885 [astro-ph.CO]} \BibitemShut {NoStop}%
\bibitem [{\citenamefont {{Rowe}}\ \emph {et~al.}(2015)\citenamefont {{Rowe}}, \citenamefont {{Jarvis}}, \citenamefont {{Mandelbaum}}, \citenamefont {{Bernstein}}, \citenamefont {{Bosch}}, \citenamefont {{Simet}}, \citenamefont {{Meyers}}, \citenamefont {{Kacprzak}}, \citenamefont {{Nakajima}}, \citenamefont {{Zuntz}}, \citenamefont {{Miyatake}}, \citenamefont {{Dietrich}}, \citenamefont {{Armstrong}}, \citenamefont {{Melchior}},\ and\ \citenamefont {{Gill}}}]{Rowe2014}%
  \BibitemOpen
  \bibfield  {author} {\bibinfo {author} {\bibfnamefont {B.~T.~P.}\ \bibnamefont {{Rowe}}}, \bibinfo {author} {\bibfnamefont {M.}~\bibnamefont {{Jarvis}}}, \bibinfo {author} {\bibfnamefont {R.}~\bibnamefont {{Mandelbaum}}}, \bibinfo {author} {\bibfnamefont {G.~M.}\ \bibnamefont {{Bernstein}}}, \bibinfo {author} {\bibfnamefont {J.}~\bibnamefont {{Bosch}}}, \bibinfo {author} {\bibfnamefont {M.}~\bibnamefont {{Simet}}}, \bibinfo {author} {\bibfnamefont {J.~E.}\ \bibnamefont {{Meyers}}}, \bibinfo {author} {\bibfnamefont {T.}~\bibnamefont {{Kacprzak}}}, \bibinfo {author} {\bibfnamefont {R.}~\bibnamefont {{Nakajima}}}, \bibinfo {author} {\bibfnamefont {J.}~\bibnamefont {{Zuntz}}}, \bibinfo {author} {\bibfnamefont {H.}~\bibnamefont {{Miyatake}}}, \bibinfo {author} {\bibfnamefont {J.~P.}\ \bibnamefont {{Dietrich}}}, \bibinfo {author} {\bibfnamefont {R.}~\bibnamefont {{Armstrong}}}, \bibinfo {author} {\bibfnamefont {P.}~\bibnamefont {{Melchior}}},\ and\ \bibinfo {author} {\bibfnamefont {M.~S.~S.}\ \bibnamefont
  {{Gill}}},\ }\href {https://doi.org/10.1016/j.ascom.2015.02.002} {\bibfield  {journal} {\bibinfo  {journal} {Astronomy and Computing}\ }\textbf {\bibinfo {volume} {10}},\ \bibinfo {pages} {121} (\bibinfo {year} {2015})},\ \Eprint {https://arxiv.org/abs/1407.7676} {arXiv:1407.7676 [astro-ph.IM]} \BibitemShut {NoStop}%
\bibitem [{\citenamefont {{Zhang}}\ \emph {et~al.}(2023{\natexlab{a}})\citenamefont {{Zhang}}, \citenamefont {{Li}}, \citenamefont {{Dalal}}, \citenamefont {{Mandelbaum}}, \citenamefont {{Strauss}}, \citenamefont {{Kannawadi}}, \citenamefont {{Miyatake}}, \citenamefont {{Nicola}}, \citenamefont {{Malag{\'o}n}}, \citenamefont {{Shirasaki}}, \citenamefont {{Sugiyama}}, \citenamefont {{Takada}},\ and\ \citenamefont {{More}}}]{Zhang2023b}%
  \BibitemOpen
  \bibfield  {author} {\bibinfo {author} {\bibfnamefont {T.}~\bibnamefont {{Zhang}}}, \bibinfo {author} {\bibfnamefont {X.}~\bibnamefont {{Li}}}, \bibinfo {author} {\bibfnamefont {R.}~\bibnamefont {{Dalal}}}, \bibinfo {author} {\bibfnamefont {R.}~\bibnamefont {{Mandelbaum}}}, \bibinfo {author} {\bibfnamefont {M.~A.}\ \bibnamefont {{Strauss}}}, \bibinfo {author} {\bibfnamefont {A.}~\bibnamefont {{Kannawadi}}}, \bibinfo {author} {\bibfnamefont {H.}~\bibnamefont {{Miyatake}}}, \bibinfo {author} {\bibfnamefont {A.}~\bibnamefont {{Nicola}}}, \bibinfo {author} {\bibfnamefont {A.~A.~P.}\ \bibnamefont {{Malag{\'o}n}}}, \bibinfo {author} {\bibfnamefont {M.}~\bibnamefont {{Shirasaki}}}, \bibinfo {author} {\bibfnamefont {S.}~\bibnamefont {{Sugiyama}}}, \bibinfo {author} {\bibfnamefont {M.}~\bibnamefont {{Takada}}},\ and\ \bibinfo {author} {\bibfnamefont {S.}~\bibnamefont {{More}}},\ }\href {https://doi.org/10.1093/mnras/stad1801} {\bibfield  {journal} {\bibinfo  {journal} {\mnras}\ }\textbf {\bibinfo {volume} {525}},\
  \bibinfo {pages} {2441} (\bibinfo {year} {2023}{\natexlab{a}})},\ \Eprint {https://arxiv.org/abs/2212.03257} {arXiv:2212.03257 [astro-ph.CO]} \BibitemShut {NoStop}%
\bibitem [{\citenamefont {{Nishizawa}}\ \emph {et~al.}(2020)\citenamefont {{Nishizawa}}, \citenamefont {{Hsieh}}, \citenamefont {{Tanaka}},\ and\ \citenamefont {{Takata}}}]{Nishizawa2020}%
  \BibitemOpen
  \bibfield  {author} {\bibinfo {author} {\bibfnamefont {A.~J.}\ \bibnamefont {{Nishizawa}}}, \bibinfo {author} {\bibfnamefont {B.-C.}\ \bibnamefont {{Hsieh}}}, \bibinfo {author} {\bibfnamefont {M.}~\bibnamefont {{Tanaka}}},\ and\ \bibinfo {author} {\bibfnamefont {T.}~\bibnamefont {{Takata}}},\ }\href {https://doi.org/10.48550/arXiv.2003.01511} {\bibfield  {journal} {\bibinfo  {journal} {arXiv e-prints}\ ,\ \bibinfo {eid} {arXiv:2003.01511}} (\bibinfo {year} {2020})},\ \Eprint {https://arxiv.org/abs/2003.01511} {arXiv:2003.01511 [astro-ph.GA]} \BibitemShut {NoStop}%
\bibitem [{\citenamefont {{Mandelbaum}}\ \emph {et~al.}(2013)\citenamefont {{Mandelbaum}}, \citenamefont {{Slosar}}, \citenamefont {{Baldauf}}, \citenamefont {{Seljak}}, \citenamefont {{Hirata}}, \citenamefont {{Nakajima}}, \citenamefont {{Reyes}},\ and\ \citenamefont {{Smith}}}]{mandelbaum2013}%
  \BibitemOpen
  \bibfield  {author} {\bibinfo {author} {\bibfnamefont {R.}~\bibnamefont {{Mandelbaum}}}, \bibinfo {author} {\bibfnamefont {A.}~\bibnamefont {{Slosar}}}, \bibinfo {author} {\bibfnamefont {T.}~\bibnamefont {{Baldauf}}}, \bibinfo {author} {\bibfnamefont {U.}~\bibnamefont {{Seljak}}}, \bibinfo {author} {\bibfnamefont {C.~M.}\ \bibnamefont {{Hirata}}}, \bibinfo {author} {\bibfnamefont {R.}~\bibnamefont {{Nakajima}}}, \bibinfo {author} {\bibfnamefont {R.}~\bibnamefont {{Reyes}}},\ and\ \bibinfo {author} {\bibfnamefont {R.~E.}\ \bibnamefont {{Smith}}},\ }\href {https://doi.org/10.1093/mnras/stt572} {\bibfield  {journal} {\bibinfo  {journal} {\mnras}\ }\textbf {\bibinfo {volume} {432}},\ \bibinfo {pages} {1544} (\bibinfo {year} {2013})},\ \Eprint {https://arxiv.org/abs/1207.1120} {arXiv:1207.1120 [astro-ph.CO]} \BibitemShut {NoStop}%
\bibitem [{\citenamefont {{Sheldon}}\ \emph {et~al.}(2004)\citenamefont {{Sheldon}}, \citenamefont {{Johnston}}, \citenamefont {{Frieman}}, \citenamefont {{Scranton}}, \citenamefont {{McKay}}, \citenamefont {{Connolly}}, \citenamefont {{Budav{\'a}ri}}, \citenamefont {{Zehavi}}, \citenamefont {{Bahcall}}, \citenamefont {{Brinkmann}},\ and\ \citenamefont {{Fukugita}}}]{sheldon2004}%
  \BibitemOpen
  \bibfield  {author} {\bibinfo {author} {\bibfnamefont {E.~S.}\ \bibnamefont {{Sheldon}}}, \bibinfo {author} {\bibfnamefont {D.~E.}\ \bibnamefont {{Johnston}}}, \bibinfo {author} {\bibfnamefont {J.~A.}\ \bibnamefont {{Frieman}}}, \bibinfo {author} {\bibfnamefont {R.}~\bibnamefont {{Scranton}}}, \bibinfo {author} {\bibfnamefont {T.~A.}\ \bibnamefont {{McKay}}}, \bibinfo {author} {\bibfnamefont {A.~J.}\ \bibnamefont {{Connolly}}}, \bibinfo {author} {\bibfnamefont {T.}~\bibnamefont {{Budav{\'a}ri}}}, \bibinfo {author} {\bibfnamefont {I.}~\bibnamefont {{Zehavi}}}, \bibinfo {author} {\bibfnamefont {N.~A.}\ \bibnamefont {{Bahcall}}}, \bibinfo {author} {\bibfnamefont {J.}~\bibnamefont {{Brinkmann}}},\ and\ \bibinfo {author} {\bibfnamefont {M.}~\bibnamefont {{Fukugita}}},\ }\href {https://doi.org/10.1086/383293} {\bibfield  {journal} {\bibinfo  {journal} {\aj}\ }\textbf {\bibinfo {volume} {127}},\ \bibinfo {pages} {2544} (\bibinfo {year} {2004})},\ \Eprint {https://arxiv.org/abs/astro-ph/0312036}
  {arXiv:astro-ph/0312036 [astro-ph]} \BibitemShut {NoStop}%
\bibitem [{\citenamefont {{Rau}}\ \emph {et~al.}(2023)\citenamefont {{Rau}}, \citenamefont {{Dalal}}, \citenamefont {{Zhang}}, \citenamefont {{Li}}, \citenamefont {{Nishizawa}}, \citenamefont {{More}}, \citenamefont {{Mandelbaum}}, \citenamefont {{Miyatake}}, \citenamefont {{Strauss}},\ and\ \citenamefont {{Takada}}}]{Rau2022}%
  \BibitemOpen
  \bibfield  {author} {\bibinfo {author} {\bibfnamefont {M.~M.}\ \bibnamefont {{Rau}}}, \bibinfo {author} {\bibfnamefont {R.}~\bibnamefont {{Dalal}}}, \bibinfo {author} {\bibfnamefont {T.}~\bibnamefont {{Zhang}}}, \bibinfo {author} {\bibfnamefont {X.}~\bibnamefont {{Li}}}, \bibinfo {author} {\bibfnamefont {A.~J.}\ \bibnamefont {{Nishizawa}}}, \bibinfo {author} {\bibfnamefont {S.}~\bibnamefont {{More}}}, \bibinfo {author} {\bibfnamefont {R.}~\bibnamefont {{Mandelbaum}}}, \bibinfo {author} {\bibfnamefont {H.}~\bibnamefont {{Miyatake}}}, \bibinfo {author} {\bibfnamefont {M.~A.}\ \bibnamefont {{Strauss}}},\ and\ \bibinfo {author} {\bibfnamefont {M.}~\bibnamefont {{Takada}}},\ }\bibfield  {journal} {\bibinfo  {journal} {\mnras}\ }\href {https://doi.org/10.1093/mnras/stad1962} {10.1093/mnras/stad1962} (\bibinfo {year} {2023}),\ \Eprint {https://arxiv.org/abs/2211.16516} {arXiv:2211.16516 [astro-ph.CO]} \BibitemShut {NoStop}%
\bibitem [{\citenamefont {Oguri}(2014)}]{CAMIRA_Oguri2014}%
  \BibitemOpen
  \bibfield  {author} {\bibinfo {author} {\bibfnamefont {M.}~\bibnamefont {Oguri}},\ }\href {https://doi.org/10.1093/mnras/stu1446} {\bibfield  {journal} {\bibinfo  {journal} {Monthly Notices of the Royal Astronomical Society}\ }\textbf {\bibinfo {volume} {444}},\ \bibinfo {pages} {147} (\bibinfo {year} {2014})},\ \Eprint {https://arxiv.org/abs/1407.4693} {arXiv:1407.4693 [astro-ph.CO]} \BibitemShut {NoStop}%
\bibitem [{\citenamefont {{Ishikawa}}\ \emph {et~al.}(2021)\citenamefont {{Ishikawa}}, \citenamefont {{Okumura}}, \citenamefont {{Oguri}},\ and\ \citenamefont {{Lin}}}]{CAMIRA3}%
  \BibitemOpen
  \bibfield  {author} {\bibinfo {author} {\bibfnamefont {S.}~\bibnamefont {{Ishikawa}}}, \bibinfo {author} {\bibfnamefont {T.}~\bibnamefont {{Okumura}}}, \bibinfo {author} {\bibfnamefont {M.}~\bibnamefont {{Oguri}}},\ and\ \bibinfo {author} {\bibfnamefont {S.-C.}\ \bibnamefont {{Lin}}},\ }\href {https://doi.org/10.3847/1538-4357/ac1f90} {\bibfield  {journal} {\bibinfo  {journal} {\apj}\ }\textbf {\bibinfo {volume} {922}},\ \bibinfo {eid} {23} (\bibinfo {year} {2021})},\ \Eprint {https://arxiv.org/abs/2103.08628} {arXiv:2103.08628 [astro-ph.GA]} \BibitemShut {NoStop}%
\bibitem [{\citenamefont {{Takahashi}}\ \emph {et~al.}(2017)\citenamefont {{Takahashi}}, \citenamefont {{Hamana}}, \citenamefont {{Shirasaki}}, \citenamefont {{Namikawa}}, \citenamefont {{Nishimichi}}, \citenamefont {{Osato}},\ and\ \citenamefont {{Shiroyama}}}]{takahashi2017}%
  \BibitemOpen
  \bibfield  {author} {\bibinfo {author} {\bibfnamefont {R.}~\bibnamefont {{Takahashi}}}, \bibinfo {author} {\bibfnamefont {T.}~\bibnamefont {{Hamana}}}, \bibinfo {author} {\bibfnamefont {M.}~\bibnamefont {{Shirasaki}}}, \bibinfo {author} {\bibfnamefont {T.}~\bibnamefont {{Namikawa}}}, \bibinfo {author} {\bibfnamefont {T.}~\bibnamefont {{Nishimichi}}}, \bibinfo {author} {\bibfnamefont {K.}~\bibnamefont {{Osato}}},\ and\ \bibinfo {author} {\bibfnamefont {K.}~\bibnamefont {{Shiroyama}}},\ }\href {https://doi.org/10.3847/1538-4357/aa943d} {\bibfield  {journal} {\bibinfo  {journal} {\apj}\ }\textbf {\bibinfo {volume} {850}},\ \bibinfo {eid} {24} (\bibinfo {year} {2017})},\ \Eprint {https://arxiv.org/abs/1706.01472} {arXiv:1706.01472 [astro-ph.CO]} \BibitemShut {NoStop}%
\bibitem [{\citenamefont {{Alam}}\ \emph {et~al.}(2015)\citenamefont {{Alam}}, \citenamefont {{Albareti}}, \citenamefont {{Allende Prieto}}, \citenamefont {{Anders}}, \citenamefont {{Anderson}}, \citenamefont {{Anderton}}, \citenamefont {{Andrews}}, \citenamefont {{Armengaud}}, \citenamefont {{Aubourg}}, \citenamefont {{Bailey}}, \citenamefont {{Basu}}, \citenamefont {{Bautista}}, \citenamefont {{Beaton}}, \citenamefont {{Beers}}, \citenamefont {{Bender}}, \citenamefont {{Berlind}}, \citenamefont {{Beutler}}, \citenamefont {{Bhardwaj}}, \citenamefont {{Bird}}, \citenamefont {{Bizyaev}}, \citenamefont {{Blake}}, \citenamefont {{Blanton}}, \citenamefont {{Blomqvist}}, \citenamefont {{Bochanski}}, \citenamefont {{Bolton}}, \citenamefont {{Bovy}}, \citenamefont {{Shelden Bradley}}, \citenamefont {{Brandt}}, \citenamefont {{Brauer}}, \citenamefont {{Brinkmann}}, \citenamefont {{Brown}}, \citenamefont {{Brownstein}}, \citenamefont {{Burden}}, \citenamefont {{Burtin}}, \citenamefont {{Busca}}, \citenamefont {{Cai}},
  \citenamefont {{Capozzi}}, \citenamefont {{Carnero Rosell}}, \citenamefont {{Carr}}, \citenamefont {{Carrera}}, \citenamefont {{Chambers}}, \citenamefont {{Chaplin}}, \citenamefont {{Chen}}, \citenamefont {{Chiappini}}, \citenamefont {{Chojnowski}}, \citenamefont {{Chuang}}, \citenamefont {{Clerc}}, \citenamefont {{Comparat}}, \citenamefont {{Covey}}, \citenamefont {{Croft}}, \citenamefont {{Cuesta}}, \citenamefont {{Cunha}}, \citenamefont {{da Costa}}, \citenamefont {{Da Rio}}, \citenamefont {{Davenport}}, \citenamefont {{Dawson}}, \citenamefont {{De Lee}}, \citenamefont {{Delubac}}, \citenamefont {{Deshpande}}, \citenamefont {{Dhital}}, \citenamefont {{Dutra-Ferreira}}, \citenamefont {{Dwelly}}, \citenamefont {{Ealet}}, \citenamefont {{Ebelke}}, \citenamefont {{Edmondson}}, \citenamefont {{Eisenstein}}, \citenamefont {{Ellsworth}}, \citenamefont {{Elsworth}}, \citenamefont {{Epstein}}, \citenamefont {{Eracleous}}, \citenamefont {{Escoffier}}, \citenamefont {{Esposito}}, \citenamefont {{Evans}},
  \citenamefont {{Fan}}, \citenamefont {{Fern{\'a}ndez-Alvar}}, \citenamefont {{Feuillet}}, \citenamefont {{Filiz Ak}}, \citenamefont {{Finley}}, \citenamefont {{Finoguenov}}, \citenamefont {{Flaherty}}, \citenamefont {{Fleming}}, \citenamefont {{Font-Ribera}}, \citenamefont {{Foster}}, \citenamefont {{Frinchaboy}}, \citenamefont {{Galbraith-Frew}}, \citenamefont {{Garc{\'\i}a}}, \citenamefont {{Garc{\'\i}a-Hern{\'a}ndez}}, \citenamefont {{Garc{\'\i}a P{\'e}rez}}, \citenamefont {{Gaulme}}, \citenamefont {{Ge}}, \citenamefont {{G{\'e}nova-Santos}}, \citenamefont {{Georgakakis}}, \citenamefont {{Ghezzi}}, \citenamefont {{Gillespie}}, \citenamefont {{Girardi}}, \citenamefont {{Goddard}}, \citenamefont {{Gontcho}}, \citenamefont {{Gonz{\'a}lez Hern{\'a}ndez}}, \citenamefont {{Grebel}}, \citenamefont {{Green}}, \citenamefont {{Grieb}}, \citenamefont {{Grieves}}, \citenamefont {{Gunn}}, \citenamefont {{Guo}}, \citenamefont {{Harding}}, \citenamefont {{Hasselquist}}, \citenamefont {{Hawley}}, \citenamefont
  {{Hayden}}, \citenamefont {{Hearty}}, \citenamefont {{Hekker}}, \citenamefont {{Ho}}, \citenamefont {{Hogg}}, \citenamefont {{Holley-Bockelmann}}, \citenamefont {{Holtzman}}, \citenamefont {{Honscheid}}, \citenamefont {{Huber}}, \citenamefont {{Huehnerhoff}}, \citenamefont {{Ivans}}, \citenamefont {{Jiang}}, \citenamefont {{Johnson}}, \citenamefont {{Kinemuchi}}, \citenamefont {{Kirkby}}, \citenamefont {{Kitaura}}, \citenamefont {{Klaene}}, \citenamefont {{Knapp}}, \citenamefont {{Kneib}}, \citenamefont {{Koenig}}, \citenamefont {{Lam}}, \citenamefont {{Lan}}, \citenamefont {{Lang}}, \citenamefont {{Laurent}}, \citenamefont {{Le Goff}}, \citenamefont {{Leauthaud}}, \citenamefont {{Lee}}, \citenamefont {{Lee}}, \citenamefont {{Licquia}}, \citenamefont {{Liu}}, \citenamefont {{Long}}, \citenamefont {{L{\'o}pez-Corredoira}}, \citenamefont {{Lorenzo-Oliveira}}, \citenamefont {{Lucatello}}, \citenamefont {{Lundgren}}, \citenamefont {{Lupton}}, \citenamefont {{Mack}}, \citenamefont {{Mahadevan}}, \citenamefont
  {{Maia}}, \citenamefont {{Majewski}}, \citenamefont {{Malanushenko}}, \citenamefont {{Malanushenko}}, \citenamefont {{Manchado}}, \citenamefont {{Manera}}, \citenamefont {{Mao}}, \citenamefont {{Maraston}}, \citenamefont {{Marchwinski}}, \citenamefont {{Margala}}, \citenamefont {{Martell}}, \citenamefont {{Martig}}, \citenamefont {{Masters}}, \citenamefont {{Mathur}}, \citenamefont {{McBride}}, \citenamefont {{McGehee}}, \citenamefont {{McGreer}}, \citenamefont {{McMahon}}, \citenamefont {{M{\'e}nard}}, \citenamefont {{Menzel}}, \citenamefont {{Merloni}}, \citenamefont {{M{\'e}sz{\'a}ros}}, \citenamefont {{Miller}}, \citenamefont {{Miralda-Escud{\'e}}}, \citenamefont {{Miyatake}}, \citenamefont {{Montero-Dorta}}, \citenamefont {{More}}, \citenamefont {{Morganson}}, \citenamefont {{Morice-Atkinson}}, \citenamefont {{Morrison}}, \citenamefont {{Mosser}}, \citenamefont {{Muna}}, \citenamefont {{Myers}}, \citenamefont {{Nandra}}, \citenamefont {{Newman}}, \citenamefont {{Neyrinck}}, \citenamefont {{Nguyen}},
  \citenamefont {{Nichol}}, \citenamefont {{Nidever}}, \citenamefont {{Noterdaeme}}, \citenamefont {{Nuza}}, \citenamefont {{O'Connell}}, \citenamefont {{O'Connell}}, \citenamefont {{O'Connell}}, \citenamefont {{Ogando}}, \citenamefont {{Olmstead}}, \citenamefont {{Oravetz}}, \citenamefont {{Oravetz}}, \citenamefont {{Osumi}}, \citenamefont {{Owen}}, \citenamefont {{Padgett}}, \citenamefont {{Padmanabhan}}, \citenamefont {{Paegert}}, \citenamefont {{Palanque-Delabrouille}}, \citenamefont {{Pan}}, \citenamefont {{Parejko}}, \citenamefont {{P{\^a}ris}}, \citenamefont {{Park}}, \citenamefont {{Pattarakijwanich}}, \citenamefont {{Pellejero-Ibanez}}, \citenamefont {{Pepper}}, \citenamefont {{Percival}}, \citenamefont {{P{\'e}rez-Fournon}}, \citenamefont {{P{\'e}rez-R{\`a}fols}}, \citenamefont {{Petitjean}}, \citenamefont {{Pieri}}, \citenamefont {{Pinsonneault}}, \citenamefont {{Porto de Mello}}, \citenamefont {{Prada}}, \citenamefont {{Prakash}}, \citenamefont {{Price-Whelan}}, \citenamefont {{Protopapas}},
  \citenamefont {{Raddick}}, \citenamefont {{Rahman}}, \citenamefont {{Reid}}, \citenamefont {{Rich}}, \citenamefont {{Rix}}, \citenamefont {{Robin}}, \citenamefont {{Rockosi}}, \citenamefont {{Rodrigues}}, \citenamefont {{Rodr{\'\i}guez-Torres}}, \citenamefont {{Roe}}, \citenamefont {{Ross}}, \citenamefont {{Ross}}, \citenamefont {{Rossi}}, \citenamefont {{Ruan}}, \citenamefont {{Rubi{\~n}o-Mart{\'\i}n}}, \citenamefont {{Rykoff}}, \citenamefont {{Salazar-Albornoz}}, \citenamefont {{Salvato}}, \citenamefont {{Samushia}}, \citenamefont {{S{\'a}nchez}}, \citenamefont {{Santiago}}, \citenamefont {{Sayres}}, \citenamefont {{Schiavon}}, \citenamefont {{Schlegel}}, \citenamefont {{Schmidt}}, \citenamefont {{Schneider}}, \citenamefont {{Schultheis}}, \citenamefont {{Schwope}}, \citenamefont {{Sc{\'o}ccola}}, \citenamefont {{Scott}}, \citenamefont {{Sellgren}}, \citenamefont {{Seo}}, \citenamefont {{Serenelli}}, \citenamefont {{Shane}}, \citenamefont {{Shen}}, \citenamefont {{Shetrone}}, \citenamefont {{Shu}},
  \citenamefont {{Silva Aguirre}}, \citenamefont {{Sivarani}}, \citenamefont {{Skrutskie}}, \citenamefont {{Slosar}}, \citenamefont {{Smith}}, \citenamefont {{Sobreira}}, \citenamefont {{Souto}}, \citenamefont {{Stassun}}, \citenamefont {{Steinmetz}}, \citenamefont {{Stello}}, \citenamefont {{Strauss}}, \citenamefont {{Streblyanska}}, \citenamefont {{Suzuki}}, \citenamefont {{Swanson}}, \citenamefont {{Tan}}, \citenamefont {{Tayar}}, \citenamefont {{Terrien}}, \citenamefont {{Thakar}}, \citenamefont {{Thomas}}, \citenamefont {{Thomas}}, \citenamefont {{Thompson}}, \citenamefont {{Tinker}}, \citenamefont {{Tojeiro}}, \citenamefont {{Troup}}, \citenamefont {{Vargas-Maga{\~n}a}}, \citenamefont {{Vazquez}}, \citenamefont {{Verde}}, \citenamefont {{Viel}}, \citenamefont {{Vogt}}, \citenamefont {{Wake}}, \citenamefont {{Wang}}, \citenamefont {{Weaver}}, \citenamefont {{Weinberg}}, \citenamefont {{Weiner}}, \citenamefont {{White}}, \citenamefont {{Wilson}}, \citenamefont {{Wisniewski}}, \citenamefont {{Wood-Vasey}},
  \citenamefont {{Ye`che}}, \citenamefont {{York}}, \citenamefont {{Zakamska}}, \citenamefont {{Zamora}}, \citenamefont {{Zasowski}}, \citenamefont {{Zehavi}}, \citenamefont {{Zhao}}, \citenamefont {{Zheng}}, \citenamefont {{Zhou}}, \citenamefont {{Zhou}}, \citenamefont {{Zou}},\ and\ \citenamefont {{Zhu}}}]{sdss_dr11}%
  \BibitemOpen
  \bibfield  {author} {\bibinfo {author} {\bibfnamefont {S.}~\bibnamefont {{Alam}}}, \bibinfo {author} {\bibfnamefont {F.~D.}\ \bibnamefont {{Albareti}}}, \bibinfo {author} {\bibfnamefont {C.}~\bibnamefont {{Allende Prieto}}}, \bibinfo {author} {\bibfnamefont {F.}~\bibnamefont {{Anders}}}, \bibinfo {author} {\bibfnamefont {S.~F.}\ \bibnamefont {{Anderson}}}, \bibinfo {author} {\bibfnamefont {T.}~\bibnamefont {{Anderton}}}, \bibinfo {author} {\bibfnamefont {B.~H.}\ \bibnamefont {{Andrews}}}, \bibinfo {author} {\bibfnamefont {E.}~\bibnamefont {{Armengaud}}}, \bibinfo {author} {\bibfnamefont {{\'E}.}~\bibnamefont {{Aubourg}}}, \bibinfo {author} {\bibfnamefont {S.}~\bibnamefont {{Bailey}}}, \bibinfo {author} {\bibfnamefont {S.}~\bibnamefont {{Basu}}}, \bibinfo {author} {\bibfnamefont {J.~E.}\ \bibnamefont {{Bautista}}}, \bibinfo {author} {\bibfnamefont {R.~L.}\ \bibnamefont {{Beaton}}}, \bibinfo {author} {\bibfnamefont {T.~C.}\ \bibnamefont {{Beers}}}, \bibinfo {author} {\bibfnamefont {C.~F.}\ \bibnamefont
  {{Bender}}}, \bibinfo {author} {\bibfnamefont {A.~A.}\ \bibnamefont {{Berlind}}}, \bibinfo {author} {\bibfnamefont {F.}~\bibnamefont {{Beutler}}}, \bibinfo {author} {\bibfnamefont {V.}~\bibnamefont {{Bhardwaj}}}, \bibinfo {author} {\bibfnamefont {J.~C.}\ \bibnamefont {{Bird}}}, \bibinfo {author} {\bibfnamefont {D.}~\bibnamefont {{Bizyaev}}}, \bibinfo {author} {\bibfnamefont {C.~H.}\ \bibnamefont {{Blake}}}, \bibinfo {author} {\bibfnamefont {M.~R.}\ \bibnamefont {{Blanton}}}, \bibinfo {author} {\bibfnamefont {M.}~\bibnamefont {{Blomqvist}}}, \bibinfo {author} {\bibfnamefont {J.~J.}\ \bibnamefont {{Bochanski}}}, \bibinfo {author} {\bibfnamefont {A.~S.}\ \bibnamefont {{Bolton}}}, \bibinfo {author} {\bibfnamefont {J.}~\bibnamefont {{Bovy}}}, \bibinfo {author} {\bibfnamefont {A.}~\bibnamefont {{Shelden Bradley}}}, \bibinfo {author} {\bibfnamefont {W.~N.}\ \bibnamefont {{Brandt}}}, \bibinfo {author} {\bibfnamefont {D.~E.}\ \bibnamefont {{Brauer}}}, \bibinfo {author} {\bibfnamefont {J.}~\bibnamefont
  {{Brinkmann}}}, \bibinfo {author} {\bibfnamefont {P.~J.}\ \bibnamefont {{Brown}}}, \bibinfo {author} {\bibfnamefont {J.~R.}\ \bibnamefont {{Brownstein}}}, \bibinfo {author} {\bibfnamefont {A.}~\bibnamefont {{Burden}}}, \bibinfo {author} {\bibfnamefont {E.}~\bibnamefont {{Burtin}}}, \bibinfo {author} {\bibfnamefont {N.~G.}\ \bibnamefont {{Busca}}}, \bibinfo {author} {\bibfnamefont {Z.}~\bibnamefont {{Cai}}}, \bibinfo {author} {\bibfnamefont {D.}~\bibnamefont {{Capozzi}}}, \bibinfo {author} {\bibfnamefont {A.}~\bibnamefont {{Carnero Rosell}}}, \bibinfo {author} {\bibfnamefont {M.~A.}\ \bibnamefont {{Carr}}}, \bibinfo {author} {\bibfnamefont {R.}~\bibnamefont {{Carrera}}}, \bibinfo {author} {\bibfnamefont {K.~C.}\ \bibnamefont {{Chambers}}}, \bibinfo {author} {\bibfnamefont {W.~J.}\ \bibnamefont {{Chaplin}}}, \bibinfo {author} {\bibfnamefont {Y.-C.}\ \bibnamefont {{Chen}}}, \bibinfo {author} {\bibfnamefont {C.}~\bibnamefont {{Chiappini}}}, \bibinfo {author} {\bibfnamefont {S.~D.}\ \bibnamefont {{Chojnowski}}},
  \bibinfo {author} {\bibfnamefont {C.-H.}\ \bibnamefont {{Chuang}}}, \bibinfo {author} {\bibfnamefont {N.}~\bibnamefont {{Clerc}}}, \bibinfo {author} {\bibfnamefont {J.}~\bibnamefont {{Comparat}}}, \bibinfo {author} {\bibfnamefont {K.}~\bibnamefont {{Covey}}}, \bibinfo {author} {\bibfnamefont {R.~A.~C.}\ \bibnamefont {{Croft}}}, \bibinfo {author} {\bibfnamefont {A.~J.}\ \bibnamefont {{Cuesta}}}, \bibinfo {author} {\bibfnamefont {K.}~\bibnamefont {{Cunha}}}, \bibinfo {author} {\bibfnamefont {L.~N.}\ \bibnamefont {{da Costa}}}, \bibinfo {author} {\bibfnamefont {N.}~\bibnamefont {{Da Rio}}}, \bibinfo {author} {\bibfnamefont {J.~R.~A.}\ \bibnamefont {{Davenport}}}, \bibinfo {author} {\bibfnamefont {K.~S.}\ \bibnamefont {{Dawson}}}, \bibinfo {author} {\bibfnamefont {N.}~\bibnamefont {{De Lee}}}, \bibinfo {author} {\bibfnamefont {T.}~\bibnamefont {{Delubac}}}, \bibinfo {author} {\bibfnamefont {R.}~\bibnamefont {{Deshpande}}}, \bibinfo {author} {\bibfnamefont {S.}~\bibnamefont {{Dhital}}}, \bibinfo {author}
  {\bibfnamefont {L.}~\bibnamefont {{Dutra-Ferreira}}}, \bibinfo {author} {\bibfnamefont {T.}~\bibnamefont {{Dwelly}}}, \bibinfo {author} {\bibfnamefont {A.}~\bibnamefont {{Ealet}}}, \bibinfo {author} {\bibfnamefont {G.~L.}\ \bibnamefont {{Ebelke}}}, \bibinfo {author} {\bibfnamefont {E.~M.}\ \bibnamefont {{Edmondson}}}, \bibinfo {author} {\bibfnamefont {D.~J.}\ \bibnamefont {{Eisenstein}}}, \bibinfo {author} {\bibfnamefont {T.}~\bibnamefont {{Ellsworth}}}, \bibinfo {author} {\bibfnamefont {Y.}~\bibnamefont {{Elsworth}}}, \bibinfo {author} {\bibfnamefont {C.~R.}\ \bibnamefont {{Epstein}}}, \bibinfo {author} {\bibfnamefont {M.}~\bibnamefont {{Eracleous}}}, \bibinfo {author} {\bibfnamefont {S.}~\bibnamefont {{Escoffier}}}, \bibinfo {author} {\bibfnamefont {M.}~\bibnamefont {{Esposito}}}, \bibinfo {author} {\bibfnamefont {M.~L.}\ \bibnamefont {{Evans}}}, \bibinfo {author} {\bibfnamefont {X.}~\bibnamefont {{Fan}}}, \bibinfo {author} {\bibfnamefont {E.}~\bibnamefont {{Fern{\'a}ndez-Alvar}}}, \bibinfo {author}
  {\bibfnamefont {D.}~\bibnamefont {{Feuillet}}}, \bibinfo {author} {\bibfnamefont {N.}~\bibnamefont {{Filiz Ak}}}, \bibinfo {author} {\bibfnamefont {H.}~\bibnamefont {{Finley}}}, \bibinfo {author} {\bibfnamefont {A.}~\bibnamefont {{Finoguenov}}}, \bibinfo {author} {\bibfnamefont {K.}~\bibnamefont {{Flaherty}}}, \bibinfo {author} {\bibfnamefont {S.~W.}\ \bibnamefont {{Fleming}}}, \bibinfo {author} {\bibfnamefont {A.}~\bibnamefont {{Font-Ribera}}}, \bibinfo {author} {\bibfnamefont {J.}~\bibnamefont {{Foster}}}, \bibinfo {author} {\bibfnamefont {P.~M.}\ \bibnamefont {{Frinchaboy}}}, \bibinfo {author} {\bibfnamefont {J.~G.}\ \bibnamefont {{Galbraith-Frew}}}, \bibinfo {author} {\bibfnamefont {R.~A.}\ \bibnamefont {{Garc{\'\i}a}}}, \bibinfo {author} {\bibfnamefont {D.~A.}\ \bibnamefont {{Garc{\'\i}a-Hern{\'a}ndez}}}, \bibinfo {author} {\bibfnamefont {A.~E.}\ \bibnamefont {{Garc{\'\i}a P{\'e}rez}}}, \bibinfo {author} {\bibfnamefont {P.}~\bibnamefont {{Gaulme}}}, \bibinfo {author} {\bibfnamefont {J.}~\bibnamefont
  {{Ge}}}, \bibinfo {author} {\bibfnamefont {R.}~\bibnamefont {{G{\'e}nova-Santos}}}, \bibinfo {author} {\bibfnamefont {A.}~\bibnamefont {{Georgakakis}}}, \bibinfo {author} {\bibfnamefont {L.}~\bibnamefont {{Ghezzi}}}, \bibinfo {author} {\bibfnamefont {B.~A.}\ \bibnamefont {{Gillespie}}}, \bibinfo {author} {\bibfnamefont {L.}~\bibnamefont {{Girardi}}}, \bibinfo {author} {\bibfnamefont {D.}~\bibnamefont {{Goddard}}}, \bibinfo {author} {\bibfnamefont {S.~G.~A.}\ \bibnamefont {{Gontcho}}}, \bibinfo {author} {\bibfnamefont {J.~I.}\ \bibnamefont {{Gonz{\'a}lez Hern{\'a}ndez}}}, \bibinfo {author} {\bibfnamefont {E.~K.}\ \bibnamefont {{Grebel}}}, \bibinfo {author} {\bibfnamefont {P.~J.}\ \bibnamefont {{Green}}}, \bibinfo {author} {\bibfnamefont {J.~N.}\ \bibnamefont {{Grieb}}}, \bibinfo {author} {\bibfnamefont {N.}~\bibnamefont {{Grieves}}}, \bibinfo {author} {\bibfnamefont {J.~E.}\ \bibnamefont {{Gunn}}}, \bibinfo {author} {\bibfnamefont {H.}~\bibnamefont {{Guo}}}, \bibinfo {author} {\bibfnamefont {P.}~\bibnamefont
  {{Harding}}}, \bibinfo {author} {\bibfnamefont {S.}~\bibnamefont {{Hasselquist}}}, \bibinfo {author} {\bibfnamefont {S.~L.}\ \bibnamefont {{Hawley}}}, \bibinfo {author} {\bibfnamefont {M.}~\bibnamefont {{Hayden}}}, \bibinfo {author} {\bibfnamefont {F.~R.}\ \bibnamefont {{Hearty}}}, \bibinfo {author} {\bibfnamefont {S.}~\bibnamefont {{Hekker}}}, \bibinfo {author} {\bibfnamefont {S.}~\bibnamefont {{Ho}}}, \bibinfo {author} {\bibfnamefont {D.~W.}\ \bibnamefont {{Hogg}}}, \bibinfo {author} {\bibfnamefont {K.}~\bibnamefont {{Holley-Bockelmann}}}, \bibinfo {author} {\bibfnamefont {J.~A.}\ \bibnamefont {{Holtzman}}}, \bibinfo {author} {\bibfnamefont {K.}~\bibnamefont {{Honscheid}}}, \bibinfo {author} {\bibfnamefont {D.}~\bibnamefont {{Huber}}}, \bibinfo {author} {\bibfnamefont {J.}~\bibnamefont {{Huehnerhoff}}}, \bibinfo {author} {\bibfnamefont {I.~I.}\ \bibnamefont {{Ivans}}}, \bibinfo {author} {\bibfnamefont {L.}~\bibnamefont {{Jiang}}}, \bibinfo {author} {\bibfnamefont {J.~A.}\ \bibnamefont {{Johnson}}},
  \bibinfo {author} {\bibfnamefont {K.}~\bibnamefont {{Kinemuchi}}}, \bibinfo {author} {\bibfnamefont {D.}~\bibnamefont {{Kirkby}}}, \bibinfo {author} {\bibfnamefont {F.}~\bibnamefont {{Kitaura}}}, \bibinfo {author} {\bibfnamefont {M.~A.}\ \bibnamefont {{Klaene}}}, \bibinfo {author} {\bibfnamefont {G.~R.}\ \bibnamefont {{Knapp}}}, \bibinfo {author} {\bibfnamefont {J.-P.}\ \bibnamefont {{Kneib}}}, \bibinfo {author} {\bibfnamefont {X.~P.}\ \bibnamefont {{Koenig}}}, \bibinfo {author} {\bibfnamefont {C.~R.}\ \bibnamefont {{Lam}}}, \bibinfo {author} {\bibfnamefont {T.-W.}\ \bibnamefont {{Lan}}}, \bibinfo {author} {\bibfnamefont {D.}~\bibnamefont {{Lang}}}, \bibinfo {author} {\bibfnamefont {P.}~\bibnamefont {{Laurent}}}, \bibinfo {author} {\bibfnamefont {J.-M.}\ \bibnamefont {{Le Goff}}}, \bibinfo {author} {\bibfnamefont {A.}~\bibnamefont {{Leauthaud}}}, \bibinfo {author} {\bibfnamefont {K.-G.}\ \bibnamefont {{Lee}}}, \bibinfo {author} {\bibfnamefont {Y.~S.}\ \bibnamefont {{Lee}}}, \bibinfo {author} {\bibfnamefont
  {T.~C.}\ \bibnamefont {{Licquia}}}, \bibinfo {author} {\bibfnamefont {J.}~\bibnamefont {{Liu}}}, \bibinfo {author} {\bibfnamefont {D.~C.}\ \bibnamefont {{Long}}}, \bibinfo {author} {\bibfnamefont {M.}~\bibnamefont {{L{\'o}pez-Corredoira}}}, \bibinfo {author} {\bibfnamefont {D.}~\bibnamefont {{Lorenzo-Oliveira}}}, \bibinfo {author} {\bibfnamefont {S.}~\bibnamefont {{Lucatello}}}, \bibinfo {author} {\bibfnamefont {B.}~\bibnamefont {{Lundgren}}}, \bibinfo {author} {\bibfnamefont {R.~H.}\ \bibnamefont {{Lupton}}}, \bibinfo {author} {\bibfnamefont {I.}~\bibnamefont {{Mack}}, \bibfnamefont {Claude~E.}}, \bibinfo {author} {\bibfnamefont {S.}~\bibnamefont {{Mahadevan}}}, \bibinfo {author} {\bibfnamefont {M.~A.~G.}\ \bibnamefont {{Maia}}}, \bibinfo {author} {\bibfnamefont {S.~R.}\ \bibnamefont {{Majewski}}}, \bibinfo {author} {\bibfnamefont {E.}~\bibnamefont {{Malanushenko}}}, \bibinfo {author} {\bibfnamefont {V.}~\bibnamefont {{Malanushenko}}}, \bibinfo {author} {\bibfnamefont {A.}~\bibnamefont {{Manchado}}},
  \bibinfo {author} {\bibfnamefont {M.}~\bibnamefont {{Manera}}}, \bibinfo {author} {\bibfnamefont {Q.}~\bibnamefont {{Mao}}}, \bibinfo {author} {\bibfnamefont {C.}~\bibnamefont {{Maraston}}}, \bibinfo {author} {\bibfnamefont {R.~C.}\ \bibnamefont {{Marchwinski}}}, \bibinfo {author} {\bibfnamefont {D.}~\bibnamefont {{Margala}}}, \bibinfo {author} {\bibfnamefont {S.~L.}\ \bibnamefont {{Martell}}}, \bibinfo {author} {\bibfnamefont {M.}~\bibnamefont {{Martig}}}, \bibinfo {author} {\bibfnamefont {K.~L.}\ \bibnamefont {{Masters}}}, \bibinfo {author} {\bibfnamefont {S.}~\bibnamefont {{Mathur}}}, \bibinfo {author} {\bibfnamefont {C.~K.}\ \bibnamefont {{McBride}}}, \bibinfo {author} {\bibfnamefont {P.~M.}\ \bibnamefont {{McGehee}}}, \bibinfo {author} {\bibfnamefont {I.~D.}\ \bibnamefont {{McGreer}}}, \bibinfo {author} {\bibfnamefont {R.~G.}\ \bibnamefont {{McMahon}}}, \bibinfo {author} {\bibfnamefont {B.}~\bibnamefont {{M{\'e}nard}}}, \bibinfo {author} {\bibfnamefont {M.-L.}\ \bibnamefont {{Menzel}}}, \bibinfo
  {author} {\bibfnamefont {A.}~\bibnamefont {{Merloni}}}, \bibinfo {author} {\bibfnamefont {S.}~\bibnamefont {{M{\'e}sz{\'a}ros}}}, \bibinfo {author} {\bibfnamefont {A.~A.}\ \bibnamefont {{Miller}}}, \bibinfo {author} {\bibfnamefont {J.}~\bibnamefont {{Miralda-Escud{\'e}}}}, \bibinfo {author} {\bibfnamefont {H.}~\bibnamefont {{Miyatake}}}, \bibinfo {author} {\bibfnamefont {A.~D.}\ \bibnamefont {{Montero-Dorta}}}, \bibinfo {author} {\bibfnamefont {S.}~\bibnamefont {{More}}}, \bibinfo {author} {\bibfnamefont {E.}~\bibnamefont {{Morganson}}}, \bibinfo {author} {\bibfnamefont {X.}~\bibnamefont {{Morice-Atkinson}}}, \bibinfo {author} {\bibfnamefont {H.~L.}\ \bibnamefont {{Morrison}}}, \bibinfo {author} {\bibfnamefont {B.}~\bibnamefont {{Mosser}}}, \bibinfo {author} {\bibfnamefont {D.}~\bibnamefont {{Muna}}}, \bibinfo {author} {\bibfnamefont {A.~D.}\ \bibnamefont {{Myers}}}, \bibinfo {author} {\bibfnamefont {K.}~\bibnamefont {{Nandra}}}, \bibinfo {author} {\bibfnamefont {J.~A.}\ \bibnamefont {{Newman}}}, \bibinfo
  {author} {\bibfnamefont {M.}~\bibnamefont {{Neyrinck}}}, \bibinfo {author} {\bibfnamefont {D.~C.}\ \bibnamefont {{Nguyen}}}, \bibinfo {author} {\bibfnamefont {R.~C.}\ \bibnamefont {{Nichol}}}, \bibinfo {author} {\bibfnamefont {D.~L.}\ \bibnamefont {{Nidever}}}, \bibinfo {author} {\bibfnamefont {P.}~\bibnamefont {{Noterdaeme}}}, \bibinfo {author} {\bibfnamefont {S.~E.}\ \bibnamefont {{Nuza}}}, \bibinfo {author} {\bibfnamefont {J.~E.}\ \bibnamefont {{O'Connell}}}, \bibinfo {author} {\bibfnamefont {R.~W.}\ \bibnamefont {{O'Connell}}}, \bibinfo {author} {\bibfnamefont {R.}~\bibnamefont {{O'Connell}}}, \bibinfo {author} {\bibfnamefont {R.~L.~C.}\ \bibnamefont {{Ogando}}}, \bibinfo {author} {\bibfnamefont {M.~D.}\ \bibnamefont {{Olmstead}}}, \bibinfo {author} {\bibfnamefont {A.~E.}\ \bibnamefont {{Oravetz}}}, \bibinfo {author} {\bibfnamefont {D.~J.}\ \bibnamefont {{Oravetz}}}, \bibinfo {author} {\bibfnamefont {K.}~\bibnamefont {{Osumi}}}, \bibinfo {author} {\bibfnamefont {R.}~\bibnamefont {{Owen}}}, \bibinfo
  {author} {\bibfnamefont {D.~L.}\ \bibnamefont {{Padgett}}}, \bibinfo {author} {\bibfnamefont {N.}~\bibnamefont {{Padmanabhan}}}, \bibinfo {author} {\bibfnamefont {M.}~\bibnamefont {{Paegert}}}, \bibinfo {author} {\bibfnamefont {N.}~\bibnamefont {{Palanque-Delabrouille}}}, \bibinfo {author} {\bibfnamefont {K.}~\bibnamefont {{Pan}}}, \bibinfo {author} {\bibfnamefont {J.~K.}\ \bibnamefont {{Parejko}}}, \bibinfo {author} {\bibfnamefont {I.}~\bibnamefont {{P{\^a}ris}}}, \bibinfo {author} {\bibfnamefont {C.}~\bibnamefont {{Park}}}, \bibinfo {author} {\bibfnamefont {P.}~\bibnamefont {{Pattarakijwanich}}}, \bibinfo {author} {\bibfnamefont {M.}~\bibnamefont {{Pellejero-Ibanez}}}, \bibinfo {author} {\bibfnamefont {J.}~\bibnamefont {{Pepper}}}, \bibinfo {author} {\bibfnamefont {W.~J.}\ \bibnamefont {{Percival}}}, \bibinfo {author} {\bibfnamefont {I.}~\bibnamefont {{P{\'e}rez-Fournon}}}, \bibinfo {author} {\bibfnamefont {I.}~\bibnamefont {{P{\'e}rez-R{\`a}fols}}}, \bibinfo {author} {\bibfnamefont {P.}~\bibnamefont
  {{Petitjean}}}, \bibinfo {author} {\bibfnamefont {M.~M.}\ \bibnamefont {{Pieri}}}, \bibinfo {author} {\bibfnamefont {M.~H.}\ \bibnamefont {{Pinsonneault}}}, \bibinfo {author} {\bibfnamefont {G.~F.}\ \bibnamefont {{Porto de Mello}}}, \bibinfo {author} {\bibfnamefont {F.}~\bibnamefont {{Prada}}}, \bibinfo {author} {\bibfnamefont {A.}~\bibnamefont {{Prakash}}}, \bibinfo {author} {\bibfnamefont {A.~M.}\ \bibnamefont {{Price-Whelan}}}, \bibinfo {author} {\bibfnamefont {P.}~\bibnamefont {{Protopapas}}}, \bibinfo {author} {\bibfnamefont {M.~J.}\ \bibnamefont {{Raddick}}}, \bibinfo {author} {\bibfnamefont {M.}~\bibnamefont {{Rahman}}}, \bibinfo {author} {\bibfnamefont {B.~A.}\ \bibnamefont {{Reid}}}, \bibinfo {author} {\bibfnamefont {J.}~\bibnamefont {{Rich}}}, \bibinfo {author} {\bibfnamefont {H.-W.}\ \bibnamefont {{Rix}}}, \bibinfo {author} {\bibfnamefont {A.~C.}\ \bibnamefont {{Robin}}}, \bibinfo {author} {\bibfnamefont {C.~M.}\ \bibnamefont {{Rockosi}}}, \bibinfo {author} {\bibfnamefont {T.~S.}\ \bibnamefont
  {{Rodrigues}}}, \bibinfo {author} {\bibfnamefont {S.}~\bibnamefont {{Rodr{\'\i}guez-Torres}}}, \bibinfo {author} {\bibfnamefont {N.~A.}\ \bibnamefont {{Roe}}}, \bibinfo {author} {\bibfnamefont {A.~J.}\ \bibnamefont {{Ross}}}, \bibinfo {author} {\bibfnamefont {N.~P.}\ \bibnamefont {{Ross}}}, \bibinfo {author} {\bibfnamefont {G.}~\bibnamefont {{Rossi}}}, \bibinfo {author} {\bibfnamefont {J.~J.}\ \bibnamefont {{Ruan}}}, \bibinfo {author} {\bibfnamefont {J.~A.}\ \bibnamefont {{Rubi{\~n}o-Mart{\'\i}n}}}, \bibinfo {author} {\bibfnamefont {E.~S.}\ \bibnamefont {{Rykoff}}}, \bibinfo {author} {\bibfnamefont {S.}~\bibnamefont {{Salazar-Albornoz}}}, \bibinfo {author} {\bibfnamefont {M.}~\bibnamefont {{Salvato}}}, \bibinfo {author} {\bibfnamefont {L.}~\bibnamefont {{Samushia}}}, \bibinfo {author} {\bibfnamefont {A.~G.}\ \bibnamefont {{S{\'a}nchez}}}, \bibinfo {author} {\bibfnamefont {B.}~\bibnamefont {{Santiago}}}, \bibinfo {author} {\bibfnamefont {C.}~\bibnamefont {{Sayres}}}, \bibinfo {author} {\bibfnamefont {R.~P.}\
  \bibnamefont {{Schiavon}}}, \bibinfo {author} {\bibfnamefont {D.~J.}\ \bibnamefont {{Schlegel}}}, \bibinfo {author} {\bibfnamefont {S.~J.}\ \bibnamefont {{Schmidt}}}, \bibinfo {author} {\bibfnamefont {D.~P.}\ \bibnamefont {{Schneider}}}, \bibinfo {author} {\bibfnamefont {M.}~\bibnamefont {{Schultheis}}}, \bibinfo {author} {\bibfnamefont {A.~D.}\ \bibnamefont {{Schwope}}}, \bibinfo {author} {\bibfnamefont {C.~G.}\ \bibnamefont {{Sc{\'o}ccola}}}, \bibinfo {author} {\bibfnamefont {C.}~\bibnamefont {{Scott}}}, \bibinfo {author} {\bibfnamefont {K.}~\bibnamefont {{Sellgren}}}, \bibinfo {author} {\bibfnamefont {H.-J.}\ \bibnamefont {{Seo}}}, \bibinfo {author} {\bibfnamefont {A.}~\bibnamefont {{Serenelli}}}, \bibinfo {author} {\bibfnamefont {N.}~\bibnamefont {{Shane}}}, \bibinfo {author} {\bibfnamefont {Y.}~\bibnamefont {{Shen}}}, \bibinfo {author} {\bibfnamefont {M.}~\bibnamefont {{Shetrone}}}, \bibinfo {author} {\bibfnamefont {Y.}~\bibnamefont {{Shu}}}, \bibinfo {author} {\bibfnamefont {V.}~\bibnamefont {{Silva
  Aguirre}}}, \bibinfo {author} {\bibfnamefont {T.}~\bibnamefont {{Sivarani}}}, \bibinfo {author} {\bibfnamefont {M.~F.}\ \bibnamefont {{Skrutskie}}}, \bibinfo {author} {\bibfnamefont {A.}~\bibnamefont {{Slosar}}}, \bibinfo {author} {\bibfnamefont {V.~V.}\ \bibnamefont {{Smith}}}, \bibinfo {author} {\bibfnamefont {F.}~\bibnamefont {{Sobreira}}}, \bibinfo {author} {\bibfnamefont {D.}~\bibnamefont {{Souto}}}, \bibinfo {author} {\bibfnamefont {K.~G.}\ \bibnamefont {{Stassun}}}, \bibinfo {author} {\bibfnamefont {M.}~\bibnamefont {{Steinmetz}}}, \bibinfo {author} {\bibfnamefont {D.}~\bibnamefont {{Stello}}}, \bibinfo {author} {\bibfnamefont {M.~A.}\ \bibnamefont {{Strauss}}}, \bibinfo {author} {\bibfnamefont {A.}~\bibnamefont {{Streblyanska}}}, \bibinfo {author} {\bibfnamefont {N.}~\bibnamefont {{Suzuki}}}, \bibinfo {author} {\bibfnamefont {M.~E.~C.}\ \bibnamefont {{Swanson}}}, \bibinfo {author} {\bibfnamefont {J.~C.}\ \bibnamefont {{Tan}}}, \bibinfo {author} {\bibfnamefont {J.}~\bibnamefont {{Tayar}}}, \bibinfo
  {author} {\bibfnamefont {R.~C.}\ \bibnamefont {{Terrien}}}, \bibinfo {author} {\bibfnamefont {A.~R.}\ \bibnamefont {{Thakar}}}, \bibinfo {author} {\bibfnamefont {D.}~\bibnamefont {{Thomas}}}, \bibinfo {author} {\bibfnamefont {N.}~\bibnamefont {{Thomas}}}, \bibinfo {author} {\bibfnamefont {B.~A.}\ \bibnamefont {{Thompson}}}, \bibinfo {author} {\bibfnamefont {J.~L.}\ \bibnamefont {{Tinker}}}, \bibinfo {author} {\bibfnamefont {R.}~\bibnamefont {{Tojeiro}}}, \bibinfo {author} {\bibfnamefont {N.~W.}\ \bibnamefont {{Troup}}}, \bibinfo {author} {\bibfnamefont {M.}~\bibnamefont {{Vargas-Maga{\~n}a}}}, \bibinfo {author} {\bibfnamefont {J.~A.}\ \bibnamefont {{Vazquez}}}, \bibinfo {author} {\bibfnamefont {L.}~\bibnamefont {{Verde}}}, \bibinfo {author} {\bibfnamefont {M.}~\bibnamefont {{Viel}}}, \bibinfo {author} {\bibfnamefont {N.~P.}\ \bibnamefont {{Vogt}}}, \bibinfo {author} {\bibfnamefont {D.~A.}\ \bibnamefont {{Wake}}}, \bibinfo {author} {\bibfnamefont {J.}~\bibnamefont {{Wang}}}, \bibinfo {author} {\bibfnamefont
  {B.~A.}\ \bibnamefont {{Weaver}}}, \bibinfo {author} {\bibfnamefont {D.~H.}\ \bibnamefont {{Weinberg}}}, \bibinfo {author} {\bibfnamefont {B.~J.}\ \bibnamefont {{Weiner}}}, \bibinfo {author} {\bibfnamefont {M.}~\bibnamefont {{White}}}, \bibinfo {author} {\bibfnamefont {J.~C.}\ \bibnamefont {{Wilson}}}, \bibinfo {author} {\bibfnamefont {J.~P.}\ \bibnamefont {{Wisniewski}}}, \bibinfo {author} {\bibfnamefont {W.~M.}\ \bibnamefont {{Wood-Vasey}}}, \bibinfo {author} {\bibfnamefont {C.}~\bibnamefont {{Ye`che}}}, \bibinfo {author} {\bibfnamefont {D.~G.}\ \bibnamefont {{York}}}, \bibinfo {author} {\bibfnamefont {N.~L.}\ \bibnamefont {{Zakamska}}}, \bibinfo {author} {\bibfnamefont {O.}~\bibnamefont {{Zamora}}}, \bibinfo {author} {\bibfnamefont {G.}~\bibnamefont {{Zasowski}}}, \bibinfo {author} {\bibfnamefont {I.}~\bibnamefont {{Zehavi}}}, \bibinfo {author} {\bibfnamefont {G.-B.}\ \bibnamefont {{Zhao}}}, \bibinfo {author} {\bibfnamefont {Z.}~\bibnamefont {{Zheng}}}, \bibinfo {author} {\bibfnamefont {X.}~\bibnamefont
  {{Zhou}}}, \bibinfo {author} {\bibfnamefont {Z.}~\bibnamefont {{Zhou}}}, \bibinfo {author} {\bibfnamefont {H.}~\bibnamefont {{Zou}}},\ and\ \bibinfo {author} {\bibfnamefont {G.}~\bibnamefont {{Zhu}}},\ }\href {https://doi.org/10.1088/0067-0049/219/1/12} {\bibfield  {journal} {\bibinfo  {journal} {\apjs}\ }\textbf {\bibinfo {volume} {219}},\ \bibinfo {eid} {12} (\bibinfo {year} {2015})},\ \Eprint {https://arxiv.org/abs/1501.00963} {arXiv:1501.00963 [astro-ph.IM]} \BibitemShut {NoStop}%
\bibitem [{\citenamefont {{Landy}}\ and\ \citenamefont {{Szalay}}(1993)}]{Landy1993}%
  \BibitemOpen
  \bibfield  {author} {\bibinfo {author} {\bibfnamefont {S.~D.}\ \bibnamefont {{Landy}}}\ and\ \bibinfo {author} {\bibfnamefont {A.~S.}\ \bibnamefont {{Szalay}}},\ }\href {https://doi.org/10.1086/172900} {\bibfield  {journal} {\bibinfo  {journal} {\apj}\ }\textbf {\bibinfo {volume} {412}},\ \bibinfo {pages} {64} (\bibinfo {year} {1993})}\BibitemShut {NoStop}%
\bibitem [{\citenamefont {{Bartelmann}}\ and\ \citenamefont {{Schneider}}(2001)}]{bartelmann2001}%
  \BibitemOpen
  \bibfield  {author} {\bibinfo {author} {\bibfnamefont {M.}~\bibnamefont {{Bartelmann}}}\ and\ \bibinfo {author} {\bibfnamefont {P.}~\bibnamefont {{Schneider}}},\ }\href {https://doi.org/10.1016/S0370-1573(00)00082-X} {\bibfield  {journal} {\bibinfo  {journal} {\physrep}\ }\textbf {\bibinfo {volume} {340}},\ \bibinfo {pages} {291} (\bibinfo {year} {2001})},\ \Eprint {https://arxiv.org/abs/astro-ph/9912508} {arXiv:astro-ph/9912508 [astro-ph]} \BibitemShut {NoStop}%
\bibitem [{\citenamefont {{Shirasaki}}\ \emph {et~al.}(2017)\citenamefont {{Shirasaki}}, \citenamefont {{Takada}}, \citenamefont {{Miyatake}}, \citenamefont {{Takahashi}}, \citenamefont {{Hamana}}, \citenamefont {{Nishimichi}},\ and\ \citenamefont {{Murata}}}]{Shirasaki2017}%
  \BibitemOpen
  \bibfield  {author} {\bibinfo {author} {\bibfnamefont {M.}~\bibnamefont {{Shirasaki}}}, \bibinfo {author} {\bibfnamefont {M.}~\bibnamefont {{Takada}}}, \bibinfo {author} {\bibfnamefont {H.}~\bibnamefont {{Miyatake}}}, \bibinfo {author} {\bibfnamefont {R.}~\bibnamefont {{Takahashi}}}, \bibinfo {author} {\bibfnamefont {T.}~\bibnamefont {{Hamana}}}, \bibinfo {author} {\bibfnamefont {T.}~\bibnamefont {{Nishimichi}}},\ and\ \bibinfo {author} {\bibfnamefont {R.}~\bibnamefont {{Murata}}},\ }\href {https://doi.org/10.1093/mnras/stx1477} {\bibfield  {journal} {\bibinfo  {journal} {\mnras}\ }\textbf {\bibinfo {volume} {470}},\ \bibinfo {pages} {3476} (\bibinfo {year} {2017})},\ \Eprint {https://arxiv.org/abs/1607.08679} {arXiv:1607.08679 [astro-ph.CO]} \BibitemShut {NoStop}%
\bibitem [{\citenamefont {{Shirasaki}}\ \emph {et~al.}(2019)\citenamefont {{Shirasaki}}, \citenamefont {{Hamana}}, \citenamefont {{Takada}}, \citenamefont {{Takahashi}},\ and\ \citenamefont {{Miyatake}}}]{Shirasaki2019}%
  \BibitemOpen
  \bibfield  {author} {\bibinfo {author} {\bibfnamefont {M.}~\bibnamefont {{Shirasaki}}}, \bibinfo {author} {\bibfnamefont {T.}~\bibnamefont {{Hamana}}}, \bibinfo {author} {\bibfnamefont {M.}~\bibnamefont {{Takada}}}, \bibinfo {author} {\bibfnamefont {R.}~\bibnamefont {{Takahashi}}},\ and\ \bibinfo {author} {\bibfnamefont {H.}~\bibnamefont {{Miyatake}}},\ }\href {https://doi.org/10.1093/mnras/stz791} {\bibfield  {journal} {\bibinfo  {journal} {\mnras}\ }\textbf {\bibinfo {volume} {486}},\ \bibinfo {pages} {52} (\bibinfo {year} {2019})},\ \Eprint {https://arxiv.org/abs/1901.09488} {arXiv:1901.09488 [astro-ph.CO]} \BibitemShut {NoStop}%
\bibitem [{\citenamefont {{Zuntz}}\ \emph {et~al.}(2015)\citenamefont {{Zuntz}}, \citenamefont {{Paterno}}, \citenamefont {{Jennings}}, \citenamefont {{Rudd}}, \citenamefont {{Manzotti}}, \citenamefont {{Dodelson}}, \citenamefont {{Bridle}}, \citenamefont {{Sehrish}},\ and\ \citenamefont {{Kowalkowski}}}]{Zuntz2015}%
  \BibitemOpen
  \bibfield  {author} {\bibinfo {author} {\bibfnamefont {J.}~\bibnamefont {{Zuntz}}}, \bibinfo {author} {\bibfnamefont {M.}~\bibnamefont {{Paterno}}}, \bibinfo {author} {\bibfnamefont {E.}~\bibnamefont {{Jennings}}}, \bibinfo {author} {\bibfnamefont {D.}~\bibnamefont {{Rudd}}}, \bibinfo {author} {\bibfnamefont {A.}~\bibnamefont {{Manzotti}}}, \bibinfo {author} {\bibfnamefont {S.}~\bibnamefont {{Dodelson}}}, \bibinfo {author} {\bibfnamefont {S.}~\bibnamefont {{Bridle}}}, \bibinfo {author} {\bibfnamefont {S.}~\bibnamefont {{Sehrish}}},\ and\ \bibinfo {author} {\bibfnamefont {J.}~\bibnamefont {{Kowalkowski}}},\ }\href {https://doi.org/10.1016/j.ascom.2015.05.005} {\bibfield  {journal} {\bibinfo  {journal} {Astronomy and Computing}\ }\textbf {\bibinfo {volume} {12}},\ \bibinfo {pages} {45} (\bibinfo {year} {2015})},\ \Eprint {https://arxiv.org/abs/1409.3409} {arXiv:1409.3409 [astro-ph.CO]} \BibitemShut {NoStop}%
\bibitem [{\citenamefont {{Lewis}}\ \emph {et~al.}(2000)\citenamefont {{Lewis}}, \citenamefont {{Challinor}},\ and\ \citenamefont {{Lasenby}}}]{Lewis2000}%
  \BibitemOpen
  \bibfield  {author} {\bibinfo {author} {\bibfnamefont {A.}~\bibnamefont {{Lewis}}}, \bibinfo {author} {\bibfnamefont {A.}~\bibnamefont {{Challinor}}},\ and\ \bibinfo {author} {\bibfnamefont {A.}~\bibnamefont {{Lasenby}}},\ }\href {https://doi.org/10.1086/309179} {\bibfield  {journal} {\bibinfo  {journal} {\apj}\ }\textbf {\bibinfo {volume} {538}},\ \bibinfo {pages} {473} (\bibinfo {year} {2000})},\ \Eprint {https://arxiv.org/abs/astro-ph/9911177} {arXiv:astro-ph/9911177 [astro-ph]} \BibitemShut {NoStop}%
\bibitem [{\citenamefont {{Mead}}\ \emph {et~al.}(2016)\citenamefont {{Mead}}, \citenamefont {{Heymans}}, \citenamefont {{Lombriser}}, \citenamefont {{Peacock}}, \citenamefont {{Steele}},\ and\ \citenamefont {{Winther}}}]{Mead2016}%
  \BibitemOpen
  \bibfield  {author} {\bibinfo {author} {\bibfnamefont {A.~J.}\ \bibnamefont {{Mead}}}, \bibinfo {author} {\bibfnamefont {C.}~\bibnamefont {{Heymans}}}, \bibinfo {author} {\bibfnamefont {L.}~\bibnamefont {{Lombriser}}}, \bibinfo {author} {\bibfnamefont {J.~A.}\ \bibnamefont {{Peacock}}}, \bibinfo {author} {\bibfnamefont {O.~I.}\ \bibnamefont {{Steele}}},\ and\ \bibinfo {author} {\bibfnamefont {H.~A.}\ \bibnamefont {{Winther}}},\ }\href {https://doi.org/10.1093/mnras/stw681} {\bibfield  {journal} {\bibinfo  {journal} {\mnras}\ }\textbf {\bibinfo {volume} {459}},\ \bibinfo {pages} {1468} (\bibinfo {year} {2016})},\ \Eprint {https://arxiv.org/abs/1602.02154} {arXiv:1602.02154 [astro-ph.CO]} \BibitemShut {NoStop}%
\bibitem [{\citenamefont {{Chisari}}\ \emph {et~al.}(2018)\citenamefont {{Chisari}}, \citenamefont {{Richardson}}, \citenamefont {{Devriendt}}, \citenamefont {{Dubois}}, \citenamefont {{Schneider}}, \citenamefont {{Le Brun}}, \citenamefont {{Beckmann}}, \citenamefont {{Peirani}}, \citenamefont {{Slyz}},\ and\ \citenamefont {{Pichon}}}]{Chisari2018}%
  \BibitemOpen
  \bibfield  {author} {\bibinfo {author} {\bibfnamefont {N.~E.}\ \bibnamefont {{Chisari}}}, \bibinfo {author} {\bibfnamefont {M.~L.~A.}\ \bibnamefont {{Richardson}}}, \bibinfo {author} {\bibfnamefont {J.}~\bibnamefont {{Devriendt}}}, \bibinfo {author} {\bibfnamefont {Y.}~\bibnamefont {{Dubois}}}, \bibinfo {author} {\bibfnamefont {A.}~\bibnamefont {{Schneider}}}, \bibinfo {author} {\bibfnamefont {A.~M.~C.}\ \bibnamefont {{Le Brun}}}, \bibinfo {author} {\bibfnamefont {R.~S.}\ \bibnamefont {{Beckmann}}}, \bibinfo {author} {\bibfnamefont {S.}~\bibnamefont {{Peirani}}}, \bibinfo {author} {\bibfnamefont {A.}~\bibnamefont {{Slyz}}},\ and\ \bibinfo {author} {\bibfnamefont {C.~o.}\ \bibnamefont {{Pichon}}},\ }\href {https://doi.org/10.1093/mnras/sty2093} {\bibfield  {journal} {\bibinfo  {journal} {\mnras}\ }\textbf {\bibinfo {volume} {480}},\ \bibinfo {pages} {3962} (\bibinfo {year} {2018})},\ \Eprint {https://arxiv.org/abs/1801.08559} {arXiv:1801.08559 [astro-ph.CO]} \BibitemShut {NoStop}%
\bibitem [{\citenamefont {{Huang}}\ \emph {et~al.}(2021)\citenamefont {{Huang}}, \citenamefont {{Eifler}}, \citenamefont {{Mandelbaum}}, \citenamefont {{Bernstein}}, \citenamefont {{Chen}}, \citenamefont {{Choi}}, \citenamefont {{Garc{\'\i}a-Bellido}}, \citenamefont {{Huterer}}, \citenamefont {{Krause}}, \citenamefont {{Rozo}} \emph {et~al.}}]{Huang2021}%
  \BibitemOpen
  \bibfield  {author} {\bibinfo {author} {\bibfnamefont {H.-J.}\ \bibnamefont {{Huang}}}, \bibinfo {author} {\bibfnamefont {T.}~\bibnamefont {{Eifler}}}, \bibinfo {author} {\bibfnamefont {R.}~\bibnamefont {{Mandelbaum}}}, \bibinfo {author} {\bibfnamefont {G.~M.}\ \bibnamefont {{Bernstein}}}, \bibinfo {author} {\bibfnamefont {A.}~\bibnamefont {{Chen}}}, \bibinfo {author} {\bibfnamefont {A.}~\bibnamefont {{Choi}}}, \bibinfo {author} {\bibfnamefont {J.}~\bibnamefont {{Garc{\'\i}a-Bellido}}}, \bibinfo {author} {\bibfnamefont {D.}~\bibnamefont {{Huterer}}}, \bibinfo {author} {\bibfnamefont {E.}~\bibnamefont {{Krause}}}, \bibinfo {author} {\bibfnamefont {E.}~\bibnamefont {{Rozo}}}, \emph {et~al.},\ }\href {https://doi.org/10.1093/mnras/stab357} {\bibfield  {journal} {\bibinfo  {journal} {\mnras}\ }\textbf {\bibinfo {volume} {502}},\ \bibinfo {pages} {6010} (\bibinfo {year} {2021})},\ \Eprint {https://arxiv.org/abs/2007.15026} {arXiv:2007.15026 [astro-ph.CO]} \BibitemShut {NoStop}%
\bibitem [{\citenamefont {{Terasawa}}\ \emph {et~al.}(2025)\citenamefont {{Terasawa}}, \citenamefont {{Li}}, \citenamefont {{Takada}}, \citenamefont {{Nishimichi}}, \citenamefont {{Tanaka}}, \citenamefont {{Sugiyama}}, \citenamefont {{Kurita}}, \citenamefont {{Zhang}}, \citenamefont {{Shirasaki}}, \citenamefont {{Takahashi}}, \citenamefont {{Miyatake}}, \citenamefont {{More}},\ and\ \citenamefont {{Nishizawa}}}]{Ryo2025}%
  \BibitemOpen
  \bibfield  {author} {\bibinfo {author} {\bibfnamefont {R.}~\bibnamefont {{Terasawa}}}, \bibinfo {author} {\bibfnamefont {X.}~\bibnamefont {{Li}}}, \bibinfo {author} {\bibfnamefont {M.}~\bibnamefont {{Takada}}}, \bibinfo {author} {\bibfnamefont {T.}~\bibnamefont {{Nishimichi}}}, \bibinfo {author} {\bibfnamefont {S.}~\bibnamefont {{Tanaka}}}, \bibinfo {author} {\bibfnamefont {S.}~\bibnamefont {{Sugiyama}}}, \bibinfo {author} {\bibfnamefont {T.}~\bibnamefont {{Kurita}}}, \bibinfo {author} {\bibfnamefont {T.}~\bibnamefont {{Zhang}}}, \bibinfo {author} {\bibfnamefont {M.}~\bibnamefont {{Shirasaki}}}, \bibinfo {author} {\bibfnamefont {R.}~\bibnamefont {{Takahashi}}}, \bibinfo {author} {\bibfnamefont {H.}~\bibnamefont {{Miyatake}}}, \bibinfo {author} {\bibfnamefont {S.}~\bibnamefont {{More}}},\ and\ \bibinfo {author} {\bibfnamefont {A.~J.}\ \bibnamefont {{Nishizawa}}},\ }\href {https://doi.org/10.1103/PhysRevD.111.063509} {\bibfield  {journal} {\bibinfo  {journal} {\prd}\ }\textbf {\bibinfo {volume} {111}},\
  \bibinfo {eid} {063509} (\bibinfo {year} {2025})},\ \Eprint {https://arxiv.org/abs/2403.20323} {arXiv:2403.20323 [astro-ph.CO]} \BibitemShut {NoStop}%
\bibitem [{\citenamefont {{Asgari}}\ \emph {et~al.}(2018)\citenamefont {{Asgari}}, \citenamefont {{Taylor}}, \citenamefont {{Joachimi}},\ and\ \citenamefont {{Kitching}}}]{Asgari2018}%
  \BibitemOpen
  \bibfield  {author} {\bibinfo {author} {\bibfnamefont {M.}~\bibnamefont {{Asgari}}}, \bibinfo {author} {\bibfnamefont {A.}~\bibnamefont {{Taylor}}}, \bibinfo {author} {\bibfnamefont {B.}~\bibnamefont {{Joachimi}}},\ and\ \bibinfo {author} {\bibfnamefont {T.~D.}\ \bibnamefont {{Kitching}}},\ }\href {https://doi.org/10.1093/mnras/sty1412} {\bibfield  {journal} {\bibinfo  {journal} {\mnras}\ }\textbf {\bibinfo {volume} {479}},\ \bibinfo {pages} {454} (\bibinfo {year} {2018})},\ \Eprint {https://arxiv.org/abs/1612.04664} {arXiv:1612.04664 [astro-ph.CO]} \BibitemShut {NoStop}%
\bibitem [{\citenamefont {{Joachimi}}\ \emph {et~al.}(2021)\citenamefont {{Joachimi}}, \citenamefont {{Lin}}, \citenamefont {{Asgari}}, \citenamefont {{Tr{\"o}ster}}, \citenamefont {{Heymans}}, \citenamefont {{Hildebrandt}}, \citenamefont {{K{\"o}hlinger}}, \citenamefont {{S{\'a}nchez}}, \citenamefont {{Wright}}, \citenamefont {{Bilicki}} \emph {et~al.}}]{Joachimi2021}%
  \BibitemOpen
  \bibfield  {author} {\bibinfo {author} {\bibfnamefont {B.}~\bibnamefont {{Joachimi}}}, \bibinfo {author} {\bibfnamefont {C.~A.}\ \bibnamefont {{Lin}}}, \bibinfo {author} {\bibfnamefont {M.}~\bibnamefont {{Asgari}}}, \bibinfo {author} {\bibfnamefont {T.}~\bibnamefont {{Tr{\"o}ster}}}, \bibinfo {author} {\bibfnamefont {C.}~\bibnamefont {{Heymans}}}, \bibinfo {author} {\bibfnamefont {H.}~\bibnamefont {{Hildebrandt}}}, \bibinfo {author} {\bibfnamefont {F.}~\bibnamefont {{K{\"o}hlinger}}}, \bibinfo {author} {\bibfnamefont {A.~G.}\ \bibnamefont {{S{\'a}nchez}}}, \bibinfo {author} {\bibfnamefont {A.~H.}\ \bibnamefont {{Wright}}}, \bibinfo {author} {\bibfnamefont {M.}~\bibnamefont {{Bilicki}}}, \emph {et~al.},\ }\href {https://doi.org/10.1051/0004-6361/202038831} {\bibfield  {journal} {\bibinfo  {journal} {\aap}\ }\textbf {\bibinfo {volume} {646}},\ \bibinfo {eid} {A129} (\bibinfo {year} {2021})},\ \Eprint {https://arxiv.org/abs/2007.01844} {arXiv:2007.01844 [astro-ph.CO]} \BibitemShut {NoStop}%
\bibitem [{\citenamefont {{Hirata}}\ and\ \citenamefont {{Seljak}}(2004)}]{Hirata2004}%
  \BibitemOpen
  \bibfield  {author} {\bibinfo {author} {\bibfnamefont {C.~M.}\ \bibnamefont {{Hirata}}}\ and\ \bibinfo {author} {\bibfnamefont {U.}~\bibnamefont {{Seljak}}},\ }\href {https://doi.org/10.1103/PhysRevD.70.063526} {\bibfield  {journal} {\bibinfo  {journal} {\prd}\ }\textbf {\bibinfo {volume} {70}},\ \bibinfo {eid} {063526} (\bibinfo {year} {2004})},\ \Eprint {https://arxiv.org/abs/astro-ph/0406275} {arXiv:astro-ph/0406275 [astro-ph]} \BibitemShut {NoStop}%
\bibitem [{\citenamefont {{Limber}}(1953)}]{Limber1953}%
  \BibitemOpen
  \bibfield  {author} {\bibinfo {author} {\bibfnamefont {D.~N.}\ \bibnamefont {{Limber}}},\ }\href {https://doi.org/10.1086/145672} {\bibfield  {journal} {\bibinfo  {journal} {\apj}\ }\textbf {\bibinfo {volume} {117}},\ \bibinfo {pages} {134} (\bibinfo {year} {1953})}\BibitemShut {NoStop}%
\bibitem [{\citenamefont {{Blazek}}\ \emph {et~al.}(2019)\citenamefont {{Blazek}}, \citenamefont {{MacCrann}}, \citenamefont {{Troxel}},\ and\ \citenamefont {{Fang}}}]{Blazek2019}%
  \BibitemOpen
  \bibfield  {author} {\bibinfo {author} {\bibfnamefont {J.~A.}\ \bibnamefont {{Blazek}}}, \bibinfo {author} {\bibfnamefont {N.}~\bibnamefont {{MacCrann}}}, \bibinfo {author} {\bibfnamefont {M.~A.}\ \bibnamefont {{Troxel}}},\ and\ \bibinfo {author} {\bibfnamefont {X.}~\bibnamefont {{Fang}}},\ }\href {https://doi.org/10.1103/PhysRevD.100.103506} {\bibfield  {journal} {\bibinfo  {journal} {\prd}\ }\textbf {\bibinfo {volume} {100}},\ \bibinfo {eid} {103506} (\bibinfo {year} {2019})},\ \Eprint {https://arxiv.org/abs/1708.09247} {arXiv:1708.09247 [astro-ph.CO]} \BibitemShut {NoStop}%
\bibitem [{\citenamefont {{McEwen}}\ \emph {et~al.}(2016)\citenamefont {{McEwen}}, \citenamefont {{Fang}}, \citenamefont {{Hirata}},\ and\ \citenamefont {{Blazek}}}]{McEwen2016}%
  \BibitemOpen
  \bibfield  {author} {\bibinfo {author} {\bibfnamefont {J.~E.}\ \bibnamefont {{McEwen}}}, \bibinfo {author} {\bibfnamefont {X.}~\bibnamefont {{Fang}}}, \bibinfo {author} {\bibfnamefont {C.~M.}\ \bibnamefont {{Hirata}}},\ and\ \bibinfo {author} {\bibfnamefont {J.~A.}\ \bibnamefont {{Blazek}}},\ }\href {https://doi.org/10.1088/1475-7516/2016/09/015} {\bibfield  {journal} {\bibinfo  {journal} {\jcap}\ }\textbf {\bibinfo {volume} {2016}},\ \bibinfo {eid} {015} (\bibinfo {year} {2016})},\ \Eprint {https://arxiv.org/abs/1603.04826} {arXiv:1603.04826 [astro-ph.CO]} \BibitemShut {NoStop}%
\bibitem [{\citenamefont {{Fang}}\ \emph {et~al.}(2017)\citenamefont {{Fang}}, \citenamefont {{Blazek}}, \citenamefont {{McEwen}},\ and\ \citenamefont {{Hirata}}}]{Fang2017}%
  \BibitemOpen
  \bibfield  {author} {\bibinfo {author} {\bibfnamefont {X.}~\bibnamefont {{Fang}}}, \bibinfo {author} {\bibfnamefont {J.~A.}\ \bibnamefont {{Blazek}}}, \bibinfo {author} {\bibfnamefont {J.~E.}\ \bibnamefont {{McEwen}}},\ and\ \bibinfo {author} {\bibfnamefont {C.~M.}\ \bibnamefont {{Hirata}}},\ }\href {https://doi.org/10.1088/1475-7516/2017/02/030} {\bibfield  {journal} {\bibinfo  {journal} {\jcap}\ }\textbf {\bibinfo {volume} {2017}},\ \bibinfo {eid} {030} (\bibinfo {year} {2017})},\ \Eprint {https://arxiv.org/abs/1609.05978} {arXiv:1609.05978 [astro-ph.CO]} \BibitemShut {NoStop}%
\bibitem [{\citenamefont {{Bridle}}\ and\ \citenamefont {{King}}(2007)}]{Bridle2007}%
  \BibitemOpen
  \bibfield  {author} {\bibinfo {author} {\bibfnamefont {S.}~\bibnamefont {{Bridle}}}\ and\ \bibinfo {author} {\bibfnamefont {L.}~\bibnamefont {{King}}},\ }\href {https://doi.org/10.1088/1367-2630/9/12/444} {\bibfield  {journal} {\bibinfo  {journal} {New Journal of Physics}\ }\textbf {\bibinfo {volume} {9}},\ \bibinfo {pages} {444} (\bibinfo {year} {2007})},\ \Eprint {https://arxiv.org/abs/0705.0166} {arXiv:0705.0166 [astro-ph]} \BibitemShut {NoStop}%
\bibitem [{\citenamefont {{Krause}}\ and\ \citenamefont {{Hirata}}(2010)}]{Krause2010}%
  \BibitemOpen
  \bibfield  {author} {\bibinfo {author} {\bibfnamefont {E.}~\bibnamefont {{Krause}}}\ and\ \bibinfo {author} {\bibfnamefont {C.~M.}\ \bibnamefont {{Hirata}}},\ }\href {https://doi.org/10.1051/0004-6361/200913524} {\bibfield  {journal} {\bibinfo  {journal} {\aap}\ }\textbf {\bibinfo {volume} {523}},\ \bibinfo {eid} {A28} (\bibinfo {year} {2010})},\ \Eprint {https://arxiv.org/abs/0910.3786} {arXiv:0910.3786 [astro-ph.CO]} \BibitemShut {NoStop}%
\bibitem [{\citenamefont {{Sugiyama}}\ \emph {et~al.}(2020)\citenamefont {{Sugiyama}}, \citenamefont {{Takada}}, \citenamefont {{Kobayashi}}, \citenamefont {{Miyatake}}, \citenamefont {{Shirasaki}}, \citenamefont {{Nishimichi}},\ and\ \citenamefont {{Park}}}]{Sugiyama2020}%
  \BibitemOpen
  \bibfield  {author} {\bibinfo {author} {\bibfnamefont {S.}~\bibnamefont {{Sugiyama}}}, \bibinfo {author} {\bibfnamefont {M.}~\bibnamefont {{Takada}}}, \bibinfo {author} {\bibfnamefont {Y.}~\bibnamefont {{Kobayashi}}}, \bibinfo {author} {\bibfnamefont {H.}~\bibnamefont {{Miyatake}}}, \bibinfo {author} {\bibfnamefont {M.}~\bibnamefont {{Shirasaki}}}, \bibinfo {author} {\bibfnamefont {T.}~\bibnamefont {{Nishimichi}}},\ and\ \bibinfo {author} {\bibfnamefont {Y.}~\bibnamefont {{Park}}},\ }\href {https://doi.org/10.1103/PhysRevD.102.083520} {\bibfield  {journal} {\bibinfo  {journal} {\prd}\ }\textbf {\bibinfo {volume} {102}},\ \bibinfo {eid} {083520} (\bibinfo {year} {2020})},\ \Eprint {https://arxiv.org/abs/2008.06873} {arXiv:2008.06873 [astro-ph.CO]} \BibitemShut {NoStop}%
\bibitem [{\citenamefont {{Fang}}\ \emph {et~al.}(2020)\citenamefont {{Fang}}, \citenamefont {{Eifler}},\ and\ \citenamefont {{Krause}}}]{Fang2020}%
  \BibitemOpen
  \bibfield  {author} {\bibinfo {author} {\bibfnamefont {X.}~\bibnamefont {{Fang}}}, \bibinfo {author} {\bibfnamefont {T.}~\bibnamefont {{Eifler}}},\ and\ \bibinfo {author} {\bibfnamefont {E.}~\bibnamefont {{Krause}}},\ }\href {https://doi.org/10.1093/mnras/staa1726} {\bibfield  {journal} {\bibinfo  {journal} {\mnras}\ }\textbf {\bibinfo {volume} {497}},\ \bibinfo {pages} {2699} (\bibinfo {year} {2020})},\ \Eprint {https://arxiv.org/abs/2004.04833} {arXiv:2004.04833 [astro-ph.CO]} \BibitemShut {NoStop}%
\bibitem [{\citenamefont {{Zehavi}}\ \emph {et~al.}(2005)\citenamefont {{Zehavi}}, \citenamefont {{Zheng}}, \citenamefont {{Weinberg}}, \citenamefont {{Frieman}}, \citenamefont {{Berlind}}, \citenamefont {{Blanton}}, \citenamefont {{Scoccimarro}}, \citenamefont {{Sheth}}, \citenamefont {{Strauss}}, \citenamefont {{Kayo}}, \citenamefont {{Suto}}, \citenamefont {{Fukugita}}, \citenamefont {{Nakamura}}, \citenamefont {{Bahcall}}, \citenamefont {{Brinkmann}}, \citenamefont {{Gunn}}, \citenamefont {{Hennessy}}, \citenamefont {{Ivezi{\'c}}}, \citenamefont {{Knapp}}, \citenamefont {{Loveday}}, \citenamefont {{Meiksin}}, \citenamefont {{Schlegel}}, \citenamefont {{Schneider}}, \citenamefont {{Szapudi}}, \citenamefont {{Tegmark}}, \citenamefont {{Vogeley}}, \citenamefont {{York}},\ and\ \citenamefont {{SDSS Collaboration}}}]{Zehavi2005}%
  \BibitemOpen
  \bibfield  {author} {\bibinfo {author} {\bibfnamefont {I.}~\bibnamefont {{Zehavi}}}, \bibinfo {author} {\bibfnamefont {Z.}~\bibnamefont {{Zheng}}}, \bibinfo {author} {\bibfnamefont {D.~H.}\ \bibnamefont {{Weinberg}}}, \bibinfo {author} {\bibfnamefont {J.~A.}\ \bibnamefont {{Frieman}}}, \bibinfo {author} {\bibfnamefont {A.~A.}\ \bibnamefont {{Berlind}}}, \bibinfo {author} {\bibfnamefont {M.~R.}\ \bibnamefont {{Blanton}}}, \bibinfo {author} {\bibfnamefont {R.}~\bibnamefont {{Scoccimarro}}}, \bibinfo {author} {\bibfnamefont {R.~K.}\ \bibnamefont {{Sheth}}}, \bibinfo {author} {\bibfnamefont {M.~A.}\ \bibnamefont {{Strauss}}}, \bibinfo {author} {\bibfnamefont {I.}~\bibnamefont {{Kayo}}}, \bibinfo {author} {\bibfnamefont {Y.}~\bibnamefont {{Suto}}}, \bibinfo {author} {\bibfnamefont {M.}~\bibnamefont {{Fukugita}}}, \bibinfo {author} {\bibfnamefont {O.}~\bibnamefont {{Nakamura}}}, \bibinfo {author} {\bibfnamefont {N.~A.}\ \bibnamefont {{Bahcall}}}, \bibinfo {author} {\bibfnamefont {J.}~\bibnamefont {{Brinkmann}}},
  \bibinfo {author} {\bibfnamefont {J.~E.}\ \bibnamefont {{Gunn}}}, \bibinfo {author} {\bibfnamefont {G.~S.}\ \bibnamefont {{Hennessy}}}, \bibinfo {author} {\bibfnamefont {{\v{Z}}.}~\bibnamefont {{Ivezi{\'c}}}}, \bibinfo {author} {\bibfnamefont {G.~R.}\ \bibnamefont {{Knapp}}}, \bibinfo {author} {\bibfnamefont {J.}~\bibnamefont {{Loveday}}}, \bibinfo {author} {\bibfnamefont {A.}~\bibnamefont {{Meiksin}}}, \bibinfo {author} {\bibfnamefont {D.~J.}\ \bibnamefont {{Schlegel}}}, \bibinfo {author} {\bibfnamefont {D.~P.}\ \bibnamefont {{Schneider}}}, \bibinfo {author} {\bibfnamefont {I.}~\bibnamefont {{Szapudi}}}, \bibinfo {author} {\bibfnamefont {M.}~\bibnamefont {{Tegmark}}}, \bibinfo {author} {\bibfnamefont {M.~S.}\ \bibnamefont {{Vogeley}}}, \bibinfo {author} {\bibfnamefont {D.~G.}\ \bibnamefont {{York}}},\ and\ \bibinfo {author} {\bibnamefont {{SDSS Collaboration}}},\ }\href {https://doi.org/10.1086/431891} {\bibfield  {journal} {\bibinfo  {journal} {\apj}\ }\textbf {\bibinfo {volume} {630}},\ \bibinfo {pages}
  {1} (\bibinfo {year} {2005})},\ \Eprint {https://arxiv.org/abs/astro-ph/0408569} {arXiv:astro-ph/0408569 [astro-ph]} \BibitemShut {NoStop}%
\bibitem [{\citenamefont {{Kaiser}}(1984)}]{Kaiser1984}%
  \BibitemOpen
  \bibfield  {author} {\bibinfo {author} {\bibfnamefont {N.}~\bibnamefont {{Kaiser}}},\ }\href {https://doi.org/10.1086/184341} {\bibfield  {journal} {\bibinfo  {journal} {\apjl}\ }\textbf {\bibinfo {volume} {284}},\ \bibinfo {pages} {L9} (\bibinfo {year} {1984})}\BibitemShut {NoStop}%
\bibitem [{\citenamefont {{Baldauf}}\ \emph {et~al.}(2010)\citenamefont {{Baldauf}}, \citenamefont {{Smith}}, \citenamefont {{Seljak}},\ and\ \citenamefont {{Mandelbaum}}}]{Baldauf2010}%
  \BibitemOpen
  \bibfield  {author} {\bibinfo {author} {\bibfnamefont {T.}~\bibnamefont {{Baldauf}}}, \bibinfo {author} {\bibfnamefont {R.~E.}\ \bibnamefont {{Smith}}}, \bibinfo {author} {\bibfnamefont {U.}~\bibnamefont {{Seljak}}},\ and\ \bibinfo {author} {\bibfnamefont {R.}~\bibnamefont {{Mandelbaum}}},\ }\href {https://doi.org/10.1103/PhysRevD.81.063531} {\bibfield  {journal} {\bibinfo  {journal} {\prd}\ }\textbf {\bibinfo {volume} {81}},\ \bibinfo {eid} {063531} (\bibinfo {year} {2010})},\ \Eprint {https://arxiv.org/abs/0911.4973} {arXiv:0911.4973 [astro-ph.CO]} \BibitemShut {NoStop}%
\bibitem [{\citenamefont {{Dodelson}}(2010)}]{Dodelson2010}%
  \BibitemOpen
  \bibfield  {author} {\bibinfo {author} {\bibfnamefont {S.}~\bibnamefont {{Dodelson}}},\ }\href {https://doi.org/10.1103/PhysRevD.82.023522} {\bibfield  {journal} {\bibinfo  {journal} {\prd}\ }\textbf {\bibinfo {volume} {82}},\ \bibinfo {eid} {023522} (\bibinfo {year} {2010})},\ \Eprint {https://arxiv.org/abs/1001.5012} {arXiv:1001.5012 [astro-ph.CO]} \BibitemShut {NoStop}%
\bibitem [{\citenamefont {{Jefferson}}\ \emph {et~al.}(2025)\citenamefont {{Jefferson}}, \citenamefont {{Omori}}, \citenamefont {{Chang}}, \citenamefont {{Agarwal}}, \citenamefont {{Zuntz}}, \citenamefont {{Asgari}}, \citenamefont {{Gatti}}, \citenamefont {{Giblin}}, \citenamefont {{H{\'e}bert}}, \citenamefont {{Jarvis}}, \citenamefont {{Pedersen}}, \citenamefont {{Prat}}, \citenamefont {{Schutt}}, \citenamefont {{Zhang}},\ and\ \citenamefont {{the LSST Dark Energy Science Collaboration}}}]{jefferson2025}%
  \BibitemOpen
  \bibfield  {author} {\bibinfo {author} {\bibfnamefont {J.}~\bibnamefont {{Jefferson}}}, \bibinfo {author} {\bibfnamefont {Y.}~\bibnamefont {{Omori}}}, \bibinfo {author} {\bibfnamefont {C.}~\bibnamefont {{Chang}}}, \bibinfo {author} {\bibfnamefont {S.}~\bibnamefont {{Agarwal}}}, \bibinfo {author} {\bibfnamefont {J.}~\bibnamefont {{Zuntz}}}, \bibinfo {author} {\bibfnamefont {M.}~\bibnamefont {{Asgari}}}, \bibinfo {author} {\bibfnamefont {M.}~\bibnamefont {{Gatti}}}, \bibinfo {author} {\bibfnamefont {B.}~\bibnamefont {{Giblin}}}, \bibinfo {author} {\bibfnamefont {C.-A.}\ \bibnamefont {{H{\'e}bert}}}, \bibinfo {author} {\bibfnamefont {M.}~\bibnamefont {{Jarvis}}}, \bibinfo {author} {\bibfnamefont {E.~M.}\ \bibnamefont {{Pedersen}}}, \bibinfo {author} {\bibfnamefont {J.}~\bibnamefont {{Prat}}}, \bibinfo {author} {\bibfnamefont {T.}~\bibnamefont {{Schutt}}}, \bibinfo {author} {\bibfnamefont {T.}~\bibnamefont {{Zhang}}},\ and\ \bibinfo {author} {\bibnamefont {{the LSST Dark Energy Science Collaboration}}},\ }\href
  {https://doi.org/10.48550/arXiv.2505.03964} {\bibfield  {journal} {\bibinfo  {journal} {arXiv e-prints}\ ,\ \bibinfo {eid} {arXiv:2505.03964}} (\bibinfo {year} {2025})},\ \Eprint {https://arxiv.org/abs/2505.03964} {arXiv:2505.03964 [astro-ph.CO]} \BibitemShut {NoStop}%
\bibitem [{\citenamefont {{Oguri}}\ \emph {et~al.}(2018)\citenamefont {{Oguri}}, \citenamefont {{Lin}}, \citenamefont {{Lin}}, \citenamefont {{Nishizawa}}, \citenamefont {{More}}, \citenamefont {{More}}, \citenamefont {{Hsieh}}, \citenamefont {{Medezinski}}, \citenamefont {{Miyatake}}, \citenamefont {{Jian}} \emph {et~al.}}]{Oguri2018_camira}%
  \BibitemOpen
  \bibfield  {author} {\bibinfo {author} {\bibfnamefont {M.}~\bibnamefont {{Oguri}}}, \bibinfo {author} {\bibfnamefont {Y.-T.}\ \bibnamefont {{Lin}}}, \bibinfo {author} {\bibfnamefont {S.-C.}\ \bibnamefont {{Lin}}}, \bibinfo {author} {\bibfnamefont {A.~J.}\ \bibnamefont {{Nishizawa}}}, \bibinfo {author} {\bibfnamefont {A.}~\bibnamefont {{More}}}, \bibinfo {author} {\bibfnamefont {S.}~\bibnamefont {{More}}}, \bibinfo {author} {\bibfnamefont {B.-C.}\ \bibnamefont {{Hsieh}}}, \bibinfo {author} {\bibfnamefont {E.}~\bibnamefont {{Medezinski}}}, \bibinfo {author} {\bibfnamefont {H.}~\bibnamefont {{Miyatake}}}, \bibinfo {author} {\bibfnamefont {H.-Y.}\ \bibnamefont {{Jian}}}, \emph {et~al.},\ }\href {https://doi.org/10.1093/pasj/psx042} {\bibfield  {journal} {\bibinfo  {journal} {\pasj}\ }\textbf {\bibinfo {volume} {70}},\ \bibinfo {eid} {S20} (\bibinfo {year} {2018})},\ \Eprint {https://arxiv.org/abs/1701.00818} {arXiv:1701.00818 [astro-ph.CO]} \BibitemShut {NoStop}%
\bibitem [{\citenamefont {{Zhang}}\ \emph {et~al.}(2023{\natexlab{b}})\citenamefont {{Zhang}}, \citenamefont {{Rau}}, \citenamefont {{Mandelbaum}}, \citenamefont {{Li}},\ and\ \citenamefont {{Moews}}}]{Zhang2023_nz}%
  \BibitemOpen
  \bibfield  {author} {\bibinfo {author} {\bibfnamefont {T.}~\bibnamefont {{Zhang}}}, \bibinfo {author} {\bibfnamefont {M.~M.}\ \bibnamefont {{Rau}}}, \bibinfo {author} {\bibfnamefont {R.}~\bibnamefont {{Mandelbaum}}}, \bibinfo {author} {\bibfnamefont {X.}~\bibnamefont {{Li}}},\ and\ \bibinfo {author} {\bibfnamefont {B.}~\bibnamefont {{Moews}}},\ }\href {https://doi.org/10.1093/mnras/stac3090} {\bibfield  {journal} {\bibinfo  {journal} {\mnras}\ }\textbf {\bibinfo {volume} {518}},\ \bibinfo {pages} {709} (\bibinfo {year} {2023}{\natexlab{b}})},\ \Eprint {https://arxiv.org/abs/2206.10169} {arXiv:2206.10169 [astro-ph.CO]} \BibitemShut {NoStop}%
\bibitem [{\citenamefont {{More}}\ \emph {et~al.}(2015)\citenamefont {{More}}, \citenamefont {{Miyatake}}, \citenamefont {{Mandelbaum}}, \citenamefont {{Takada}}, \citenamefont {{Spergel}}, \citenamefont {{Brownstein}},\ and\ \citenamefont {{Schneider}}}]{more2015}%
  \BibitemOpen
  \bibfield  {author} {\bibinfo {author} {\bibfnamefont {S.}~\bibnamefont {{More}}}, \bibinfo {author} {\bibfnamefont {H.}~\bibnamefont {{Miyatake}}}, \bibinfo {author} {\bibfnamefont {R.}~\bibnamefont {{Mandelbaum}}}, \bibinfo {author} {\bibfnamefont {M.}~\bibnamefont {{Takada}}}, \bibinfo {author} {\bibfnamefont {D.~N.}\ \bibnamefont {{Spergel}}}, \bibinfo {author} {\bibfnamefont {J.~R.}\ \bibnamefont {{Brownstein}}},\ and\ \bibinfo {author} {\bibfnamefont {D.~P.}\ \bibnamefont {{Schneider}}},\ }\href {https://doi.org/10.1088/0004-637X/806/1/2} {\bibfield  {journal} {\bibinfo  {journal} {\apj}\ }\textbf {\bibinfo {volume} {806}},\ \bibinfo {eid} {2} (\bibinfo {year} {2015})},\ \Eprint {https://arxiv.org/abs/1407.1856} {arXiv:1407.1856 [astro-ph.CO]} \BibitemShut {NoStop}%
\bibitem [{\citenamefont {{Newman}}(2008)}]{Newman2008}%
  \BibitemOpen
  \bibfield  {author} {\bibinfo {author} {\bibfnamefont {J.~A.}\ \bibnamefont {{Newman}}},\ }\href {https://doi.org/10.1086/589982} {\bibfield  {journal} {\bibinfo  {journal} {\apj}\ }\textbf {\bibinfo {volume} {684}},\ \bibinfo {pages} {88} (\bibinfo {year} {2008})},\ \Eprint {https://arxiv.org/abs/0805.1409} {arXiv:0805.1409 [astro-ph]} \BibitemShut {NoStop}%
\bibitem [{\citenamefont {{Hildebrandt}}\ \emph {et~al.}(2017)\citenamefont {{Hildebrandt}}, \citenamefont {{Viola}}, \citenamefont {{Heymans}}, \citenamefont {{Joudaki}}, \citenamefont {{Kuijken}}, \citenamefont {{Blake}}, \citenamefont {{Erben}}, \citenamefont {{Joachimi}}, \citenamefont {{Klaes}}, \citenamefont {{Miller}} \emph {et~al.}}]{Hildebrandt2017}%
  \BibitemOpen
  \bibfield  {author} {\bibinfo {author} {\bibfnamefont {H.}~\bibnamefont {{Hildebrandt}}}, \bibinfo {author} {\bibfnamefont {M.}~\bibnamefont {{Viola}}}, \bibinfo {author} {\bibfnamefont {C.}~\bibnamefont {{Heymans}}}, \bibinfo {author} {\bibfnamefont {S.}~\bibnamefont {{Joudaki}}}, \bibinfo {author} {\bibfnamefont {K.}~\bibnamefont {{Kuijken}}}, \bibinfo {author} {\bibfnamefont {C.}~\bibnamefont {{Blake}}}, \bibinfo {author} {\bibfnamefont {T.}~\bibnamefont {{Erben}}}, \bibinfo {author} {\bibfnamefont {B.}~\bibnamefont {{Joachimi}}}, \bibinfo {author} {\bibfnamefont {D.}~\bibnamefont {{Klaes}}}, \bibinfo {author} {\bibfnamefont {L.}~\bibnamefont {{Miller}}}, \emph {et~al.},\ }\href {https://doi.org/10.1093/mnras/stw2805} {\bibfield  {journal} {\bibinfo  {journal} {\mnras}\ }\textbf {\bibinfo {volume} {465}},\ \bibinfo {pages} {1454} (\bibinfo {year} {2017})},\ \Eprint {https://arxiv.org/abs/1606.05338} {arXiv:1606.05338 [astro-ph.CO]} \BibitemShut {NoStop}%
\bibitem [{\citenamefont {{DESI Collaboration}}\ \emph {et~al.}(2025)\citenamefont {{DESI Collaboration}}, \citenamefont {{Karim}}, \citenamefont {{Adame}}, \citenamefont {{Aguado}}, \citenamefont {{Aguilar}}, \citenamefont {{Ahlen}}, \citenamefont {{Alam}}, \citenamefont {{Aldering}}, \citenamefont {{Alexander}}, \citenamefont {{Alfarsy}}, \citenamefont {{Allen}}, \citenamefont {{Allende Prieto}}, \citenamefont {{Alves}}, \citenamefont {{Anand}}, \citenamefont {{Andrade}}, \citenamefont {{Armengaud}}, \citenamefont {{Avila}}, \citenamefont {{Aviles}}, \citenamefont {{Awan}}, \citenamefont {{Bailey}}, \citenamefont {{Baleato Lizancos}}, \citenamefont {{Ballester}}, \citenamefont {{Bault}}, \citenamefont {{Bautista}}, \citenamefont {{BenZvi}}, \citenamefont {{Silva}}, \citenamefont {{Bermejo-Climent}}, \citenamefont {{Beutler}}, \citenamefont {{Bianchi}}, \citenamefont {{Blake}}, \citenamefont {{Blum}}, \citenamefont {{Bolton}}, \citenamefont {{Bonici}}, \citenamefont {{Brieden}}, \citenamefont {{Brodzeller}},
  \citenamefont {{Brooks}}, \citenamefont {{Buckley-Geer}}, \citenamefont {{Burtin}}, \citenamefont {{Canning}}, \citenamefont {{Carnero Rosell}}, \citenamefont {{Carr}}, \citenamefont {{Carrilho}}, \citenamefont {{Casas}}, \citenamefont {{Castander}}, \citenamefont {{Cereskaite}}, \citenamefont {{Cervantes-Cota}}, \citenamefont {{Chaussidon}}, \citenamefont {{Chaves-Montero}}, \citenamefont {{Chen}}, \citenamefont {{Chen}}, \citenamefont {{Claybaugh}}, \citenamefont {{Cole}}, \citenamefont {{Cooper}}, \citenamefont {{Cousinou}}, \citenamefont {{Cuceu}}, \citenamefont {{Davis}}, \citenamefont {{Dawson}}, \citenamefont {{de Belsunce}}, \citenamefont {{de la Cruz}}, \citenamefont {{de la Macorra}}, \citenamefont {{de Mattia}}, \citenamefont {{Deiosso}}, \citenamefont {{Della Costa}}, \citenamefont {{Demina}}, \citenamefont {{Demirbozan}}, \citenamefont {{DeRose}}, \citenamefont {{Dey}}, \citenamefont {{Dey}}, \citenamefont {{Ding}}, \citenamefont {{Ding}}, \citenamefont {{Doel}}, \citenamefont {{Douglass}},
  \citenamefont {{Dowicz}}, \citenamefont {{Ebina}}, \citenamefont {{Edelstein}}, \citenamefont {{Eisenstein}}, \citenamefont {{Elbers}}, \citenamefont {{Emas}}, \citenamefont {{Escoffier}}, \citenamefont {{Fagrelius}}, \citenamefont {{Fan}}, \citenamefont {{Fanning}}, \citenamefont {{Fawcett}}, \citenamefont {{Fern{\'a}ndez-Garc{\'\i}a}}, \citenamefont {{Ferraro}}, \citenamefont {{Findlay}}, \citenamefont {{Font-Ribera}}, \citenamefont {{Forero-Romero}}, \citenamefont {{Forero-S{\'a}nchez}}, \citenamefont {{Frenk}}, \citenamefont {{G{\"a}nsicke}}, \citenamefont {{Galbany}}, \citenamefont {{Garc{\'\i}a-Bellido}}, \citenamefont {{Garcia-Quintero}}, \citenamefont {{Garrison}}, \citenamefont {{Gazta{\~n}aga}}, \citenamefont {{Gil-Mar{\'\i}n}}, \citenamefont {{Gnedin}}, \citenamefont {{Gontcho}}, \citenamefont {{Gonzalez-Morales}}, \citenamefont {{Gonzalez-Perez}}, \citenamefont {{Gordon}}, \citenamefont {{Graur}}, \citenamefont {{Green}}, \citenamefont {{Gruen}}, \citenamefont {{Gsponer}}, \citenamefont
  {{Guandalin}}, \citenamefont {{Gutierrez}}, \citenamefont {{Guy}}, \citenamefont {{Hahn}}, \citenamefont {{Han}}, \citenamefont {{Han}}, \citenamefont {{He}}, \citenamefont {{Herrera-Alcantar}}, \citenamefont {{Honscheid}}, \citenamefont {{Hou}}, \citenamefont {{Howlett}}, \citenamefont {{Huterer}}, \citenamefont {{Ir{\v{s}}i{\v{c}}}}, \citenamefont {{Ishak}}, \citenamefont {{Jacques}}, \citenamefont {{Jimenez}}, \citenamefont {{Jing}}, \citenamefont {{Joachimi}}, \citenamefont {{Joudaki}}, \citenamefont {{Joyce}}, \citenamefont {{Jullo}}, \citenamefont {{Juneau}}, \citenamefont {{Kara{\c{c}}ayl{\i}}}, \citenamefont {{Karim}}, \citenamefont {{Kehoe}}, \citenamefont {{Kent}}, \citenamefont {{Khederlarian}}, \citenamefont {{Kirkby}}, \citenamefont {{Kisner}}, \citenamefont {{Kitaura}}, \citenamefont {{Kizhuprakkat}}, \citenamefont {{Kong}}, \citenamefont {{Koposov}}, \citenamefont {{Kremin}}, \citenamefont {{Krolewski}}, \citenamefont {{Lahav}}, \citenamefont {{Lai}}, \citenamefont {{Lamman}}, \citenamefont
  {{Lan}}, \citenamefont {{Landriau}}, \citenamefont {{Lang}}, \citenamefont {{Lange}}, \citenamefont {{Lasker}}, \citenamefont {{Le Goff}}, \citenamefont {{Le Guillou}}, \citenamefont {{Leauthaud}}, \citenamefont {{Levi}}, \citenamefont {{Li}}, \citenamefont {{Li}}, \citenamefont {{Lodha}}, \citenamefont {{Lokken}}, \citenamefont {{Luo}}, \citenamefont {{Magneville}}, \citenamefont {{Manera}}, \citenamefont {{Manser}}, \citenamefont {{Margala}}, \citenamefont {{Martini}}, \citenamefont {{Maus}}, \citenamefont {{McCullough}}, \citenamefont {{McDonald}}, \citenamefont {{Medina}}, \citenamefont {{Medina-Varela}}, \citenamefont {{Meisner}}, \citenamefont {{Mena-Fern{\'a}ndez}}, \citenamefont {{Menegas}}, \citenamefont {{Mezcua}}, \citenamefont {{Miquel}}, \citenamefont {{Montero-Camacho}}, \citenamefont {{Moon}}, \citenamefont {{Moustakas}}, \citenamefont {{Mu{\~n}oz-Guti{\'e}rrez}}, \citenamefont {{Mu{\~n}oz-Santos}}, \citenamefont {{Myers}}, \citenamefont {{Myles}}, \citenamefont {{Nadathur}}, \citenamefont
  {{Najita}}, \citenamefont {{Napolitano}}, \citenamefont {{Newman}}, \citenamefont {{Nikakhtar}}, \citenamefont {{Nikutta}}, \citenamefont {{Niz}}, \citenamefont {{Noriega}}, \citenamefont {{Padmanabhan}}, \citenamefont {{Paillas}}, \citenamefont {{Palanque-Delabrouille}}, \citenamefont {{Palmese}}, \citenamefont {{Pan}}, \citenamefont {{Pan}}, \citenamefont {{Parkinson}}, \citenamefont {{Peacock}}, \citenamefont {{Percival}}, \citenamefont {{P{\'e}rez-Fern{\'a}ndez}}, \citenamefont {{P{\'e}rez-R{\`a}fols}},\ and\ \citenamefont {{Peterson}}}]{desi_dr1}%
  \BibitemOpen
  \bibfield  {author} {\bibinfo {author} {\bibnamefont {{DESI Collaboration}}}, \bibinfo {author} {\bibfnamefont {M.~A.}\ \bibnamefont {{Karim}}}, \bibinfo {author} {\bibfnamefont {A.~G.}\ \bibnamefont {{Adame}}}, \bibinfo {author} {\bibfnamefont {D.}~\bibnamefont {{Aguado}}}, \bibinfo {author} {\bibfnamefont {J.}~\bibnamefont {{Aguilar}}}, \bibinfo {author} {\bibfnamefont {S.}~\bibnamefont {{Ahlen}}}, \bibinfo {author} {\bibfnamefont {S.}~\bibnamefont {{Alam}}}, \bibinfo {author} {\bibfnamefont {G.}~\bibnamefont {{Aldering}}}, \bibinfo {author} {\bibfnamefont {D.~M.}\ \bibnamefont {{Alexander}}}, \bibinfo {author} {\bibfnamefont {R.}~\bibnamefont {{Alfarsy}}}, \bibinfo {author} {\bibfnamefont {L.}~\bibnamefont {{Allen}}}, \bibinfo {author} {\bibfnamefont {C.}~\bibnamefont {{Allende Prieto}}}, \bibinfo {author} {\bibfnamefont {O.}~\bibnamefont {{Alves}}}, \bibinfo {author} {\bibfnamefont {A.}~\bibnamefont {{Anand}}}, \bibinfo {author} {\bibfnamefont {U.}~\bibnamefont {{Andrade}}}, \bibinfo {author}
  {\bibfnamefont {E.}~\bibnamefont {{Armengaud}}}, \bibinfo {author} {\bibfnamefont {S.}~\bibnamefont {{Avila}}}, \bibinfo {author} {\bibfnamefont {A.}~\bibnamefont {{Aviles}}}, \bibinfo {author} {\bibfnamefont {H.}~\bibnamefont {{Awan}}}, \bibinfo {author} {\bibfnamefont {S.}~\bibnamefont {{Bailey}}}, \bibinfo {author} {\bibfnamefont {A.}~\bibnamefont {{Baleato Lizancos}}}, \bibinfo {author} {\bibfnamefont {O.}~\bibnamefont {{Ballester}}}, \bibinfo {author} {\bibfnamefont {A.}~\bibnamefont {{Bault}}}, \bibinfo {author} {\bibfnamefont {J.}~\bibnamefont {{Bautista}}}, \bibinfo {author} {\bibfnamefont {S.}~\bibnamefont {{BenZvi}}}, \bibinfo {author} {\bibfnamefont {L.~B.~e.}\ \bibnamefont {{Silva}}}, \bibinfo {author} {\bibfnamefont {J.~R.}\ \bibnamefont {{Bermejo-Climent}}}, \bibinfo {author} {\bibfnamefont {F.}~\bibnamefont {{Beutler}}}, \bibinfo {author} {\bibfnamefont {D.}~\bibnamefont {{Bianchi}}}, \bibinfo {author} {\bibfnamefont {C.}~\bibnamefont {{Blake}}}, \bibinfo {author} {\bibfnamefont
  {R.}~\bibnamefont {{Blum}}}, \bibinfo {author} {\bibfnamefont {A.~S.}\ \bibnamefont {{Bolton}}}, \bibinfo {author} {\bibfnamefont {M.}~\bibnamefont {{Bonici}}}, \bibinfo {author} {\bibfnamefont {S.}~\bibnamefont {{Brieden}}}, \bibinfo {author} {\bibfnamefont {A.}~\bibnamefont {{Brodzeller}}}, \bibinfo {author} {\bibfnamefont {D.}~\bibnamefont {{Brooks}}}, \bibinfo {author} {\bibfnamefont {E.}~\bibnamefont {{Buckley-Geer}}}, \bibinfo {author} {\bibfnamefont {E.}~\bibnamefont {{Burtin}}}, \bibinfo {author} {\bibfnamefont {R.}~\bibnamefont {{Canning}}}, \bibinfo {author} {\bibfnamefont {A.}~\bibnamefont {{Carnero Rosell}}}, \bibinfo {author} {\bibfnamefont {A.}~\bibnamefont {{Carr}}}, \bibinfo {author} {\bibfnamefont {P.}~\bibnamefont {{Carrilho}}}, \bibinfo {author} {\bibfnamefont {L.}~\bibnamefont {{Casas}}}, \bibinfo {author} {\bibfnamefont {F.~J.}\ \bibnamefont {{Castander}}}, \bibinfo {author} {\bibfnamefont {R.}~\bibnamefont {{Cereskaite}}}, \bibinfo {author} {\bibfnamefont {J.~L.}\ \bibnamefont
  {{Cervantes-Cota}}}, \bibinfo {author} {\bibfnamefont {E.}~\bibnamefont {{Chaussidon}}}, \bibinfo {author} {\bibfnamefont {J.}~\bibnamefont {{Chaves-Montero}}}, \bibinfo {author} {\bibfnamefont {S.}~\bibnamefont {{Chen}}}, \bibinfo {author} {\bibfnamefont {X.}~\bibnamefont {{Chen}}}, \bibinfo {author} {\bibfnamefont {T.}~\bibnamefont {{Claybaugh}}}, \bibinfo {author} {\bibfnamefont {S.}~\bibnamefont {{Cole}}}, \bibinfo {author} {\bibfnamefont {A.~P.}\ \bibnamefont {{Cooper}}}, \bibinfo {author} {\bibfnamefont {M.~C.}\ \bibnamefont {{Cousinou}}}, \bibinfo {author} {\bibfnamefont {A.}~\bibnamefont {{Cuceu}}}, \bibinfo {author} {\bibfnamefont {T.~M.}\ \bibnamefont {{Davis}}}, \bibinfo {author} {\bibfnamefont {K.~S.}\ \bibnamefont {{Dawson}}}, \bibinfo {author} {\bibfnamefont {R.}~\bibnamefont {{de Belsunce}}}, \bibinfo {author} {\bibfnamefont {R.}~\bibnamefont {{de la Cruz}}}, \bibinfo {author} {\bibfnamefont {A.}~\bibnamefont {{de la Macorra}}}, \bibinfo {author} {\bibfnamefont {A.}~\bibnamefont {{de
  Mattia}}}, \bibinfo {author} {\bibfnamefont {N.}~\bibnamefont {{Deiosso}}}, \bibinfo {author} {\bibfnamefont {J.}~\bibnamefont {{Della Costa}}}, \bibinfo {author} {\bibfnamefont {R.}~\bibnamefont {{Demina}}}, \bibinfo {author} {\bibfnamefont {U.}~\bibnamefont {{Demirbozan}}}, \bibinfo {author} {\bibfnamefont {J.}~\bibnamefont {{DeRose}}}, \bibinfo {author} {\bibfnamefont {A.}~\bibnamefont {{Dey}}}, \bibinfo {author} {\bibfnamefont {B.}~\bibnamefont {{Dey}}}, \bibinfo {author} {\bibfnamefont {J.}~\bibnamefont {{Ding}}}, \bibinfo {author} {\bibfnamefont {Z.}~\bibnamefont {{Ding}}}, \bibinfo {author} {\bibfnamefont {P.}~\bibnamefont {{Doel}}}, \bibinfo {author} {\bibfnamefont {K.}~\bibnamefont {{Douglass}}}, \bibinfo {author} {\bibfnamefont {M.}~\bibnamefont {{Dowicz}}}, \bibinfo {author} {\bibfnamefont {H.}~\bibnamefont {{Ebina}}}, \bibinfo {author} {\bibfnamefont {J.}~\bibnamefont {{Edelstein}}}, \bibinfo {author} {\bibfnamefont {D.~J.}\ \bibnamefont {{Eisenstein}}}, \bibinfo {author} {\bibfnamefont
  {W.}~\bibnamefont {{Elbers}}}, \bibinfo {author} {\bibfnamefont {N.}~\bibnamefont {{Emas}}}, \bibinfo {author} {\bibfnamefont {S.}~\bibnamefont {{Escoffier}}}, \bibinfo {author} {\bibfnamefont {P.}~\bibnamefont {{Fagrelius}}}, \bibinfo {author} {\bibfnamefont {X.}~\bibnamefont {{Fan}}}, \bibinfo {author} {\bibfnamefont {K.}~\bibnamefont {{Fanning}}}, \bibinfo {author} {\bibfnamefont {V.~A.}\ \bibnamefont {{Fawcett}}}, \bibinfo {author} {\bibfnamefont {E.}~\bibnamefont {{Fern{\'a}ndez-Garc{\'\i}a}}}, \bibinfo {author} {\bibfnamefont {S.}~\bibnamefont {{Ferraro}}}, \bibinfo {author} {\bibfnamefont {N.}~\bibnamefont {{Findlay}}}, \bibinfo {author} {\bibfnamefont {A.}~\bibnamefont {{Font-Ribera}}}, \bibinfo {author} {\bibfnamefont {J.~E.}\ \bibnamefont {{Forero-Romero}}}, \bibinfo {author} {\bibfnamefont {D.}~\bibnamefont {{Forero-S{\'a}nchez}}}, \bibinfo {author} {\bibfnamefont {C.~S.}\ \bibnamefont {{Frenk}}}, \bibinfo {author} {\bibfnamefont {B.~T.}\ \bibnamefont {{G{\"a}nsicke}}}, \bibinfo {author}
  {\bibfnamefont {L.}~\bibnamefont {{Galbany}}}, \bibinfo {author} {\bibfnamefont {J.}~\bibnamefont {{Garc{\'\i}a-Bellido}}}, \bibinfo {author} {\bibfnamefont {C.}~\bibnamefont {{Garcia-Quintero}}}, \bibinfo {author} {\bibfnamefont {L.~H.}\ \bibnamefont {{Garrison}}}, \bibinfo {author} {\bibfnamefont {E.}~\bibnamefont {{Gazta{\~n}aga}}}, \bibinfo {author} {\bibfnamefont {H.}~\bibnamefont {{Gil-Mar{\'\i}n}}}, \bibinfo {author} {\bibfnamefont {O.~Y.}\ \bibnamefont {{Gnedin}}}, \bibinfo {author} {\bibfnamefont {S.~G.~A.}\ \bibnamefont {{Gontcho}}}, \bibinfo {author} {\bibfnamefont {A.~X.}\ \bibnamefont {{Gonzalez-Morales}}}, \bibinfo {author} {\bibfnamefont {V.}~\bibnamefont {{Gonzalez-Perez}}}, \bibinfo {author} {\bibfnamefont {C.}~\bibnamefont {{Gordon}}}, \bibinfo {author} {\bibfnamefont {O.}~\bibnamefont {{Graur}}}, \bibinfo {author} {\bibfnamefont {D.}~\bibnamefont {{Green}}}, \bibinfo {author} {\bibfnamefont {D.}~\bibnamefont {{Gruen}}}, \bibinfo {author} {\bibfnamefont {R.}~\bibnamefont {{Gsponer}}},
  \bibinfo {author} {\bibfnamefont {C.}~\bibnamefont {{Guandalin}}}, \bibinfo {author} {\bibfnamefont {G.}~\bibnamefont {{Gutierrez}}}, \bibinfo {author} {\bibfnamefont {J.}~\bibnamefont {{Guy}}}, \bibinfo {author} {\bibfnamefont {C.}~\bibnamefont {{Hahn}}}, \bibinfo {author} {\bibfnamefont {J.~J.}\ \bibnamefont {{Han}}}, \bibinfo {author} {\bibfnamefont {J.}~\bibnamefont {{Han}}}, \bibinfo {author} {\bibfnamefont {S.}~\bibnamefont {{He}}}, \bibinfo {author} {\bibfnamefont {H.~K.}\ \bibnamefont {{Herrera-Alcantar}}}, \bibinfo {author} {\bibfnamefont {K.}~\bibnamefont {{Honscheid}}}, \bibinfo {author} {\bibfnamefont {J.}~\bibnamefont {{Hou}}}, \bibinfo {author} {\bibfnamefont {C.}~\bibnamefont {{Howlett}}}, \bibinfo {author} {\bibfnamefont {D.}~\bibnamefont {{Huterer}}}, \bibinfo {author} {\bibfnamefont {V.}~\bibnamefont {{Ir{\v{s}}i{\v{c}}}}}, \bibinfo {author} {\bibfnamefont {M.}~\bibnamefont {{Ishak}}}, \bibinfo {author} {\bibfnamefont {A.}~\bibnamefont {{Jacques}}}, \bibinfo {author} {\bibfnamefont
  {J.}~\bibnamefont {{Jimenez}}}, \bibinfo {author} {\bibfnamefont {Y.~P.}\ \bibnamefont {{Jing}}}, \bibinfo {author} {\bibfnamefont {B.}~\bibnamefont {{Joachimi}}}, \bibinfo {author} {\bibfnamefont {S.}~\bibnamefont {{Joudaki}}}, \bibinfo {author} {\bibfnamefont {R.}~\bibnamefont {{Joyce}}}, \bibinfo {author} {\bibfnamefont {E.}~\bibnamefont {{Jullo}}}, \bibinfo {author} {\bibfnamefont {S.}~\bibnamefont {{Juneau}}}, \bibinfo {author} {\bibfnamefont {N.~G.}\ \bibnamefont {{Kara{\c{c}}ayl{\i}}}}, \bibinfo {author} {\bibfnamefont {T.}~\bibnamefont {{Karim}}}, \bibinfo {author} {\bibfnamefont {R.}~\bibnamefont {{Kehoe}}}, \bibinfo {author} {\bibfnamefont {S.}~\bibnamefont {{Kent}}}, \bibinfo {author} {\bibfnamefont {A.}~\bibnamefont {{Khederlarian}}}, \bibinfo {author} {\bibfnamefont {D.}~\bibnamefont {{Kirkby}}}, \bibinfo {author} {\bibfnamefont {T.}~\bibnamefont {{Kisner}}}, \bibinfo {author} {\bibfnamefont {F.~S.}\ \bibnamefont {{Kitaura}}}, \bibinfo {author} {\bibfnamefont {N.}~\bibnamefont
  {{Kizhuprakkat}}}, \bibinfo {author} {\bibfnamefont {H.}~\bibnamefont {{Kong}}}, \bibinfo {author} {\bibfnamefont {S.~E.}\ \bibnamefont {{Koposov}}}, \bibinfo {author} {\bibfnamefont {A.}~\bibnamefont {{Kremin}}}, \bibinfo {author} {\bibfnamefont {A.}~\bibnamefont {{Krolewski}}}, \bibinfo {author} {\bibfnamefont {O.}~\bibnamefont {{Lahav}}}, \bibinfo {author} {\bibfnamefont {Y.}~\bibnamefont {{Lai}}}, \bibinfo {author} {\bibfnamefont {C.}~\bibnamefont {{Lamman}}}, \bibinfo {author} {\bibfnamefont {T.~W.}\ \bibnamefont {{Lan}}}, \bibinfo {author} {\bibfnamefont {M.}~\bibnamefont {{Landriau}}}, \bibinfo {author} {\bibfnamefont {D.}~\bibnamefont {{Lang}}}, \bibinfo {author} {\bibfnamefont {J.~U.}\ \bibnamefont {{Lange}}}, \bibinfo {author} {\bibfnamefont {J.}~\bibnamefont {{Lasker}}}, \bibinfo {author} {\bibfnamefont {J.~M.}\ \bibnamefont {{Le Goff}}}, \bibinfo {author} {\bibfnamefont {L.}~\bibnamefont {{Le Guillou}}}, \bibinfo {author} {\bibfnamefont {A.}~\bibnamefont {{Leauthaud}}}, \bibinfo {author}
  {\bibfnamefont {M.~E.}\ \bibnamefont {{Levi}}}, \bibinfo {author} {\bibfnamefont {S.}~\bibnamefont {{Li}}}, \bibinfo {author} {\bibfnamefont {T.~S.}\ \bibnamefont {{Li}}}, \bibinfo {author} {\bibfnamefont {K.}~\bibnamefont {{Lodha}}}, \bibinfo {author} {\bibfnamefont {M.}~\bibnamefont {{Lokken}}}, \bibinfo {author} {\bibfnamefont {Y.}~\bibnamefont {{Luo}}}, \bibinfo {author} {\bibfnamefont {C.}~\bibnamefont {{Magneville}}}, \bibinfo {author} {\bibfnamefont {M.}~\bibnamefont {{Manera}}}, \bibinfo {author} {\bibfnamefont {C.~J.}\ \bibnamefont {{Manser}}}, \bibinfo {author} {\bibfnamefont {D.}~\bibnamefont {{Margala}}}, \bibinfo {author} {\bibfnamefont {P.}~\bibnamefont {{Martini}}}, \bibinfo {author} {\bibfnamefont {M.}~\bibnamefont {{Maus}}}, \bibinfo {author} {\bibfnamefont {J.}~\bibnamefont {{McCullough}}}, \bibinfo {author} {\bibfnamefont {P.}~\bibnamefont {{McDonald}}}, \bibinfo {author} {\bibfnamefont {G.~E.}\ \bibnamefont {{Medina}}}, \bibinfo {author} {\bibfnamefont {L.}~\bibnamefont
  {{Medina-Varela}}}, \bibinfo {author} {\bibfnamefont {A.}~\bibnamefont {{Meisner}}}, \bibinfo {author} {\bibfnamefont {J.}~\bibnamefont {{Mena-Fern{\'a}ndez}}}, \bibinfo {author} {\bibfnamefont {A.}~\bibnamefont {{Menegas}}}, \bibinfo {author} {\bibfnamefont {M.}~\bibnamefont {{Mezcua}}}, \bibinfo {author} {\bibfnamefont {R.}~\bibnamefont {{Miquel}}}, \bibinfo {author} {\bibfnamefont {P.}~\bibnamefont {{Montero-Camacho}}}, \bibinfo {author} {\bibfnamefont {J.}~\bibnamefont {{Moon}}}, \bibinfo {author} {\bibfnamefont {J.}~\bibnamefont {{Moustakas}}}, \bibinfo {author} {\bibfnamefont {A.}~\bibnamefont {{Mu{\~n}oz-Guti{\'e}rrez}}}, \bibinfo {author} {\bibfnamefont {D.}~\bibnamefont {{Mu{\~n}oz-Santos}}}, \bibinfo {author} {\bibfnamefont {A.~D.}\ \bibnamefont {{Myers}}}, \bibinfo {author} {\bibfnamefont {J.}~\bibnamefont {{Myles}}}, \bibinfo {author} {\bibfnamefont {S.}~\bibnamefont {{Nadathur}}}, \bibinfo {author} {\bibfnamefont {J.}~\bibnamefont {{Najita}}}, \bibinfo {author} {\bibfnamefont {L.}~\bibnamefont
  {{Napolitano}}}, \bibinfo {author} {\bibfnamefont {J.~A.}\ \bibnamefont {{Newman}}}, \bibinfo {author} {\bibfnamefont {F.}~\bibnamefont {{Nikakhtar}}}, \bibinfo {author} {\bibfnamefont {R.}~\bibnamefont {{Nikutta}}}, \bibinfo {author} {\bibfnamefont {G.}~\bibnamefont {{Niz}}}, \bibinfo {author} {\bibfnamefont {H.~E.}\ \bibnamefont {{Noriega}}}, \bibinfo {author} {\bibfnamefont {N.}~\bibnamefont {{Padmanabhan}}}, \bibinfo {author} {\bibfnamefont {E.}~\bibnamefont {{Paillas}}}, \bibinfo {author} {\bibfnamefont {N.}~\bibnamefont {{Palanque-Delabrouille}}}, \bibinfo {author} {\bibfnamefont {A.}~\bibnamefont {{Palmese}}}, \bibinfo {author} {\bibfnamefont {J.}~\bibnamefont {{Pan}}}, \bibinfo {author} {\bibfnamefont {Z.}~\bibnamefont {{Pan}}}, \bibinfo {author} {\bibfnamefont {D.}~\bibnamefont {{Parkinson}}}, \bibinfo {author} {\bibfnamefont {J.}~\bibnamefont {{Peacock}}}, \bibinfo {author} {\bibfnamefont {W.~J.}\ \bibnamefont {{Percival}}}, \bibinfo {author} {\bibfnamefont {A.}~\bibnamefont
  {{P{\'e}rez-Fern{\'a}ndez}}}, \bibinfo {author} {\bibfnamefont {I.}~\bibnamefont {{P{\'e}rez-R{\`a}fols}}},\ and\ \bibinfo {author} {\bibfnamefont {P.}~\bibnamefont {{Peterson}}},\ }\href@noop {} {\bibfield  {journal} {\bibinfo  {journal} {arXiv e-prints}\ ,\ \bibinfo {eid} {arXiv:2503.14745}} (\bibinfo {year} {2025})},\ \Eprint {https://arxiv.org/abs/2503.14745} {arXiv:2503.14745 [astro-ph.CO]} \BibitemShut {NoStop}%
\bibitem [{\citenamefont {{Jarvis}}\ \emph {et~al.}(2021)\citenamefont {{Jarvis}}, \citenamefont {{Bernstein}}, \citenamefont {{Amon}}, \citenamefont {{Davis}}, \citenamefont {{L{\'e}get}}, \citenamefont {{Bechtol}}, \citenamefont {{Harrison}}, \citenamefont {{Gatti}}, \citenamefont {{Roodman}}, \citenamefont {{Chang}}, \citenamefont {{Chen}}, \citenamefont {{Choi}}, \citenamefont {{Desai}}, \citenamefont {{Drlica-Wagner}}, \citenamefont {{Gruen}}, \citenamefont {{Gruendl}}, \citenamefont {{Hernandez}}, \citenamefont {{MacCrann}}, \citenamefont {{Meyers}}, \citenamefont {{Navarro-Alsina}}, \citenamefont {{Pandey}}, \citenamefont {{Plazas}}, \citenamefont {{Secco}}, \citenamefont {{Sheldon}}, \citenamefont {{Troxel}}, \citenamefont {{Vorperian}}, \citenamefont {{Wei}}, \citenamefont {{Zuntz}}, \citenamefont {{Abbott}}, \citenamefont {{Aguena}}, \citenamefont {{Allam}}, \citenamefont {{Avila}}, \citenamefont {{Bhargava}}, \citenamefont {{Bridle}}, \citenamefont {{Brooks}}, \citenamefont {{Carnero Rosell}},
  \citenamefont {{Carrasco Kind}}, \citenamefont {{Carretero}}, \citenamefont {{Costanzi}}, \citenamefont {{da Costa}}, \citenamefont {{De Vicente}}, \citenamefont {{Diehl}}, \citenamefont {{Doel}}, \citenamefont {{Everett}}, \citenamefont {{Flaugher}}, \citenamefont {{Fosalba}}, \citenamefont {{Frieman}}, \citenamefont {{Garc{\'\i}a-Bellido}}, \citenamefont {{Gaztanaga}}, \citenamefont {{Gerdes}}, \citenamefont {{Gutierrez}}, \citenamefont {{Hinton}}, \citenamefont {{Hollowood}}, \citenamefont {{Honscheid}}, \citenamefont {{James}}, \citenamefont {{Kent}}, \citenamefont {{Kuehn}}, \citenamefont {{Kuropatkin}}, \citenamefont {{Lahav}}, \citenamefont {{Maia}}, \citenamefont {{March}}, \citenamefont {{Marshall}}, \citenamefont {{Melchior}}, \citenamefont {{Menanteau}}, \citenamefont {{Miquel}}, \citenamefont {{Ogando}}, \citenamefont {{Paz-Chinch{\'o}n}}, \citenamefont {{Rykoff}}, \citenamefont {{Sanchez}}, \citenamefont {{Scarpine}}, \citenamefont {{Schubnell}}, \citenamefont {{Serrano}}, \citenamefont
  {{Sevilla-Noarbe}}, \citenamefont {{Smith}}, \citenamefont {{Suchyta}}, \citenamefont {{Swanson}}, \citenamefont {{Tarle}}, \citenamefont {{Varga}}, \citenamefont {{Walker}}, \citenamefont {{Wester}}, \citenamefont {{Wilkinson}},\ and\ \citenamefont {{DES Collaboration}}}]{jarvis2021}%
  \BibitemOpen
  \bibfield  {author} {\bibinfo {author} {\bibfnamefont {M.}~\bibnamefont {{Jarvis}}}, \bibinfo {author} {\bibfnamefont {G.~M.}\ \bibnamefont {{Bernstein}}}, \bibinfo {author} {\bibfnamefont {A.}~\bibnamefont {{Amon}}}, \bibinfo {author} {\bibfnamefont {C.}~\bibnamefont {{Davis}}}, \bibinfo {author} {\bibfnamefont {P.~F.}\ \bibnamefont {{L{\'e}get}}}, \bibinfo {author} {\bibfnamefont {K.}~\bibnamefont {{Bechtol}}}, \bibinfo {author} {\bibfnamefont {I.}~\bibnamefont {{Harrison}}}, \bibinfo {author} {\bibfnamefont {M.}~\bibnamefont {{Gatti}}}, \bibinfo {author} {\bibfnamefont {A.}~\bibnamefont {{Roodman}}}, \bibinfo {author} {\bibfnamefont {C.}~\bibnamefont {{Chang}}}, \bibinfo {author} {\bibfnamefont {R.}~\bibnamefont {{Chen}}}, \bibinfo {author} {\bibfnamefont {A.}~\bibnamefont {{Choi}}}, \bibinfo {author} {\bibfnamefont {S.}~\bibnamefont {{Desai}}}, \bibinfo {author} {\bibfnamefont {A.}~\bibnamefont {{Drlica-Wagner}}}, \bibinfo {author} {\bibfnamefont {D.}~\bibnamefont {{Gruen}}}, \bibinfo {author}
  {\bibfnamefont {R.~A.}\ \bibnamefont {{Gruendl}}}, \bibinfo {author} {\bibfnamefont {A.}~\bibnamefont {{Hernandez}}}, \bibinfo {author} {\bibfnamefont {N.}~\bibnamefont {{MacCrann}}}, \bibinfo {author} {\bibfnamefont {J.}~\bibnamefont {{Meyers}}}, \bibinfo {author} {\bibfnamefont {A.}~\bibnamefont {{Navarro-Alsina}}}, \bibinfo {author} {\bibfnamefont {S.}~\bibnamefont {{Pandey}}}, \bibinfo {author} {\bibfnamefont {A.~A.}\ \bibnamefont {{Plazas}}}, \bibinfo {author} {\bibfnamefont {L.~F.}\ \bibnamefont {{Secco}}}, \bibinfo {author} {\bibfnamefont {E.}~\bibnamefont {{Sheldon}}}, \bibinfo {author} {\bibfnamefont {M.~A.}\ \bibnamefont {{Troxel}}}, \bibinfo {author} {\bibfnamefont {S.}~\bibnamefont {{Vorperian}}}, \bibinfo {author} {\bibfnamefont {K.}~\bibnamefont {{Wei}}}, \bibinfo {author} {\bibfnamefont {J.}~\bibnamefont {{Zuntz}}}, \bibinfo {author} {\bibfnamefont {T.~M.~C.}\ \bibnamefont {{Abbott}}}, \bibinfo {author} {\bibfnamefont {M.}~\bibnamefont {{Aguena}}}, \bibinfo {author} {\bibfnamefont
  {S.}~\bibnamefont {{Allam}}}, \bibinfo {author} {\bibfnamefont {S.}~\bibnamefont {{Avila}}}, \bibinfo {author} {\bibfnamefont {S.}~\bibnamefont {{Bhargava}}}, \bibinfo {author} {\bibfnamefont {S.~L.}\ \bibnamefont {{Bridle}}}, \bibinfo {author} {\bibfnamefont {D.}~\bibnamefont {{Brooks}}}, \bibinfo {author} {\bibfnamefont {A.}~\bibnamefont {{Carnero Rosell}}}, \bibinfo {author} {\bibfnamefont {M.}~\bibnamefont {{Carrasco Kind}}}, \bibinfo {author} {\bibfnamefont {J.}~\bibnamefont {{Carretero}}}, \bibinfo {author} {\bibfnamefont {M.}~\bibnamefont {{Costanzi}}}, \bibinfo {author} {\bibfnamefont {L.~N.}\ \bibnamefont {{da Costa}}}, \bibinfo {author} {\bibfnamefont {J.}~\bibnamefont {{De Vicente}}}, \bibinfo {author} {\bibfnamefont {H.~T.}\ \bibnamefont {{Diehl}}}, \bibinfo {author} {\bibfnamefont {P.}~\bibnamefont {{Doel}}}, \bibinfo {author} {\bibfnamefont {S.}~\bibnamefont {{Everett}}}, \bibinfo {author} {\bibfnamefont {B.}~\bibnamefont {{Flaugher}}}, \bibinfo {author} {\bibfnamefont {P.}~\bibnamefont
  {{Fosalba}}}, \bibinfo {author} {\bibfnamefont {J.}~\bibnamefont {{Frieman}}}, \bibinfo {author} {\bibfnamefont {J.}~\bibnamefont {{Garc{\'\i}a-Bellido}}}, \bibinfo {author} {\bibfnamefont {E.}~\bibnamefont {{Gaztanaga}}}, \bibinfo {author} {\bibfnamefont {D.~W.}\ \bibnamefont {{Gerdes}}}, \bibinfo {author} {\bibfnamefont {G.}~\bibnamefont {{Gutierrez}}}, \bibinfo {author} {\bibfnamefont {S.~R.}\ \bibnamefont {{Hinton}}}, \bibinfo {author} {\bibfnamefont {D.~L.}\ \bibnamefont {{Hollowood}}}, \bibinfo {author} {\bibfnamefont {K.}~\bibnamefont {{Honscheid}}}, \bibinfo {author} {\bibfnamefont {D.~J.}\ \bibnamefont {{James}}}, \bibinfo {author} {\bibfnamefont {S.}~\bibnamefont {{Kent}}}, \bibinfo {author} {\bibfnamefont {K.}~\bibnamefont {{Kuehn}}}, \bibinfo {author} {\bibfnamefont {N.}~\bibnamefont {{Kuropatkin}}}, \bibinfo {author} {\bibfnamefont {O.}~\bibnamefont {{Lahav}}}, \bibinfo {author} {\bibfnamefont {M.~A.~G.}\ \bibnamefont {{Maia}}}, \bibinfo {author} {\bibfnamefont {M.}~\bibnamefont {{March}}},
  \bibinfo {author} {\bibfnamefont {J.~L.}\ \bibnamefont {{Marshall}}}, \bibinfo {author} {\bibfnamefont {P.}~\bibnamefont {{Melchior}}}, \bibinfo {author} {\bibfnamefont {F.}~\bibnamefont {{Menanteau}}}, \bibinfo {author} {\bibfnamefont {R.}~\bibnamefont {{Miquel}}}, \bibinfo {author} {\bibfnamefont {R.~L.~C.}\ \bibnamefont {{Ogando}}}, \bibinfo {author} {\bibfnamefont {F.}~\bibnamefont {{Paz-Chinch{\'o}n}}}, \bibinfo {author} {\bibfnamefont {E.~S.}\ \bibnamefont {{Rykoff}}}, \bibinfo {author} {\bibfnamefont {E.}~\bibnamefont {{Sanchez}}}, \bibinfo {author} {\bibfnamefont {V.}~\bibnamefont {{Scarpine}}}, \bibinfo {author} {\bibfnamefont {M.}~\bibnamefont {{Schubnell}}}, \bibinfo {author} {\bibfnamefont {S.}~\bibnamefont {{Serrano}}}, \bibinfo {author} {\bibfnamefont {I.}~\bibnamefont {{Sevilla-Noarbe}}}, \bibinfo {author} {\bibfnamefont {M.}~\bibnamefont {{Smith}}}, \bibinfo {author} {\bibfnamefont {E.}~\bibnamefont {{Suchyta}}}, \bibinfo {author} {\bibfnamefont {M.~E.~C.}\ \bibnamefont {{Swanson}}},
  \bibinfo {author} {\bibfnamefont {G.}~\bibnamefont {{Tarle}}}, \bibinfo {author} {\bibfnamefont {T.~N.}\ \bibnamefont {{Varga}}}, \bibinfo {author} {\bibfnamefont {A.~R.}\ \bibnamefont {{Walker}}}, \bibinfo {author} {\bibfnamefont {W.}~\bibnamefont {{Wester}}}, \bibinfo {author} {\bibfnamefont {R.~D.}\ \bibnamefont {{Wilkinson}}},\ and\ \bibinfo {author} {\bibnamefont {{DES Collaboration}}},\ }\href {https://doi.org/10.1093/mnras/staa3679} {\bibfield  {journal} {\bibinfo  {journal} {\mnras}\ }\textbf {\bibinfo {volume} {501}},\ \bibinfo {pages} {1282} (\bibinfo {year} {2021})},\ \Eprint {https://arxiv.org/abs/2011.03409} {arXiv:2011.03409 [astro-ph.IM]} \BibitemShut {NoStop}%
\bibitem [{\citenamefont {{Zhang}}\ \emph {et~al.}(2023{\natexlab{c}})\citenamefont {{Zhang}}, \citenamefont {{Almoubayyed}}, \citenamefont {{Mandelbaum}}, \citenamefont {{Meyers}}, \citenamefont {{Jarvis}}, \citenamefont {{Kannawadi}}, \citenamefont {{Schmitz}}, \citenamefont {{Guinot}},\ and\ \citenamefont {{LSST Dark Energy Science Collaboration}}}]{zhang2023a}%
  \BibitemOpen
  \bibfield  {author} {\bibinfo {author} {\bibfnamefont {T.}~\bibnamefont {{Zhang}}}, \bibinfo {author} {\bibfnamefont {H.}~\bibnamefont {{Almoubayyed}}}, \bibinfo {author} {\bibfnamefont {R.}~\bibnamefont {{Mandelbaum}}}, \bibinfo {author} {\bibfnamefont {J.~E.}\ \bibnamefont {{Meyers}}}, \bibinfo {author} {\bibfnamefont {M.}~\bibnamefont {{Jarvis}}}, \bibinfo {author} {\bibfnamefont {A.}~\bibnamefont {{Kannawadi}}}, \bibinfo {author} {\bibfnamefont {M.~A.}\ \bibnamefont {{Schmitz}}}, \bibinfo {author} {\bibfnamefont {A.}~\bibnamefont {{Guinot}}},\ and\ \bibinfo {author} {\bibnamefont {{LSST Dark Energy Science Collaboration}}},\ }\href {https://doi.org/10.1093/mnras/stac3350} {\bibfield  {journal} {\bibinfo  {journal} {\mnras}\ }\textbf {\bibinfo {volume} {520}},\ \bibinfo {pages} {2328} (\bibinfo {year} {2023}{\natexlab{c}})},\ \Eprint {https://arxiv.org/abs/2205.07892} {arXiv:2205.07892 [astro-ph.CO]} \BibitemShut {NoStop}%
\bibitem [{\citenamefont {{Heymans}}\ \emph {et~al.}(2021)\citenamefont {{Heymans}}, \citenamefont {{Tr{\"o}ster}}, \citenamefont {{Asgari}}, \citenamefont {{Blake}}, \citenamefont {{Hildebrandt}}, \citenamefont {{Joachimi}}, \citenamefont {{Kuijken}}, \citenamefont {{Lin}}, \citenamefont {{S{\'a}nchez}}, \citenamefont {{van den Busch}} \emph {et~al.}}]{Heymans2021}%
  \BibitemOpen
  \bibfield  {author} {\bibinfo {author} {\bibfnamefont {C.}~\bibnamefont {{Heymans}}}, \bibinfo {author} {\bibfnamefont {T.}~\bibnamefont {{Tr{\"o}ster}}}, \bibinfo {author} {\bibfnamefont {M.}~\bibnamefont {{Asgari}}}, \bibinfo {author} {\bibfnamefont {C.}~\bibnamefont {{Blake}}}, \bibinfo {author} {\bibfnamefont {H.}~\bibnamefont {{Hildebrandt}}}, \bibinfo {author} {\bibfnamefont {B.}~\bibnamefont {{Joachimi}}}, \bibinfo {author} {\bibfnamefont {K.}~\bibnamefont {{Kuijken}}}, \bibinfo {author} {\bibfnamefont {C.-A.}\ \bibnamefont {{Lin}}}, \bibinfo {author} {\bibfnamefont {A.~G.}\ \bibnamefont {{S{\'a}nchez}}}, \bibinfo {author} {\bibfnamefont {J.~L.}\ \bibnamefont {{van den Busch}}}, \emph {et~al.},\ }\href {https://doi.org/10.1051/0004-6361/202039063} {\bibfield  {journal} {\bibinfo  {journal} {\aap}\ }\textbf {\bibinfo {volume} {646}},\ \bibinfo {eid} {A140} (\bibinfo {year} {2021})},\ \Eprint {https://arxiv.org/abs/2007.15632} {arXiv:2007.15632 [astro-ph.CO]} \BibitemShut {NoStop}%
\bibitem [{\citenamefont {{Sugiyama}}\ \emph {et~al.}(2022)\citenamefont {{Sugiyama}}, \citenamefont {{Takada}}, \citenamefont {{Miyatake}}, \citenamefont {{Nishimichi}}, \citenamefont {{Shirasaki}}, \citenamefont {{Kobayashi}}, \citenamefont {{Mandelbaum}}, \citenamefont {{More}}, \citenamefont {{Takahashi}}, \citenamefont {{Osato}} \emph {et~al.}}]{Sugiyama2022}%
  \BibitemOpen
  \bibfield  {author} {\bibinfo {author} {\bibfnamefont {S.}~\bibnamefont {{Sugiyama}}}, \bibinfo {author} {\bibfnamefont {M.}~\bibnamefont {{Takada}}}, \bibinfo {author} {\bibfnamefont {H.}~\bibnamefont {{Miyatake}}}, \bibinfo {author} {\bibfnamefont {T.}~\bibnamefont {{Nishimichi}}}, \bibinfo {author} {\bibfnamefont {M.}~\bibnamefont {{Shirasaki}}}, \bibinfo {author} {\bibfnamefont {Y.}~\bibnamefont {{Kobayashi}}}, \bibinfo {author} {\bibfnamefont {R.}~\bibnamefont {{Mandelbaum}}}, \bibinfo {author} {\bibfnamefont {S.}~\bibnamefont {{More}}}, \bibinfo {author} {\bibfnamefont {R.}~\bibnamefont {{Takahashi}}}, \bibinfo {author} {\bibfnamefont {K.}~\bibnamefont {{Osato}}}, \emph {et~al.},\ }\href {https://doi.org/10.1103/PhysRevD.105.123537} {\bibfield  {journal} {\bibinfo  {journal} {\prd}\ }\textbf {\bibinfo {volume} {105}},\ \bibinfo {eid} {123537} (\bibinfo {year} {2022})},\ \Eprint {https://arxiv.org/abs/2111.10966} {arXiv:2111.10966 [astro-ph.CO]} \BibitemShut {NoStop}%
\bibitem [{\citenamefont {{Hartlap}}\ \emph {et~al.}(2007)\citenamefont {{Hartlap}}, \citenamefont {{Simon}},\ and\ \citenamefont {{Schneider}}}]{Hartlap2007}%
  \BibitemOpen
  \bibfield  {author} {\bibinfo {author} {\bibfnamefont {J.}~\bibnamefont {{Hartlap}}}, \bibinfo {author} {\bibfnamefont {P.}~\bibnamefont {{Simon}}},\ and\ \bibinfo {author} {\bibfnamefont {P.}~\bibnamefont {{Schneider}}},\ }\href {https://doi.org/10.1051/0004-6361:20066170} {\bibfield  {journal} {\bibinfo  {journal} {\aap}\ }\textbf {\bibinfo {volume} {464}},\ \bibinfo {pages} {399} (\bibinfo {year} {2007})},\ \Eprint {https://arxiv.org/abs/astro-ph/0608064} {arXiv:astro-ph/0608064 [astro-ph]} \BibitemShut {NoStop}%
\bibitem [{\citenamefont {{Park}}\ and\ \citenamefont {{Rozo}}(2020)}]{Park2020}%
  \BibitemOpen
  \bibfield  {author} {\bibinfo {author} {\bibfnamefont {Y.}~\bibnamefont {{Park}}}\ and\ \bibinfo {author} {\bibfnamefont {E.}~\bibnamefont {{Rozo}}},\ }\href {https://doi.org/10.1093/mnras/staa2647} {\bibfield  {journal} {\bibinfo  {journal} {\mnras}\ }\textbf {\bibinfo {volume} {499}},\ \bibinfo {pages} {4638} (\bibinfo {year} {2020})},\ \Eprint {https://arxiv.org/abs/1907.05798} {arXiv:1907.05798 [astro-ph.CO]} \BibitemShut {NoStop}%
\bibitem [{\citenamefont {{Ferreira}}\ \emph {et~al.}(2021)\citenamefont {{Ferreira}}, \citenamefont {{Zhang}}, \citenamefont {{Chen}}, \citenamefont {{Dodelson}},\ and\ \citenamefont {{LSST Dark Energy Science Collaboration}}}]{Ferreira2021}%
  \BibitemOpen
  \bibfield  {author} {\bibinfo {author} {\bibfnamefont {T.}~\bibnamefont {{Ferreira}}}, \bibinfo {author} {\bibfnamefont {T.}~\bibnamefont {{Zhang}}}, \bibinfo {author} {\bibfnamefont {N.}~\bibnamefont {{Chen}}}, \bibinfo {author} {\bibfnamefont {S.}~\bibnamefont {{Dodelson}}},\ and\ \bibinfo {author} {\bibnamefont {{LSST Dark Energy Science Collaboration}}},\ }\href {https://doi.org/10.1103/PhysRevD.103.103535} {\bibfield  {journal} {\bibinfo  {journal} {\prd}\ }\textbf {\bibinfo {volume} {103}},\ \bibinfo {eid} {103535} (\bibinfo {year} {2021})},\ \Eprint {https://arxiv.org/abs/2010.15986} {arXiv:2010.15986 [astro-ph.CO]} \BibitemShut {NoStop}%
\bibitem [{\citenamefont {{Schaye}}\ \emph {et~al.}(2010)\citenamefont {{Schaye}}, \citenamefont {{Dalla Vecchia}}, \citenamefont {{Booth}}, \citenamefont {{Wiersma}}, \citenamefont {{Theuns}}, \citenamefont {{Haas}}, \citenamefont {{Bertone}}, \citenamefont {{Duffy}}, \citenamefont {{McCarthy}},\ and\ \citenamefont {{van de Voort}}}]{Schaye2010}%
  \BibitemOpen
  \bibfield  {author} {\bibinfo {author} {\bibfnamefont {J.}~\bibnamefont {{Schaye}}}, \bibinfo {author} {\bibfnamefont {C.}~\bibnamefont {{Dalla Vecchia}}}, \bibinfo {author} {\bibfnamefont {C.~M.}\ \bibnamefont {{Booth}}}, \bibinfo {author} {\bibfnamefont {R.~P.~C.}\ \bibnamefont {{Wiersma}}}, \bibinfo {author} {\bibfnamefont {T.}~\bibnamefont {{Theuns}}}, \bibinfo {author} {\bibfnamefont {M.~R.}\ \bibnamefont {{Haas}}}, \bibinfo {author} {\bibfnamefont {S.}~\bibnamefont {{Bertone}}}, \bibinfo {author} {\bibfnamefont {A.~R.}\ \bibnamefont {{Duffy}}}, \bibinfo {author} {\bibfnamefont {I.~G.}\ \bibnamefont {{McCarthy}}},\ and\ \bibinfo {author} {\bibfnamefont {F.~o.}\ \bibnamefont {{van de Voort}}},\ }\href {https://doi.org/10.1111/j.1365-2966.2009.16029.x} {\bibfield  {journal} {\bibinfo  {journal} {\mnras}\ }\textbf {\bibinfo {volume} {402}},\ \bibinfo {pages} {1536} (\bibinfo {year} {2010})},\ \Eprint {https://arxiv.org/abs/0909.5196} {arXiv:0909.5196 [astro-ph.CO]} \BibitemShut {NoStop}%
\bibitem [{\citenamefont {{McCarthy}}\ \emph {et~al.}(2017)\citenamefont {{McCarthy}}, \citenamefont {{Schaye}}, \citenamefont {{Bird}},\ and\ \citenamefont {{Le Brun}}}]{mccarthy2017}%
  \BibitemOpen
  \bibfield  {author} {\bibinfo {author} {\bibfnamefont {I.~G.}\ \bibnamefont {{McCarthy}}}, \bibinfo {author} {\bibfnamefont {J.}~\bibnamefont {{Schaye}}}, \bibinfo {author} {\bibfnamefont {S.}~\bibnamefont {{Bird}}},\ and\ \bibinfo {author} {\bibfnamefont {A.~M.~C.}\ \bibnamefont {{Le Brun}}},\ }\href {https://doi.org/10.1093/mnras/stw2792} {\bibfield  {journal} {\bibinfo  {journal} {\mnras}\ }\textbf {\bibinfo {volume} {465}},\ \bibinfo {pages} {2936} (\bibinfo {year} {2017})},\ \Eprint {https://arxiv.org/abs/1603.02702} {arXiv:1603.02702 [astro-ph.CO]} \BibitemShut {NoStop}%
\bibitem [{\citenamefont {{Le Brun}}\ \emph {et~al.}(2014)\citenamefont {{Le Brun}}, \citenamefont {{McCarthy}}, \citenamefont {{Schaye}},\ and\ \citenamefont {{Ponman}}}]{LeBrun2014}%
  \BibitemOpen
  \bibfield  {author} {\bibinfo {author} {\bibfnamefont {A.~M.~C.}\ \bibnamefont {{Le Brun}}}, \bibinfo {author} {\bibfnamefont {I.~G.}\ \bibnamefont {{McCarthy}}}, \bibinfo {author} {\bibfnamefont {J.}~\bibnamefont {{Schaye}}},\ and\ \bibinfo {author} {\bibfnamefont {T.~J.}\ \bibnamefont {{Ponman}}},\ }\href {https://doi.org/10.1093/mnras/stu608} {\bibfield  {journal} {\bibinfo  {journal} {\mnras}\ }\textbf {\bibinfo {volume} {441}},\ \bibinfo {pages} {1270} (\bibinfo {year} {2014})},\ \Eprint {https://arxiv.org/abs/1312.5462} {arXiv:1312.5462 [astro-ph.CO]} \BibitemShut {NoStop}%
\end{thebibliography}%

\end{document}